\documentclass[10pt,epsfig]{report}
\usepackage{graphicx}
\usepackage{amssymb}
\usepackage{amsmath}
\usepackage{pifont}
\usepackage{epsfig}
\newcommand{\pho}{\phantom{1}}
\newcommand{\beq}{\begin{equation}}
\newcommand{\eeq}{\end{equation}}
\newcommand{\dtp}{\frac{d^3p}{(2\pi)^3}}
\newcommand{\dfp}{\frac{d^4p}{(2\pi)^4}}
\newcommand{\dfk}{\frac{d^4k}{(2\pi)^4}}
\newcommand{\phm}{\phantom{-}}
\newcommand{\mi}{\mathit}
\newcommand{\rb}{\rho_B}
\newcommand{\MeV}{{\rm MeV}}
\newcommand{\fm}{{\rm fm}}
\newcommand{\eps}{\varepsilon}
\newcommand{\ct}{s_{22}}
\newcommand{\cf}{s_{55}}
\newcommand{\cs}{s_{77}}
\newcommand{\xup}{x_{uu}^+}
\newcommand{\xum}{x_{uu}^-}
\newcommand{\xsp}{x_{ss}^+}
\newcommand{\xsm}{x_{ss}^-}
\newcommand{\xusp}{x_{us}^+}
\newcommand{\xusm}{x_{us}^-}
\newcommand{\yu}{y_{uu}}
\newcommand{\ys}{y_{ss}}
\newcommand{\yus}{y_{us}}
\newcommand{\dt}{|\Delta_2|}
\newcommand{\df}{|\Delta_5|}
\newcommand{\delsl}{\partial\hspace{-1.95mm}/}
\newcommand{\psl}{p\hspace{-1.5mm}/}
\newcommand{\Qsl}{Q\hspace{-2.5mm}/}
\newcommand{\unity}{1\hspace{-1.3mm}1}
\newcommand{\ave}[1]{\langle{#1}\rangle}
\newcommand{\bra}[1]{\langle{#1}|}
\newcommand{\ket}[1]{|{#1}\rangle}
\newcommand{\tr}[1]{{\rm Tr}\,[{#1}]}
\newcommand{\eq}[1]{Eq.~(\ref{#1})}
\newcommand{\eqs}[1]{Eqs.~(\ref{#1})}
\newcommand{\fig}[1]{Fig.~\ref{#1}}
\newcommand{\tab}[1]{Table~\ref{#1}}
\newcommand{\sect}[1]{Sec.~\ref{#1}}
\newcommand{\chap}[1]{Chap.~\ref{#1}}
\newcommand{\doublespace}{\renewcommand{\baselinestretch}{1.6}\large\normalsize}
\begin {document}
%

\pagestyle{empty} 
\begin{center}
\doublespace
\phantom{.} \vspace{40mm}
\begin{huge}
{\bf NJL-model analysis \\[2mm]
of dense quark matter}\\
\end{huge}
\vskip 20mm
Michael Buballa\\[10mm]
{\small{\it Institut f\"ur Kernphysik, Technische Universit\"at Darmstadt,\\
                 Schlossgartenstr. 9, D-64289 Darmstadt, Germany}}
\end{center}
\vskip 55mm
{\small e-mail:\quad  michael.buballa@physik.tu-darmstadt.de\\
phone:\quad +49\; 6151\;16\;3272 \qquad
fax:\; +49\; 6151\;16\;6076}

\newpage


\pagestyle{empty} 
\abstract{

Investigations of deconfined quark matter within NJL-type models are reviewed,
focusing on the regime of low temperatures and ``moderate'' densities,
which is not accessible by perturbative QCD. 
Central issue is the interplay between chiral symmetry restoration and the 
formation of color superconducting phases.
In order to lay a solid ground for this analysis, we begin 
with a rather detailed discussion of two and three-flavor NJL models
and their phase structure, neglecting the possibility of diquark pairing
in a first step. 
An important aspect of this part is a comparison with the MIT bag model. 
The NJL model is also applied to investigate the possibility of absolutely
stable strange quark matter. 
In the next step the formalism is extended to include diquark condensates. 
We discuss     
the role and mutual influence of several conventional and less 
conventional quark-antiquark and diquark condensates.
As a particularly interesting example, we analyze a spin-1 diquark
condensate as a possible pairing channel for those quarks which are
left over from the standard spin-0 condensate. 
For three-flavor systems,
we find that a self-consistent calculation of the strange quark mass, 
together with the diquark condensates, is crucial for a realistic
description of the 2SC-CFL phase transition.
We also study the effect of neutrality constraints
which are of relevance for compact stars.
Both, homogeneous and mixed, neutral phases are constructed. 
Although neutrality constraints generally tend to disfavor the 2SC phase
we find that this phase is again stabilized by the large values of
the dynamical strange quark mass which follow from the self-consistent
treatment.
Finally, we combine our solutions with existing hadronic equations of 
state to investigate the existence of quark matter cores in 
neutron stars.  


}

\newpage
\pagestyle{plain}
\pagenumbering{arabic}
\setcounter{page}{1}

\tableofcontents

\chapter{Introduction}
\pagestyle{plain}
\label{intro}

\begin{figure}
\begin{center}
\epsfig{file=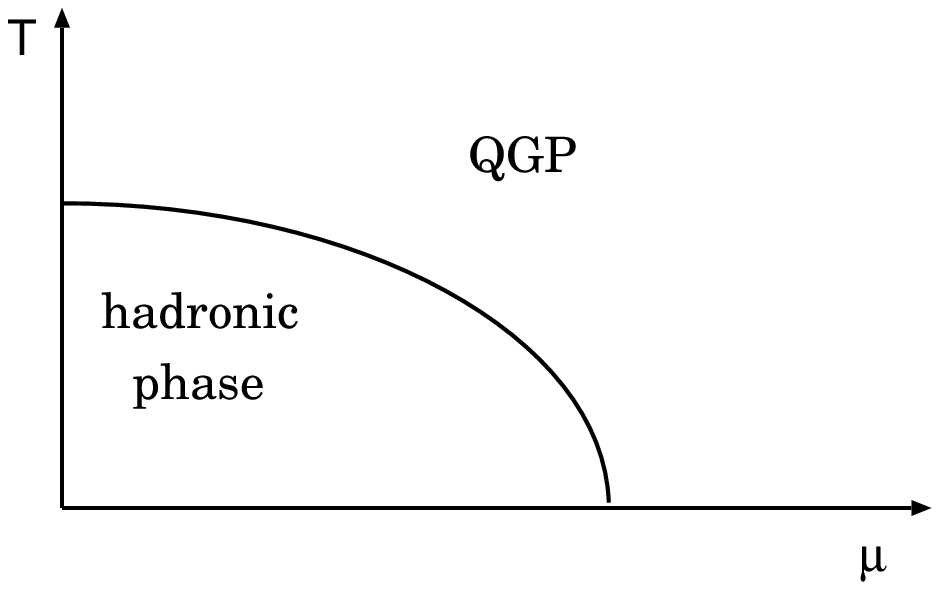,width=6.5cm}\hspace{1.0cm}
\epsfig{file=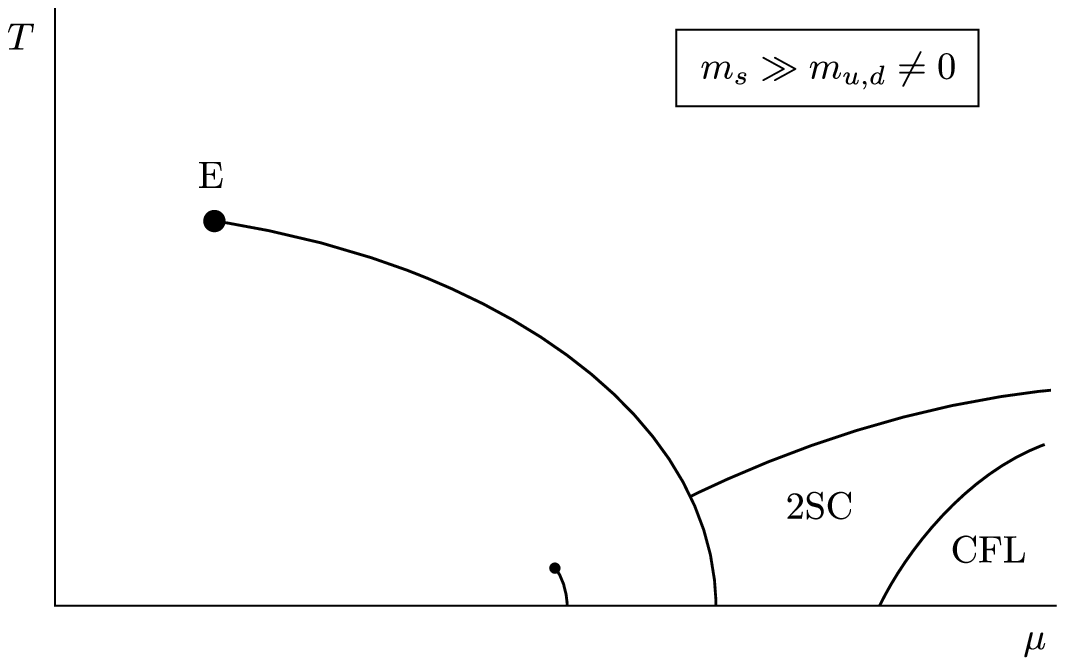,width=6.5cm}\\
\epsfig{file=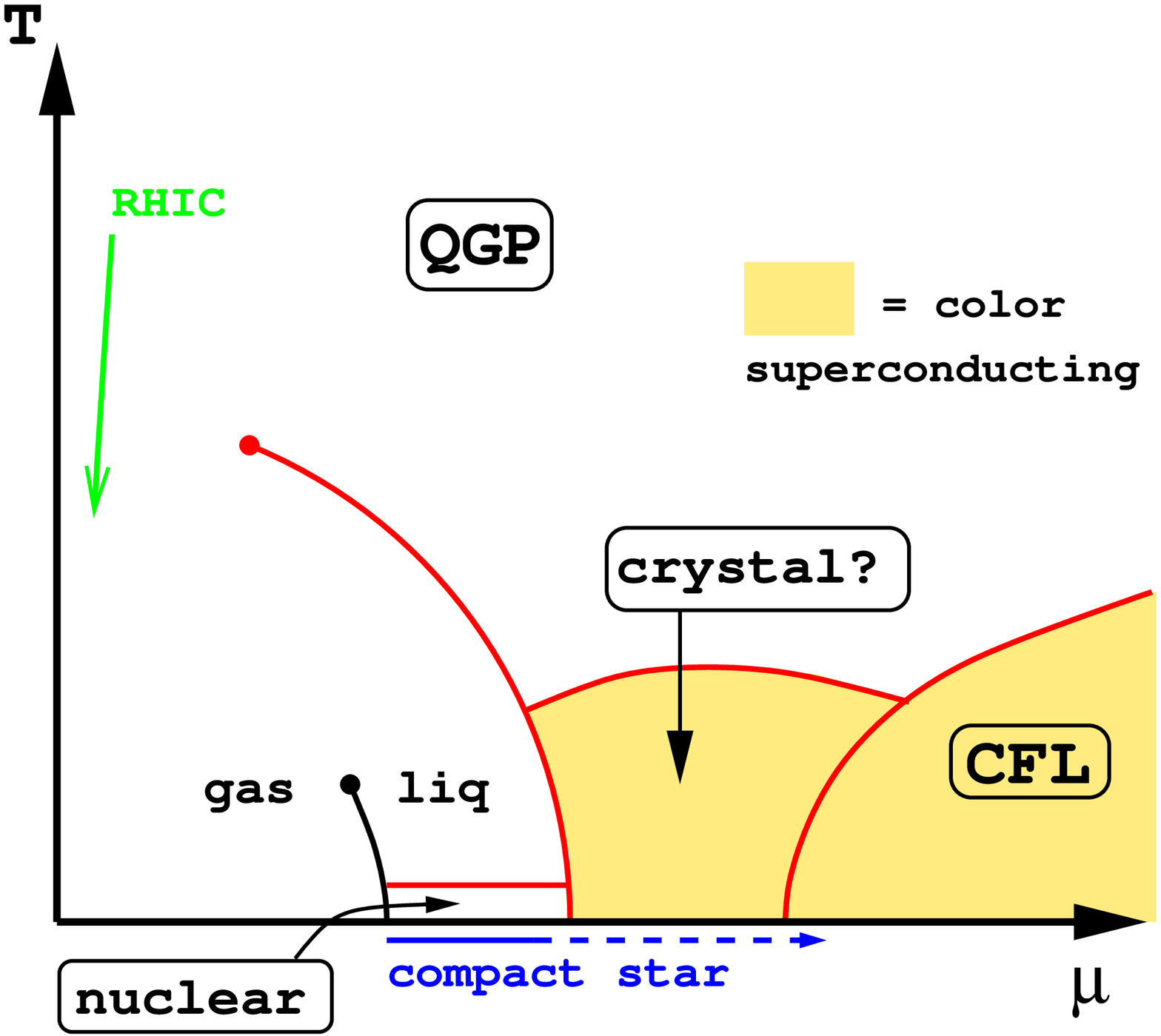,width=6.5cm}\hspace{1.0cm}
\epsfig{file=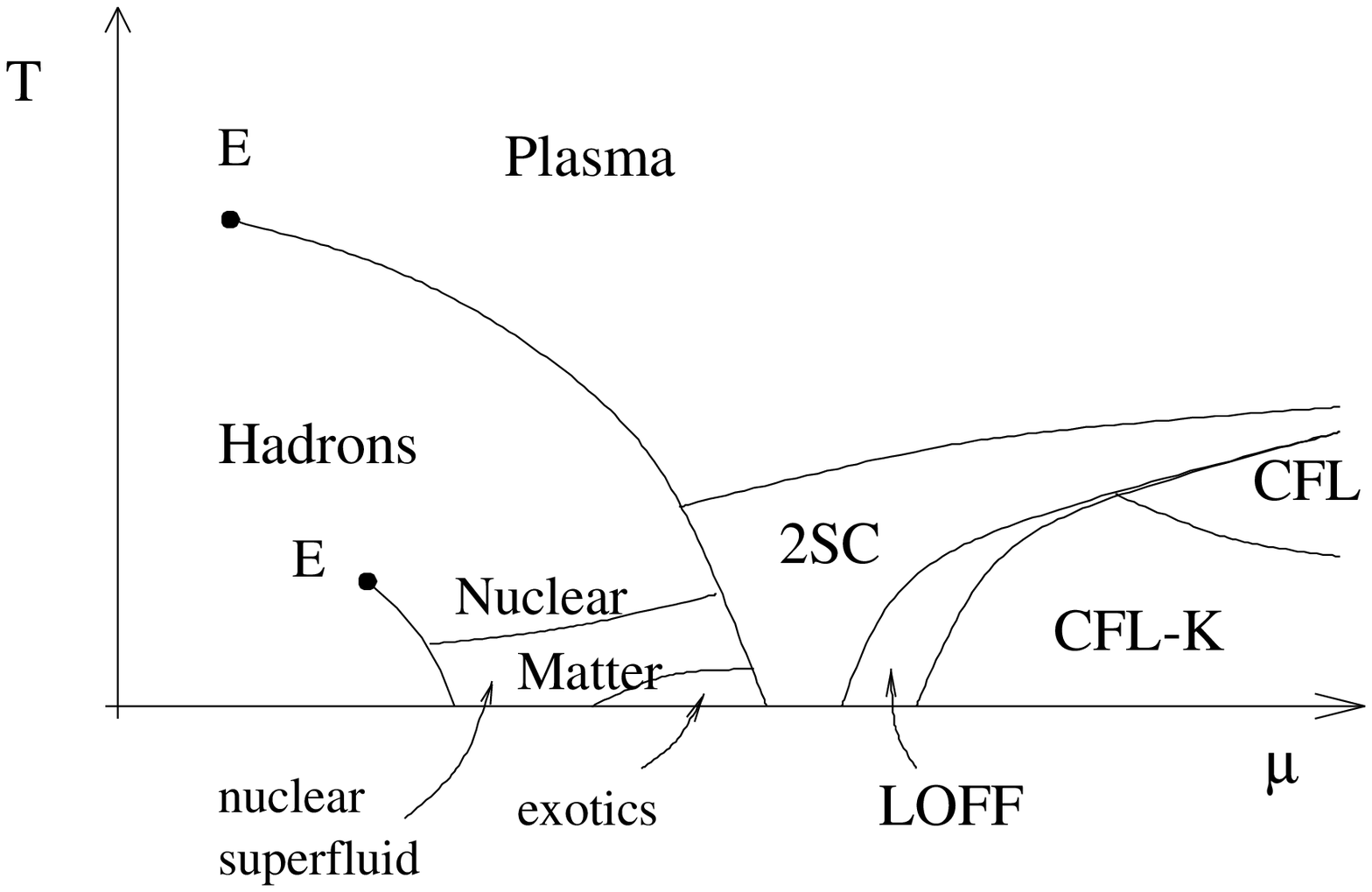,width=6.5cm}
\end{center}
\vspace{-0.5cm}
\caption{\small Schematic QCD phase diagrams in the chemical 
         potential--temperature plane. 
         Upper left: generic phase diagram of the 
         ``pre-color superconductivity era'', see, e.g., 
         Refs.~\cite{CGS86,MeOr96}.  
         The other diagrams are taken from the literature.
         Upper right: Rajagopal (1999)~\cite{Raja99}.
         Lower left: Alford (2003)~\cite{Al03}.
         Lower right: Sch\"afer (2003)~\cite{Schaefer}.   
}
\label{figschemphase}
\end{figure}

Exploring the phase structure of quantum chromodynamics (QCD)
is certainly one of the most exciting topics in the field of strong 
interaction physics.
Already in the 70s, rather soon after it had become clear that hadrons 
consist of confined quarks and gluons, it was argued that the latter
should become deconfined at high temperature or density when 
the hadrons strongly overlap and loose their 
individuality~\cite{CoPe75,CaPa75}.
In this picture, there are thus two distinct phases,
the ``hadronic phase'' where quarks and gluons are confined,
and the so-called quark-gluon plasma (QGP) where they are deconfined.
This scenario is illustrated in the upper left panel of \fig{figschemphase}
by a schematic phase diagram in the plane of (quark number) chemical 
potential and temperature.
A diagram of this type has essentially been drawn already in 
Ref.~\cite{CaPa75} and can be found, e.g., in Refs.~\cite{CGS86,MeOr96}.

In nature, the QGP surely existed in the early universe, a few microseconds 
after the Big Bang when the temperature was very high.
It is less clear whether deconfined quark matter also exists in the 
relatively cold but dense centers of neutron stars. 
Experimentally, the creation and identification of the QGP
is the ultimate goal of ultra-relativistic heavy-ion collisions.
First indications of success have been reported
in press releases at CERN (SPS)~\cite{SpS} and BNL (RHIC)~\cite{RHIC},
although the interpretation of the data is still under debate.
There is little doubt that the QGP will be created at the Large Hadron
Collider (LHC), which is currently being built at CERN.

At least on a schematic level, the phase diagram shown in the
upper left panel of \fig{figschemphase} remained the standard picture
for about two decades. In particular the possibility of having more than 
one deconfined phase was not taken into account. 
Although Cooper pairing in cold, dense quark matter
(``color superconductivity'') had been mentioned already in 1975~\cite{CoPe75} 
and had further been worked out in Refs.~\cite{Ba77,Fr78,BaLo84},
the relevance of this idea for the QCD phase diagram was widely ignored
until the end of the 90s.
At that time, new approaches to color superconductivity revealed that 
the related gaps in the fermion spectrum could 
be of the order of 100~MeV~\cite{ARW98,RSSV98}, much larger than expected 
earlier. 
Since larger gaps are related to larger critical temperatures,
this would imply a sizeable extention of the color superconducting
region into the temperature direction. 
Hence, in addition to the two standard phases, there should be
a non-negligible region in the QCD phase diagram
where strongly interacting matter is a color superconductor. 
(For reviews on color superconductivity, see 
Refs.~\cite{RaWi00,Alford,Schaefer,Rischkereview}.)

Once color superconductivity was on the agenda, the door was open for
many new possibilities.
This is illustrated by the remaining three phase diagrams of
\fig{figschemphase}, which are taken from the literature. 
It is expected that at large chemical potentials up, down, and strange quarks 
are paired in a so-called color-flavor locked (CFL) condensate~\cite{ARW99}.
However, this might become unfavorable at lower densities, where the strange 
quarks are suppressed by their mass. It is thus possible that in some
intermediate regime there is a second color superconducting phase (2SC)
where only up and down quarks are paired.
This scenario is depicted in the upper right diagram of \fig{figschemphase},
taken from Ref.~\cite{Raja99}.
More recently, further phases, like three-flavor color superconductors
with condensed kaons (CFL-K)~\cite{Sch00c,BS02,KR02}
or crystalline color superconductors (``LOFF phase'') \cite{ABR01,BR02}
have been suggested, which might partially (lower right diagram 
\cite{Schaefer}) or even completely (lower left diagram \cite{Al03}) 
replace the 2SC phase.

\fig{figschemphase}, which is only an incomplete compilation of recent
suggestions, illustrates the potential richness of the phase structure, 
which has not been appreciated for a long time.
At the same time, it makes obvious that the issue is not at all settled. 
Note that all phase diagrams shown in the figure are only ``schematic'',
i.e., educated guesses, based on certain theoretical results or arguments.
In this situation, and since exact results from QCD are rather limited,  
model calculations may provide a useful tool to test the robustness of
these ideas and to develop new ones.

In the present report we discuss the phase diagram and related issues
which result from studies with Nambu--Jona-Lasinio (NJL) type models.
These are schematic models with point-like quark- (anti-)quark vertices,
but no gluons.  
As a consequence, NJL-type models have several well-known shortcomings,
most important, they do not have the confinement property of QCD. 
This is certainly a major drawback in the hadronic phase,
where constituent quarks are not the proper quasi-particle degrees of
freedom. At high temperatures, confinement becomes less relevant
but obviously a realistic description of the quark-{\it gluon} plasma
requires explicit gluon degrees of freedom. 
On the other hand, the use of NJL-type models seems to be justified
-- at least on a schematic level -- to study cold deconfined quark matter
where both, confinement and gluon degrees of freedom, are of minor
importance.  

In any case, every model calculation should be confronted with the ``facts'',
as far as available. 
To that end, we briefly list the main features of QCD in \sect{qcd}
and summarize what is currently known about the QCD phase diagram in 
\sect{facts}.
The present work will then be outlined in more details in \sect{outline}.

\section{Basics of QCD}
\label{qcd}

Quantum chromodynamics is defined by the Lagrangian~\cite{Wein73,FGL73}
\beq
    {\cal L}_{QCD} \;=\; \bar q \,(\,i\gamma^\mu D_\mu - \hat m\,)\,q 
    \;-\; \frac{1}{4}\, G^{a\,\mu\nu}\, G^a_{\;\mu\nu}~,
\label{Lqcd}
\eeq
where $q$ denotes a quark field with six flavor ($u,d,s,c,b,t$) and three 
color degrees of freedom, and $\hat m = {\rm diag}_f(m_u, m_d, \dots)$
is the corresponding mass matrix in flavor space. 
The covariant derivative
\beq
    D_\mu \;=\; \partial_\mu \;-\;i g \,\frac{\lambda^a}{2}\,A_\mu^a
\eeq
is related to the gluon field $A_\mu^a$, and
\beq
    G^a_{\;\mu\nu} \;=\; \partial_\mu\,A^a_\nu \;-\;\partial_\nu\,A^a_\mu
    \;+\; g\,f^{abc}\,A_\mu^b\,A_\nu^c
\label{Gamn}
\eeq
is the gluon field strength tensor. $\lambda^a$ and $f^{abc}$ denote the
generators of $SU(3)$ (Gell-Mann matrices) and the corresponding
antisymmetric structure constants, respectively. 
$g$ is the QCD coupling constant. 

The QCD Lagrangian is by construction symmetric under $SU(3)$ gauge
transformations in color space. 
Because of the non-Abelian character of the gauge group, underlined
by the presence of the $f^{abc}$ term in \eq{Gamn}, the theory has
several non-trivial features which are not present in Abelian gauge
theories, like quantum electrodynamics:

\begin{itemize}
\item ${\cal L}_{QCD}$ contains gluonic self-couplings (three- and
      four-gluon vertices), i.e., gluons carry color.

\item QCD is an asymptotically free theory~\cite{GW73,Pol73}, i.e., 
      the coupling becomes weak at short distances or, equivalently, 
      large Euclidean momenta $Q$. To one-loop order, 
\beq
      \alpha_s(Q^2) \;\equiv\; \frac{g^2(Q^2)}{4\pi}\;=\;
      \frac{4\pi}{(11 - \frac{2}{3} N_f) \ln{(Q^2/\Lambda_{QCD}^2)}}~,
\label{alpharun}
\eeq
      where $N_f$ is the number of relevant flavors and 
      $\Lambda_{QCD}$ is the QCD scale parameter which can be
      determined, e.g., by fitting \eq{alpharun} (or improved versions
      thereof) to experimental data at large $Q^2$. In this way, one
      finds $\Lambda_{QCD} \simeq 200$~MeV for five flavors
      in the $\overline{MS}$ scheme. 
 
      \eq{alpharun} is the fundamental basis for the perturbative 
      treatment of QCD in the high-momentum regime. 
      For instance at $Q^2=M_Z^2$ one finds $\alpha_s \simeq 0.12$.
      (For a recent overview, see Ref.~\cite{alphas}.)

\item In turn, \eq{alpharun} implies that the coupling becomes strong
      at low momenta. In particular, perturbative QCD is not 
      applicable to describe hadrons with masses below $\sim 2$~GeV.
      This may or may not be related to the phenomenon of ``confinement'',
      i.e., to the empirical fact that colored objects, like
      quarks and gluons, do not exist as physical degrees of freedom in
      vacuum. 
      There are interesting attempts to relate confinement to particular
      topological objects in the QCD vacuum, like monopoles
      or vortices, 
      but it is fair to say that confinement is not yet fully understood
      (see, e.g., Refs.~\cite{Gr03,Ri03} and references therein). 
\end{itemize}
 
Another important feature of ${\cal L}_{QCD}$ is its (approximate)
chiral symmetry, i.e., its symmetry under global
$SU(N_f)_L \times SU(N_f)_R$ transformations. This is equivalent to be
invariant under global vector and axial-vector $SU(N_f)$ transformations,
\beq
    SU(N_f)_V: \quad q \;\rightarrow\; 
    \exp{(i \theta^V_a \tau_a)}\;q~, \qquad
    SU(N_f)_A: \quad q \;\rightarrow\; 
    \exp{(i \theta^A_a \gamma_5\tau_a)}\;q~, 
\label{chiral}
\eeq
where $\tau_a$ are the generators of flavor $SU(N_f)$.
These symmetries would be exact in the limit of $N_f$ massless flavors.
(For $SU(N_f)_V$ it is sufficient to have $N_f$ {\it degenerate} 
flavors.)
In reality all quarks have non-vanishing masses.
Still, chiral symmetry is a 
useful concept in the up/down sector ($N_f=2$) and even, although
with larger deviations, when strange quarks are 
included as well ($N_f=3$). 
The sector of heavy quarks
(charm, bottom, top) is governed by the opposite limit,
corresponding to an expansion in inverse quark masses,
but this will not be of further interest for us.

The $SU(N_f)_V$ is also an (approximate) symmetry of the
QCD vacuum, reflected by the existence of nearly degenerate
$SU(N_f)$ multiplets in the hadron spectrum.
If this was also true for the axial symmetry each hadron should have
an approximately degenerate ``chiral partner'' of opposite parity.
Since this is not the case, one concludes that chiral symmetry is 
spontaneously broken in vacuum.
This is closely related to the existence of a non-vanishing
quark condensate $\ave{\bar q q}$, which is not invariant under
$SU(N_f)_A$ and therefore often deals as an order parameter for
spontaneous chiral symmetry breaking\footnote{
Note, however, that $\ave{\bar q q} = 0$ does not necessarily mean that 
chiral symmetry is restored since it could still be broken by
other condensates. This is for instance the case for three massless
flavors in the CFL phase, see \sect{cflprop}.\label{foot1}}.

Another hint for the spontaneously broken chiral symmetry is the 
low mass of the pion which comes about quite naturally if the pions are
interpreted as the corresponding Goldstone bosons in the two-flavor case.
If chiral symmetry was exact on the Lagrangian level (``chiral limit'')
they would be massless, while the small but finite pion mass reflects
the explicit symmetry breaking through the up and down quark masses.
In the analogous way, the pseudoscalar meson octet corresponds to
the Goldstone bosons in the three-flavor case. 
The Goldstone bosons obey several low-energy theorems
which provide the basis for chiral perturbation theory 
($\chi$PT)~\cite{gasserleutwyler,Ecker,BKM95}.
Unlike ordinary perturbation theory, $\chi PT$ corresponds to an expansion
in quark masses and momenta and can be applied in regions where
$\alpha_s$ is large. 

Since the perturbative vacuum is chirally symmetric for massless quarks,
one expects that chiral symmetry gets restored at high temperature.
This would also be the case at high density if the matter was in 
a trivial rather than in a color superconducting state (see footnote
\ref{foot1}).
The ``partial restoration'' of chiral symmetry, i.e., the in-medium 
reduction of $|\ave{\bar q q}|$ at small temperatures or densities
can be studied within $\chi$PT.
For two flavors one finds to leading order in temperature and 
density~\cite{GeLe89,CFG92} 
\beq
     \frac{\ave{\bar q q}_{T,\rho_B}}{\ave{\bar q q}_{0,0}} \;=\;
     1 \;-\; \frac{T^2}{8f_\pi^2}  
     \;-\; \frac{\sigma_{\pi N}}{f_\pi^2 m_\pi^2}\,\rho_B \;+\; \dots~,
\label{ChPTrho}
\eeq
where $f_\pi = 92.4$~MeV is the pion decay constant 
and $\sigma_{\pi N} \simeq $~45~MeV is the pion-nucleon sigma term.  
The two correction terms on the r.h.s. describe the effect of 
thermally excited non-interacting pions and of nucleons, respectively.
(Interaction effects are of higher order.) 
To be precise, the $T^2$ behavior is only correct in the chiral limit,
i.e., for massless pions, which are otherwise exponentially suppressed.
Since in the hadronic phase
the physical pion mass can never be assumed to be small against
the temperature, \eq{ChPTrho} is more of theoretical rather than practical
interest. Systematic mass corrections are of course possible~\cite{GeLe89}.
It remains at least qualitatively correct that pions,
being the lightest hadrons, dominate the low-temperature behavior. 

On the classical level, ${\cal L}_{QCD}$ is also invariant under global
\beq
    U_A(1):\quad
    q \rightarrow \exp{(i \alpha \gamma_5)}\;q~,
\eeq
in the limit of massless quarks. 
This symmetry is, however, broken on the quantum level 
and therefore not a real symmetry of QCD~\cite{tHooft}. 
Most prominently, this is reflected by the relatively heavy
$\eta'$ meson which should be much lighter (lighter than the $\eta$-meson)
if it was a Goldstone boson of a spontaneously broken symmetry. 

In this context instantons play a crucial role. 
These are semiclassical objects 
which originate in the fact that the classical Yang-Mills action
gives rise to an infinite number of topologically distinct degenerate
vacuum solutions.  
Instantons correspond to tunneling events between these vacua.    
When quarks are included, the instantons mediate an interaction
which is $SU(N_f)_L \times SU(N_f)_R$ symmetric, but explicitly
breaks the $U_A(1)$ symmetry (``'t Hooft interaction'')~\cite{tHooft}.
In particular it is repulsive in the $\eta'$-channel.

In the instanton liquid model~\cite{instliq}, the gauge field contribution 
to the QCD partition function is replaced by an
ensemble of instantons, characterized be a certain size and density
distribution in Euclidean space. 
It turned out that hadronic correlators obtained in this way
agree remarkably well with the ``exact'' solutions on the lattice
(see Ref.~\cite{SS98} for review).

\section{The QCD phase diagram}
\label{facts}

We have pointed out that the phase diagrams presented in 
\fig{figschemphase} are schematic conjectures, which are 
constrained only by a relatively small number of 
safe theoretical or empirical facts.  
Before discussing them in more detail,
we should note that in general we are not restricted to a single
chemical potential, but there is a chemical potential for each 
conserved quantity. Hence, the QCD phase diagram has in general
more dimensions than shown in \fig{figschemphase}.
If we talk about one chemical potential only, we thus have to
specify under which conditions this is fixed.
For instance, it is often simply assumed that the chemical potential
is the same for all quark species.
On the other hand, in a neutron star we should consider neutral matter
in beta equilibrium whereas in heavy-ion collisions we should 
conserve isospin and strangeness.
This means, the different sources of information we are going to 
discuss below describe different phase diagrams 
(or different slices of the complete multi-dimensional phase diagram)
as far as they correspond to different physical situations.
Moreover, the results are often obtained in or extrapolated to 
certain unphysical limits, like vanishing or unrealistically large quark
masses or the neglect of electromagnetism.

Direct empirical information about the phase structure of strongly interacting
matter is basically restricted to two points at zero temperature,
both belonging to the ``hadronic phase'' where quarks and
gluons are confined and chiral symmetry is spontaneously broken.
The first one corresponds to the vacuum, i.e., $\mu = 0$,
the second to nuclear matter at saturation density
(baryon density $\rho_B = \rho_0 \simeq 0.17~\fm^{-3}$),
which can be inferred from systematics of atomic 
nuclei\footnote{The numbers quoted in Ref.~\cite{BrMa90} are
a binding energy of $(16\pm 1)$~MeV per nucleon and a Fermi momentum of
$k_F = (1.35 \pm 0.05)$~fm$^{-1}$.}
Since the binding energy of nuclear matter is about $E_b \simeq$~16~MeV per
nucleon, it follows a baryon number chemical potential of
$\mu_B = m_N - E_b \simeq$~923~MeV, corresponding to a quark number chemical
potential $\mu = \mu_B/3 \simeq$~308~MeV. 
Unless there exists a so-far unknown exotic state which is bound more
strongly (like absolutely stable strange quark matter~\cite{Bodmer,Witten}),  
this point marks the onset of dense matter, i.e., the entire regime 
at $T=0$ and $\mu <$~308~MeV belongs to the vacuum\footnote{The
authors of Ref.~\cite{HJSSV} distinguish between ``QCD'', which is a 
theoretical object with strong interactions only, and ``QCD+'',
which corresponds to the real world with electromagnetic effects included.
In this terminology, our discussion refers to QCD.
In QCD+ the ground state of matter is solid iron, i.e., a crystal of iron 
nuclei and electrons. Here the onset takes place at $\mu_B \simeq$~930~MeV
and the density is about 13 orders of magnitude smaller than in 
symmetric nuclear matter.}.
The onset point is also part of a first-order phase boundary which 
separates a hadron gas at lower chemical potentials from a hadronic liquid at
higher chemical potentials if one moves to finite temperature. 
This phase boundary is expected to end in a critical endpoint,
which one tries to identify within multifragmentation experiments.
Preliminary results seem to indicate a corresponding critical temperature
of about 15~MeV~\cite{Nato02}.
Of course both, the gas and the liquid are part of the hadronic
phase. They have been mentioned for completeness but are not subject
of this report. There might be further hadronic phases, e.g., due to 
the onset of hyperons or to superfluidity\footnote{
We are using the word ``phase'' in a rather lose sense.
If there is a first-order phase boundary which ends in a critical endpoint
one can obviously go around this point without meeting a singularity.
Hence, in a strict sense, both sides of the ``phase boundary'' belong 
to the same phase. However, from a practical point of view it often makes
sense to be less strict, since the properties of matter sometimes
change rather drastically {\it across} the boundary (see, e.g., liquid 
water and vapor).}.

Obtaining empirical information about the QGP is the general aim of
ultra-relativistic heavy-ion collisions. As mentioned earlier, there are
some indications that this phase has indeed been reached at 
SPS~\cite{SpS} and at RHIC~\cite{RHIC}. 
There are also claims that the chemical freeze-out points, which are
determined in a thermal-model fit to the measured particle ratios,
must be very close to the phase boundary~\cite{BMSW04}.
However, since the system cannot be investigated under static conditions
but only integrating along a trajectory in the phase diagram,
an interpretation of the results without theoretical guidance
is obviously very difficult. 

There is also only little hope that information about 
color superconducting phases can be obtained from ultra-relativistic 
heavy-ion collisions, which
are more suited to study high temperatures rather than high densities.
For instance, at chemical freeze-out one finds 
$T\simeq 125$~MeV and $\mu_B \simeq 540$~MeV at AGS and  
$T\simeq 165$~MeV and $\mu_B \simeq 275$~MeV for  
Pb-Pb collisions at SPS~\cite{BMHS99}.
Even though the CBM project at the future GSI machine is designed to
reach higher densities~\cite{GSI}, it is very unlikely that the corresponding
temperatures are low enough to allow for diquark condensation. 
Of course, no final statement can be made as long as reliable predictions
for the related critical temperatures are missing.

On the theoretical side, most of our present knowledge about the QCD 
phase structure comes from ab-initio Monte Carlo calculations on the
lattice (see Refs. \cite{LaPh03,KaLa03} for recent reviews).
For a long time, these were restricted to zero chemical potential, 
i.e., to the temperature axis of the phase diagram, and
only recently some progress has been made in handling non-zero
chemical potentials. 
Before summarizing the main results, let us mention that twenty
years ago, the chiral phase transition at finite temperature has  
been analyzed on a more general basis, applying universality arguments
related to the symmetries of the problem~\cite{PiWi84}.
It was found that the order of the phase transition depends on
the number of light flavors: 
If the strange quark is heavy, the phase transition is second order
for massless up and down quarks and becomes a smooth cross-over if
up and down have non-vanishing masses.
On the other hand, if $m_s$ is small enough the phase transition 
becomes first-order.
The question which scenario corresponds to the physical quark masses
cannot be decided on symmetry arguments but must be worked out 
quantitatively. 
In principle, this can be done on the lattice. In practice, there is
the problem that it is not yet possible to perform lattice calculations 
with realistic up and down quark masses. The calculations are therefore
performed with relatively large masses and have to be extrapolated 
down to the physical values. Although this imposes some uncertainty,
the result is that the transition at $\mu = 0$ is most likely a 
cross-over~\cite{LaPh03,KaLa03}.

While there is thus no real phase transition, the cross-over is
sufficiently rapid that the definition of a transition temperature 
makes sense. This can be defined as the maximum of the chiral 
susceptibility, which is proportional to the slope of the quark condensate.
One finds a transition temperature of about 170~MeV.
It is remarkable that the susceptibility related to the Polyakov loop
-- which deals as order parameter of the deconfinement transition -- 
peaks at the same temperature, i.e., chiral and deconfinement transition 
coincide.
It is usually expected that this is a general feature which also holds
at finite chemical potential, but this is not clear.

When extrapolated to the chiral limit, the critical temperature is 
found to be $(173 \pm 8)$~MeV for two flavors and $(154 \pm 8)$~MeV
for three flavors~\cite{KLP01}. 
This is considerably lower than in the pure $SU(3)$ gauge theory
without quarks where $T_c = (269 \pm 1)$~MeV~\cite{LaPh03,KaLa03}.
According to the symmetry arguments mentioned above, the phase
transition in the two-flavor case is expected to belong to the 
$O(4)$ universality class, thus having $O(4)$ critical exponents.
Present lattice results seem to be consistent with this, 
but they are not yet precise enough to rule out other possibilities.

The extension of lattice analyses to (real) non-zero chemical potentials 
is complicated by the fact that in this case the fermion determinant
in the QCD partition function becomes complex. As a consequence 
the standard statistical weight for the importance sampling is no longer 
positive definite which spoils the convergence of the procedure.
Quite recently, several methods have been developed which allow
to circumvent this problem, at least for not too large chemical
potentials ($\mu/T \lesssim 1$).
One possibility is to perform a Taylor expansion
in terms of $\mu/T$ and to evaluate the corresponding coefficients
at $\mu=0$~\cite{AEHKKLS,EAHK03}. 
The second method is a reweighting technique where the ratio of the
fermion determinants at $\mu \neq 0$ and at $\mu=0$ is taken as a
part of the operator which is then averaged over an ensemble 
produced at $\mu=0$~\cite{FoKa02a,FoKa02,FoKa04}.
The third way is to perform a calculation at imaginary chemical 
potentials~\cite{HLP01,dFPh02,DELo03}.
In this case the fermion determinant remains real and the ensemble 
averaging can be done in the standard way. 
The results are then parametrized in terms of simple functions 
and analytically continued to real chemical potentials.

These methods have been applied to study the behavior of the phase
boundary for non-zero $\mu$. 
From the Taylor expansion one finds for the curvature of the phase boundary 
at $\mu=0$~\cite{AEHKKLS},
$T_c (d^2T_c/d\mu^2)|_{\mu=0} = -0.14 \pm 0.06$, i.e.,
$T_c(\mu) \simeq T_c(0) -(0.07 \pm 0.03) \mu^2/T_c(0)$.
Within error bars this result is consistent with the two other methods. 
However, all these calculations have been performed with relatively
large quark masses, and the curvature is expected to become larger
for smaller masses. 

Another important result is the lattice determination of a critical 
endpoint~\cite{EAHK03,FoKa02,FoKa04}.
We have seen that at $\mu=0$ the phase transition should be second order
for two massless flavors and most likely is a rapid cross-over for
realistic quark masses.  
On the other hand, it has been argued some time ago that at low temperatures 
and large chemical potentials the phase transition is probably 
first order.
Hence, for two massless flavors there should be a 
``tricritical point'', where the second-order phase boundary
turns into a first-order one~\cite{HJSSV,CDGP89,AsYa89}.
Similarly, for realistic quark masses one expects a first-order phase
boundary at large $\mu$, which ends in a (second-order) critical
endpoint, as indicated in the three last phase diagrams 
in \fig{figschemphase}.

It has been suggested that this point could possibly be detected in
heavy-ion experiments through event-by-event fluctuations in the 
multiplicities $N_\pi$ and mean transverse momenta $p_T^\pi$ of charged pions.
These fluctuations should arise as a result of critical fluctuations in 
the vicinity of the endpoint~\cite{SRS98,SRS99}.
To that end $N_\pi$ and $p_T^\pi$ should be measured as a function of
a control parameter $x$ 
which determines the trajectory of the evolving system in the phase diagram.
This parameter could be, e.g., the beam energy or the centrality of the 
collision. If for some value of $x$ the trajectory comes very close to the
critical point this should show up as a maximum in the above
event-by-event fluctuations. In this case the system is expected to 
freeze out close to the critical point because of critical slowing down. 
For a more detailed discussion of possible signatures under realistic
conditions (including finite size and finite time effects), see
Ref.~\cite{SRS99}.

On the lattice, the position of the endpoint has been determined first
by Fodor and Katz, employing the reweighting method~\cite{FoKa02}.
The result was $T = (160 \pm 3.5)$~MeV and 
$\mu_B = 3\mu = (725 \pm 35)$~MeV.
However, these calculation suffered from the fact that they have been 
performed on a rather small lattice with relatively large up and down quark 
masses. 
More recently, the authors have performed an improved calculation on a
larger lattice and with physical quark masses, shifting the endpoint to
$T = (162 \pm 2)$~MeV and  $\mu_B = (360 \pm 40)$~MeV. 
This result is consistent with estimates using the Taylor expansion method,
indicating a critical endpoint at $\mu/T \sim 1$, i.e., 
$\mu_B \sim 3\,T$~\cite{EAHK03}.

The techniques described above do not allow to study
the expected transition to color superconducting phases at large
chemical potentials but low temperatures.
So far, the only controlled way to investigate these phases is a
weak-coupling expansion, which becomes possible at very large 
densities because of asymptotic freedom.
These analyses show that strongly interacting matter at asymptotic
densities is indeed a color superconductor, which for three flavors
is in the CFL phase~\cite{Sch00b,EHHS00}.
Unfortunately, the weak-coupling expansion breaks down at densities 
several orders of magnitude higher than what is of ``practical''
relevance~\cite{RaSh00}, e.g., for the interior of compact stars,
and it can, of course, not be applied to study the hadron-quark 
phase transition itself.

\section{Scope and outline of this report}
\label{outline}

The above discussion has shown that the detailed phase structure
of strongly interacting matter at (not asymptotically) high density
and low temperature is largely unknown. 
In particular, there is practically no exact information about the
density region just above the hadron-quark phase transition
which might be relevant for the interiors of compact stars. 

In this situation models may play an important role in developing
and testing new ideas on a semi-quantitative basis and checking
the robustness of older ones.
Since models are simpler than the fundamental theory (QCD)
-- otherwise they are useless --
they often allow for studying more complex situations
than accessible by the latter.
The price for this is, of course, a reduced predictive power
due to dependencies on model parameters or certain approximation
schemes.
The results should therefore always be confronted with 
model independent statements or empirical facts, as far as available.
In turn, models can help to interprete the latter where no other
theory is available.
Also, ``model independent results'' are often derived by an expansion
in parameters which are assumed to be small. Thus, although mathematically
rigorous, they do not necessarily describe the real physical situation.
Here models can give hints about the validity of these assumptions
or even uncover further assumptions which are hidden.

An example which will be one of the central points in this report is the
effect of the strange quark mass $M_s$ on the phase structure.
Starting from the idealized case
of three massless flavors, this is often studied within an
expansion in terms of $M_s/\mu$, assuming that the strange quark
mass is much smaller than the chemical potential.
Obviously, this can always be done for large enough values of $\mu$,
but the expansion eventually breaks down, when $\mu$ becomes of the order 
of $M_s$.
However, the crucial point is that $M_s$ itself should be considered
to be a $\mu$-dependent effective (``constituent'') quark mass. 
This does not only imply that the expansion breaks down earlier 
than one might naively expect (because $M_s$ can be considerably 
larger than the perturbative quark mass $m_s$ which is listed in the 
particle data book), but also that effects due to the 
$\mu$ dependence of $M_s$ are missed completely. 
In particular, there can be strong discontinuities in $M_s$ across
first-order phase boundaries which, of course, cannot be described
by a Taylor expansion.

In the present work we investigate this kind of questions within
models of the Nambu--Jona-Lasinio type, i.e., schematic quark models 
with simple four-fermion (sometimes six-fermion) interactions.
Historically, the NJL model~\cite{NJL1,NJL2} has been introduced to describe 
spontaneous chiral symmetry breaking in vacuum {\it in analogy} to the BCS 
mechanism for superconductivity~\cite{BCS} and has later been extended to 
study its restoration at non-vanishing temperatures or densities. 
On the other hand, NJL-type models are straight-forwardly used 
in the original BCS sense to calculate color superconducting pairing gaps.
In fact, the renewed interest in color superconductivity was
caused by analyses in such models~\cite{ARW98,RSSV98}.
For the obvious next step, namely considering chiral quark condensates and
diquark condensates simultaneously and studying their competition and 
mutual influence, NJL-type models are therefore the natural 
choice~\cite{BeRa99}.
Going further,
the relatively simple interaction allows to attack quite involved
problems. For instance, in order to describe the transition from
two- to three-flavor color superconductors including dynamical mass
effects, we have to allow for six different condensates which can
be all different if we consider beta equilibrated matter.  

The details of the results of these investigations
are of course model dependent. 
In particular, it is not clear whether the model parameters,
which are usually fitted to vacuum properties, 
can still be applied at large densities.
In fact, it seems to be quite natural that the four-point couplings 
are $\mu$ and $T$ dependent quantities, just like the effective quark 
masses we compute. 
Nevertheless, at least qualitatively, our analysis 
can give important hints which of the so-far neglected 
effects could be important and which are indeed negligible. 
In the first place it should therefore be viewed in 
conjunction with and relative to other approaches. 

There are few exceptions, where we make definite predictions for
-- in principle -- observable quantities, like absolutely stable
strange quark matter or quark matter cores in neutron stars. 
Here, although we try to estimate the robustness of
the results with respect to the parameters, we are forced to 
assume that these are at least not {\it completely} different from the
vacuum fit. In these cases the arguments could be turned around:
If at some stage the predictions turn out to be wrong, this would
mean that the model parameters must change drastically. 

The focus of the present report is an NJL-model study of the
phase diagram at large densities,
with special emphasis on color superconducting phases and their properties.
We begin, however, with a rather detailed discussion of the model
and its phase structure {\it without} taking into account diquark pairing.
Besides defining the basic concepts of the model, this is done because we
think that in order to understand the influence of color superconductivity
one should know how the model behaves without.
In particular, some detailed knowledge about the mechanism of the chiral 
phase transition in the model will be helpful when this 
is combined with the pairing transition.  
This pre-discussion will also allow us to point out the main limitations
of the model. 
Many of them, like artifacts of missing confinement or of the 
mean-field treatment, are not restricted to the NJL model, 
and this discussion could also be useful for other approaches.
\\[2mm]
The remainder of this report is organized as follows. 

In Chap.~\ref{bag} we review the thermal properties of the NJL model,
concentrating on color non-superconducting quark matter with two flavors.
Besides setting up the formalism, a central aspect will be a comparison 
with the MIT bag model, which is often used to describe dense matter.
In Chap.~\ref{njl3} the model is extended to three quark flavors.
As a first application, we investigate the possibility
of absolutely stable strange quark matter.  

Color superconducting phases are included in the 
Chapters~\ref{tsc}-\ref{stars}.
In Chap.~\ref{tsc} we again restrict ourselves to two-flavor systems.
We begin with a general overview about the basic concepts of 
color superconductivity and briefly discuss other approaches.
We then extend our formalism to include diquark pairing and discuss
the role and mutual influence of several conventional and less 
conventional quark-antiquark and diquark condensates.
A point of particular interest will be the discussion of a spin-1 diquark
condensate as a possible pairing channel for those quarks which are
left over from the standard spin-0 condensate. 
Next, in Chap.~\ref{3sc} we consider color superconductivity in a
three-flavor system. Central issue will be the description of the
2SC-CFL phase transition, taking into account $\mu$ and $T$-dependent
constituent quark masses together with the diquark condensates.
For simplicity, we consider a single chemical potential for all quarks.
This restriction is relaxed in Chap.~\ref{neut}, where we construct
neutral quark matter in beta equilibrium which is of possible
relevance for compact stars. To that end we have to introduce up to
four independent chemical potentials. 
Both, homogeneous and mixed neutral phases are discussed. 
In Chap.~\ref{stars} we use the resulting homogeneous quark matter
equation of state to investigate the possibility of a quark matter
core in neutron stars. In order to model the hadronic phase
we take existing hadronic equations of state from the literature. 
Finally, in Chap.~\ref{conclusions} we summarize what we have done
and, more important, what remains to be done.

\chapter{NJL-model description of color non-superconducting two-flavor
         quark matter}
\label{bag}

In this chapter we give a general introduction to the use of 
Nambu--Jona-Lasinio (NJL) type models for analyzing quark matter at
non-zero density or temperature. To that end, we concentrate on the 
-- technically simpler -- two-flavor version of the model and
neglect the possibility of color superconducting phases. 
The NJL model will be introduced in \sect{njl}
where we briefly summarize its vacuum properties, before the
analysis is extended to hot and dense matter in \sect{njlthermo}.
Of course, this is not meant to be exhaustive, and the interested reader 
is referred to Refs.~\cite{vogl,klevansky,hatsuda,ripka} for reviews. 
Here, our main intention is to lay a solid 
ground for our later investigations, including a critical discussion of 
the limits of the model. 

A central aspect of the present chapter will be a comparison with the
MIT bag model, which is the most frequently used model to describe 
quark-gluon matter at large temperature or density.  
Originally, both models have been developed to analyze hadron properties.
Focusing on two different aspects of QCD, they are almost complementary:
Whereas the MIT bag model is based on a phenomenological realization
of confinement, the main characteristics of the NJL model is chiral 
symmetry and its spontaneous breakdown in vacuum. 
On the other hand the NJL model does not confine and the
MIT bag model violates chiral symmetry. Nevertheless, the models behave 
quite similarly when they are employed to calculate the equation of state of
deconfined quark matter at high density.
In this regime the essential feature of both models is the existence
of a non-vanishing vacuum pressure (``bag constant'') whereas confinement
and chiral symmetry are of course less important in the deconfined,
chirally restored regime\footnote{In fact, the MIT bag model is chirally 
symmetric in the thermodynamic limit, because chiral symmetry is broken 
only at the bag surface.}.
In the MIT bag model the bag constant is an external 
parameter, while in the NJL model it is dynamically generated.
In order to work out this correspondence in some detail, 
we begin with a brief summary of the basic features of the MIT bag model.

\section{MIT bag model}
\label{mit}

The MIT bag model has been suggested in the mid-seventies as 
a microscopic model for hadrons~\cite{MIT1,MIT2,MIT3}. 
At that time, QCD had already been formulated
and the MIT bag model was one of the first quark models
where the notions of confinement and asymptotic freedom 
have been implemented in a constitutive way\footnote{ 
Strictly speaking, rather than on asymptotic freedom, 
the model was based on the experimental fact of Bjorken scaling.
Since QCD was not yet generally accepted,
at least at the beginning~\cite{MIT1}, the model was presented
in a more general way, referring to QCD only as one of several 
possibilities. 
}.

In the MIT bag model, hadrons consist of free (or only weakly interacting)
quarks which are confined to a finite region of space: the ``bag''.
The confinement is not a dynamical result of the underlying theory,
but put in by hand, imposing the appropriate boundary conditions\footnote{ 
However, if the quarks are coupled to a non-Abelian gauge
field, one finds the non-trivial result that 
the boundary conditions can only be fulfilled if the system
is a color singlet~\cite{MIT1}.
}.
The bag is stabilized by a term of the form $g^{\mu\nu} B$
which is added to the energy-momentum tensor {\it inside} the bag. 
Recalling the energy-momentum tensor of a perfect fluid in its
rest frame, 
\beq
    T^{\mu\nu}_{fluid} \;=\; {\rm diag}(\epsilon, p, p ,p)~,
\eeq
the bag constant $B$ is immediately interpreted as positive contribution
to the energy density $\epsilon$ and a negative contribution to the 
pressure $p$ inside the bag. Equivalently, we may attribute a term
$-g^{\mu\nu} B$ to the region {\it outside} the bag. This leads to the
picture of a non-trivial vacuum with a negative energy density 
$\epsilon_\mi{vac} = -B$ and a positive pressure $p_\mi{vac} = +B$. 
The stability of the hadron then results from balancing this 
positive vacuum pressure with the pressure caused by the
quarks inside the bag. 

The MIT bag model says nothing about the origin
of the non-trivial vacuum, but treats $B$ as a free parameter.   
Evaluating the energy-momentum tensor in QCD, one finds
\beq
    B_\mi{QCD} \;=\; -\frac{1}{4}\,\ave{T_\mu^\mu}
    \;=\; \frac{11-\frac{2}{3}N_f}{32}\,
    \frac{\alpha_s}{\pi}\,\ave{G_a^{\;\mu\nu} G_{a\,\mu\nu}}
    \;-\; \frac{1}{4} \sum_f m_f \, \ave{\bar q_f q_f}~,
\label{BQCD}
\eeq
which is dominated by the contribution of the gluon condensate
(first term on the r.h.s.). 
In the second term $q_f$ denotes a quark with flavor $f$, and $m_f$ 
is the corresponding current quark mass. 
Employing the QCD sum-rule result of Ref.~\cite{VZS78}, Shuryak obtained
$B_\mi{QCD} \approx$~455~MeV$/\fm^3$~\cite{Shu78}, while a modern value
of the gluon condensate would yield a somewhat larger result.
In any case, as we will see below, this is much larger than the values
of $B$ one obtains in a typical bag model fit.

\subsection{Hadron properties}

Assuming a static spherical bag of radius $R$, 
the mass of a hadron in the MIT bag model is given by the sum~\cite{MIT3}
\beq
    E_\mi{BM} = \frac{4\pi}{3} B R^3 - \frac{z_0}{R} + \frac{1}{R} \sum_q x_q 
              + E_\mi{pert}~. 
\label{EMIT}
\eeq
The first term on the r.h.s. corresponds to the volume energy, 
required to replace the non-trivial vacuum by the trivial one inside the bag.
The second term was introduced in Ref.~\cite{MIT3} 
to parametrize the finite part of the zero-point energy of the bag.
The constant $z_0$ was treated as a free parameter,
whose theoretical determination was left for future work. 
In a later analysis, however, it turned out that not all singularities 
arising from the zero-point energy could be absorbed in a 
renormalization of the model parameters, like the bag constant~\cite{BeHa76}
(also see \cite{Jaffe03} for a recent discussion).
Therefore the definition of the finite part is ambiguous and
$z_0$ remained an undetermined fitting parameter in the literature.  
The third term  in \eq{EMIT} is the (rest + kinetic) energy of the
quarks. For massless quarks in the lowest $j = 1/2$ state one finds
$x_q = 2.04$ as solutions of the eigenvalue problem.
Finally, $E_{pert}$ corresponds to perturbative corrections due to
lowest-order gluon exchange. This term gives rise, e.g., to the 
$N-\Delta$ mass splitting.

\begin{table}[b!]
\begin{center}
\begin{tabular}{|l| c c c c c | c c|}
\hline
&&&&&&&\\[-3mm]
 Fit & $m_u$ [MeV] & $m_s$ [MeV] & $B$ [MeV/fm$^3$] & $z_0$  
 & $\alpha_s$ & $m_\pi$ [MeV] & $R_N$ [fm]
\\[1mm]
\hline
&&&&&&&\\[-3mm]
\cite{MIT3} &   0 & 279 &  57.5 & 1.84 & 2.2            & 280         & 1.0 \\
\cite{MIT3} & 108 & 353 &  31.8 & 1.95 & 3.0            & 175         & 1.1 \\
\cite{CHP83}&   0 & 288 & 351.7 & 0.00 & $<1$ (running) & (tachyonic) & 0.6 \\
\cite{SaTh95}&  5 & 354 &  44.7 & 1.17 &   0            & not given   & 1.0 \\
\cite{SaTh95}&  5 & 356 & 161.5 & 2.04 &   0            & not given   & 0.6 \\
\hline
\end{tabular}
\end{center}
\caption{\small Bag-model parameters obtained from fits to light
hadron spectra: quark masses $m_u = m_d$ and $m_s$, 
bag constant $B$, parameter for the zero-point energy $z_0$, and
strong coupling constant $\alpha_s$. In the last two columns we list
the resulting pion mass and the bag radius of the nucleon.
The first two lines correspond to the original fit of the MIT 
group~\cite{MIT3}, while the other fits have been performed later
within partially modified models (see text).}
\label{tabmitfit}
\end{table}  

\eq{EMIT} contains the following parameters:
The bag constant $B$, the parameter $z_0$, the quark masses 
(entering $x_q$), and the strong coupling constant $\alpha_s$ 
(entering $E_{pert}$). 
The bag radius $R$ is not a parameter but is separately fixed for each
hadron to minimize its mass. 
The parameters of the original fit~\cite{MIT3} are listed in
Table~\ref{tabmitfit} (first two lines). They have been adjusted to fit
the masses of the nucleon, the $\Delta$, the $\Omega^-$, and the
$\omega$-meson. The light quark masses ($m_u = m_d$) have not been
fitted, but have been set to zero (fit A) and to 108~MeV (fit B) to
test the sensitivity of the fit to its variation. 
With these parameters the authors
of Ref.~\cite{MIT3} obtained a good overall fit of the light
hadron spectra (baryon octet and decuplet, and vector meson nonet), 
magnetic moments, and charge radii.

There are, however, a couple of well-known problems: 
Since chiral symmetry is explicitly broken on the bag surface
and the $U_A(1)$ anomaly is not included, the pion mass comes out too
large, whereas the $\eta'$ is too light. Also, the values of $\alpha_s$ 
needed to reproduce the $N-\Delta$ mass splitting are extremely large and 
obviously inconsistent with the idea that the corrections are perturbative.  
Finally, the nucleon radii of 1.0 - 1.1~fm, as shown in Table~\ref{tabmitfit},
would mean that the bags overlap with each other in a nucleus, 
which is clearly inconsistent with the success of meson-exchange models.   

Some of these problems are related to each other. 
For instance, in the so-called chiral bag models, where chiral symmetry
is restored by coupling external pion fields (introduced as elementary
fields) to the bag surface~\cite{ChTh75}, the bag radii can be considerably 
smaller than in the MIT bag (``little bag''~\cite{BrRh79} $R_N \sim 0.3$~fm,
``cloudy bag''~\cite{CBM} $R_N \sim 0.8$~fm). 
This also leads to smaller values of $\alpha_s$ needed to fit the
$N-\Delta$ mass splitting~\cite{BrRh79,CBM}.
In the chiral bag models the nucleon mass is the sum of the bag contribution
and a self-energy contribution from the pion cloud. Since the latter is
negative, the former can be (much) larger than the physical nucleon mass.
This is the reason why the radii can be smaller than in the MIT fit.
In fact, many observables are quite insensitive to the bag radius, which
can even be taken to be zero (``Cheshire cat principle'', for review see, 
e.g., Ref.~\cite{Cheshi}).

\eq{EMIT} contains effects of spurious c.m.\@ motion, which can approximately
be projected out if one replaces $E_\mi{BM}$ by the mass
$M_{BM} = (E_\mi{BM}^2 - \sum_q (x_q/R)^2)^{1/2}$~\cite{cm1,cm2,CHP83}.
This correction has not been performed in the original MIT fit~\cite{MIT3}
but is standard in modern bag model calculations.
It has the appreciable effect that the bag radius is reduced by about 30\%
and might also remove some of the other problems discussed above. 
In fact, instead of a too large pion mass, the authors of Ref.~\cite{CHP83}
obtain $m_\pi^2$ slightly negative, and reasonable values for $R_N$ and
$\alpha_s$, once c.m.\@ corrections are included
(third line of \tab{tabmitfit}). 
Note, however, that this model also differs from the original MIT bag 
model in the treatment of the perturbative corrections, bringing in 
additional parameters, while the parameter $z_0$ was not employed in the fit.
On the other hand, the authors of Ref.~\cite{SaTh95} did not include
perturbative corrections and restricted their fit to the baryon octet. 
They also considered $\rho$ and $\omega$ mesons, but to that end
they introduced independent zero-point energies as additional parameters.
In this way the (baryonic) zero-point energy was left undetermined and
could be employed to vary the bag radius at will (cf. last two lines 
of \tab{tabmitfit}).

For the thermodynamic description of hot or dense quark-gluon matter,
many problems discussed above are not of direct relevance (see next section).
We have mentioned them, however, because they have a strong effect on
the parameter fit, and thereby {\it indirectly} on the thermodynamics.
In this context, the most important parameter is the bag constant. 
As obvious from \tab{tabmitfit}, the fitted values of $B$ vary by more
than a factor of six, although they are all
smaller than the value derived from the gluon condensate, \eq{BQCD}. 
From a modern point of view it is also remarkable that the strange
quark mass $m_s$ in all fits listed in the table is 
considerably larger than today accepted values of its ``current quark mass''
(see \tab{tabnjl3fit}).
Later we will see that similar masses appear in 
NJL-model quark matter at comparable densities.

\subsection{Thermodynamics}
\label{mitthermo}

When we consider a large number of quarks and gluons in a large MIT bag
we can replace the exact solutions of the boundary problem by plane waves,
while zero-point energy and c.m.\@ correction terms drop out.
Hence the energy density $\epsilon = E_\mi{BM}/V$ 
at temperature $T$ and a set of chemical potentials $\{\mu_f\}$ reduces to
\beq
    \epsilon(T,\{\mu_f\}) = B 
   + \epsilon_\mi{free}(T,\{\mu_f\}) + \epsilon_\mi{pert}(T,\{\mu_f\})~,
\eeq
where $\epsilon_\mi{free}$ is the energy density of a free relativistic gas
of quarks, antiquarks, and gluons, while $\epsilon_\mi{pert}$ corresponds
to perturbative corrections. 
Equivalently, we can write for the pressure
\beq
    p(T,\{\mu_f\}) = -B + p_\mi{free}(T,\{\mu_f\})
    + p_\mi{pert}(T,\{\mu_f\})~.
\eeq
Here the earlier mentioned role of $B$ as a negative pressure inside the 
bag relative to the non-trivial vacuum is obvious. 
The free part is given by
\begin{alignat}{1}
     p_\mi{free}(T,\{\mu_f\}) = 6\,&T \sum_{f} \int \dtp\,\Big\{
      \ln{[1+\exp{(-\frac{1}{T}(E_{p,f} - \mu_f))}]}  
    + \ln{[1+\exp{(-\frac{1}{T}(E_{p,f} + \mu_f))}]}\Big\} \nonumber\\
     - 16\,&T \int \dtp\, \ln [1 -\exp (-\frac{p}{T})]~,
\label{pfree}
\end{alignat} 
where the second integral corresponds to the gluons (2 spin and 8 color
degrees of freedom)
and the first integral corresponds to the quarks and antiquarks
(2 spin and 3 color degrees of freedom and a sum over flavors).
$E_{p,f} = (p^2 + m_f^2)^{1/2}$ is the on-shell energy of a quark of flavor $f$
with three-momentum $p$.

In the following we consider the case of two massless quark flavors 
with a common chemical potential $\mu$. 
The integrals in \eq{pfree} are then readily evaluated
and the bag model pressure becomes
\beq
    p(T,\mu) = -B + 37\,\frac{\pi^2}{90}\,T^4 + \mu^2T^2 +
    \frac{\mu^4}{2\pi^2}~,
\label{pbm}
\eeq 
where we have neglected the perturbative contribution.
The famous factor 37 = 16+21 in front of the $T^4$ term is the sum of the 16
gluonic degrees of freedom and the product of the 24 quark and antiquark 
degrees of freedom with a factor $7/8$ due to Fermi statistics. 

In order to be stable, the pressure in the quark-gluon phase must 
not be negative. For $\mu = 0$ this leads to a minimum temperature 
\beq
    T_0 = \Big(\frac{90}{37\pi^2}\,B\Big)^{1/4}~,
\eeq
while for $T = 0$ we obtain a minimum chemical potential
\beq
    \mu_0 = \Big(2\pi^2\,B\Big)^{1/4}~.
\eeq
It is interesting that $T_0$ has already been derived by the MIT group 
in their first paper about the bag model~\cite{MIT1}. There $T_0$ was 
identified with the ``limiting temperature'' of a highly excited hadron, 
i.e., the temperature a hadronic bag must not exceed in order to remain 
stable. Thus, although formally equivalent to the above derivation of
$T_0$, the perspective is rather opposite. 

\begin{figure}
\begin{center}
\epsfig{file=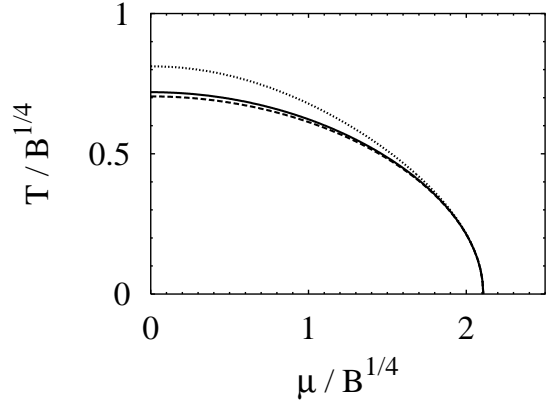,width=8.0cm}
\end{center}
\vspace{-0.5cm}
\caption{\small Lines of zero pressure in a two-flavor bag model
         with non-interacting massless quarks and gluons (dashed)
         and with quarks only (dotted). The solid line indicates the
         phase boundary, separating the  ``hadronic phase'', described 
         by a gas of non-interacting massless pions, from the 
         quark-gluon phase.
}
\label{figbag}
\end{figure}

For arbitrary chemical potentials smaller than $\mu_0$ one can easily 
solve for the temperature $T(\mu)$ for which the pressure vanishes.
The result is displayed in \fig{figbag} (dashed line).
For later purposes we also show the corresponding curve one obtains 
with quark degrees of freedom only (dotted). In this case the
factor 37 in \eq{pbm} is replaced by 21. Accordingly the minimum 
temperature is enhanced by a factor $(37/21)^{1/4} = 1.15$, while
the minimum chemical potential remains unchanged.  

\eq{pbm} is the most simple example for a bag-model description of the
quark-gluon plasma. In order to construct a phase transition, we also 
need an equation of state for the hadronic phase.
Usually, the latter is not described within the bag model as well
but taken from models with hadronic degrees of freedom.
As a prototype, we consider a gas of non-interacting 
massless pions, which should dominate the low-temperature $\mu = 0$-regime.
The resulting phase boundary, i.e., the line where $p$ becomes
equal to the pressure of the pion gas, is indicated by the solid line in
\fig{figbag}. Since the pions have an isospin degeneracy of three, the
pressure difference is given by \eq{pbm}, but with 37 replaced by
37-3=34. Hence, the critical temperature at $\mu = 0$ is given by 
$T_c = (37/34)^{1/4}~T_0 = 1.02~T_0$.
    
So far we have not set the absolute scale in our phase diagram.
Taking the bag constant from the original MIT fit, 
$B = 57.5$~MeV/fm$^3$ ($B^{1/4} = 145$~MeV)~\cite{MIT3}, we obtain 
$T_c = 104$~MeV, whereas in order to get the lattice value, 
$T_c \simeq 170$~MeV, we need a seven times larger bag constant
$B \simeq 400$~MeV/fm$^3$ ($B^{1/4} \simeq 235$~MeV).
For $\mu_0$, which in the present model is the critical chemical potential 
for the phase transition at $T=0$ , this variation of
$B$ leads to values roughly ranging from 300 to 500~MeV.

For several reasons, however, it is clear that a model of this type is too
simplistic:

\begin{itemize}

\item
Since mesons are not sensitive to the (quark number) chemical potential
they do not influence the phase transition in the large-$\mu$ low-$T$ regime. 
Thus, for a more realistic description, we need baryonic degrees of freedom.
In the above example, it would be most natural to add the contribution 
of a free nucleon gas to the hadronic phase.
It is a well-known fact, however, that this would lead to the unphysical
situation that the hadronic phase ``wins'' at large chemical potentials.
This is obvious from the contribution of a fermion of type $i$
to the pressure at $T=0$,
\beq
    p_i(T=0,\mu_i) \;=\; \frac{g_i}{24\pi^2}\,\mu_i^4 \;+\; \dots~,
\eeq
where the ellipsis indicates corrections due to the fermion mass
which can be neglected at very large chemical potentials.
Here $g_i$ is a degeneracy factor which is three times larger for
quarks than for nucleons because of color. On the other hand, since
nucleons consist of three valence quarks, their chemical potential  
is three times larger than the quark number chemical potential.
Altogether, this means $p_q : p_N = 1 : 27$, which leads to the above
mentioned unphysical result.

\item
As we have seen, the phase transition at $\mu = 0$ is a result of the 
larger number of quark-gluon degrees of freedom (37) compared with 
the hadronic ones (3 for the pion gas), i.e., the larger coefficient
of the $T^4$-term. This necessarily means, that the phase transition is
first-order with a latent heat per volume
\beq
    \frac{\Delta Q}{V} = (37 - 3)\,\frac{4\pi^2}{90}\,T_c^4~.
\eeq
This is in strong contrast to the universality arguments mentioned in
the Introduction, according to which we would expect a 
second-order phase transition in QCD with two massless flavors~\cite{PiWi84}.
It is possible to reduce $\Delta Q$ by introducing additional 
hadrons, but obviously it is very difficult to get $\Delta Q = 0$
without inhibiting the phase transition.
\end{itemize}

It is quite plausible that these problems can be traced back to the hybrid 
nature of the above model, i.e., the fact that the hadronic and the 
quark-gluon degrees of freedom are not derived from the same Lagrangian.
In a more consistent picture one should start from a gas of hadronic bags
to describe the hadronic phase.
Since the bags have finite sizes which can only be reduced at the 
expense of energy, it is obvious that the system will not stay in the
hadronic phase up to arbitrarily large densities.  
In the most naive picture the bags would simply unite
to form a uniform phase when the average quark number density exceeds 
the density inside a single bag. (In fact, it is more
difficult to prevent the bags from forming one large bag already at
low densities. For this one would need to introduce a negative surface 
tension or some repulsive force between separate bags.)
In this way also the connection between the chiral and the deconfinement
phase transition appears quite natural if one attributes
the non-trivial vacuum outside the bags to the spontaneous breakdown of
chiral symmetry, which is restored inside the bags.
We will come back to this point of view in \sect{njlbag}.

While qualitatively the picture drawn in the previous paragraph looks
quite attractive, its quantitative realization is of course very difficult.
An interesting step in this direction is the quark-meson coupling model
where nuclear matter and even finite nuclei are described by MIT bags
which interact by exchange of (elementary) scalar and vector 
mesons~\cite{SaTh95,Gui88,SaTh94,STT96}.
A somewhat ``cheaper'' alternative is to employ hybrid models with
finite volume corrections on the hadronic side (see, e.g., Ref.~\cite{BJBP93}).

\subsection{Equation of state at zero temperature}
\label{miteos}

Since we are mostly interested in quark matter at low temperature
we would like to discuss a few more details of the bag model equation
of state in the zero temperature limit. This will also provide a
basis for our later comparison with the NJL model.

For $T=0$, \eq{pfree} simplifies, and the total pressure is given by 
\beq
     p(\{\mu_f\}) \;=\; -B \;+\;
     \frac{3}{\pi^2} \sum_{f}\,\int_0^{p_F^f} dp\;p^2\,
     (\mu_f - E_{p,f}) \;+\; p_\mi{pert}(\{\mu_f\})~,
\label{OmegaBM}
\eeq
where $p_F^f = \theta(\mu_f - m_f) (\mu_f^2 - m_f^2)^{1/2}$
denotes the Fermi momentum of flavor $f$. 
For simplicity, we have dropped the temperature argument.

\begin{figure}
\begin{center}
\epsfig{file=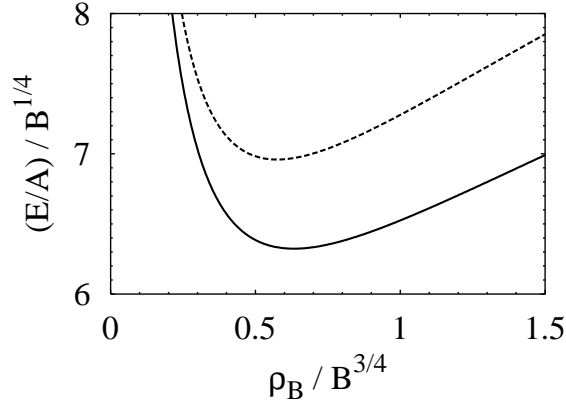,width=8.0cm}
\end{center}
\vspace{-0.5cm}
\caption{\small Energy per baryon number $E/A$ in the bag model
as a function of baryon number
density $\rho_B$. Solid line: $\alpha_s = 0$. Dashed line: $\alpha_s = 0.5$.
}
\label{figeabag}
\end{figure}

In the following, we concentrate again on the case of a uniform chemical
potential $\mu$ for two massless flavors. Neglecting the
perturbative term and applying standard thermodynamic relations, 
we obtain for the pressure, energy density, and quark number density
\beq
    p(\mu) = -B + \frac{\mu^4}{2\pi^2}~,\quad
    \epsilon(\mu) = B + \frac{3\mu^4}{2\pi^2}~,\quad
    n(\mu) = \frac{2\mu^3}{\pi^2}~.
\label{pen}
\eeq
This is simply a free gas behavior, modified by the bag constant.
From these expressions we immediately get for the function $\epsilon(p)$,
which, e.g., determines the mass-radius relation of neutron stars,
\beq
    \epsilon(p) = 3 p + 4B~.
\label{eosbag}
\eeq
Another quantity of interest, which follows from \eq{pen}, 
is the energy per particle $E/N$ as a function of density,
\beq
    \frac{E}{N}\,(n) \;=\; \frac{\epsilon}{n}\,(n) \;=\; \frac{B}{n} + 
    \frac{3}{4}\,\Big( \frac{\pi^2}{2}\, n \Big)^{\frac{1}{3}}~.
\eeq
For later convenience, we rephrase this as the energy per baryon number $A$
in terms of the baryon number density $\rho_B = n/3$,
\beq
    \frac{E}{A}\,(\rb) \;=\; \frac{\epsilon}{\rb}\,(\rb) \;=\; \frac{B}{\rb} + 
    \frac{9}{4}\,\Big( \frac{3\pi^2}{2}\, \rb \Big)^{\frac{1}{3}}~.
\label{eabag}
\eeq
This function is plotted in \fig{figeabag} (solid line).
Its general structure is easily understood if we recall that
$\rho_B = A/V$. Thus the first term on the r.h.s. is just the volume
energy of the MIT bag, while the second term reflects the $1/R$ behavior 
of the quark energy in \eq{EMIT}. Of course, the coefficient of the latter
depends on the number of occupied states and therefore it is different
in the thermodynamic limit from the case of a single hadron.

Because of the interplay of the two terms, $E/A$
diverges for both, $\rb \rightarrow 0$ and $\rb \rightarrow \infty$,
and has a minimum at
\beq
    \Big(\frac{E}{A}\Big)_\mi{min} \;=\;
     3\,(2\pi^2\,B)^{\frac{1}{4}} \;\equiv\;3\,\mu_0
\quad \text{at} \quad
    \rb \;=\; \frac{1}{3}\,\Big(\frac{2}{\pi^2}\Big)^{\frac{1}{4}}
     \,(4B)^{\frac{3}{4}} \;=:\;\rho^*~.
\eeq
Note that this is just the point where the pressure vanishes, reflecting
the general thermodynamic relation
\beq
    \frac{\partial}{\partial \rb}\,\Big(\frac{\epsilon}{\rb}\Big) \;=\;
    \frac{p}{\rb^2}~.
\label{eader}
\eeq
The physical meaning of this relation becomes clear if we consider 
a finite lump of quark matter (large enough that the thermodynamic treatment 
is valid). For $\rb < \rho^*$ (i.e., $\mu < \mu_0$), the pressure is 
negative and the lump shrinks, thereby increasing the density. 
On the other hand, for $\rb > \rho^*$ ($\mu > \mu_0$), the pressure is 
positive and the lump tends to expand, unless this is prevented by external
forces. Hence, the only stable point is  $\rb = \rho^*$, where the pressure
vanishes. In a canonical (instead of grand canonical) treatment, which
would be more appropriate for this example with fixed particle number,
this stability becomes manifest as a minimum in the energy.
(Recall that $T=0$ and hence the Helmholtz free energy $F = E - TS = E$.) 

The absolute scale of \fig{figeabag} is again set by the value of the bag
constant. 
As discussed above, varying $B$ between the two ``extreme'' values, 
$B = 57.5$~MeV/fm$^3$ from the original MIT fit and
$B = 400$~MeV/fm$^3$ from fitting  
$T_c \simeq 170$~MeV within the simple hybrid model, leads to values of
$\mu_0$ between about 300 and 500~MeV. This corresponds to $(E/A)_{min}$
ranging from about 900~MeV at  $\rb = 1.4\,\rho_0$ to 1500~MeV at 
$\rb = 6.5\,\rho_0$. 
Of course, values for $(E/A)_{min}$ lower than the energy per nucleon
in atomic nuclei must be excluded, since otherwise nuclei should be
able to decay into a large bag of deconfined quarks. We will come back 
to this in \sect{sqm}.  

Finally, we discuss briefly the influence of perturbative
corrections, which we have neglected so far.
To order $\alpha_s$ the free-gas terms in \eq{pen} are scaled by a
factor~\cite{Farhi},
\beq
    p(\mu) = -B + \gamma\,\frac{\mu^4}{2\pi^2}~,\quad
    \epsilon(\mu) = B +  \gamma\,\frac{3\mu^4}{2\pi^2}~,\quad
    n(\mu) = \gamma\,\frac{2\mu^3}{\pi^2}~;\qquad
    \gamma := (1 - 2\,\frac{\alpha_s}{\pi})~.
\eeq
Obviously, this correction makes only sense for $\alpha_s \ll \pi/2$.
This demonstrates again that the value of the original MIT fit, 
$\alpha_s = 2.2$, is completely out of range for a perturbative treatment.
Note that it would even change the sign of the free-gas terms. 

As a consequence of the correction, the chemical potential $\mu_0$ at
which the pressure vanishes and thus $(E/A)_\mi{min}$
are enhanced by a factor $\gamma^{-1/4}$,
whereas the corresponding baryon number density $\rho^*$ is reduced
by a factor $\gamma^{1/4}$. Accordingly, the minimum of $E/A$ is shifted
to a larger value at a lower density. 
This is illustrated by the dashed line in \fig{figeabag}, which corresponds
to $\alpha_s = 0.5$. 
On the other hand, the relation between $\epsilon$ and $p$, \eq{eosbag}, 
remains unaffected by the correction.

\section{Nambu--Jona-Lasinio model in vacuum}
\label{njl}

As pointed out earlier, the Nambu--Jona-Lasinio (NJL) model is to
some extent complementary to the MIT bag model. 
Historically, it goes back to two papers by Nambu and
Jona-Lasinio in 1961~\cite{NJL1,NJL2}, i.e., to a time when 
QCD and even quarks were still unknown. 
In its original version, the NJL model was therefore a model of
interacting {\it nucleons}, and obviously, confinement
-- the main physics input of the MIT bag model -- was not an issue. 
On the other hand, even in the pre-QCD era there
were already indications for the existence of a (partially) conserved
axial vector current (PCAC), i.e., chiral symmetry.
Since (approximate) chiral symmetry implies (almost) massless fermions
on the Lagrangian level, the problem was to find a mechanism which
explains the large nucleon mass without destroying the symmetry. 
It was the pioneering idea of Nambu and Jona-Lasinio that the mass gap
in the Dirac spectrum of the nucleon can be generated quite analogously
to the energy gap of a superconductor in BCS theory, which has been
developed a few years earlier~\cite{BCS}. 
To that end they introduced a Lagrangian for a nucleon field
$\psi$ with a point-like, chirally symmetric  four-fermion 
interaction~\cite{NJL2},
\beq
    {\cal L} = \bar\psi (i\delsl - m) \psi
    + G\,\Big\{(\bar\psi\psi)^2 + (\bar\psi i\gamma_5\vec\tau\psi)^2 \Big\}~.
\label{LNJL}
\eeq
Here $m$ is a small bare mass of the nucleon, $\vec\tau$ is a Pauli matrix
acting in isospin space, and $G$ a dimensionful coupling constant.
As we will discuss in more detail in \sect{njlmesons}, the self-energy 
induced by the interaction generates an effective mass $M$ which can
be considerably larger than $m$ and stays large, even when $m$ is taken
to zero (``chiral limit''). At the same time there are light 
collective nucleon-antinucleon excitations which become massless in the 
chiral limit: The pion emerges as the Goldstone boson of the spontaneously 
broken chiral symmetry. 
In fact, this discovery was an important milestone on the way to the
general derivation of the Goldstone theorem in the same year~\cite{Goldstone}.

After the development of QCD, the NJL model was reinterpreted as a
schematic quark model~\cite{Kleinert,Volkov,HaKu84}. At that point, 
of course, the lack
of confinement became a problem, severely limiting the applicability
of the model. On the other hand, there are many situations where chiral
symmetry is the relevant feature of QCD, confinement being less important.
The most prominent example is again the Goldstone nature of the pion. 
In this aspect the NJL model is superior to the MIT bag, which, as we have 
seen, fails to explain the low pion mass.

End of the nineties, a third era of the NJL model began when the model
was employed to study color superconducting phases in deconfined 
quark matter. 
There, by definition, lack of confinement is again of minor relevance. 
As outlined in the Introduction, color superconductivity will be 
discussed in great detail in \chap{tsc} and thereafter. 

After the reinterpretation of the NJL model as a quark model, many authors
kept the original form of the Lagrangian, \eq{LNJL}, with $\psi$ now 
being a quark field with two flavor and three color degrees of freedom.
However, this choice is not unique and we can write down 
many other chirally symmetric interaction terms.
For instance, from a modern point of view, local 4-point interactions between 
quarks in a two-flavor system can be thought of to be abstracted from
instanton induced interactions~\cite{tHooft}. In this case the interaction
Lagrangian should have the form
\beq 
{\cal L}_{inst} = G\,\Big\{ ({\bar q}q)^2 - ({\bar q}\,\vec\tau q)^2 
- ({\bar q}\,i\gamma_5 q)^2 + ({\bar q}\,i\gamma_5\vec\tau q)^2 \Big\}~. 
\label{Linst}
\eeq 
Here and in the following we denote quark fields by $q$.
In general we will call all these models ``NJL-type models'' 
(or just ``NJL models'') as long as they describe quarks
interacting via 4-point vertices (or sometimes higher $n$-point vertices).

\subsection{Constituent quarks and mesons}
\label{njlmesons}

In this section we briefly review the vacuum properties of quarks and mesons
described within the NJL model. For simplicity, we restrict this discussion
to the standard NJL-Lagrangian, \eq{LNJL}, for quarks with two flavor and
three color degrees of freedom. Most of this can easily be generalized
to other NJL-type Lagrangians with two degenerate flavors.
The discussion of the three-flavor case, which has some additional features,
is deferred to the next chapter.

\begin{figure}
\begin{center}
\epsfig{file=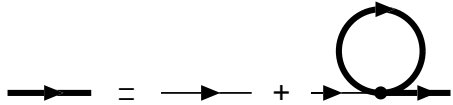,width=6.0cm}
\end{center}
\vspace{-0.5cm}
\caption{\small  Dyson equation for the quark propagator in Hartree
         approximation. The bare (dressed) propagator is denoted by the
         thin (bold) line.}
\label{fignjlgap}
\end{figure}

In most publications (including the original papers by Nambu and 
Jona-Lasinio~\cite{NJL1,NJL2}) the quark self-energy which arises from the 
interaction term has been calculated within Hartree or Hartree-Fock
approximation. The corresponding Dyson equation is depicted in 
\fig{fignjlgap}.
Since in this approximation the self-energy is local, it only gives rise
to a constant shift in the quark mass,
\beq
    M \;=\; m \,+\, 2i\,G \int \dfp \; {\rm Tr}\,S(p)~.
\label{njlgap1}
\eeq
Here
$S(p) = (\psl - M +i\varepsilon)^{-1}$ is the {\it dressed} quark propagator,
underlining the non-perturbative character of the approximation.
The trace is to be taken in color, flavor, and Dirac space.
One finds
\beq
    M \;=\; m \,+\, 8N_f N_c\,G\,i\int \dfp \, \frac{M}{p^2 - M^2 + i\eps}~,
\label{njlgap2}
\eeq
where $N_f=2$ and $N_c=3$ are the number of flavors and colors, respectively.
For a sufficiently strong coupling $G$,
this allows for a non-trivial solution $M \neq m$,
even in the chiral limit $m = 0$, producing a gap of $\Delta E = 2M$
in the quark spectrum. In analogy to BCS theory, \eq{njlgap2}
is therefore often referred to as ``gap equation''.
$M$ is often called ``constituent quark mass''.
A closely related quantity is the quark condensate, which is generally
given by
\beq
    \ave{\bar q q} \;=\; - i \int \dfp \; {\rm Tr}\,S(p)~,  
\label{njlqbarqgen}
\eeq
and thus in the present case
\beq
    \ave{\bar q q} \;=\; - \frac{M-m}{2G}~.  
\label{njlqbarqh}
\eeq

\begin{figure}
\begin{center}
\epsfig{file=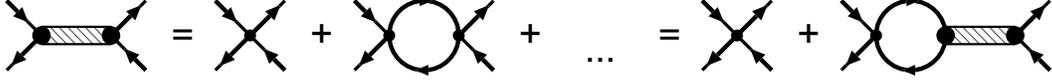,width=14.0cm}
\end{center}
\vspace{-0.5cm}
\caption{\small Bethe-Salpeter equation for quark-antiquark T-matrix 
        (``meson propagators'', shaded boxes) in RPA.
        The solid lines correspond to the dressed quark propagators in Hartree 
        approximation (\fig{fignjlgap}).}       
\label{fignjlrpa}
\end{figure}

Iterating the 4-point vertex as shown in \fig{fignjlrpa} yields the 
quark-antiquark T-matrix in random phase approximation (RPA),
\beq
    T_M(q^2) \;=\; \frac{2G}{1 - 2G \,\Pi_M(q^2)}~,
\eeq
where 
\beq
    \Pi_M(q^2) \;=\; i \int \dfp \tr{{\cal O}_M\,S(p+q)\,{\cal O}_M\,S(p)}
\label{rpapol}
\eeq
is the quark-antiquark polarization in the channel with the quantum numbers
$\{M\}$. For the Lagrangian \eq{LNJL} we have the sigma channel 
(${\cal O}_\sigma = \unity$) and three pion channels 
(${\cal O}_{\pi_a} = i\gamma_5 \tau_a$, $a=1,2,3$). 
Evaluating the traces and employing the gap equation for $M\neq 0$ one finds
\begin{alignat}{1}
\Pi_\sigma(q^2) \;&=\; \frac{1}{2G}\Big(1-\frac{m}{M}\Big) 
- \frac{1}{2}(q^2 - 4M^2)\,I(q^2)~,
\nonumber\\
\Pi_{\pi_a}(q^2) \;&=\; \frac{1}{2G}\Big(1-\frac{m}{M}\Big) 
- \frac{1}{2}q^2\,I(q^2)~,
\label{sigpipol}
\end{alignat}
where
\beq
    I(q^2) \;=\; 4N_f N_c\,i\int\dfp\frac{1}{[(p+q)^2 - M^2 + i\eps]
    [p^2 - M^2 + i\eps]}~.
\label{Iq}
\eeq
In order to determine meson properties one interpretes $T_M$
as an effective meson exchange between the external quark legs in
\fig{fignjlrpa} and parametrizes the pole structure as
\beq
    T_M(q^2) \;=\; \frac{-g_{Mqq}^2}{q^2 - m_M^2}~.
\eeq
Thus
\beq
    1 - 2G \,\Pi_M(q^2=m_M^2) = 0 \quad \text{and} \quad
    g_{Mqq}^{-2} = \frac{d \Pi_M}{d q^2}\Big|_{q^2=m_M^2}  
~.    
\label{rpamass}
\eeq
Using \eq{sigpipol}, one immediately finds that $m_\pi = 0$ if $m = 0$,
in accordance with the Goldstone theorem. For $m \neq 0$, $m_\pi$ also
becomes non-zero (see below). 

\begin{figure}
\begin{center}
\epsfig{file=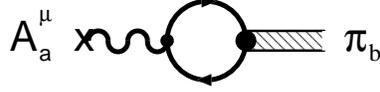,width= 5.0cm}
\end{center}
\vspace{-0.5cm}
\caption{\small One-pion-to-vacuum matrix element in RPA, giving rise to the 
         weak pion decay: A pion with isospin index $b$ is coupled via a quark
         loop to an axial current with isospin index $a$.}       
\label{fignjlfpi}
\end{figure}

The pion decay constant can be obtained from the one-pion-to-vacuum
matrix element visualized in \fig{fignjlfpi},
\beq
    f_\pi\,q^\mu\,\delta_{ab} \;=\; g_{\pi qq} \int \dfp 
    \tr{\gamma^\mu\gamma_5\frac{\tau_a}{2}\,S(p+q)\,i\gamma_5\tau_b\,S(p)}~.
\label{njlfpi}
\eeq
It is straight forward to show that in the chiral limit the generalized
Goldberger-Treiman relation~\cite{GT},
\beq
    g_{\pi qq}\,f_\pi \;=\; M \;+\; {\cal O}(m)
\eeq
holds. Moreover, in first non-vanishing order in $m$, the pion mass
satisfies the Gell-Mann Oakes Renner relation~\cite{GOR},
\beq
   f_\pi^2 m_\pi^2 \;=\; - m \,\ave{\bar qq} \;+\; {\cal O}(m^2)~.
\eeq 

In the brief discussion presented above, we have ignored several problems:

\begin{itemize}

\item In general, the gap equation has more than one solution. 
      For instance, in the chiral limit, $m=0$, there is always a 
      trivial solution $M=0$, but there can be non-trivial solutions
      $M = \pm M_0 \neq 0$ as well. In this case, one has to find out
      which solution minimizes the vacuum energy.

      One way is to go back to the underlying mechanism
      of the gap equation, which is a Bogoliubov-Valatin rotation
      of the fields. Starting point is a variational ansatz for
      the non-trivial vacuum of the model~\cite{ARW98,NJL1,klevansky},
\beq
      \ket{vac} \;=\; \prod_{\vec p,s,f,c} \Big[\cos{\theta_s(\vec p)}
       + e^{i\xi_s(\vec p)}\,\sin{\theta_s(\vec p)}\, 
       b^\dagger(\vec p,s,f,c)\,  
       d^\dagger(-\vec p,-s,f,c) \Big] \ket{0}~,
       \label{njlvac}
\eeq
      where $\ket{0}$ is the perturbative vacuum, and $b(\vec p,s,f,c)$
      and $d(\vec p,s,f,c)$ are the corresponding annihilation operators 
      for a quark or antiquark, respectively, with momentum $\vec p$, 
      helicity $s$, flavor $f$, and color $c$. 
      According to this ansatz, $\ket{vac}$ is a coherent
      state composed of quark-antiquark pairs with zero total momentum.
      This underlines once more the analogy to the BCS
      ground state. Minimizing the ground state energy, 
\beq
      {\cal E}_\mi{vac}[\theta_s(\vec p),\xi_s(\vec p)] = 
      \bra{vac} \hat H \ket{vac}~,
\eeq
      ($\hat H =$ Hamiltonian of the model)
      with respect to variations of the functions $\theta_s(\vec p)$
      and $\xi_s(\vec p)$
      leads to a self-consistency equation, which is equivalent to
      the Hartree-Fock gap equation discussed above. (For the difference
      between Hartree and Hartree-Fock, see below.)
      It turns out that the vacuum energy is minimized by the solution
      with the largest $M$. In the chiral limit this means that the
      non-trivial solution is stable whenever it exists. 

      An alternative way to calculate the ground state energy 
      will be presented in \sect{njlthermo} in the context of the 
      thermodynamics of the model. There we will discuss further details.

\item The NJL model is not renormalizable. Since the above expressions
      contain divergent integrals, e.g., \eqs{njlgap2}, (\ref{Iq}), and
      (\ref{njlfpi}), we have to specify how to regularize these
      divergencies. This prescription is then part of the model.
      There are several regularization schemes which have been used
      in the literature, and each of them have certain advantages
      and disadvantages~\cite{klevansky,ripka}.
      When the model is applied to thermodynamics, most authors
      prefer to regularize the integrals by a (sharp or smooth) 3-momentum
      cut-off. Besides being relatively simple, this has the advantage 
      that it preserves the analytical structure, necessary, e.g., for the 
      analytical continuation of functions given on imaginary 
      Matsubara frequencies. Of course, 3-momentum cut-offs
      violate the Lorentz covariance of the model. It is often argued
      that this problem is less severe at finite temperature or density 
      where manifest covariance is anyway broken by the medium. 
      Although this argument is questionable, since it makes a
      difference whether the symmetry is broken by physical effects
      or by hand, it is perhaps true that a 3-momentum cut-off has 
      the {\it least} impact on the medium parts of the regularized integrals,
      in particular at $T=0$. This will become more clear in \sect{njlthermo}.
      In this report we will therefore regularize the model using a 
      sharp 3-momentum cut-off $\Lambda$, unless stated otherwise.         

\item We already mentioned that the NJL model does not confine.
      Formally, this is reflected by the fact that the integral
      $I(q^2)$, and hence the polarization functions $\Pi_M(q^2)$,
      get an imaginary part above the $q\bar q$-threshold, i.e.,
      for $q^2 > 4 M^2$. As a consequence, mesons with a mass larger
      than $2M$ have a finite width, which indicates that they are 
      unstable against decay into a quark-antiquark pair.
      The pion is of course not affected by this problem.  
      However, as can be seen from \eqs{sigpipol} and (\ref{rpamass}),
      $m_\sigma = 2M$ if $m=0$ and it moves above the threshold 
      if $m>0$. If vector mesons are included, it depends on the parameters
      whether they have masses above or below the $q\bar q$ threshold,
      while axial vector mesons always decay into $q\bar q$ pairs in the
      model.
     
\item The formulae given above correspond to the Hartree approximation 
      and to RPA without Pauli exchange terms, respectively. 
      However, because of the local 4-point interaction, exchange
      diagrams can always be cast in the form of direct diagrams
      via a Fierz transformation (see App.~\ref{fierz}). This means, 
      the Hartree-Fock
      approximation is equivalent to the Hartree approximation  with
      appropriately redefined coupling constants. 
      In this sense, Hartree is as good as Hartree-Fock, as long as 
      the interaction terms in the Lagrangian are not fixed by
      some underlying theory.  

      Extensions of the approximation scheme beyond Hartree-Fock + RPA
      are much more difficult. This topic will briefly be discussed in
      \sect{nc}.
\end{itemize}

\subsection{Parameter fit}
\label{njlfit}

As a basis for the subsequent discussions we perform a first parameter
fit for the simple two-flavor model, \eq{LNJL}, within the 
Hartree + RPA scheme. As mentioned above, we will regularize the integrals
by a sharp 3-momentum cut-off. We thus have three parameters,
the bare quark mass $m$, the coupling constant $G$, and the cut-off $\Lambda$.
These parameters are usually fixed by fitting the pion mass, the pion decay 
constant, and the quark condensate. 
Whereas the pion mass, 
$m_\pi = 135.0$~MeV~\cite{PDG}\footnote{This corresponds to the mass of the 
$\pi^0$, which is not affected by ${\cal O}(\alpha)$ electromagnetic 
corrections~\cite{Dashen}. For the NJL model, this was explicitly
proven in Ref.~\cite{DLT92}. Of course, having the other uncertainties
in mind, fitting $m_\pi$ to the charged pion mass would not cause
any practical difference.}
and the pion decay constant, $f_\pi = 92.4 \pm 0.2 $~MeV~\cite{Holstein},
are known quite accurately, the uncertainties for
the quark condensate are rather large. Limits extracted from sum rules
are 190~MeV~$\lesssim -\ave{\bar uu}^{1/3} \lesssim$~ 260~MeV
at a renormalization scale of 1~GeV~\cite{dosch}, while 
lattice calculations yield 
$\ave{\bar uu}^{1/3} = -(231 \pm 4 \pm 8 \pm 6)$~MeV~\cite{giusti}.

\begin{figure}
\begin{center}
\epsfig{file=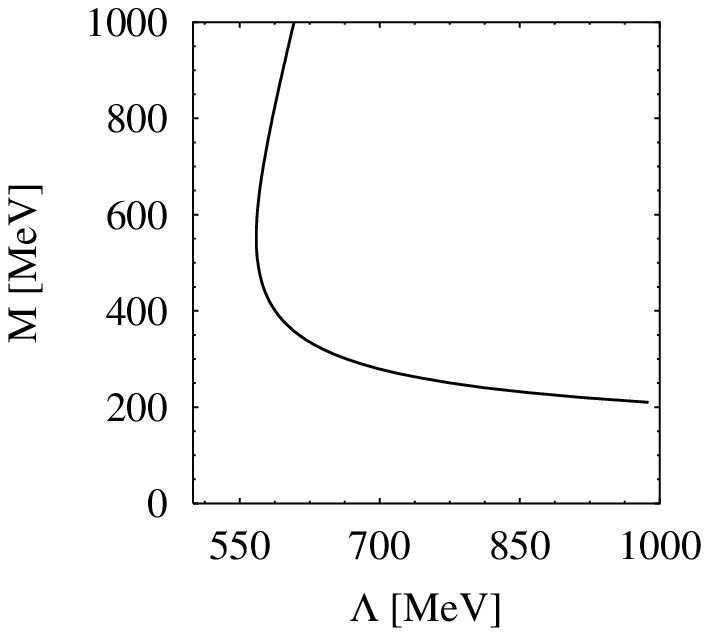,width=7.0cm}\hspace{0.5cm}
\epsfig{file=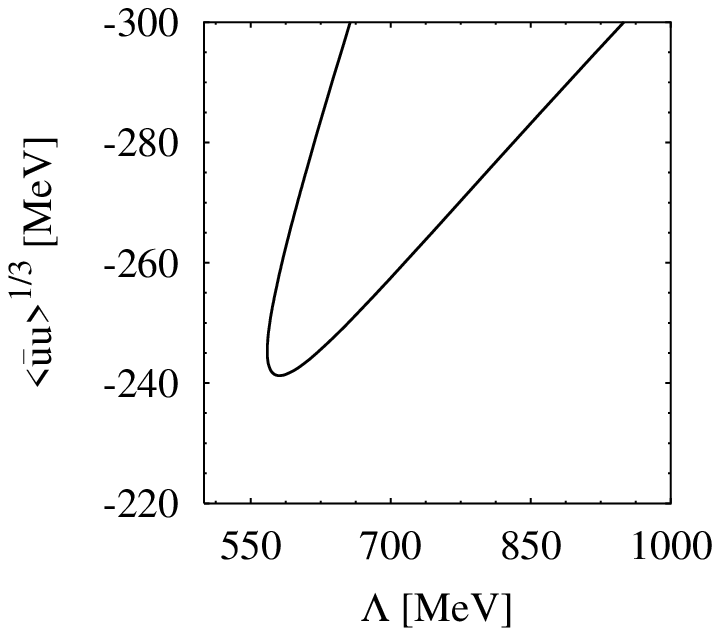,width=7.0cm}
\end{center}
\vspace{-0.5cm}
\caption{\small Constituent quark mass (left) and quark condensate (right)
                as functions of the 3-momentum cut-off for fixed
                $f_\pi = 92.4$~MeV and $m_\pi = 135$~MeV.}
\label{fignjl2fit}
\end{figure}

In this situation we first fix $G$ and $m$
for arbitrary values of $\Lambda$ by fitting $f_\pi$ and $m_\pi$ 
to their empirical values.
The corresponding solutions of the gap equation are displayed in 
\fig{fignjl2fit}. In the left panel the constituent mass $M$
is shown as a function of the cut-off. Obviously, $\Lambda$ must be 
larger than some critical value (about 568~MeV~$\hat =$~6.1~$f_\pi$)
to find a solution. Above this value, there is a ``low-mass'' branch
($M \lesssim 550$~MeV) and a ``high-mass'' branch of solutions. 
This kind of behavior is typical for the model and is also found
within other regularization schemes~\cite{ripka}.

On the r.h.s. of \fig{fignjl2fit} we show the corresponding values of
the quark condensate. We see that the model (together with the used
regularization scheme) cannot accommodate values of
$-\ave{\bar uu}^{1/3}$ smaller than 240~MeV.
Taking the upper limit of Ref.~\cite{dosch}, 
$-\ave{\bar uu}^{1/3} \lesssim$~ 260~MeV,
we find $\Lambda \lesssim$~ 720~MeV for the low-mass branch and 
$\Lambda \lesssim$~ 585~MeV for the high-mass branch.
This restricts the constituent mass to lie between about 270 and 800~MeV.
Taking the upper limit of $-\ave{\bar uu}^{1/3}$ to be 250~MeV, as suggested
by the results of Ref.~\cite{giusti}, would constrain $M$ to an interval
between about 300 and 640~MeV.

Four parameter sets, more or less representing this interval, are listed
in \tab{tabnjl2fit}.
In the last column we also list the corresponding ``bag constants'',
i.e., the energy gain per volume, due to the formation of the 
non-trivial vacuum state. 
Obviously, the uncertainty in $M$ (caused by the uncertainty in
the quark condensate and the peculiar behavior of having two solutions
for the same value of $\ave{\bar uu}$) leads to a big uncertainty in
the bag constant and thereby in the thermodynamic behavior of the 
model. This will be discussed in more detail in the next section.

\begin{table}[b!]
\begin{center}
\begin{tabular}{|c| c c c | c c c|}
\hline
&&&&&&\\[-3mm]
 set & $\Lambda$ [MeV] & $G\Lambda^2$ & $m$ [MeV] & $M$ [MeV]  
 & $\ave{\bar uu}^{1/3}$ [MeV] & $B$ [MeV/fm$^3$]
\\[1mm]
\hline
&&&&&&\\[-3mm]
1 & 664.3 & 2.06 & 5.0 & 300 & -250.8 & {\phantom 1}76.3 \\
2 & 587.9 & 2.44 & 5.6 & 400 & -240.8 & 141.4 \\
3 & 569.3 & 2.81 & 5.5 & 500 & -242.4 & 234.1 \\
4 & 568.6 & 3.17 & 5.1 & 600 & -247.5 & 356.1 \\
\hline
\end{tabular}
\end{center}
\caption{\small Model parameters (3-momentum cut-off $\Lambda$,
coupling constant $G$, and current quark mass $m$) and related
quantities (constituent quark mass $M$, quark condensate $\ave{\bar uu}$,
and bag constant $B$) for the two-flavor NJL model,
\eq{LNJL}, treated in Hartree + RPA approximation.
The parameters have been determined fitting the pion decay constant
and the pion mass to their empirical values, $f_\pi = 92.4$~MeV and
$m_\pi = 135.0$~MeV. The definition of the bag constant is given in
\sect{njlpot}, \eq{njlb}.}
\label{tabnjl2fit}
\end{table}  

\section{Non-zero densities and temperatures}
\label{njlthermo}

Soon after the reinterpretation of the NJL model as an effective quark model
it has been employed to study quark and meson properties
in hot or dense matter~\cite{BMZ87,HaKu87a}. 

Applying standard techniques of thermal field theory~\cite{Kapusta}
it is straight forward to evaluate the quark loop which enters the 
gap equation or the mesonic polarization diagrams at non-vanishing 
temperature or chemical potential.
The results have basically the same structure as the vacuum expressions,
but are modified by thermal occupation numbers.  
For instance, the gap equation \eq{njlgap2} becomes  
\beq
    M \;=\; m \,+\,4 N_f N_c\,G \int \dtp\,\frac{M}{E_{p}}\,
         \Big(1 - n_{p}(T,\mu)- {\bar n}_{p}(T,\mu)\Big)~, 
\label{njlgaptmu}
\eeq
where $E_{p} = \sqrt{{\vec p}^{\,2} + M^2}$ is the on-shell energy of 
the quark, self-consistently evaluated for the constituent mass $M$ 
which solves the equation\footnote{We will not use a special notation,
like, e.g., $M^*$, to indicate in-medium quantities, since most quantities 
in this report correspond to non-zero temperature or density. 
Instead, we will sometimes indicate vacuum quantities by
the suffix ``$vac$'', if necessary.}. 
$ n_{p}$ and ${\bar n}_{p}$ are Fermi occupation numbers of quarks 
and antiquarks, respectively,
\beq
    n_{p}(T,\mu)\;=\; \frac{1}{e^{(E_{p}-\mu)/T} + 1}~,\qquad
    {\bar n}_{p}(T,\mu) \;=\; \frac{1}{e^{(E_{p}+\mu)/T} + 1}~.
\label{fermiocc}
\eeq
They are related to the total quark number density in the standard way,
\beq
    n(T,\mu)
    = 2 N_f N_c \int \dtp \Big( n_{p}(T,\mu)
       - {\bar n}_{p}(T,\mu) \Big)~.
\label{densmu}
\eeq
For $T = \mu = 0$,  we have $n_{p} = \bar n_{p} = 0$ 
and \eq{njlgaptmu} becomes identical to \eq{njlgap2} if there
the integration over $p_0$ is turned out.

\begin{figure}
\begin{center}
\epsfig{file=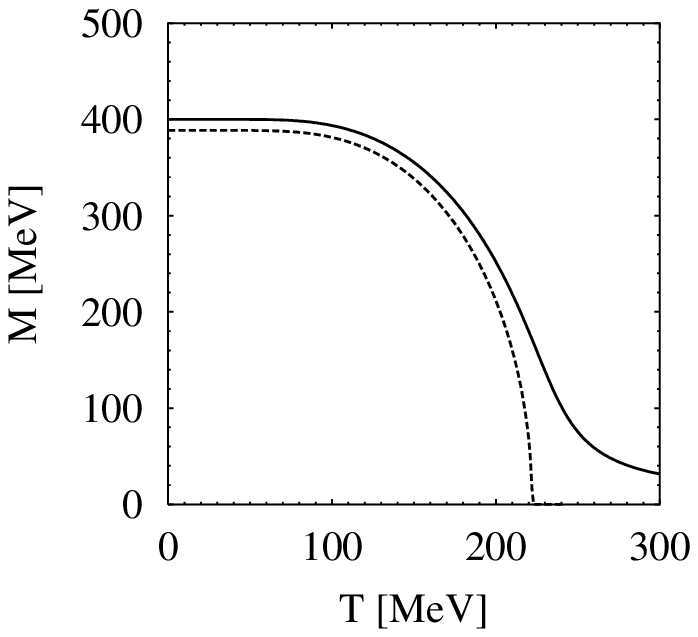,width=7.0cm}\hspace{0.5cm}
\epsfig{file=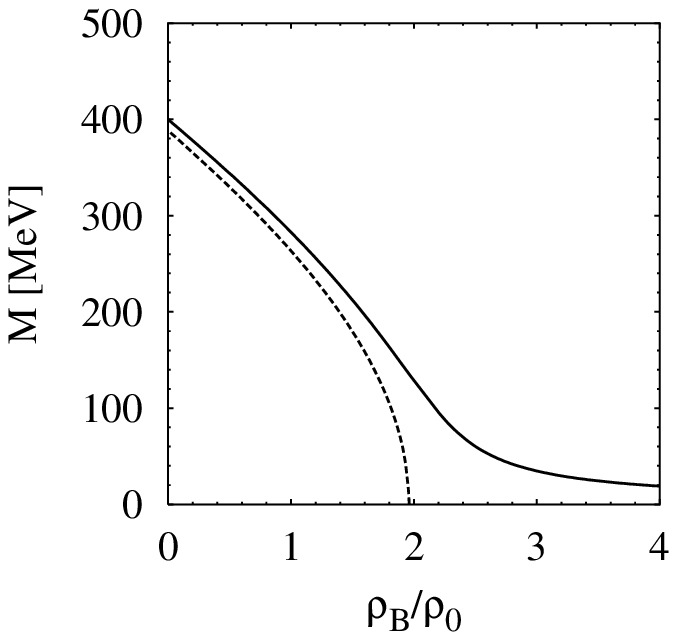,width=7.0cm}
\end{center}
\vspace{-0.5cm}
\caption{\small Constituent quark mass at zero density as a function of
         temperature (left) and at zero temperature as a function of 
         baryon number density in units of nuclear matter density 
         $\rho_0 = 0.17~{\rm fm}^{-3}$ (right). Solid lines: 
         parameters of set 2 of \tab{tabnjl2fit}, dashed lines:
         same parameters, but with $m=0$ (chiral limit).}
\label{figgaptmu}
\end{figure}

In medium, the occupation numbers are non-zero and reduce the value 
of the constituent mass. 
For large temperatures or densities the factor $(1 - n_{p}
- {\bar n_{p}})$
goes to zero, and $M$ approaches the value of the current mass $m$. 
This is illustrated in \fig{figgaptmu} for zero density and non-zero 
temperatures (left panel) and for zero temperature and non-zero densities 
(right panel). The solid lines indicate the (maximal) solutions of the 
gap equation for parameter set 2 of \tab{tabnjl2fit}. Taking the chiral
limit we arrive at the dashed lines. In this case \eq{njlgaptmu} does
no longer support non-trivial solutions for temperatures larger than
$T_c = 222$~MeV or densities larger than $\rho_c = 2.0 \rho_0$.
(For a detailed discussion of the critical line in the density-temperature 
plane, see, e.g., \cite{SKP99}.) 
Note that the critical temperature at zero density is quite large 
compared with the lattice value of about 170 MeV,
while the critical density at zero temperature seems to be rather low.
As we will discuss in \sect{njlt},
this is a typical feature of an NJL mean-field calculation, which is 
mainly due to the fact that the phase transition is driven by the wrong
degrees of freedom (unconfined quarks). 

In the beginning, the smooth behavior of the constituent quark mass as
a function of temperature or density lead several authors to believe
that the phase transition is second order~\cite{BMZ87}.
However, in order to decide on the order of the phase transition it is 
not sufficient to solve the gap equation. As we have seen in vacuum,
the gap equation does not have a unique solution
and one should look for the solution with the lowest energy.
Similarly, at non-vanishing temperature or chemical potential,
one should minimize the (grand canonical) thermodynamic potential.
In this way it was revealed by Asakawa and Yazaki that the phase transition
can indeed be first order, at least at low temperatures~\cite{AsYa89}. 
In the following we will basically adopt their method to calculate
the thermodynamic potential.

\subsection{Thermodynamic potential}
\label{njlpot}

We consider a two-flavor NJL-type Lagrangian of the form 
\beq
    {\cal L} \;=\; {\bar q}(i\delsl - m) q
    \;+\; G_S\,\Big[({\bar q} q)^2 + ({\bar q}i\gamma_5\vec\tau q)^2 \Big]
    - G_V\,({\bar q}\gamma^\mu q)^2 + \dots~.
\label{LNJLSV}
\eeq
To keep the discussion rather general we have not restricted ourselves
to the standard NJL interaction term in the scalar and pseudoscalar-isovector 
channels, but we have added explicitly a term in the vector-isoscalar
channel. It is known, e.g., from the Walecka model~\cite{Walecka}, 
that this channel
is quite important at non-zero densities. In principle we allow for 
further channels (indicated by the ellipsis), which, however, do not
contribute at mean-field level as long as we have only one common
quark chemical potential.

The thermodynamic potential per volume $V$ at temperature $T$ and quark 
chemical potential $\mu$ is defined as 
\beq
    \Omega(T,\mu) \;=\; -\frac{T}{V}\,\ln{\cal Z}
                  \;=\; -\frac{T}{V}\,\ln{{\mathbf{Tr}}\;
                  \exp{\Big( -\frac{1}{T}
                  \int d^3 x\,({\cal H} \,-\, \mu\;q^\dagger q)\Big)}}~, 
\label{Omegadef}
\eeq 
where ${\cal H}$ is the Hamiltonian density and $\mathbf{Tr}$ a
functional trace 
over all states of the system, i.e., spin, flavor, color and momentum. 
${\cal Z}$ is the grand canonical partition function.

To calculate $\Omega$ in mean-field (Hartree) approximation we
could in principle proceed in the way sketched in \sect{njlmesons},
i.e., we define a non-trivial ground state via a Bogoliubov rotation 
and then evaluate the free energy in this state.
This method would have the advantage that it contains the full
information about the structure of the ground state.
On the other hand, the derivations become quite involved
and extensions to include several condensates at the same time
are very difficult. 
Therefore we follow Ref.~\cite{AsYa89} where an equivalent but much
simpler method has been applied. To that end we 
consider two non-vanishing ``condensates'',
\beq
     \phi \;=\; \ave{\bar qq} \quad \text{and} \quad 
        n \;=\; \ave{q^\dagger q} \;\equiv\; \ave{\bar q \gamma^0 q}~,
\eeq 
i.e., the quark condensate and the total quark number density.
As long as we assume that --
apart from chiral symmetry and Lorentz invariance (which is explicitly 
broken by the chemical potential) --
all symmetries of the Lagrangian remain intact, these are the only allowed 
expectation values which are bilinear in the quark fields.
(Later we will encounter many other condensates, due to both, explicit and 
spontaneous symmetry breaking.)

Next, we linearize the interaction terms of ${\cal L}$ in the presence of
$\phi$ and $n$\footnote{
A more formal, but essentially equivalent, method is to bosonize the model. 
In that context, $\phi$ and $n$ emerge as auxiliary Bose fields 
which are introduced in the framework of a Hubbard-Stratonovich 
transformation.},
\beq
   ({\bar q} q)^2 \;\simeq\; 2\phi\;{\bar q} q \,-\, \phi^2~, \qquad
   ({\bar q} \gamma^\mu q)^2 \;\simeq\; 2n\;q^\dagger q \,-\, n^2~,
\eeq
where terms quadratic in the fluctuations, like $(\bar q q - \phi)^2$,
have been neglected. In particular terms in channels without condensate, 
like $({\bar q}i\gamma_5\vec\tau q)^2$ or the space components 
in the vector vertex, drop out.
In this approximation,
\begin{alignat}{1}
    {\cal L} \,+\, \mu\;q^\dagger q \;=&\; {\bar q}(i\delsl - m + 2G_S\,\phi)q
     \,+\, (\mu - 2G_V\,n)\;q^\dagger q \,-\, G_S\,\phi^2 + G_V\,n^2
\nonumber\\
 \;=&\; {\bar q}(i\delsl - M)q
     \,+\, \tilde\mu\;q^\dagger q \,-\, 
    \frac{(M-m)^2}{4 G_S} - \frac{(\mu-\tilde\mu)^2}{4 G_V}~,
\label{LNJLlin}
\end{alignat}
where we have introduced the constituent mass $M$ and the renormalized chemical
potential $\tilde\mu$,
\beq
    M \;=\; m \;-\; 2G_s\,\phi~,\qquad \tilde\mu \;=\; \mu \;-\; 2G_V\,n~.
\label{Mmuren}
\eeq
This means, apart from constant (i.e., field independent) terms,
which give trivial contributions to the r.h.s. of \eq{Omegadef},
the problem is equivalent to a system of non-interacting particles 
with mass $M$ at chemical potential $\tilde\mu$. 
Hence, the mean-field thermodynamic potential takes the form
\beq
    \Omega(T,\mu;M,\tilde\mu) \;=\; \Omega_M(T,\tilde\mu) \;+\; 
    \frac{(M-m)^2}{4 G_S} - \frac{(\mu-\tilde\mu)^2}{4 G_V} \;+\; const.~,
\label{OmegaMF}
\eeq
with the free Fermi-gas contribution 
\beq
    \Omega_M(T,\tilde\mu)
    \;=\; -T \sum_n \int \frac{d^3p}{(2\pi)^3} \;
    {\rm Tr}\; \ln \Big(\frac{1}{T}\,S^{-1}(i\omega_n, \vec p)
    \Big)~.
\label{Omegatr}
\eeq
Here $S^{-1}(p) = \psl - \tilde\mu\gamma^0 - M$ is the inverse fermion
propagator at chemical potential $\tilde\mu$ which has to be evaluated
at fermionic Matsubara frequencies, $p^0 = i\omega_n = (2n+1)\pi T$.

The further evaluation of $\Omega_M$ can be found in textbooks~\cite{Kapusta},
but for later comparison we summarize the main steps:
The trace is to be taken in color, flavor, and Dirac space. Using
\beq
     {\rm Tr}\; \ln (\Qsl - M) \;=\; \ln \; {\rm Det}(\Qsl - M) \;=\;
     2 N_f N_c \ln (Q^2 - M^2)~,
\eeq
this is readily done. Then, after some reordering, one can apply 
the relation~\cite{Kapusta}
\beq
     T \sum_n \ln\Big(\frac{1}{T^2}(\omega_n^2 + \lambda_k^2)\Big) \;=\;
     \lambda_k \;+\; 2  T \ln(1+e^{-\lambda_k/T})
\label{Matsu}
\end{equation}
to turn out the Matsubara sum. One finally gets
\begin{alignat}{1}
    \Omega_M(T,\tilde\mu) = - 2 N_f N_c \int \dtp\,\Big\{
    E_{p} &+ T\,\ln{\Big(1 + \exp{(-\frac{E_{p}-\tilde\mu}{T})}\Big)}
\nonumber \\
        &+ T\,\ln{\Big(1 + \exp{(-\frac{E_{p}+\tilde\mu}{T})}\Big)}\Big\}~.
\label{OmegaM}
\end{alignat}
Note that $\Omega$ is physically meaningful only up to a constant, as
indicated in \eq{OmegaMF}.

Until this point, the result for $\Omega$ depends on $M$ and $\tilde\mu$,
i.e., on our choice of $\phi$ and $n$. On the other hand,
in a thermodynamically consistent treatment, $\phi$ and $n$ should 
follow from $\Omega$ as
\beq
    \phi \;=\; \frac{\partial\Omega}{\partial m} \quad\text{and}\quad 
    n \;=\; -\frac{\partial\Omega}{\partial\mu}~. 
\label{thcons}
\eeq
Writing $M = M(m,T,\mu)$ and $\tilde\mu = \tilde\mu(m,T,\mu)$ and applying the chain rule
we get from \eq{OmegaMF} 
\beq 
    \frac{\partial\Omega}{\partial m} \;=\; \phi \,+\, 
    \frac{\delta\Omega}{\delta M} \frac{\partial M}{\partial m} \,+\, 
    \frac{\delta\Omega}{\delta \tilde\mu} 
    \frac{\partial \tilde\mu}{\partial m}~,\qquad\qquad
    \frac{\partial\Omega}{\partial \mu} \;=\; -n \,+\, 
    \frac{\delta\Omega}{\delta M} \frac{\partial M}{\partial \mu} \,+\, 
    \frac{\delta\Omega}{\delta \tilde\mu} 
    \frac{\partial \tilde\mu}{\partial \mu}~,
\eeq
where we have used \eq{Mmuren} to replace the {\it explicit} derivatives
by $\phi$ and $-n$, respectively. 
Thus, to be consistent with \eq{thcons} the {\it implicit} contributions
have to vanish. This is obviously fulfilled if
\beq
    \frac{\delta\Omega}{\delta M} \;=\;
    \frac{\delta\Omega}{\delta\tilde\mu} \;=\; 0~,
\label{station}
\eeq
i.e., the stationary points of $\Omega$ with respect to $M$ and $\tilde\mu$
are automatically thermodynamically consistent.
Explicitly, one gets
\beq
    \frac{\delta\Omega}{\delta M} \;=\;\frac{M-m}{2 G_S}  \,-\,
    2 N_f N_c \int \dtp\,\frac{M}{E_{p}}\,
         \Big(1 - n_{p}(T,\tilde\mu)- {\bar n}_{p}(T,\tilde\mu)\Big)
    \;=\; 0
\label{njlgapM}
\eeq
and
\beq
    \frac{\delta\Omega}{\delta\tilde\mu} \;=\;\frac{\mu-\tilde\mu}{2 G_V}  
    \,-\,
    2 N_f N_c \int \dtp\,
         \Big(n_{p}(T,\tilde\mu)- {\bar n}_{p}(T,\tilde\mu)\Big)
    \;=\; 0~,\hspace{10mm}
\label{njlgapmu}
\eeq
This is a coupled set of self-consistency equations for 
$M$ and $\tilde\mu$, which generalizes the gap equation 
(\ref{njlgaptmu}) to $G_V \neq 0$.
In fact, for $G_V \rightarrow 0$, \eq{njlgapmu} yields $\tilde\mu = \mu$
and \eq{njlgapM} goes over into  \eq{njlgaptmu}.
In general, $\tilde\mu \neq \mu$ and \eqs{njlgapM} and (\ref{njlgapmu}) have 
to be solved simultaneously.
If there is more than one solution, the stable one is the solution which
corresponds to the lowest value of $\Omega$. 

In the following, we will restrict ourselves to repulsive vector interactions,
$G_V \geq 0$. In this case $\tilde\mu$ is uniquely determined by 
\eq{njlgapmu} for given values of $T$, $\mu$, and $M$. 
To see this we rewrite this equation in the form 
\beq
    \mu \;=\; \tilde\mu \,+\,
    4 N_f N_c\,G_V \int \dtp\,
         \Big(n_{p}(T,\tilde\mu)- {\bar n}_{p}(T,\tilde\mu)\Big)~,
\eeq
which formally defines $\mu$ as a function of $\tilde\mu$. 
It is easy to verify that this function is strictly rising.
Therefore it can be inverted and $\tilde\mu$ is in turn a strictly
rising function of $\mu$. In particular we find that $\tilde\mu(\mu=0)=0$,
which also implies $n = 0$ (see \eq{Mmuren}).

Another important observation is that $M$, if written as a function of 
temperature and {\it density}, does not depend on the vector coupling $G_V$.
This follows from the fact that, according to \eqs{njlgapmu} and
(\ref{Mmuren}), the density is given by
\beq
    n(T,\tilde\mu)
    = 2 N_f N_c \int \dtp \Big( n_{p}(T,\tilde\mu)
       - {\bar n}_{p}(T,\tilde\mu) \Big)~,
\label{densmutilde}
\eeq
i.e., just like in a free quark gas, \eq{densmu}, but with $\mu$ replaced
by $\tilde\mu$. This equation can be inverted to calculate 
$\tilde\mu$ for given $T$ and $n$, except for $T=0$ and $n=0$ which can be
satisfied by any value of $\tilde\mu$ with $|\tilde\mu| \leq |M|$.
However, in that case the occupation functions $n_{p}$ and
${\bar n}_{p}$ are identically zero, which means that in any case
$n_{p}$ and ${\bar n}_{p}$, and thus all ingredients of the gap
equation~(\ref{njlgapM}) are uniquely determined if the density is known.  

Having found a pair of solutions $M$ and $\tilde\mu$, other thermodynamic
quantities can be obtained in the standard way.
Since the system is uniform, pressure and energy density are given by
\beq
    p(T,\mu) \;=\; -\Omega(T,\mu;M,\tilde\mu)~,\qquad
    \epsilon(T,\mu) \;=\; -p(T,\mu) \,+\,T\,s(T,\mu)
    \,+\,\mu\,n(T,\mu)~.
\eeq
The density $n$ is given by \eq{thcons}
while the entropy density is $s = -\partial\Omega/\partial T$.
As customary, we choose the irrelevant constant in 
\eq{OmegaMF} such that $p$ and $\epsilon$ vanish in vacuum, i.e.,
we choose  
\beq
    \Omega(0,0;M_{vac},0) \;=\; 0~.
\label{Omega0}
\eeq
Here $M_{vac}$ corresponds to the {\it stable} solution for $M$ 
at $T=\mu=0$.
We may also define the ``bag constant'',
\beq
     B \;=\; \Omega(0,0;m,0) - \Omega(0,0;M_{vac},0) \;=\; \Omega(0,0;m,0)~,
\label{njlb}
\eeq
where the second equality follows from our particular choice, \eq{Omega0},
whereas the first equality is the more general expression. 
Like in the bag model, $B$ describes the pressure difference between the
trivial and the non-trivial vacuum, but it is not an input parameter of
the model, but a dynamical consequence of the interaction, leading to
vacuum masses $M_{vac}\neq m$.
Note that, except in the chiral limit, $M=m$ is not a solution of the gap 
equations, i.e., not even an unstable one, but corresponds to the
perturbative vacuum.

In \tab{tabnjl2fit} we have listed the values of $B$ for the various
parameter sets.
Obviously, $B$ is extremely sensitive to the parameters,
ranging from 76.3~MeV/fm$^3$ for $M_{vac} = 300$~MeV to
356.1~MeV/fm$^3$ for $M_{vac} = 600$~MeV, i.e.,
more or less covering the same region as the bag-model fits listed in
\tab{tabmitfit}. 
It can be shown that for $\Lambda^2 \gg M_{vac}^2$ the following
relation holds in the chiral limit~\cite{Blotz},
\beq
    B \;=\; \frac{1}{2}\,M_{vac}^2\,f_\pi^2 \,+\, 
    \frac{N_f N_c}{32\pi^2}\,M_{vac}^4 
    \,+\, {\cal O}(M_{vac}^2/\Lambda^2)~.
\label{Bapp}
\eeq
In practice this formula works rather well, even for $M_{vac}$
of the same order as the cut-off. Thus the strong parameter dependence
of $B$ can mainly be attributed to the $M_{vac}^4$-term in \eq{Bapp}.

\subsection{Chiral phase transition and stable
            quark matter solutions at zero temperature}
\label{njlt0}

In the limit $T\rightarrow 0$, the thermal factors in \eq{OmegaM} go 
over into step functions and the mean-field thermodynamic potential
\eq{OmegaMF} becomes
\beq
\Omega(0,\mu;M,\tilde\mu) =\;  - 2 N_f N_c \int \dtp\,\Big\{
    E_{p} \,+\,(\tilde\mu-E_{p})\,\theta(\tilde\mu-E_{p})\Big\}
\;+\;\frac{(M-m)^2}{4 G_S} \;-\; \frac{(\mu-\tilde\mu)^2}{4 G_V} 
\;+\; \mi{const}.~,
\label{OmegaMF0}
\eeq
where we have assumed $\mu\geq 0$ and thus $\tilde\mu\geq 0$.
When we also take $\mu\rightarrow 0$ and use the fact that
in this case $\tilde\mu$ has to vanish as well,
we obtain the vacuum thermodynamic potential
\beq
    \Omega_{vac}(M) \;:=\;
    \Omega(0,0;M,0) \;=\;  - 2 N_f N_c \int \dtp\,E_{p}
    \;+\; \frac{(M-m)^2}{4 G_S} \;+\; \mi{const}.~.
\label{Omegavac}
\eeq
Here we can nicely see, that the spontaneous symmetry breaking in vacuum
comes about through the interplay between the negative contribution
from the Dirac sea (first term on the r.h.s.), which
favors large values of $M^2$, and the positive field energy of the
condensate (second term) which favors
values of $M$ close to the current mass $m$.
We should keep in mind that the integral, which would be strongly
divergent otherwise, is regularized by a cut-off. 
One can easily check that it rises quadratically with $M$ for small values
of $M$ and logarithmically if $M$ is large. Thus for large $M$, the
positive $(M-m)^2$ term always wins, whereas for small $M$ the over-all
behavior depends on the size of the coupling constant. 

\begin{figure}
\begin{center}
\epsfig{file=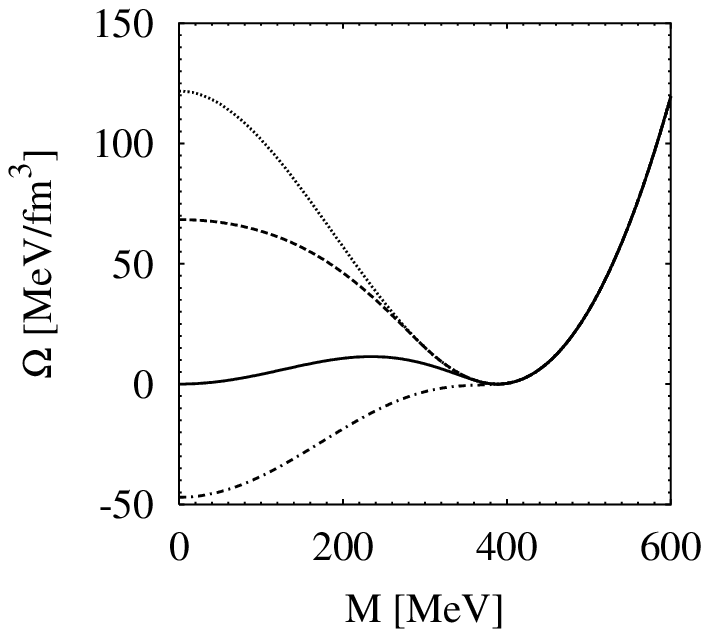,width=7.0cm}\hspace{0.5cm}
\epsfig{file=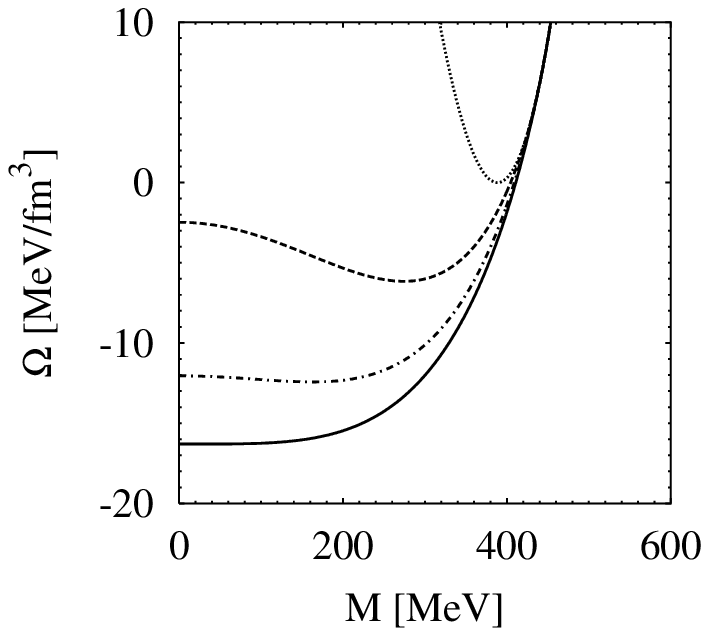,width=7.0cm}
\end{center}
\vspace{-0.5cm}
\caption{\small Mean-field thermodynamic potential as a function of the 
         auxiliary variable $M$ (``constituent quark mass'') for parameter
         set 2 of \tab{tabnjl2fit}, but with $m=0$ (chiral limit).
         For each value of $M$, the gap equation (\ref{njlgapmu}) has been
         solved to eliminate the auxiliary variable $\tilde\mu$.
         All functions are symmetric in $M$, but only the positive part
         is shown.
         Left: $G_V=0$ and chemical potentials
         $\mu=0$ (dotted), $\mu = 300$~MeV (dashed),  
         $\mu = \mu_c = 368.6$~MeV (solid), and $\mu = 400$~MeV (dash-dotted).
         Right: $G_V=G_S$ and chemical potentials  
         $\mu=0$ (dotted), $\mu = 430$~MeV (dashed), $\mu = 440$~MeV 
         (dash-dotted), and $\mu = \mu_c = 444.3$~MeV (solid).
         The vacuum result (dotted) is identical to that for $G_V=0$.}
\label{fignjlomega}
\end{figure}

An example for the vacuum thermodynamic potential as a function of $M$
is shown in the left panel of \fig{fignjlomega} (dotted line). We have 
used parameter set 2 of \tab{tabnjl2fit}, but in the chiral limit ($m=0$).
In this case one obtains a vacuum mass $M_{vac} = 388.5$~MeV, which
corresponds to a minimum $\Omega_{vac}$, while the trivial solution
$M=0$ is a maximum. 

For $\mu>0$ (but still $T=0$), $\Omega$ gets modified by the term
\beq
    \delta\Omega_{med}(0,\mu;M,\tilde\mu) \;=\;  - 2 N_f N_c \int \dtp\,
    (\tilde\mu-E_{p})\,\theta(\tilde\mu-E_{p})
    \;-\; \frac{(\mu-\tilde\mu)^2}{4 G_V}~.
\label{Omegamed}
\eeq
Unlike the vacuum part, this term is finite, even without regularization
because the integral is cut off by the step function at the Fermi momentum
$p_F = \theta(\tilde\mu-M)\,\sqrt{\tilde\mu^2-M^2}$.
Thus, as long as $p_F<\Lambda$, $\delta\Omega_{med}$ is not affected by
the cut-off. Taking a typical value, $\Lambda=600$~MeV, this corresponds
to a baryon number density of about $11 \rho_0$.
This was what we had in mind, when we said that a sharp 3-momentum cut-off
is probably the least severe regularization of medium integrals. 
(Note, however, that the cut-off does have an impact on the medium 
contributions at finite $T$ or in color superconducting phases,
when the Fermi surface is smeared out.)

For $\tilde\mu\leq M$, $p_F=0$ and the integral vanishes.
In this case the second term in \eq{Omegamed} yields the stationary solution 
$\tilde\mu=\mu$, i.e., it vanishes, too\footnote{Note that for repulsive
vector interactions, $G_V>0$, the stationary solution corresponds to a
{\it maximum} of $\Omega$ with respect to $\tilde\mu$. This phenomenon
is well-known, e.g., from the Walecka model~\cite{Walecka}.
It means that the condition $\delta\Omega/\delta\tilde\mu = 0$ must 
not be viewed as a variational principle, but as a constraint: Values of 
$\tilde\mu$ which do not fulfill this condition are not thermodynamically
consistent and should be discarded.}. Since $\tilde\mu$ is a strictly
rising function of $\mu$ and vice versa, we conclude that   
$\tilde\mu=\mu$ for all $\mu\leq M$ and $\delta\Omega_{med}$ vanishes
in this regime. 
From a physical point of view, this makes sense: At $T=0$ the
chemical potential corresponds to the Fermi energy of the system. 
As long as this is smaller than the constituent quark mass, no quark state
can be populated, i.e., the density remains zero. Since 
$n = -\partial\Omega/\partial\mu$, this implies that $\Omega$ remains
unchanged\footnote{Although one might think that this argument is only 
valid for thermodynamic consistent points, it applies to $\Omega$ at any 
fixed $M\geq\mu$ because $\delta\Omega_{med}$ does not ``know'' whether 
or not $M$ corresponds to a stationary point of the total thermodynamic 
potential.}. Moreover, according to \eq{Mmuren}, $\tilde\mu = \mu$
for $n=0$.

\begin{figure}
\begin{center}
\epsfig{file=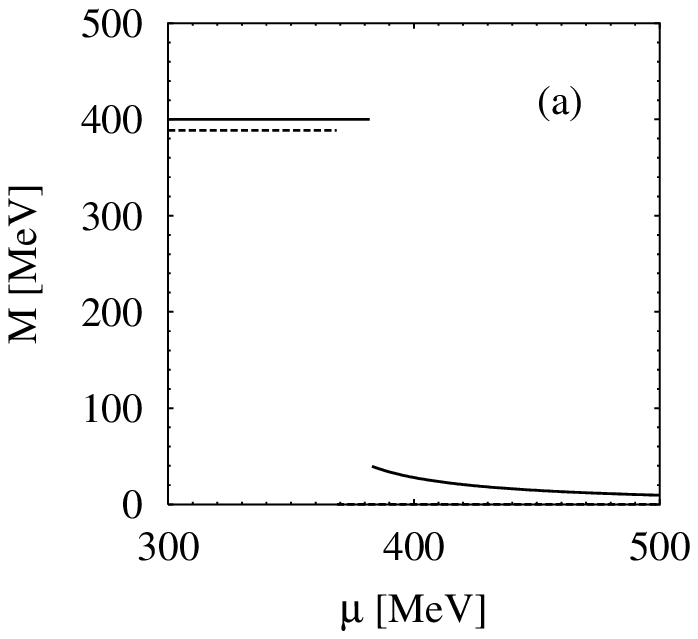,width=7.0cm}\hspace{0.5cm}
\epsfig{file=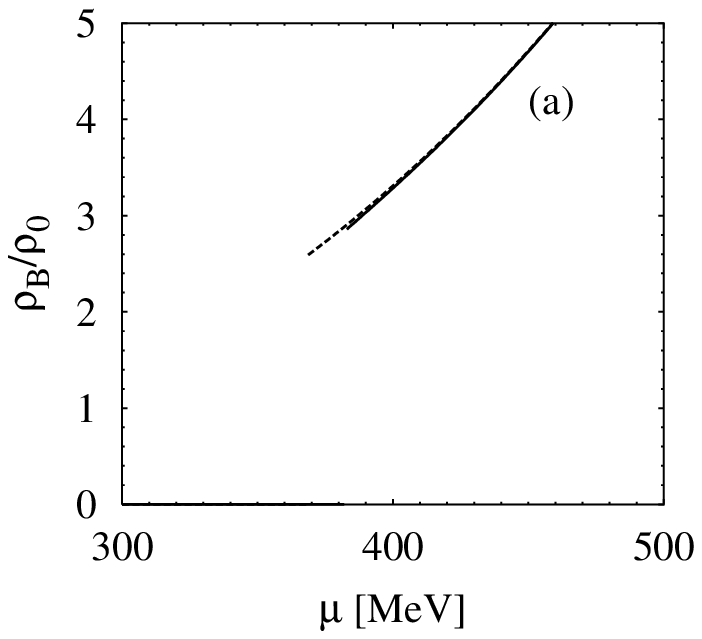,width=7.0cm}
\epsfig{file=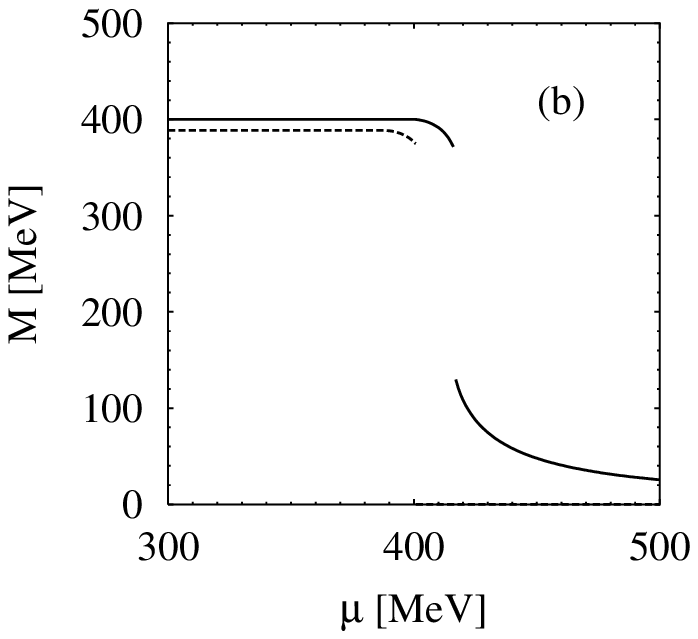,width=7.0cm}\hspace{0.5cm}
\epsfig{file=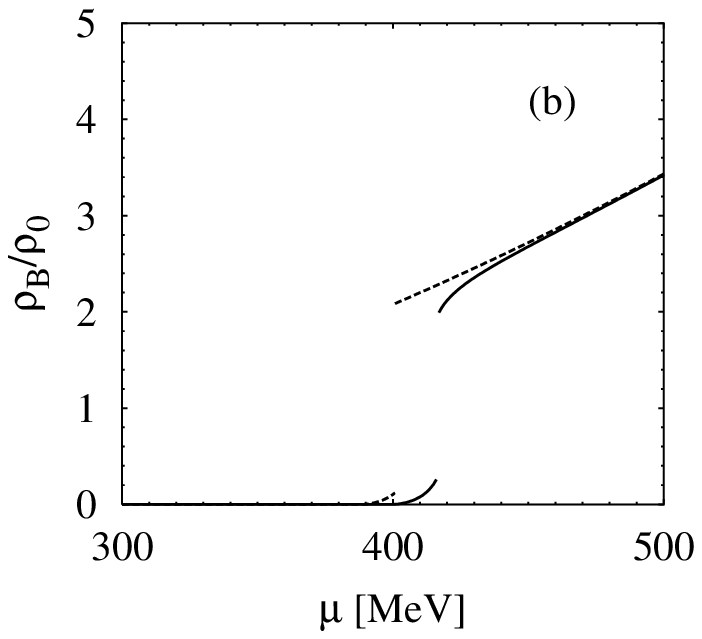,width=7.0cm}
\epsfig{file=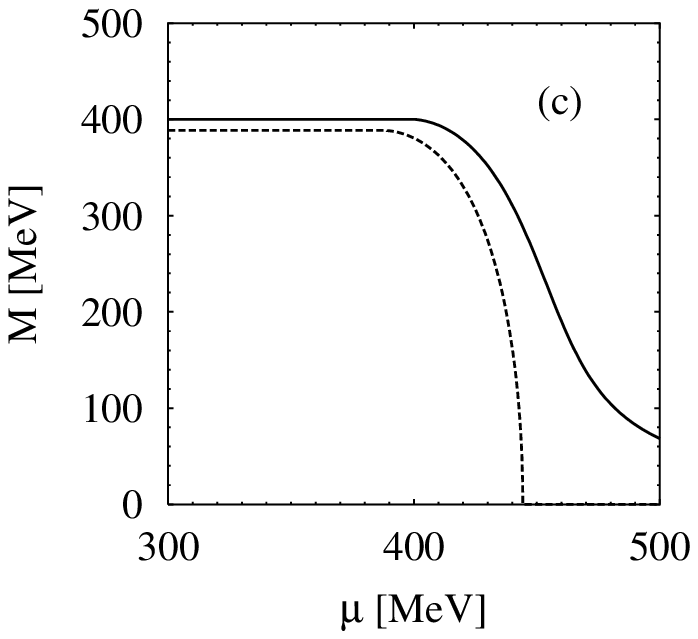,width=7.0cm}\hspace{0.5cm}
\epsfig{file=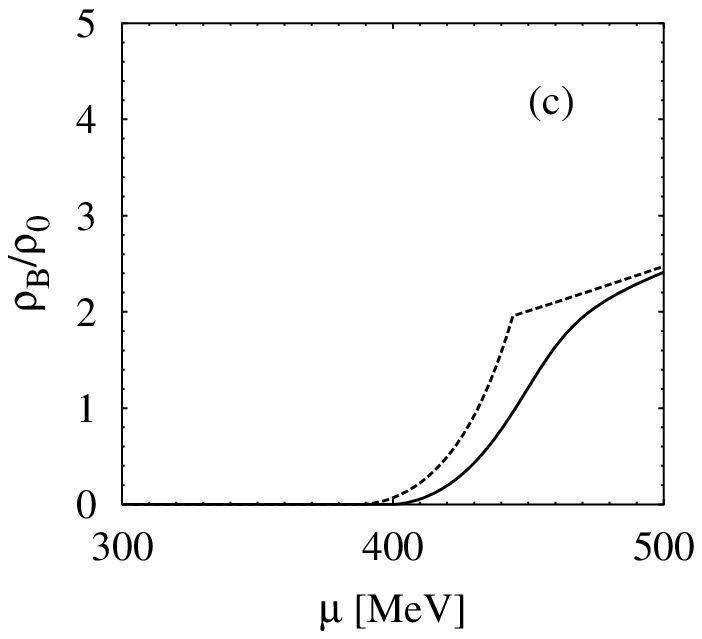,width=7.0cm}
\end{center}
\vspace{-0.5cm}
\caption{\small Masses (left) and baryon number densities (right) as 
functions of the chemical potential $\mu$, illustrating the three types
of phase transitions in the NJL model at $T=0$. The solid lines correspond
to parameter set 2  of \tab{tabnjl2fit}, the dashed lines to the chiral
limit. The vector coupling is $G_V=0$ (a),  $G_V=0.5~G_S$ (b), and
$G_V=G_S$ (c).}
\label{figphasetypes}
\end{figure}

For $\mu>M$, $\delta\Omega_{med}$ does not vanish and leads to a reduction
of $\Omega$, favoring small values of $M$. 
In the chiral limit this eventually leads
to a restoration of chiral symmetry at some critical chemical potential
$\mu_c$. Above this value, the absolute minimum of the thermodynamic
potential corresponds to $M=0$. 
It turns out that there are three different ways how the 
restored phase can be reached in the model~\cite{Bu96,BuOe98}.
These scenarios are illustrated in \fig{figphasetypes} where the constituent
quark masses (left panels) and the densities (right panel) are displayed
as functions of $\mu$. 
The plots are based on calculations with parameter set 2 of \tab{tabnjl2fit}
and different values of the vector coupling constant $G_V$:
$G_V = 0$ in the upper line (``case (a)''),
$G_V = 0.5~G_S$ in the second line (``case (b)''), and
$G_V = G_S$ in the lower line (``case (c)'').
The dashed lines
correspond to the chiral limit, while for the solid lines we used 
the current mass $m=5.6$~MeV, as given in the table. 
In that case, chiral symmetry gets of course never restored exactly, 
but the main points discussed below remain the same.  

\begin{itemize}

\item[(a)] First-order phase transition at $\mu_c < M_{vac}$: 

If the reduction of the thermodynamic potential at low masses grows 
fast enough with $\mu$, it may happen that the phase transition takes
place at a critical potential $\mu_c$ which is smaller than the vacuum mass
$M_{vac}$ (see left panel of \fig{fignjlomega} for illustration).
Since $\delta\Omega_{med} = 0$ for all $M\geq\mu$, this means
that in the vicinity of the vacuum minimum, $M=M_{vac}$, the thermodynamic
potential has still its vacuum form. In particular, the solution at
$M=M_{vac}$ itself still exists and still corresponds to zero density.
Thus, if this case is realized, there is a strong first-order phase 
transition from the vacuum solution $M=M_{vac}$ into the chirally 
restored phase with $M=0$ (or $M=$~``small'' if we are not in the 
chiral limit). At the same time the density jumps from
zero to a relatively large value.
Apart from the vacuum, there is no stable solution with broken
chiral symmetry and no other stable solution with a smaller density
than the critical one.

\item[(b)] First-order phase transition at $\mu_c > M_{vac}$: 

This case is similar to case (a), but with a slower reduction of the 
thermodynamic potential at low masses, such that the phase transition
takes only place at a critical potential $\mu_c > M_{vac}$.
This means, there is an interval $M_{vac} < \mu < \mu_c$, where the
system is still in the chirally broken phase, but $\delta\Omega_{med}$ is
already non-zero at $M=M_{vac}$ and shifts the minimum to lower 
values and its location to lower masses. Thus, in this interval,
the constituent mass goes smoothly down and the density smoothly rises
with $\mu$. Eventually, at $\mu=\mu_c$ the phase transition takes place
and constituent mass and density show a similar discontinuous behavior 
as in case (a).   

\item[(c)] Second-order phase transition ($\mu_c > M_{vac}$): 

Unlike in (a) and (b) it is also possible that $\delta\Omega_{med}$
does not produce an extra minimum at $M=0$ (in the chiral limit), 
and the ``old'' minimum moves all the way down to zero when $\mu$ is 
increased to sufficiently high values (see right panel of \fig{fignjlomega}).
As pointed out earlier, in the chiral limit $\Omega$ is symmetric in $M$.
Thus below $\mu_c$ there are two degenerate minima with opposite sign.
At $\mu=\mu_c$ they merge and turn the maximum at $M=0$ into a minimum.

Like in case (b) the constituent mass drops smoothly, beginning at
$\mu = M_{vac}$, but this time there is no discontinuity at any higher
value of $\mu$. In the chiral limit, the second-order phase transition
manifests itself in a discontinuous derivative of the mass and the
density as a function of $\mu$. For $m\neq 0$ there is only a cross-over,
and all variables vary smoothly.

\end{itemize}

In principle, one could imagine further scenarios.
For instance, there could be a discontinuous jump not directly into 
the restored phase but to a solution with a finite constituent quark mass,
which eventually goes to zero at higher chemical potential. Such a behavior 
would imply that the thermodynamic potential develops another minimum at 
finite $M$ which is different from the vacuum one. Inspecting the structure
of $\Omega_{vac}$ and $\delta\Omega_{med}$, this seems to be difficult to 
realize although we have not proven it rigorously. In any case, we have 
not found such solutions. This can be different, however, if we go beyond
mean-field approximation. In this context it is interesting that a behavior
of the above type has recently been found within a renormalization group
analysis of the quark-meson coupling model~\cite{SchW04}.  

It should be reminded that, although the functions $M(\mu)$ and $\rho_B(\mu)$
depend on the vector coupling $G_V$ and are therefore different for
the three examples shown in \fig{figphasetypes}, the function $M(\rho_B)$ is 
$G_V$-independent, as we have seen earlier.
Thus if we plot the masses given in \fig{figphasetypes} for the cases
(a), (b), and (c) against the respective densities they all fall on the
same lines (one for the chiral limit and one for $m=5.6$~MeV)
which agree with the functions plotted in the right panel of \fig{figgaptmu}.
However, $G_V$ does influence the {\it stability} of the solutions. 
Whereas in case (c) all points shown in \fig{figgaptmu} are stable solutions,
this is not the case for (a) and (b). 

As demonstrated by the numerical examples, all three cases, (a), (b), and (c),
can be realized within the NJL model, depending on the choice of the 
parameters. However, from a physical point of view it is clear that the cases 
(b) and (c) are unrealistic, because both of them predict the existence
of a low-density phase of homogeneously distributed constituent quarks.
This reflects the missing confinement of the model and
obviously any prediction based on these solutions, like medium 
modifications of RPA mesons, should be taken with great care.   

Case (a) is special, because it does {\it not} predict a stable quark phase
at low density. The first-order phase transition directly from the vacuum
phase to quark matter with baryon number density $\rho^*$ implies that any 
homogeneous quark distribution of density $0 < \rho_B < \rho^*$ is unstable 
against separation into a mixed phase consisting of quark matter with density 
$\rho^*$ and vacuum.
Hence, instead of a homogeneous quark gas at low densities, in this case
the model predicts the existence of quark ``droplets'' which are 
self-bound in vacuum~\cite{Bu96}.
 
\begin{figure}
\begin{center}
\epsfig{file=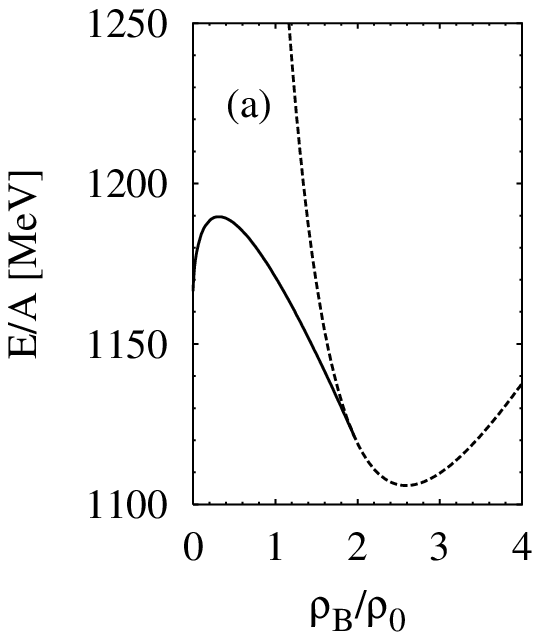,width=4.9cm}
\epsfig{file=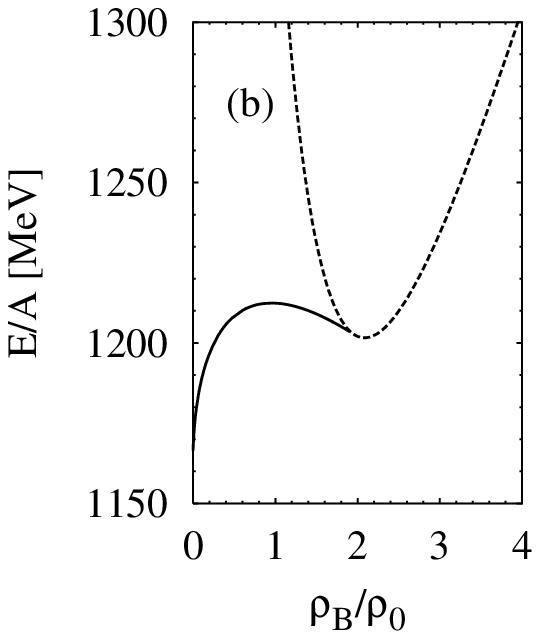,width=4.9cm}
\epsfig{file=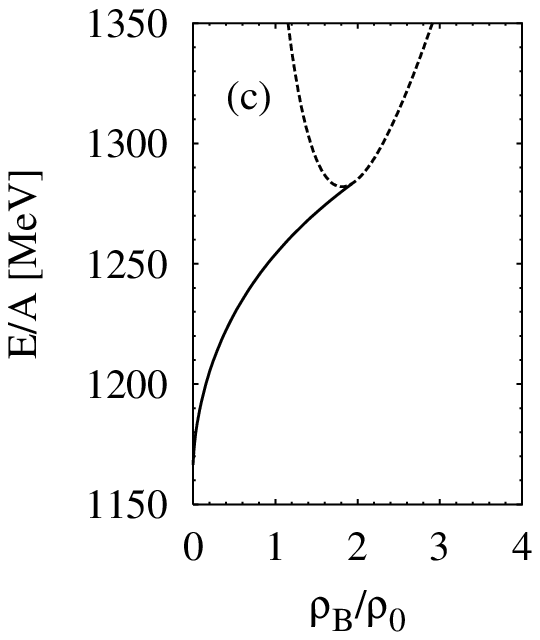,width=4.9cm}
\end{center}
\vspace{-0.5cm}
\caption{\small Energy per baryon number for the three cases (a), (b), and (c)
as functions of the baryon number density $\rho_B$.
The parameters are the same as in \fig{figphasetypes} with $m=0$. 
The solid lines correspond to the massive solutions of the gap equation,
the dashed lines to the massless ones.
Similar figures have been shown in Ref.~\cite{Bu96} for slightly
different parameters.} 
\label{fignjlea}
\end{figure}

To work this out more clearly, let us discuss the behavior of the
energy per baryon, 
\beq
    \frac{E}{A} \;=\; \frac{\epsilon}{\rho_B} \;=\;
    - \frac{p}{\rho_B} \;+\; 3\mu
\eeq
as a function of density.
In \fig{fignjlea} this is displayed for the three previous examples in
the chiral limit. The dashed lines correspond to the massless solutions
of the gap equation, the solid lines to the massive ones. As we have seen
in \fig{figgaptmu}, the latter only exist for densities 
$\rho_B < 2.0\rho_0$ for these parameters.
Note that at fixed density the massive solutions, whenever they exist, are
energetically favored.

The basic features of the curves can be understood by inspecting 
the points of zero pressure:

\begin{itemize}

\item The branch of the massive solutions starts with the non-trivial
      vacuum point which by definition has zero pressure. Approaching
      this point from above, $\rho_B\rightarrow 0^+$,
      we have $\mu \rightarrow M_{vac}$ while the pressure 
      is proportional to $\rho_B^{5/3}$~\cite{Kapusta}.
      Hence $E/A \rightarrow 3M_{vac}$ in this limit.
      On the other hand, applying \eq{eader}, the derivative diverges
      at this point.

\item For the massless solutions the pressure becomes zero at some
      chemical potential $\mu_0$ corresponding to a non-vanishing density.
      Hence there is a minimum with $E/A = 3\mu_0$.   

      In case (a), $\mu_0$ is identical to the critical chemical potential 
      $\mu_c$
      for the chiral phase transition and is smaller than $M_{vac}$.
      Therefore the minimum is an absolute minimum and corresponds to
      the state of self-bound quark matter mentioned above. 

      In the cases (b) and (c), $\mu_0 > M_{vac}$, and the point of
      lowest $E/A$ is reached for $\rho_B\rightarrow 0$. 
      This means, if not prohibited by external forces, the quark matter
      favors to become infinitely dilute. This is of course unrealistic. 

\item In the cases (a) and (b), there is a maximum in the thermodynamic
      potential which separates the massive from the massless minimum
      (see \fig{fignjlomega}).
      At some chemical potential $\mu_1$ the pressure of this maximum
      becomes zero. This leads to a maximum in the massive branch of
      $E/A$. As a consequence the minimum of the massless branch 
      corresponds to metastable quark matter in case (b). 
\end{itemize}

\begin{figure}
\begin{center}
\epsfig{file=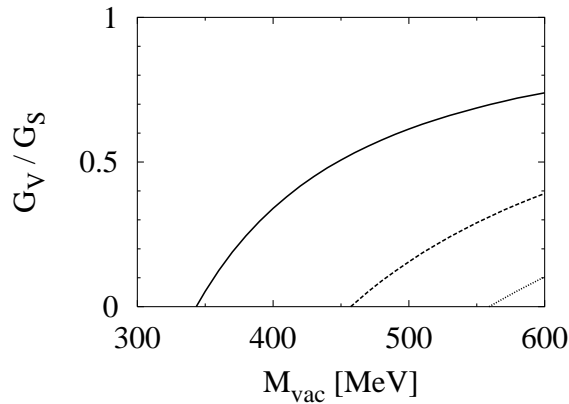,width=8.0cm}
\end{center}
\vspace{-0.5cm}
\caption{\small Lines of constant binding energy per quark, $E_b$, for fixed
         $f_\pi = 92.4$~MeV, $m=0$, and varying constituent quark masses
         $M_{vac}$ and ratios of the vector and scalar coupling constants,
         $G_V/G_S$: $E_b = 0$ (solid), $E_b = 50$~MeV (dashed), and
         $E_b = 100$~MeV (dotted). Adapted from Ref.~\cite{BuOe98}.} 
\label{figebind}
\end{figure}

It is obvious that the repulsive vector interaction disfavors the existence
of bound quark matter and therefore case (a) is only realized
for not too large values of $G_V$. 
Similarly, the attractive scalar interaction favors  the existence
of bound quark matter. In turn, if the attraction is too weak, 
case (a) is not even realized for vanishing $G_V$.
This is illustrated in \fig{figebind} where lines of constant binding
energy per quark, $E_b = M_{vac} - (E/(3A))_{min}$ are displayed in the 
$G_V/G_S - M_{vac}$ space. The calculations have been performed in the
chiral limit with $G_S$ and $\Lambda$ chosen to reproduce 
$f_\pi = 92.4$~MeV, see \fig{fignjl2fit}\footnote{We have ignored
that $f_\pi$ should be somewhat smaller in the chiral limit. 
Also, from physical arguments it might be 
reasonable to consider vector-isovector and axial vector-isovector
interactions with the same coupling strength $G_V$ as in the
vector-isoscalar channel. In this case, the pion decay constant 
gets modified by the order of 10\%. These details, 
which have not been taken into account in the parameter fixing of 
\fig{figebind}, should, however, not change the overall picture. 
}. We find that there is no bound matter for
$M_{vac} \lesssim 343$~MeV. 

This agrees well with the following simple estimate. For $M=m=0$ the 
thermodynamic potential \eq{OmegaMF0} is readily evaluated and one can 
derive $E/A$ for the massless solutions analytically. One finds
\beq
\Big(\frac{E}{A}\Big)_{M=0} \;=\; \frac{B}{\rho_B} \;+\; \frac{9}{4}
\left(\frac{3\pi^2}{2} \,\rho_B \right)^{1/3} \;+\; 9\,G_V \rho_B~.
\label{eanjl0}
\eeq
Minimizing this formula with respect to the density one finds for the
minimum 
\beq
\Big(\frac{E}{A}\Big)_{M=0,min} \;\simeq\; 
3\,\left(2\pi^2\, B \right)^{1/4} \; [ 1 + 2\,G_V 
(\frac{2B}{\pi^2})^{1/2} ] \;,
\label{eanjl0min}
\eeq
where terms of order $G_V^2$ have been neglected.
To be bound, this should be smaller than $3~M_{vac}$. 
If we use the approximate formula, \eq{Bapp}, for the bag constant
we find that for $G_V=0$
this is the case if $M_{vac} \gtrsim 4 f_\pi$,
in reasonable agreement with our numerical findings. 
Reinserting this into \eq{Bapp} we find that the minimal bag constant
which allows for bound quark matter in the NJL model is given by
$B \gtrsim$ 125 MeV fm$^{-3}$, i.e.,
about twice the original MIT value~\cite{MIT3}. 

Recently, it has been shown that the stability of NJL quark matter can 
increase if the matter is exposed to large magnetic fields~\cite{EbKl03}.
In this case, even stable quark droplets consisting of massive quarks
are possible. However, these effects require magnetic fields of the order 
$10^{19}$ Gauss, which is unlikely to be realized even in magnetars. 

We should recall, that we have not yet included diquark condensates,
i.e., color superconductivity. In our later analysis,
this will be an additional source of attraction which enhances the binding.

\subsection{Comparison with the bag model}
\label{njlbag}

The bound quark matter solutions discussed in the end of the previous
section show great similarities with the bag model equation of state.
In fact, for $G_V=0$ \eq{eanjl0} is identical to the bag-model relation,
\eq{eabag}. This is easily understood: Since \eq{eanjl0} 
describes the energy per baryon number in the chirally restored phase,
the quark condensate $\phi$ is equal to zero. Thus, if $G_V=0$
the system is not coupled to any mean field, and energy and pressure
are those of a free fermion gas, shifted by the bag constant,
\eq{njlb}, because the zero-points have been calibrated to the 
non-trivial vacuum. In other words, the quark matter phase we have 
described in this way is completely trivial. What is non-trivial,
is the vacuum. 

For $G_V\neq 0$, the chirally restored phase becomes non-trivial as
well. Nevertheless, at least qualitatively the effect of the vector
coupling is similar to the perturbative corrections in the bag model, 
shifting the minimum of $E/A$ to larger values and lower densities. 
(The quantitative behavior is, however, different: Whereas in \eq{eanjl0}
the correction term is of the order $G_V\rho_B$,  
in the bag model it is of the order $\alpha_s \rho_B^{1/3}$,
as required by dimensions.)

In spite of the arguments above, the great similarity of the NJL-model
and bag-model equations of state might be surprising, since the NJL model
does not confine the quarks, whereas the bag model is confining by 
construction. The resolution is, of course, that \eq{eanjl0}
is only valid for the massless solutions (the dashed lines
in \fig{fignjlea}). For these solutions, $E/A$ diverges in the limit 
$\rho_B \rightarrow 0$. This could indeed be interpreted as ``confinement''
in the sense, that for a fixed number of massless quarks an infinite
amount of energy would be needed to increase the bag radius to infinity.
However, in the NJL model this is not the whole story.
Here at low densities the quarks have the possibility to acquire a mass,
and for these solutions only a finite amount of energy ($M_{vac}$ times 
the number of quarks in the ``bag'') is required in the zero-density
limit: Whereas the massless quarks, just like the bag-model quarks,
are restricted to the chirally restored phase, the massive NJL-model quarks 
are permitted to enter the non-trivial vacuum. Therefore, in the zero-density
limit, instead of paying an infinite amount of energy to transform the 
vacuum, one only needs a finite amount to transform the quarks.
Thus, unfortunately, the same mechanism which gives a microscopic
explanation of the bag pressure -- chiral symmetry breaking --
prevents it from confining the quarks in the model.

Away from the chiral limit, the NJL model equation of state 
always differs from the bag model one. This is because for $m \neq 0$
chiral symmetry gets completely restored only asymptotically\footnote{In 
the NJL model with sharp 3-momentum cut-off,
``asymptotically'' means, when the Fermi momentum becomes equal to the
cut-off. At this point one gets $\phi=0$ and $M=m$. 
However, as we have seen, this only happens at very large densities,
well above the phase transition. 
This is different in the so-called ``scaled NJL model''~\cite{scaledNJL}.
In this model, the cut-off is taken to be proportional to a dilaton field
in order to maintain scale invariance. As a consequence, the cut-off 
drops discontinuously at first-order phase boundaries. Since it drops 
easily below the Fermi momentum, this often limits
the applicability of the model to the chirally broken phase~\cite{JVdB01}.}.
As we have seen in \fig{figgaptmu}, $M$ is a density dependent function
which  can stay relatively large up to rather high densities.
Thus, whereas in a bag model the quarks have just those masses which have 
been given to them and which are usually identified with the current
masses, the quarks in bound NJL matter can have considerably higher
masses. For instance, for parameter set 2 and $G_V=0$ we find bound 
quark matter with $M=40$~MeV, much larger than the current quark mass 
$m=5.6$~MeV.

For the discussion it is often useful to introduce an
``effective bag constant''. One possibility is to write
the energy density of the NJL model in the form
\beq
    \epsilon(\rho_B) \;=\; \epsilon_\mi{free}(\rho_B;M(\rho_B)) 
    \;+\; B_\mi{eff}(\rho_B)~,
\label{Beff}
\eeq
where 
\beq
    \epsilon_{free}(\rho_B;M(\rho_B)) \;=\; \frac{N_f N_c}{\pi^2} 
    \int_0^{p_F} dp\,p^2\,\sqrt{p^2 + M^2(\rho_B)}
\eeq 
is the pressure of a free gas
of quarks with mass $M(\rho_B)$ (the density dependent constituent quark
mass of the NJL model) at baryon number density $\rho_B = (N_f/3\pi^2)\,p_F^3$.

\begin{figure}
\begin{center}
\epsfig{file=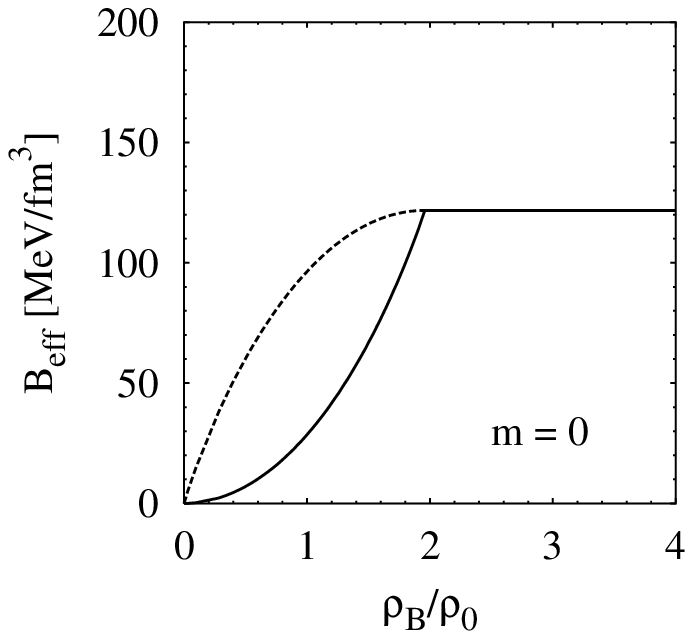,width=7.0cm}\hspace{0.5cm}
\epsfig{file=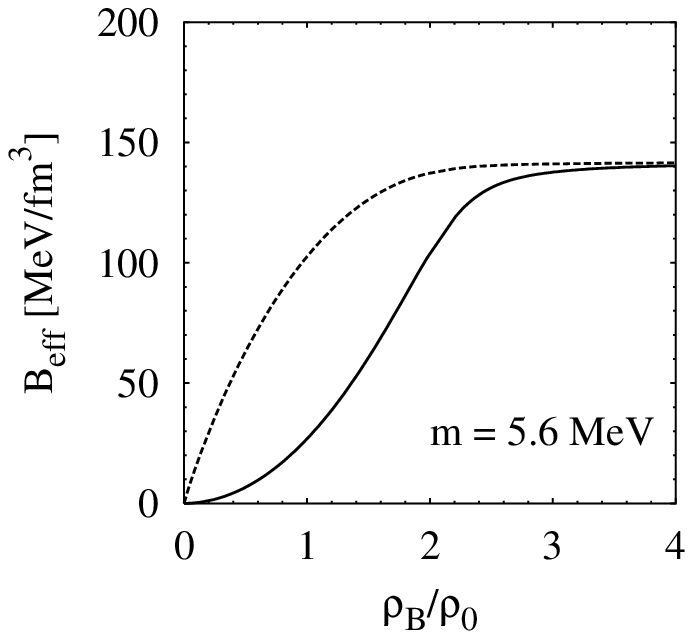,width=7.0cm}
\end{center}
\vspace{-0.5cm}
\caption{\small Effective bag constants $B_\mi{eff}$ (\eq{Beff}, solid lines) 
and $B_{eff}'$ (\eq{Beffp}, dashed lines) as functions of the baryon number 
density $\rho_B$. Right: Parameters of set 2 of \tab{tabnjl2fit} and $G_V=0$.
Left: The same, but in the chiral limit.} 
\label{figbeffnjl}
\end{figure}

$B_\mi{eff}$ as a function of $\rho_B$ is plotted in \fig{figbeffnjl}
(solid lines). The calculations have been performed using parameter set 2
of \tab{tabnjl2fit} (right panel) and the corresponding chiral limit (left
panel). In both cases the vector coupling has been set equal to zero.
The general behavior of the results can be understood as follows:
Since $\epsilon =  \epsilon_\mi{free} = 0$ at $\rho_B = 0$,
$B_\mi{eff}$ has to vanish at this point.
On the other hand, when chiral symmetry is restored, $\epsilon$ behaves
like $\epsilon_\mi{free}$, but shifted by the bag constant $B$.
Thus, in the chiral limit, $B_\mi{eff}$ rises from zero to $B$ and 
then stays constant.
For $m\neq0$, $B_\mi{eff}$ behaves similarly, but since there is no
complete restoration of chiral symmetry, the curve is smoother
and $B_\mi{eff} = B$ is reached only asymptotically. 
  
In contrast to our results, it is sometimes argued that the bag constant
should {\it decrease} with density, see, e.g., Refs.~\cite{JiJe96,BBSS02,Ag03}.
At first sight, this seems to be
natural, because in \eq{njlb} we defined the bag constant to be 
the pressure difference between the non-trivial vacuum and the trivial
vacuum at $M=m$, which goes away upon chiral restoration. 
However, the physical meaning of $B_\mi{eff}$ as defined in \eq{Beff}
is different: Here the system is interpreted as a gas of quasi quarks with 
mass $M(\rho_B)$ in a vacuum with completely or partially restored symmetry. 
$B_\mi{eff}$ is the energy per volume which is needed for this (partial) 
restoration. 
Thus at low densities where only a small step towards symmetry restoration
has been done, only a low ``price'' has to be paid, whereas  
the total amount of restoration energy, $B_\mi{eff} = B$, is only due at
high densities. Of course, from a practical point of view, it still could 
make sense to use effective bag constants which decrease with density in 
order to parametrize missing physics, like short-range
repulsion or excluded volume effects on the hadronic side. 
However, the naive argument that generalized bag models with 
decreasing bag constants are more physical because this accounts for
chiral symmetry restoration is not correct. 

In the above discussion we have identified two sources for the
deviation of the NJL model equation of state from the bag model one:
the density dependent quark masses and the density dependent
effective bag constant.
To see the net effect, we could alternatively start from a free gas of 
current quarks and attribute the entire interaction effects to the effective 
bag constant,
\beq
    \epsilon(\rho_B) \;=\; \epsilon_\mi{free}(\rho_B;m) 
    \;+\; B_\mi{eff}'(\rho_B)~.
\label{Beffp}
\eeq
$B_\mi{eff}'$ is also displayed in \fig{figbeffnjl} (dashed lines).
Again, at zero density, $\epsilon$ and $\epsilon_\mi{free}$ vanish, 
and therefore $B_\mi{eff}'$ vanishes as well. On the other hand, at 
high densities, $M \rightarrow m$  and therefore 
$B_\mi{eff} \rightarrow B_\mi{eff}'\rightarrow B$.
Between these two extremes, $B_\mi{eff}$ and $B_\mi{eff}'$ can differ
considerably. As one can see in the figure, $B_\mi{eff}'$ stays longer
close to the asymptotic value $B$ than $B_\mi{eff}$ does. This means,
the approximation of the NJL energy density by a bag model one with
constant mass $m$ and bag constant $B$ works better than one would
naively expect if one looks at $M$ and $B_\mi{eff}$ as functions of
the density.

In some models, density dependent bag constants are 
introduced {\it by hand} using some ad-hoc 
parametrization, e.g.,~\cite{JiJe96,BBSS02}.
In these cases one has to be careful not to violate thermodynamic
consistency. For instance, if we start from \eq{Beffp}, there is an
extra contribution to the chemical potential 
$\mu_B = \partial\epsilon/\partial \rho_B$ due to the density dependence
of the bag constant and hence the pressure is {\it not} given by 
the bag model expression. 
This problem will not affect us, because we will never use effective
bag constants to calculate other quantities, but only in order to 
interprete the results (which are consistently derived from the
thermodynamic potential).
Related to the above problem, there is of course some arbitrariness
in the definition of $B_\mi{eff}$. For instance, we could have started
from the bag model expression for the pressure, instead of the energy 
density. The results would be somewhat different,
but the qualitative features of \fig{figbeffnjl} would change only little.

Because of the great similarities between NJL model and bag model equations
of state, it is tempting to identify the bound-matter solutions of the 
NJL model with baryons, at least in a very schematic sense~\cite{ARW98,Bu96}.
In the previous section we pointed out that the scenarios (b) and (c) for
the chiral phase transition are unrealistic because they predict the existence
of a homogeneous gas of constituent quarks at low densities. This
contradicts confinement.  
This problem does not arise in case (a) where the dilute-gas solutions are 
unstable against phase separation, leading to droplets of dense quark
matter in the chirally restored phase surrounded by vacuum. 
Clearly, at least for two flavors, this scenario would be unrealistic
as well, unless we adopt the above interpretation of these droplets 
as baryons. 

Therefore let us neglect for a while that we have solved a thermodynamic
problem for infinite homogeneous mean fields and assume that the solutions
can be extrapolated down to three quarks in a sphere of radius $R$.
Taking the bound-matter solution of parameter set 2 with baryon number
density $\rho^* = 2.8\,\rho_0$ we obtain a reasonable bag radius 
$R = (4\pi\rho_B/3)^{-1/3} = 0.8$~fm.
(For parameter set 4, we find $\rho^* = 5.8\,\rho_0$ corresponding
to $R = 0.6$~fm.)
We may also calculate the $\bar qq$-content of the bag, which is 
defined as the difference of the quark condensate in the bag and in
the vacuum, integrated over the bag volume,
\beq
    \bra{bag}\,\bar qq\,\ket{bag} \;=\; \frac{1}{\rho^*}\,
    (\phi|_{\rho_B=\rho^*} \,-\, \phi|_{\rho_B=0})~.
\label{qbqbag}
\eeq
For parameter set 2 we find $\bra{bag}\,\bar qq\,\ket{bag} \simeq 7$.
This corresponds to a ``sigma term'' 
$\sigma_{\pi, bag} = m \, \bra{bag}\,\bar qq\,\ket{bag} \simeq 39$~MeV,
not too far away from the most commonly quoted value 
$\sigma_{\pi N} \simeq 45$~MeV,
extracted from $\pi N$ scattering data~\cite{GLS91}\footnote{Note, however, 
that much larger values are found in some more recent 
analyses~\cite{BuMe00}.}.  
For parameter set 4 we find lower values,
$\bra{bag}\,\bar qq\,\ket{bag} \simeq 4$ and $\sigma_{\pi, bag} \simeq 20$~MeV,
mostly because of the smaller bag volume. 
On the other hand, if we include vector interactions, the density of
the bound quark matter becomes smaller and, consequently, the bag radii,
$\bra{bag}\,\bar qq\,\ket{bag}$, and $\sigma_{\pi, bag}$ get larger.

Our method to determine a sigma term within the 
NJL model is rather different from that of Ref.~\cite{LKW92}. In that
reference a pion-quark sigma term, $\sigma_{\pi q}$ was extracted from
the constituent quark masses in vacuum via the Feynman-Hellmann theorem.
The authors could show
that $\sigma_{\pi q}$ governs the low-density behavior of $\ave{\bar qq}$ 
in the NJL model in the same way as $\sigma_{\pi N}$ in chiral
perturbation theory, \eq{ChPTrho}. 
Identifying $\sigma_{\pi N} = 3\sigma_{\pi q}$ they obtained
$\sigma_{\pi N} = 32$~MeV. Although this is a reasonable number, 
the description of low-density nuclear matter by a low-density quark gas 
remains questionable.
Comparison with our method might shed some light on this puzzle:
Whereas in Ref.~\cite{LKW92} the sigma term is proportional to the 
derivative of the quark condensate at $\rho_B = 0$, the value obtained
from \eq{qbqbag}
is proportional to the slope of a straight line connecting $\ave{\bar qq}$
at $\rho_B = 0$ with $\ave{\bar qq}$ at $\rho_B = \rho^*$.
Although this is not exactly the same, both numbers are quite similar,
as one can see, e.g., from the density dependence of the constituent
quark mass shown in \fig{figgaptmu} (Connect the points at 
$\rho_B = 0$ and $\rho_B = \rho^* = 2.8~\rho_0$
on the solid line of the right panel by a straight line and compare the 
slope with the slope of the solid line at  $\rho_B = 0$.).
Thus, if we believe that our method gives the correct value of the 
sigma term, the method of Ref.~\cite{LKW92} should also work rather well.
 
On the other hand, it is clear that our identification of the ``droplets'' of 
bound quark matter in the NJL model with baryons, i.e., our extrapolation 
from homogeneous infinite matter to three-quark system is too simplistic.
Obviously, a realistic modeling of baryons requires to start from
three valence quarks and to abandon the simplification of space-independent 
mean fields. In fact, much better jobs in this direction 
have already been done, describing baryons as chiral quark 
solitons~\cite{ripka,Chr96,ARW96} or solving a Fadeev equation for three
constituent quarks~\cite{IBY93,HaKr95}.
Of course, even these approaches cannot explain
why three-quark systems are favored against larger multi-quark
clusters, or, in other words, why nuclei consist of nucleons instead
of being a single large bag. This would require a better 
understanding of confinement and the inclusion of repulsive
short-range interactions which prevent the three-quark bags from
merging. 

Anyway, it is obvious that a realistic description of nuclei or nuclear 
matter with quark degrees of freedom is not possible within mean-field 
approximation. The bound quark-matter solutions 
of the NJL model describe at most some fictitious state of 
matter, which one would find if the bag pressure was the only relevant
binding force. Once confining forces and residual interactions beyond
a homogeneous mean field are taken into account, the quark-matter
solutions become unstable and decay into baryonic matter.
In this context it is remarkable that all bound quark-matter solutions
we have found above have $E/A \gtrsim 3 \times 343$~MeV,
i.e., at least 90~MeV above the nucleon mass.

As long as these mechanisms are not understood, the best way to describe
hadronic matter is to start from phenomenological hadronic interactions.
Eventually, at higher densities, homogeneous quark matter should become
favored. Thus, a more pragmatic procedure would be to employ the NJL
model only at high densities and to use a hadronic equation of state
for the hadronic phase. This will be done in Chap.~\ref{stars} where
we investigate the structure of compact stars.
On the other hand, for more schematic discussions of the phase diagram
it is often more appealing to have a single model for all phases.
An NJL mean-field description of the hadronic phase could then be 
acceptable, if one stays aware of the limitations of the model.
Particular caution is in order at finite temperature, where effects
of unconfined quarks in the ``hadronic phase'' are unavoidable.
This will briefly be discussed in the next section.

\subsection{Phase diagram}
\label{njlt}

Applying the formalism developed in \sect{njlpot}, 
it is straight forward to extend our numerical studies to non-vanishing
temperatures and to investigate the chiral phase diagram in the
$T-\mu$ plane. This has first been done by Asakawa and Yazaki~\cite{AsYa89},
followed by many others. 

\begin{figure}
\begin{center}
\epsfig{file=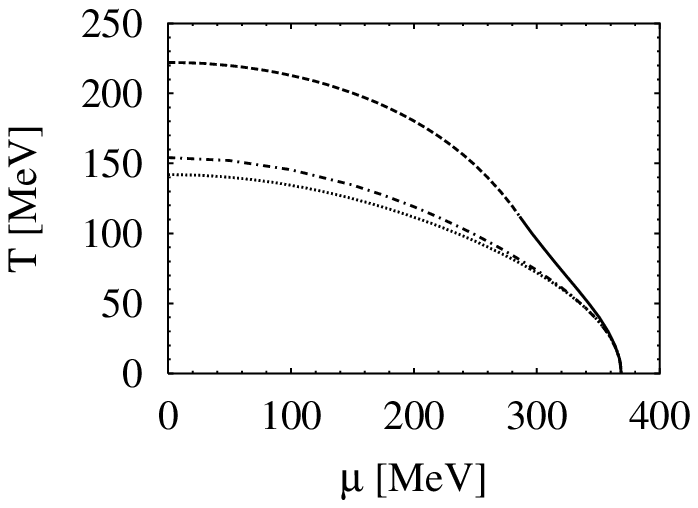,width=7.0cm}\hspace{0.5cm}
\epsfig{file=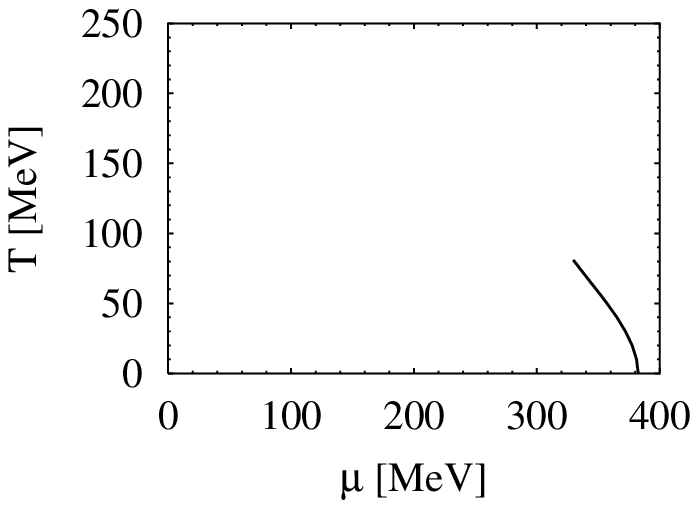,width=7.0cm}
\end{center}
\vspace{-0.5cm}
\caption{\small NJL phase diagram for parameter set 2 and $G_V=0$.
Left: Phase diagram in the chiral limit.
The first and second order phase boundaries are indicated by the solid
or dashed line, respectively. The dash-dotted line indicates the location
of massless solutions with vanishing pressure.
The dotted line corresponds to the zero-pressure line in a bag model
with the same bag constant and quark degrees of freedom only.
Right: Phase diagram for $m = 5.6$~MeV.
} 
\label{fignjl2phase}
\end{figure}

A typical phase diagram obtained in this way is shown in \fig{fignjl2phase}. 
The calculations have been performed with parameter set 2 of \tab{tabnjl2fit} 
and $G_V=0$. The phase diagram in the left panel corresponds to the 
chiral limit, $m = 0$. First and second-order phase boundaries are indicated 
by a solid or dashed line, respectively. We know already that for the present 
parameters the phase transition is first-order at $T=0$. 
On the other hand, along the $T$-axis, i.e., at $\mu = 0$,
the phase transition is second-order.
Hence, as argued in the Introduction, there must be a tricritical point
at some intermediate temperature, 
where the first-order phase boundary turns into a second-order one. 
In the present example, this point is located at $\mu = 286$~MeV and 
$T = 112$~MeV.

With finite quark masses, chiral symmetry is never restored exactly.
Therefore, at high temperatures and low chemical potentials,
instead of second-order phase transitions, we only have smooth
cross-overs where the quark condensate gets rapidly (but continuously)
reduced. In this case, the first-order
phase boundary ends in a second-order endpoint.
This is shown in the right panel of \fig{fignjl2phase}.
For the position of the endpoint we find $\mu = 330$~MeV and 
$T = 81$~MeV.

At first sight, \fig{fignjl2phase} seems to be in qualitative agreement
with common wisdom about the QCD phase diagram for two light or massless
flavors. A closer inspection, however, reveals severe problems.
Being a mean-field calculation, the phase transition is driven by
quark and antiquark degrees of freedom. In particular the ``hadronic
phase'', i.e., the phase with broken chiral symmetry is described as
a gas of constituent quarks, instead of mesons and baryons.
The only exception is the $\mu$-axis at $T=0$. There, as discussed 
in \sect{njlt0}, we have a phase transition of ``type (a)'', i.e., the
``hadronic phase'' is identical to the vacuum. 
This scenario does no longer exist at finite temperature where the
occupation numbers $n_p$ and $\bar n_p$ are always non-zero.
This underlines the difference between bound and confined quark 
matter: The finite binding energy for the self-bound
solutions at $T=0$ cannot prevent the evaporation of constituent quarks
at arbitrary small (but non-zero) temperatures.

It is therefore not surprising that the NJL model results are at variance
with several aspects of the QCD phase diagram which we have discussed
in the Introduction: 

\begin{itemize}
\item Without major modifications (like introducing temperature dependent
      coupling constants) the NJL model gives a rather poor description 
      of the lattice results at $\mu = 0$. In particular the critical
      temperature is typically too large. 
\item The curvature of the phase boundary at $\mu=0$ is also larger than
      the lattice result~\cite{AEHKKLS},
      $T_c (d^2T_c/d\mu^2)|_{\mu=0} = -0.14\pm 0.06$.
      We find -0.40 in the chiral limit and about the same number for 
      $m = 5.6$~MeV if we define the cross-over line by the 
      inflection points $\partial^2 M/\partial T^2 = 0$. 
\item The temperature of the critical endpoint is considerably smaller
      and the chemical potential is larger than for the lattice
      point of Fodor and Katz~\cite{FoKa04} who find $T = (162 \pm 2)$~MeV 
      and $\mu_B = (360 \pm 40)$~MeV, i.e., $\mu \simeq 120$~MeV.
\item In a mean-field calculation, one finds of course mean-field
      critical exponents, rather than $O(4)$. 
\item At $\mu = 0$ and low temperature, the model is not in agreement
      with chiral perturbation theory, because the pionic degrees of
      freedom are not taken into account in mean-field approximation.
      This will be discussed in more details in \sect{nc}.
\end{itemize}

Of course, the value of the critical temperature at $\mu = 0$ depends on the 
parameters. For parameter set 2 (in the chiral limit) we find $T_c = 222$~MeV, 
as we have already seen in \fig{figgaptmu}. If we take parameter set 1 we get 
a more reasonable value, $T_c = 177$~MeV, but at the same time $\mu_c$ at 
$T=0$ becomes unrealistically small: We find $\mu_c = 305$~MeV, i.e., 
the baryon chemical potential $\mu_B = 915$~MeV is less than the nuclear 
matter value.
Note that the large ratio between $T_c$ at $\mu = 0$ and
$\mu_c$ at $T=0$  
gives also a natural explanation for the too large curvature
of the phase boundary, since on average the boundary must be steeper
than for smaller ratios.

To shed some light on the possible sources of this behavior,
we compare the NJL phase boundary with the line of zero pressure in a 
bag model with quark degrees of freedom only and the same bag constant
(left panel of \fig{fignjl2phase}, dotted line).
Basically, this line corresponds to the dotted line in \fig{figbag}. 
At $T=0$, as discussed in \sect{njlbag}, both models agree.
However, at small chemical potentials, i.e., at relatively high temperatures, 
the two lines become quite different. To some extent, this is due to the 
regularization which cuts off the high momenta in the NJL-model calculation. 
Therefore the pressure of the massless solutions is somewhat smaller than
in the bag model and the line of zero pressure is shifted to higher 
temperatures (dash-dotted line). The remaining difference to the dashed or 
solid line must then be  attributed to the presence of unconfined constituent 
quarks in the chirally broken phase which add to the pressure in that phase
and thereby shift $T_c$ to higher values.
Also note that the bag-model $T_c$ becomes further reduced if gluons are
included (see \fig{figbag}).

This comparison suggests that the large critical temperatures in the NJL 
model at small chemical potentials are mainly due to unphysical effects,
namely cut-off effects, missing gluonic degrees of freedom in the 
``QGP phase'', and unconfined quarks in the ``hadronic phase''.
Although the latter might partially account for the effect of the missing 
hadronic degrees of freedom, it is clear that quantitative predictions
of the model, e.g., about the position of the critical endpoint,
should not be trusted. 
We should also recall that the introduction of a repulsive 
vector interaction in the NJL model weakens -- and finally removes -- 
the first-order phase  transition. We could thus move the endpoint 
to even lower temperatures until it vanishes completely.
Therefore, any agreement of NJL and lattice results in this point would
be accidental.

Before closing these critical reflections, we should note that the flaws 
listed above are rather general consequences of missing confinement,
together with the mean-field treatment and are not restricted to NJL-type
models. As pointed out before, it is quite obvious that, starting from quark 
degrees of freedom, one cannot get a realistic picture of the hadronic phase 
in mean-field approximation, no matter how sophisticated the interaction.

\subsection{$1/N_c$ corrections}
\label{nc}

As an example for the shortcomings of the mean-field approximation
and in order to illustrate how these problems are (partially) removed
beyond mean field, 
we briefly discuss the temperature dependence of the quark condensate
within a $1/N_c$ expansion scheme. 
More detailed discussions about the use of this scheme and
other methods beyond mean field in the NJL model can be found in
Refs.~\cite{OBW01,Oertel} and references therein. 

According to chiral perturbation theory, the low-temperature and density
behavior of the quark condensate in the chiral limit is given by \eq{ChPTrho}.
Hence, to leading order in $T$ and at zero density, the change of the 
condensate should be proportional to $T^2/(8 f_\pi^2)$. As mentioned in the
Introduction, this can be attributed to the thermal excitation of massless 
pions.  This is at variance with the NJL mean field, where the heat bath
consists entirely of constituent quarks and antiquarks. However, since
these are exponentially suppressed because of their mass, the quark condensate
changes only very little at low temperatures~\cite{ripka,LKW92}.  
This can be seen in \fig{figqbqt} where $\ave{\bar qq}$, normalized to
its value at $T=0$, is plotted as a function of temperature. 
The calculations have been performed in the chiral limit. The mean-field
result is indicated by the dashed lines\footnote{\fig{figqbqt} is based on 
the results of Refs.~\cite{OBW01,Oertel} where the quark loops have been 
regularized using the Pauli-Villars scheme. However, this is an irrelevant 
detail for the present discussion. 
For a sharp $O(3)$ cut-off, the mean-field behavior is readily read off
in \fig{figgaptmu} since in the chiral limit the mean-field
quark condensate is directly proportional to the constituent quark 
mass, see \eq{LNJLlin}.}.

A method to include the effect of thermal pions in a systematic (and
symmetry conserving) way, is an expansion in powers of the inverse
number of colors, $1/N_c$. To that end, one assigns a factor $1/N_c$
to the NJL coupling constants. 
Since the closed quark loops yield a factor $N_c$, the quark propagator
in Hartree approximation (\fig{fignjlgap}) is of the order unity, while the 
exchange of an RPA meson (\fig{fignjlrpa}) is of the order $1/N_c$.
Similarly, the quark condensate is of the order $N_c$ and 
$f_\pi$ is of the order $\sqrt{N_c}$.

\begin{figure}
\begin{center}
\epsfig{file=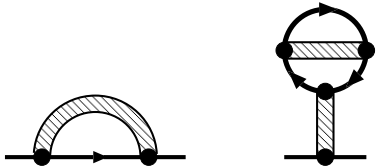,width=5.0cm}
\end{center}
\vspace{-0.5cm}
\caption{\small Correction terms of order $1/N_c$ to the quark
                self-energy. The solid lines and shaded boxes symbolize
                quark propagators and RPA meson propagators, respectively.} 
\label{fignc}
\end{figure}

Following these rules, one can construct two self-energy diagrams which 
contribute to the order $1/N_c$ to the quark propagator. They are shown in 
\fig{fignc}. Note that, unlike the Hartree term, these self-energy
contributions must not be iterated, as this would be of higher order.
The $1/N_c$ corrected quark condensate is then obtained as an integral
over the trace of the $1/N_c$ corrected quark propagator, just as in
\eq{njlqbarqgen}. 

The result is indicated by the solid lines in \fig{figqbqt}.
Because of the meson loops, the $1/N_c$ correction terms
are sensitive to thermally excited pions and therefore indeed cause
a $T^2$-behavior at small temperatures.
A careful examination of the corresponding diagrams yields~\cite{OBW01,Oertel}
\beq
    \ave{\bar qq}(T) \;=\; \ave{\bar qq}_\mi{vac} \;-\; 
    \ave{\bar qq}_\mi{vac}^{(0)}\,\frac{T^2}{8f_\pi^{(0)\;2}} \;+\; \dots~,
\label{qbqtapp}
\eeq    
where the suffix $(0)$ indicates quantities in leading order in $1/N_c$.
Since $f_\pi^{(0)\;2}$ is of the order $1/N_c$, \eq{qbqtapp} corresponds
to a consistent expansion of the $\chi PT$ result, \eq{ChPTrho}, at 
zero density to next-to-leading order in $1/N_c$. 
To illustrate the quality of this expansion, the r.h.s.\@ of \eq{qbqtapp} 
is also displayed in \fig{figqbqt} (dotted lines). As one can see, the
agreement with the solid line is excellent for $T \lesssim 80$~MeV.

\begin{figure}
\begin{center}
\epsfig{file=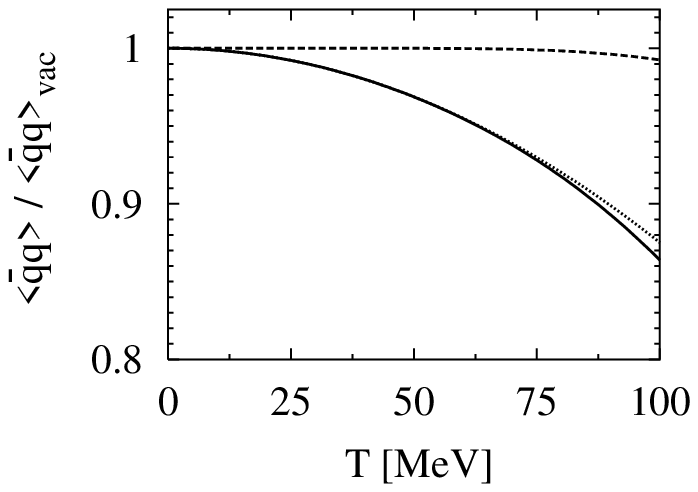,width=7.0cm}\hspace{0.5cm}
\epsfig{file=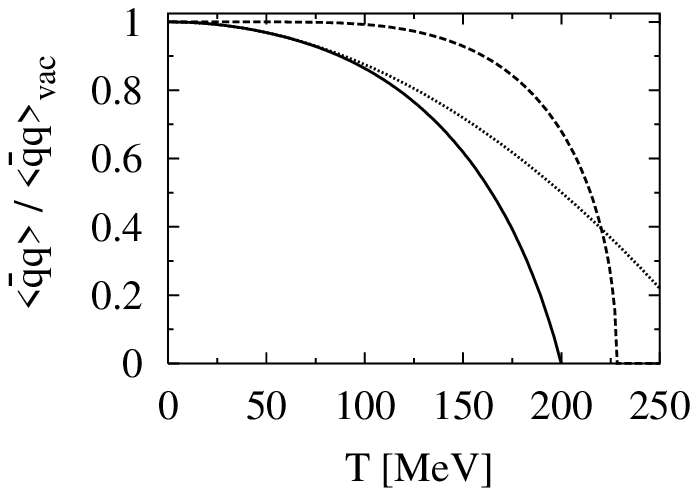,width=7.0cm}
\end{center}
\vspace{-0.5cm}
\caption{\small Quark condensate in the chiral limit, normalized to its
         vacuum value, as a function of temperature: Hartree
         approximation (dashed line) and with $1/N_c$ corrections included
         (solid). The dotted line indicates a temperature expansion of
         the $1/N_c$ corrected result up to order $T^2$ 
         (see \eq{qbqtapp}).
         The right figure has been adapted from Ref.~\cite{OBW01}. 
         The left figure is just an enlarged detail of the right one.} 
\label{figqbqt}
\end{figure}

In spite of its successes, the $1/N_c$-expansion scheme has also its
limitations.
First, it should be noted that the magnitude of the $1/N_c$-correction
terms is not uniquely determined by the leading-order parameters. 
Instead, since the NJL model is non-renormalizable, new parameters
appear at each order. 
In the model described above, the meson loops have been regularized
by an independent cut-off parameter which has been fixed by fitting
the width of the $\rho$ meson~\cite{OBW00}.

For our purposes, the most severe problem is the fact that it is a
perturbative scheme  (as pointed out before, the diagrams shown in
\fig{fignc} must not be iterated) and therefore, it cannot be 
applied to the phase transition. 
A non-perturbative, but still symmetry conserving extention of the 
NJL model beyond mean field -- the so-called meson-loop expansion (MLA) --
has been developed in Refs.~\cite{DSST95} and \cite{nikolov}.
Within this scheme one also finds the correct low-temperature behavior
of the quark condensate~\cite{OBW01,Oertel,FlBr96} and it is possible
to construct a phase transition. Unfortunately, the latter turns out to
be first-order~\cite{OBW01,Oertel,FlBr96}, which is likely to be an
artifact of the approximation scheme.

We should also note that, while we have added the relevant degrees of
freedom for low-temperature physics, we still need to include nucleons
for a correct description of the low-density regime. Staying within
the NJL model this means that one first has to solve a Fadeev equation.
Work in this direction has been performed in Refs.~\cite{BeTh01,BHIT03}. 

Finally there remains the problem that the unphysical degrees of freedom
-- quarks and antiquarks -- are not removed from the hadronic phase.
Because of the relatively large
constituent mass, they are suppressed at low temperatures,
but since they have a large degeneracy factor, they become dominant at
higher $T$. In \fig{figqbqt} this is the case above $\simeq 100$~MeV.

\section{Asymmetric matter}
\label{flamix}

So far we have restricted our analysis to the case of a uniform chemical
potential which is the same for up and down quarks. Together with the
isospin symmetry in the Lagrangian, i.e., the assumption 
$m_u = m_d \equiv m$, this implies that all quantities related to up
and down quarks, in particular constituent masses, quark condensates, 
and densities, are equal for both flavors. 
However, there are many situations in nature, where the numbers of up
and down quarks are not equal. 
Neutron stars, for instance, must be electrically neutral to a very
high degree. Therefore, if the core of a neutron star consists
of deconfined up and down quarks, the number of down quarks must be
about twice as large as the number of up quarks to ensure
neutrality. (There are also electrons, but as we will see
later on, in chemical equilibrium their fraction is very small.) 
Similarly, all heavy nuclei have an excess of neutrons over protons,
which translates into an excess of down quarks over up quarks.
Thus, if a quark-gluon plasma is formed in the collision of two
neutron-rich nuclei, one would expect that it contains more down quarks 
than up quarks.

In order to describe these situations properly, we have to allow
for different chemical potentials, $\mu_u$ and $\mu_d$,
for up and down quarks, respectively, 
\beq
\mu_u = \mu + \delta\mu~,\quad \mu_d = \mu - \delta\mu~.
\label{mudef}
\eeq
As before, $\mu = \mu_B/3$ is the chemical potential related to the total 
quark number density $n = n_u + n_d$. $\delta\mu$ is related to 
$n_u - n_d$ and in this way to the isospin density\footnote{The 
isospin density is defined as $n_I = \frac{1}{2}(n_u-n_d)$. 
This implies $\mu_I = 2 \delta\mu$.}

The introduction of a new chemical potential adds a new axis to the 
phase diagram of strong interactions.
Although most theoretical works describe the ``standard'' $\mu-T$
phase diagram with $\delta\mu = 0$, some authors have also studied other
projections. For instance the case $\mu = 0$ but 
$\delta\mu \neq 0$~\cite{SoSt01} is particularly interesting since it can 
be studied on the lattice~\cite{KoSi02}.
On the other hand, for the application to neutron star interiors one can
often neglect temperature effects, whereas $\mu \neq 0$ and $\delta\mu \neq 0$.
This case (extended to three flavors) will be one of the main tasks of 
the present work.

In this section, however, we want to discuss the effect of a 
non-vanishing, but constant $\delta\mu$ on the structure of the 
$\mu$-$T$ phase 
diagram, which might be the most interesting case for the interpretation
of heavy-ion collisions, although the isospin chemical potential is
not strictly constant along the trajectory of the process.
This case has been studied within a random matrix
model~\cite{KTV03} and within an NJL-type model~\cite{TK03}.
The authors of these references reported the interesting result that,
instead one, they found two first-order phase transitions at
low temperature and high baryon chemical potential and thus two 
second-order endpoints.
More recently, this result has been confirmed by further studies within 
the NJL model~\cite{BCPR04} and within a QCD-like model 
(``ladder QCD'')~\cite{BCPR03}. 
Since second-order endpoints, as discussed in the Introduction,
are potentially detectable in heavy-ion
collisions~\cite{SRS98,SRS99}, this could have important consequences.

In the references above an interaction was chosen, where the up and down
quarks completely decouple, i.e., the presence of up quarks has no influence
on the down quarks and vice versa. From this point of view, the fact that
there are two phase transitions -- one for up quarks and one for down quarks --
is almost trivial. It is known, however, that instanton induced interactions,
like \eq{Linst}, mix different flavors. 
One can therefore ask, whether the existence of two phase boundaries
persist, when instanton-type interactions are present,
together with non-flavor mixing interactions.
This has been investigated in Ref.~\cite{FBO03}, which we discuss in
the following.  
 
Starting point is the Lagrangian
\beq
{\cal L} = {\cal L}_0 + {\cal L}_1 + {\cal L}_2~, 
\label{Lflamix}
\eeq 
which contains a free part
\beq 
{\cal L}_0 = {\bar q} ( i \delsl - m ) q~, 
\eeq 
and two different interaction parts~\cite{klevansky,AsYa89},
\beq 
{\cal L}_1 = G_1 \Big\{ ({\bar q}
q)^2 + ({\bar q}\,\vec\tau q)^2 + ({\bar q}\,i\gamma_5 q)^2 + ({\bar
q}\,i\gamma_5\vec\tau q)^2 \Big\} 
\label{Lflamix1}
\eeq 
and 
\beq 
{\cal L}_2 = G_2
\Big\{ ({\bar q} q)^2 - ({\bar q}\,\vec\tau q)^2 - ({\bar
q}\,i\gamma_5 q)^2 + ({\bar q}\,i\gamma_5\vec\tau q)^2 \Big\}~. 
\label{Lflamix2}
\eeq
${\cal L}_2$ is identical to the instanton induced (``'t Hooft'')
interaction ${\cal L}_{inst}$, \eq{Linst}, whereas the standard
NJL Lagrangian, \eq{LNJL}, is recovered when we choose
$G_1 = G_2 = G/2$. Both terms are invariant under
$SU(2)_L\times SU(2)_R\times U(1)$ transformations. ${\cal L}_1$
exhibits an additional $U_A(1)$ symmetry which is explicitly broken by
${\cal L}_2$. 

To obtain the mean-field thermodynamic potential $\Omega(T,\mu,\delta\mu)$
we can basically apply the same techniques as before. 
Since isospin symmetry is broken by a non-vanishing $\delta\mu$,
we assume the existence of two generally different quark condensates
\beq 
\phi_u = \ave{\bar u u} \qquad \text{and} \qquad
\phi_d = \ave{\bar d d}~, 
\eeq 
and linearize the Lagrangian in the presence of these condensates. 
In principle, we should also allow for non-vanishing expectation values
with pionic quantum numbers, $\pi_a = \ave{\bar q i\gamma_5\tau_a q}$,
to describe a possible pion condensation. Indeed,
for $\mu=0$, it can be shown that these condensates become non-zero if
$\mu_I = 2|\delta\mu|$ exceeds the pion mass~\cite{MFrank,SoSt01}. 
However, as we will see below, for our present purpose it is sufficient
to restrict the model to lower values of $|\delta\mu|$ and we can 
safely assume $\pi_a=0$. In this way we get 
\beq
     \Omega(T,\mu_u,\mu_d;\phi_u,\phi_d) = \sum_{f=u,d} \Omega_{M_f}(T,\mu_f)
     + 2\,G_1\,(\phi_u^2 + \phi_d^2) + 4\,G_2\phi_u\phi_d~,
\label{flamixomega}
\eeq
where $\Omega_{M_f}(T,\mu_f)$ corresponds to the contribution of a gas
of quasiparticles of flavor $f$,
\begin{alignat}{2}
    \Omega_{M_f}(T,\mu_f) = -2N_c \int\dtp \,\Big\{
    E_{p,f} &+ T\,\ln{\Big(1 + \exp{(-\frac{1}{T}(E_{p,f}-\mu_f))}\Big)} \nonumber \\
        &+ T\,\ln{\Big(1 + \exp{(-\frac{1}{T}(E_{p,f}+\mu_f))}\Big)}\Big\}
\label{flamixom0}
\end{alignat}
The constituent quark masses are now given by
\beq
     M_i = m_i - 4\,G_1\,\phi_i - 4\,G_2\,\phi_j~,\qquad i\neq j \in \{u,d\}~,
\label{flamixmasses}
\eeq
i.e., in general $M_u \neq M_d$.
In order to determine the physical solutions, we have again to look for the 
stationary points of the thermodynamic potential, this time
with respect to the two condensates $\phi_u$ and $\phi_d$.  This leads to 
the standard expression for the quark condensates
\beq 
\phi_f = - 2N_c \int \dtp\,
\frac{M_f}{E_{p,f}}\Big\{ 1 - n_{p,f}(T,\mu_f) 
- \bar{n}_{p,f}(T,\mu_f) \Big\}~.
\label{phif}
\eeq
When these are inserted into \eq{flamixmasses}, we obtain
a set of two coupled gap equations for $M_u$ and $M_d$ which have to be 
solved self-consistently.

Note that the condensate $\phi_f$ only depends on the constituent mass
$M_f$ of the same flavor, whereas the constituent mass for one
flavor depends in general on both condensates, and therefore the two
flavors are coupled. 
If we switch off the ``instanton part'' ${\cal L}_2$, i.e.,
$G_2 = 0$, the two flavors decouple. In this case $M_i$ depends only
on the condensate of the same flavor $\phi_i$ and the mixed
contribution to $\Omega$ (last term of Eq.~(\ref{flamixomega})) vanishes. 
This limit corresponds basically to the case studied in Ref.~\cite{TK03}.
In the opposite limit, i.e., $G_1 = 0$, we have ``maximal'' mixing: 
The constituent mass of flavor $i$ only depends on the condensate $\phi_j$ 
with $i \neq j$. It is also interesting that for $G_1 = G_2$, i.e.,
for the original NJL Lagrangian \eq{LNJL} we always get $M_u = M_d$, 
even for large $\delta\mu$. 

To study the effects of flavor mixing, let us now write
\beq G_1 = (1-\alpha)\,G_0~,\qquad
G_2 = \alpha\,G_0~, 
\eeq
and calculate the phase diagram for fixed $G_0$ but different values of 
$\alpha$. The degree of flavor mixing is thereby controlled by the
particular value of $\alpha$ while the values of the vacuum
constituent quark masses $M_{vac}$ are kept constant.

For our numerical studies we use the parameters $m_u
= m_d = 6$~MeV, $\Lambda = 590$~MeV, and $G_0\Lambda^2 = 2.435$~\cite{FBO03}. 
They are close to set 2 of \tab{tabnjl2fit} and yield 
$M_{vac} = 400$~MeV, $m_\pi = 140.2$~MeV, 
$f_\pi = 92.6$~MeV and $\ave{\bar u u} = (-241.5 \,\mathrm{MeV})^3$. 

\begin{figure}[t]
\begin{center}
\includegraphics*[bb = 108 380 485 519,height = 5.5cm]{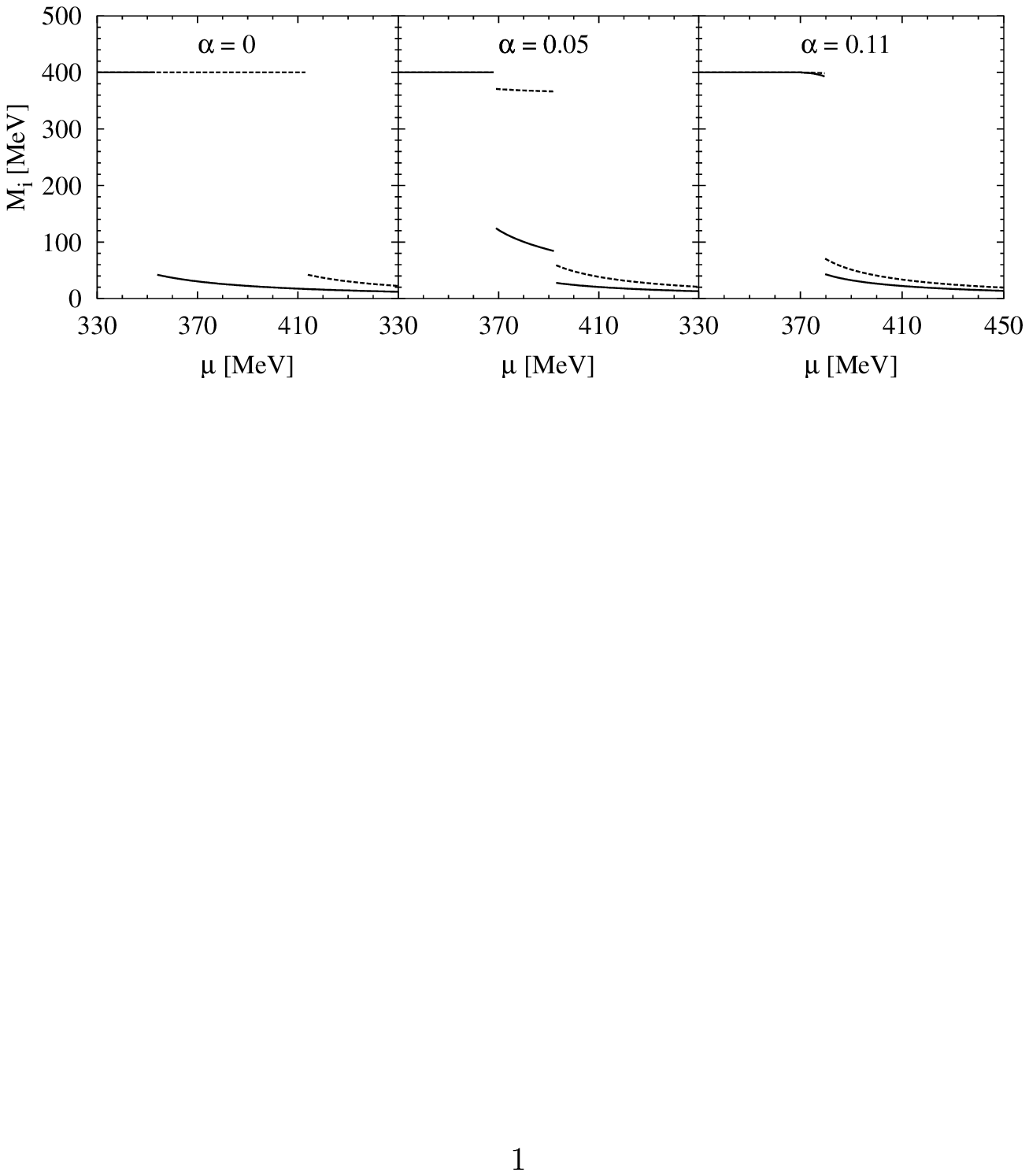}
\caption{\small Constituent quark masses $M_u$ (solid) and $M_d$ (dashed)
at $T=0$ 
as functions of quark number chemical potential $\mu$ for $\delta\mu = 30$ MeV 
and  $\alpha = 0$ (left), $\alpha = 0.05$ (center), and $\alpha = 0.11$ 
(right). From Ref.~\cite{FBO03}. 
}
\label{figflamixmass}
\end{center}
\end{figure}

We begin our discussion with the results at $T = 0$. 
In \fig{figflamixmass} we display the values of $M_u$ and $M_d$ as
functions of the quark number chemical potential $\mu$ for fixed  
$\delta\mu = 30$~MeV\footnote{Following common practice 
(e.g., Refs.~\cite{TK03,Bedaque02,KYT02})
we take a positive value of $\delta\mu$, although for the description of
heavy-ion collisions $\delta\mu<0$ would be more appropriate. However,
since changing the sign of $\delta\mu$ does only interchange the roles of up 
and down quarks, this does not alter our conclusions.}. 
The left panel corresponds to $\alpha = 0$.
We observe two distinct phase transitions at $\mu = 353$~MeV for
the up quarks and at $\mu = 413$~MeV for the down quarks. This behavior
is easily understood when we recall that at $\alpha = 0$ the up and
down quark contributions to the thermodynamic potential decouple.
Hence, if we had plotted $M_u$ and $M_d$, in terms of the corresponding
{\it flavor} chemical potential $\mu_u$ and $\mu_d$, respectively,
we would have found two identical functions with a phase transition at
$\mu_f = 383$~MeV. This is basically the result reported in 
Refs.~\cite{KTV03,TK03}.  

Now we study the influence of a non-vanishing flavor mixing.
In the central panel of \fig{figflamixmass} we show the behavior of the 
constituent quark masses for $\alpha = 0.05$.
The situation remains qualitatively unchanged, i.e.,
we still find two distinct phase transitions.
However, because $M_d$ now also depends on $\phi_u$ (and thus on
$M_u$), and vice versa, both constituent masses drop at both critical
chemical potentials.  
Moreover, this small amount of flavor mixing already diminishes the
difference between the two critical quark number chemical potentials
considerably.
Finally, for $\alpha$ larger than a critical value  of $0.104$
we find only one {\it single} first-order phase transition.
This is illustrated in the right panel of  \fig{figflamixmass},
which corresponds to $\alpha = 0.11$.

\begin{figure}[t]
\begin{center}
\includegraphics*[bb = 108 380 485 519,height = 5.5cm]{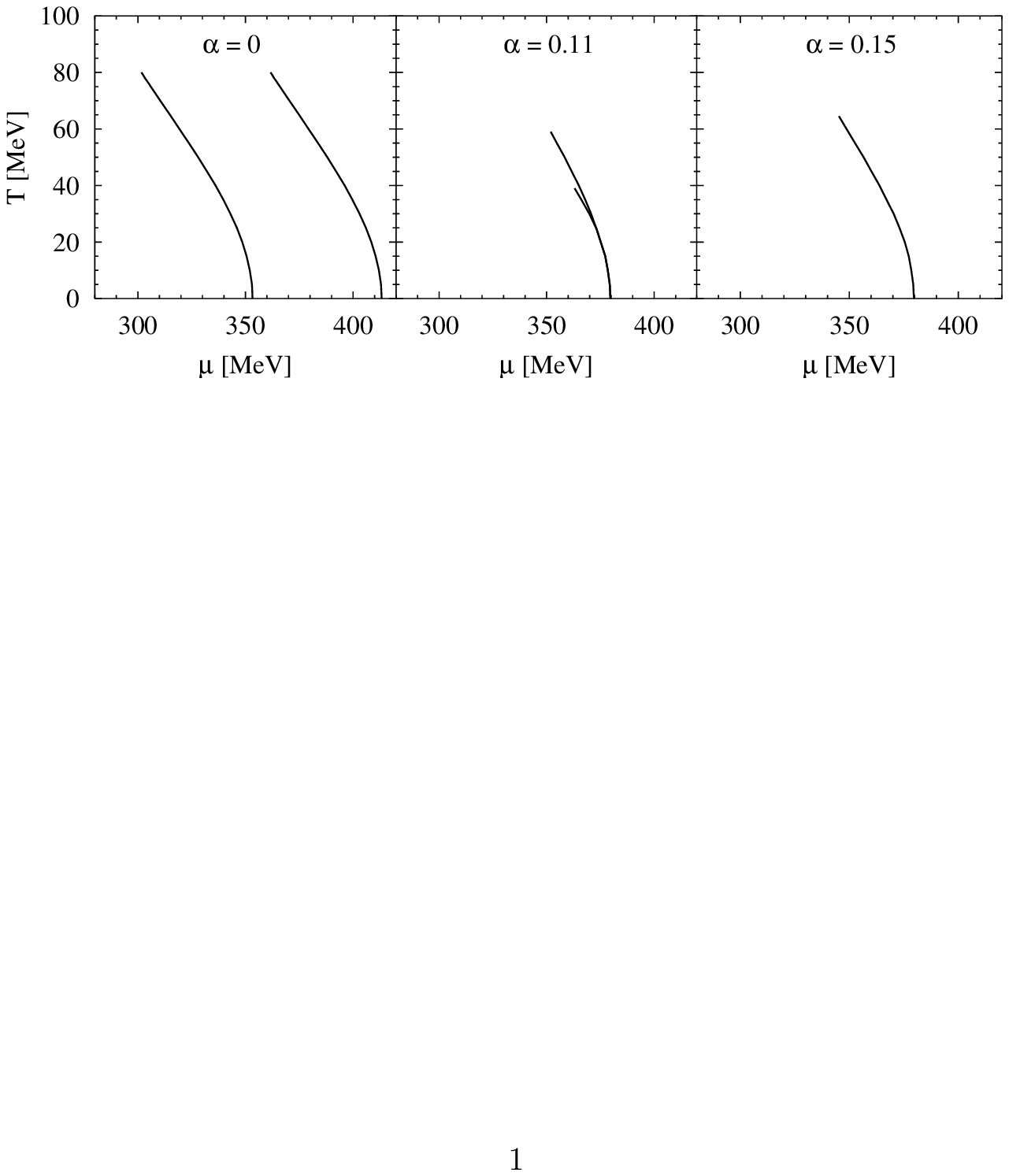}
\caption{\small Phase diagrams in the $\mu$-$T$-plane for 
$\delta\mu = 30$~MeV and $\alpha = 0$ (left), $\alpha = 0.11$ (center),
and $\alpha = 0.15$ (right). The lines correspond to first-order
phase boundaries which end in second-order endpoints.
From Ref.~\cite{FBO03}.}
\label{figflamixphase}
\end{center}
\end{figure}

Next, we extend our analysis to non-vanishing temperature.
The phase diagrams in the $\mu$-$T$ plane for fixed $\delta\mu = 30$~MeV
and three different values of $\alpha$ are shown in \fig{figflamixphase}.
At $\alpha = 0$ (left panel) we qualitatively reproduce the results 
reported in Refs.~\cite{KTV03,TK03}, i.e., two separate first-order phase 
boundaries which end in two second-order endpoints.
Again, since for $\alpha = 0$ the up and down quarks
decouple, we would obtain two identical phase diagrams
if we plotted the phase structure of flavor $f$ in the $\mu_f$-$T$ plane.
In the central panel of \fig{figflamixphase} we consider $\alpha = 0.11$,
i.e., slightly larger than the critical value $\alpha^c(T=0) = 0.104$
for a single phase transition at $T=0$.
Accordingly, there is only one phase boundary at low temperatures.
However, at $T = 25$~MeV it splits into two lines which end at two 
different second-order endpoints. 
The two branches are very close to each other though, and already
at $\alpha = 0.12$ we find only one phase boundary with a single endpoint.
This is illustrated by the diagram on the right, which corresponds
to $\alpha = 0.15$.

In our example a rather small amount of flavor mixing is sufficient
to remove the existence of the second phase transition:
Of course, there {\it must} be a single phase transition at $\alpha = 0.5$,
where $M_u$ and $M_d$ are equal (see \eq{flamixmasses}). 
(This was the case studied in Ref.~\cite{KYT02}.) However, the critical
value $\alpha^c \simeq 0.12$ we found in our example is much smaller.
At $T=0$, a rough, but perhaps more general estimate for the critical
$\alpha$ can be obtained from the observation that the phase transition 
takes place when 
the chemical potential of quark $f$ comes close to its constituent
mass, i.e., $\mu_f \approx M_i$. Applying this condition to the 
up quark we expect the first phase transition to take place
at $\mu_u \approx M_{vac}$, i.e, at $\mu \approx M_{vac} - \delta\mu$.
At this point $M_u$ drops and, according to \eq{flamixmasses},
$M_d$ drops as well. Neglecting the current quark mass, we find  
\beq
    M_d \;\approx\; -(1-\alpha)\,4G_0\, \phi_d 
        \;\lesssim\; (1-\alpha)\, M_\mathit{vac}~.
\label{md}
\eeq
If this value becomes smaller than the
value of $\mu_d$, we expect the down quarks to exhibit a phase
transition as well. Hence, we estimate 
\beq 
\alpha^c(T=0) \lesssim \frac{ 2\delta\mu}{M_\mathit{vac}}~.  
\label{ac}
\eeq
Note that this estimate would not be affected by a possible restoration
of the $U_A(1)$ symmetry at the phase boundary. 
Obviously, if $G_2$ goes to zero, $M_d$ would drop as well.

For our example, \eq{ac} gives $\alpha^c(T=0) \lesssim 0.15$.
Comparing this value with the numerical result $\alpha^c(T=0) = 0.104$, we
see that \eq{ac} is a quite conservative estimate. This is easily
understood, since in the second step of \eq{md} we have neglected
the fact, that $\phi_d$ also becomes smaller.  
Our estimate does also not include the observation, that
the critical chemical potential for the first phase transition {\it rises}
with $\alpha$.
In any case, we have to admit that our arguments cannot explain 
quantitatively why \eq{ac} seems to hold even for temperatures approaching
the critical endpoint where the quark masses do no longer drop discontinuously.

At this point one can ask, which value of $\alpha$ is ``realistic''.
All observables we have used so far to fix the parameters
($f_\pi$, $m_\pi$, and $\ave{\bar q q}$) do not depend on $\alpha$.
However, as already mentioned, for $\alpha = 0$ the interaction would
be symmetric under $U_A(1)$ transformations and consequently there 
would be another pseudo Goldstone boson, namely an isoscalar pseudoscalar 
particle, degenerate with the pions. This is of course unrealistic.
Turning on $\alpha$, the $U_A(1)$ symmetry becomes explicitly broken 
(in addition to the mass term in the Lagrangian) and the mass of the 
isoscalar meson is shifted upwards. 
Since in a pure two-flavor world this particle should be identified with 
the $\eta$ meson, one way to fix $\alpha$ is to fit the
physical $\eta$ mass. In this way one finds $\alpha = 0.11$.

However, the description of the $\eta$ meson without strange quarks 
is not very realistic. For a better way to determine $\alpha$ 
we should therefore refer to the three-flavor NJL model. For three flavors the instanton-induced
interaction which plays the role of ${\cal L}_2$ is a six-point interaction
(in general, for $N_f$ flavors it is a $2N_f$-point interaction),
and gives rise to the $\eta-\eta'$ splitting. More details of the
three-flavor NJL model will be discussed in the next chapter.
Here we just refer to \eq{njlmasses3} for the constituent quark masses.
When we compare this equation with \eq{flamixmasses} we can identify 
$G_1 = G$ and $G_2 = -\frac{1}{2} K \phi_s$ and thus
\beq
    \alpha = \frac{- K \phi_s}{2G - K \phi_s}~.
\label{alphaeff}
\eeq
The parameters $G$ and $K$ and the quark condensate
$\phi_s$ have been determined by several authors by fitting the 
masses of the pseudoscalar octet and are listed in \tab{tabnjl3fit}. 
If we take, for instance, the values of set RKH~\cite{Rehberg}, 
we find $\alpha \simeq 0.21$.
For the parameters of set HK~\cite{hatsuda} we get a somewhat smaller
value, $\alpha \simeq 0.16$. 
On the other hand, the success of the instanton liquid model to describe
vacuum correlators~\cite{SS98} would suggest that ${\cal L}_2$ is the 
dominant part of our Lagrangian, i.e., $\alpha\simeq 1$.
In all these cases we would find only one phase transition for 
$\delta\mu = 30$~MeV. (For the pure
$SU(2)$ fit of the $\eta$ meson we would just be at the intermediate case
depicted in the central panel of \fig{figflamixphase}.)

Typical values of $|\delta\mu|$ in heavy-ion collisions are likely to be
smaller than that. 
A simple estimate, assuming the density ratio  
$n_u : n_d = 290 : 334$ as in $^{208}$Pb, and the approximate relation
$n_u : n_d \approx (\mu_u : \mu_d)^3$ yields $\delta\mu \approx -10$~MeV
for $\mu = 400$~MeV.
Empirically, one finds $\delta\mu = -2.5$~MeV, at chemical freeze-out
for Pb-Pb collisions at SPS~\cite{BMHS99} and $\delta\mu = -6$~MeV for 
Si+Au collisions at AGS~\cite{heppediplom}.)
This would mean, that it is very unlikely to ``see'' two phase boundaries
in heavy-ion collisions.

Of course, before drawing quantitative conclusions,
we should recall the shortcomings of the model. 
As pointed out earlier, the description of the ``hadronic phase'' as a gas 
of quarks, rather than hadrons, is unrealistic and 
any prediction of the critical endpoint(s)
in non-confining mean-field models should not be trusted.
However, keeping this in mind, our results
show that flavor-mixing effects cannot be
neglected in the discussion of the phase diagram. The very existence
of these effects is related to instantons and the $U_A(1)$-anomaly of
QCD. Of course their magnitude is a matter of debate,
but they are likely to cancel the interesting phenomena discussed in
Refs.~\cite{KTV03,TK03}.

\chapter{Three-flavor systems}
\label{njl3}

In this chapter we extend our analysis to three quark flavors.
While most features of the two-flavor NJL model which we have discussed 
in the previous chapter remain qualitatively unaffected,
the main difference comes about from the fact that
the mass of the strange quark cannot be chosen equal to the 
non-strange quark mass in realistic calculations. 
This means we have to deal with an explicitly broken $SU(3)$ symmetry,
and thus 
$\ave{\bar ss} \neq \ave{\bar uu}$, even for equal chemical potentials.
In particular the chiral limit is in general not a good approximation
to the model with realistic masses. Therefore the finite-mass effects 
we have already encountered in the two-flavor case become much more
pronounced in the three-flavor model.

\section{Vacuum properties}
\label{njl3vac}

\subsection{Lagrangian}
\label{njl3L}

The three-flavor version of the NJL model has been developed in the mid
80s~\cite{EbRe86,BJM87,HaKu87b} and has been investigated by many authors 
since then.
The most commonly used Lagrangian reads~\cite{Rehberg,Kunihiro}
\beq
     {\cal L} \;=\; \bar q ( i \delsl \,-\, {\hat m}) q
     \;+\; {\cal L}_\mi{sym} \;+\; {\cal L}_\mi{det}~,
\label{LNJL3}
\eeq
where $q = (u,d,s)^T$ denotes a quark field with three flavors, and
$\hat m = diag_f(m_u, m_d, m_s)$ is the corresponding mass matrix.
Throughout this report we will assume isospin symmetry on the Lagrangian
level, $m_u = m_d$, whereas $m_s$ will in general be
different, thus explicitly breaking $SU(3)$-flavor symmetry. 
The Lagrangian contains two independent interaction terms which are 
given by
\beq
{\cal L}_\mi{sym} \;=\; G \sum_{a=0}^8 \Big[\,(\bar q \tau_a q)^2 + 
                 (\bar q\, i\gamma_5\tau_a q)^2\,\Big] 
\label{LNJL4}
\eeq
and
\beq
{\cal L}_\mi{det} \;=\;-\; K \,\Big[ \,{\det}_f (\bar q(1+\gamma_5) q) 
                         + {\det}_f (\bar q(1-\gamma_5) q) \,\Big]~.
\label{LNJL6}
\eeq
These terms may be seen as the three-flavor version of \eqs{Lflamix1} 
and (\ref{Lflamix2}) and can straight forwardly be generalized to 
to any number of flavors, $N_f$. 
${\cal L}_\mi{sym}$ is then a $U(N_f)_L \times U(N_f)_R$ 
symmetric 4-point interaction, where $\tau_a$, $a=1,\dots,(N_f^2-1)$ denote 
the generators of $SU(N_f)$, while $\tau_0$ is proportional
to the unit matrix. All $\tau_a$ are normalized such that 
${\rm tr}\,\tau_a\tau_b = 2\delta_{ab}$. For two flavors, this
is fulfilled by $\tau_0 = \unity_f$ and the Pauli matrices $\tau_a$, 
$a=1,2,3$. Then ${\cal L}_\mi{sym}$ becomes identical to ${\cal L}_1$
in \eq{Lflamix1} with $G_1=G$. 
Now, for three flavors, $\tau_0 = \sqrt{2/3}\,\unity_f$
and $\tau_a$, $a=1,\dots,8$ are the eight Gell-Mann matrices\footnote{
Traditionally, the Gell-Mann matrices 
are denoted by $\lambda_a$. However, for later convenience we reserve
this symbol for $SU(3)$-color generators and keep the notation
$\tau_a$ for $SU(3)$-flavor generators.}.

\begin{figure}
\begin{center}
\epsfig{file=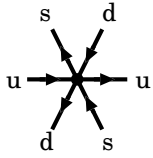,width= 2.0cm}
\end{center}
\vspace{-0.5cm}
\caption{\small Flavor structure of the six-point vertex 
('t Hooft interaction).}       
\label{figsixv}
\end{figure}

The second term, ${\cal L}_\mi{det}$, corresponds to the 't Hooft 
interaction and is a determinant in flavor space.
This means, it is a maximally flavor-mixing $2N_f$-point interaction,
involving an incoming and an outgoing quark of each flavor.
Thus, for two flavors, ${\cal L}_\mi{det}$ is a four-point interaction and 
one can easily check that it is equal to ${\cal L}_2$ in \eq{Lflamix2} 
with $G_2 = -K$.
For three flavors we have a six-point interaction of the form
\beq
    {\det}_f\,(\bar q \,{\cal O}\, q) \;:=\; \sum_{i,j,k} \varepsilon_{ijk}\;
    (\bar u \,{\cal O}\, q_i) \, (\bar d\,{\cal O}\,q_j) 
    \, (\bar s\, {\cal O}\, q_k)~,
\eeq 
where $i,j,k$ are flavor indices (see \fig{figsixv}). 

${\cal L}_\mi{det}$ is $SU(N_f)_L \times SU(N_f)_R$ 
symmetric, but it breaks the $U_A(1)$ symmetry which was left unbroken
by ${\cal L}_\mi{sym}$.
It thus translates the $U_A(1)$ anomaly, which in QCD arises 
at quantum level from the gluon sector, to a tree-level interaction in 
a pure quark model. As discussed earlier, this term is phenomenologically
important to get the correct mass splitting of the $\eta$ and $\eta'$ mesons.
In the chiral limit
($m_u=m_d=m_s=0$), the $\eta'$ mass is lifted to a finite value by
${\cal L}_\mi{det}$, while the other pseudoscalar mesons, including the
$\eta$, remain massless.

Like in the two-flavor case, there are many other terms which are consistent
with the symmetries and which could be added to the Lagrangian.
For instance, in Ref.~\cite{LKW92}
vector and axial-vector four-point interactions have been taken into account 
in addition to the terms given above.
A complete list of possible four-point and six-point terms is discussed in 
Ref.~\cite{KLVW90}.
For simplicity, however, we will restrict ourselves to the Lagrangian
defined above.

\subsection{Gap equation and meson spectrum}
\label{njl3gap}

\begin{figure}
\begin{center}
\epsfig{file=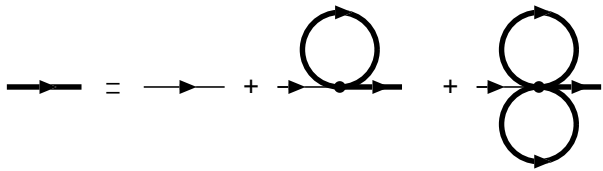,width= 8.0cm}
\end{center}
\vspace{-0.5cm}
\caption{\small Dyson equation for the quark propagator in the three-flavor
         NJL model (Hartree approximation).}       
\label{fignjl3gap}
\end{figure}

Apart from the explicit $SU(3)$-flavor breaking by the strange quark mass,
the main complication of the three-flavor version of the NJL model
as compared with the two-flavor case is caused by the six-point vertices
which arise from the 't Hooft interaction. 
For the gap equation this means that there is an additional term involving
two quark loops. This gives rise to the equation 
\beq
     M_i = m_i - 4\,G\,\phi_i +2 \,K\,\phi_j\phi_k~,
     \qquad (i,j,k) \;=\; \text{any permutation of} \;\, (u,d,s)~,
\label{njlmasses3}
\eeq   
which is diagrammatically shown in \fig{fignjl3gap}. It contains a 
non-flavor mixing term proportional to the coupling constant $G$ and
a flavor mixing term proportional to the coupling constant $K$. 

\begin{figure}
\begin{center}
\epsfig{file=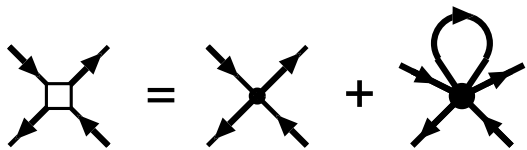,width= 7.0cm}
\end{center}
\vspace{-0.5cm}
\caption{\small Effective four-point vertex.}       
\label{figsixpoint}
\end{figure}

For describing mesons one first constructs effective four-point vertices
as a sum of the genuine four-point vertices and the six-point vertices 
with one closed loop, see \fig{figsixpoint}. These effective four-point
vertices are then used as scattering kernels in the Bethe-Salpeter
equation, \fig{fignjlrpa}. Although technically more involved, mostly
because of the unequal strange and non-strange quark masses 
(leading, among other things, to octet-singlet mixing in the 
$\eta$-$\eta'$ subspace), the basic
mechanism is the same and will not be presented here. 
For further details see, e.g., Refs.~\cite{Rehberg,KLVW90}. 

The model specified in \sect{njl3L} contains five parameters:
the bare masses $m_s$ and $m_u$, 
the coupling constants $G$ and $K$, and the cut-off $\Lambda$.
Thus, compared with the simplest version of the two-flavor model,
\eq{LNJL}, we have two more parameters: $m_s$ and the six-point coupling.  
(As we have seen in \sect{flamix}, \eq{LNJL} corresponds to a particular
choice of the more general Lagrangian \eq{Lflamix}. From this point of view,
there is even only one additional parameter.) 
On the other hand there are at least three additional observables,
namely the masses of the pseudoscalar mesons $K$, $\eta$, and $\eta'$.
Therefore, in the three-flavor model, the parameters are in principle
much better constrained than in the two-flavor model. 
In fact, whereas for two flavors we had to invoke the poorly known 
quark condensate, we now have five well-known observables, $f_\pi$, $m_\pi$,
$m_K$, $m_\eta$, and $m_{\eta'}$, and one would naively expect that one
can uniquely fix the five parameters. 

It turns out that this is not quite the case. 
In \tab{tabnjl3fit} we have listed
three different parameter sets taken from the literature, together
with related quantities in the quark and meson sectors. 
The first two sets correspond to the fits of Rehberg, Klevansky, and
H\"ufner (RKH)~\cite{Rehberg} and of Hatsuda and Kunihiro 
(HK)~\cite{hatsuda}\footnote{The parameters of Ref.~\cite{hatsuda} are almost,
but not exactly the same as the parameters of the earlier fit by 
Kunihiro~\cite{Kunihiro}.}, respectively. 
Instead of fitting five mesonic quantities, 
both groups have set $m_u$ to a value of 5.5~MeV, 
taken from Refs.~\cite{gasserleutwyler,MNS90}, and fixed the remaining 
four parameters by fitting $f_\pi$, $m_\pi$, $m_K$, and $m_{\eta'}$ to their 
empirical values. In this way the mass of the $\eta$-meson is
underestimated by 6\% in RKH and 11\% in HK.

\begin{table}[h!]
\begin{center}
\begin{tabular}{|l| c c c |c|}
\hline
&&&&\\[-3mm]
    & RKH \cite{Rehberg} & HK \cite{hatsuda} & LKW \cite{LKW92} 
    & empirical~\cite{PDG}
\\[1mm]
\hline
&&&&\\[-3mm]
$\Lambda$ [MeV] &  602.3 & 631.4 & 750 &
\\
$G\Lambda^2$    & 1.835 & 1.835 & 1.82 &
\\
$K\Lambda^5$    & 12.36 & 9.29 & 8.9 &
\\
$m_{u,d}$ [MeV] & 5.5  & 5.5 & 3.6 & 3.5 - 7.5
\\
$m_s$ [MeV]     & 140.7 & 135.7 & 87 & 110 - 210
\\
$G_V/G$ & --- & --- & 1.1 &
\\
\hline
$M_{u,d}$ [MeV] & 367.7 & 335 & 361 &
\\
$M_s$ [MeV] & 549.5 & 527 & 501 &
\\
$(\ave{\bar uu})^{1/3}$ [MeV] & -241.9 & -246.9& -287 &
\\
$(\ave{\bar ss})^{1/3}$ [MeV] & -257.7 & -267.0& -306 &
\\
$B$ [MeV/fm$^3$] & 291.7 & 295.5 & 350.0 &
\\
\hline 
$f_\pi$ [MeV] & 92.4 & 93.0 & 93 &  92.4~\cite{Holstein}
\\
$m_\pi$ [MeV] & 135.0 & 138 & 139 & 135.0, 139.6
\\
$m_K$ [MeV]   & 497.7 & 496 & 498 & 493.7, 497.7
\\
$m_\eta$ [MeV] & 514.8 & 487 & 519 & 547.3
\\
$m_{\eta'}$ [MeV] & 957.8 & 958 & 963 & 957.8
\\
$m_{\rho,\omega} $ [MeV]  & --- & --- & 765 & 771.1, 782.6
\\
$m_{K^*}$ [MeV] & --- & --- & 864 & 891.7, 896.1
\\
$m_\phi$ [MeV]  & --- & --- & 997 & 1019.5
\\
\hline
\end{tabular}
\end{center}
\caption{\small Three sets of parameters and related quark and meson
         properties in the three-flavor NJL model. 
         The empirical quark masses listed in Ref.~\cite{PDG} have been 
         rescaled to a renormalization scale of 1~GeV by multiplying them 
         by 1.35~\cite{PDG}. The values given for the light quarks 
         correspond to the average $(m_u+m_d)/2$.}
\label{tabnjl3fit}
\end{table}  

However, although both groups apparently used the same prescription,
the resulting parameter sets are not identical.
The most striking difference is found in the six-point coupling.
Comparing the dimensionless combinations $K\Lambda^5$, the RKH value
is more than 30\% larger than the HK value. (If we compare the values
of $K$ the difference is even 70\%.)
The reason for this discrepancy lies in the different treatment of the 
$\eta'$-meson~\cite{Costa}. 
Because of its large mass ($m_{\eta'}=958$~MeV), in both cases the $\eta'$ 
is above the threshold for $q\bar q$-decay, and the $q\bar q$-polarization
diagram receives an unphysical imaginary part. 
In RKH this was accepted as an unavoidable feature of the NJL model and the
authors defined the $\eta'$-mass as the real part of the corresponding
pole in the complex plane. In HK, on the contrary, the imaginary part of
the polarization function has been discarded by hand and only the real part
was retained in order to determine $m_{\eta'}$. 
Of course, since the real part is linked to the imaginary part via
dispersion relations, this prescription does not completely remove the 
unphysical effects. This leaves a general uncertainty which is reflected 
in the difference between the parameter sets RKH and HK. 

As a third example, the parameters of Lutz, Klimt, and Weise 
(LKW)~\cite{LKW92} are
also shown in \tab{tabnjl3fit}. In addition to \eqs{LNJL4} and (\ref{LNJL6}),
the authors considered a vector and axial-vector interaction term,
which enabled them to fit the vector-meson nonet ($\rho$, $\omega$,
$K^*$, and $\phi$) as well. In the pseudo-scalar meson sector they
obtained similar results as RKH and HK. However, because the longitudinal 
part of the axial-vector interaction mixes with the pseudoscalar interaction
(``$\pi a_1$-mixing''), this sector is not independent of the vector
coupling constant $G_V$. As a consequence, LKW find a relatively large 
cut-off and relatively small bare quark masses. Moreover, the six-point 
coupling is even weaker than in HK.   

In spite of these differences, it remains generally true that the parameters
of the three-flavor model are much better constrained than in the 
two-flavor case. In fact, the resulting constituent quark masses in vacuum
are quite similar for the three parameter sets, ranging from about
335~ to 370~MeV for up and down quarks and from 500 to 550~MeV for
strange quarks. The same is true for the bag constants, which are
defined in an analogous way to the two-flavor case (see below). 
They are almost equal for RKH and 
HK, while the value for LKW is about 20\% larger.
In the numerical calculations presented in this work, we will mostly
employ the parameters of RKH.

\section{Thermodynamics}
\label{njl3thermo}

\subsection{Formalism}
\label{njl3thform}

The formalism of Secs. \ref{njlpot} and \ref{flamix}
is straight-forwardly generalized to the three-flavor model.
Allowing for three independent chemical potentials $\mu_f$ for the three
flavors $f=u,d,s$, the mean-field thermodynamic potential in the 
presence of the quark condensates $\phi_f = \ave{\bar q_f q_f}$ reads
\beq
     \Omega(T,\{\mu_f\}; \{\phi_f\}) 
     = \sum_{f=u,d,s} \Omega_{M_f}(T,\mu_f)
     + 2\,G\,(\phi_u^2 + \phi_d^2 + \phi_s^2) - 4\,K\phi_u\phi_d\phi_s
     \;+\; \mi{const}.~,
\label{njl3omega}
\eeq
where $\Omega_{M_f}(T,\mu_f)$ is given in \eq{flamixom0} and corresponds
to the contribution of a gas of quasiparticles with mass $M_f$.
The latter is related to the various $\phi_i$ via \eq{njlmasses3}.
Again, the thermodynamically consistent solutions correspond to the 
stationary points of $\Omega$, where $\delta\Omega/\delta\phi_f = 0$.
One finds that $\phi_f$ is given by \eq{phif}, which now has 
to be evaluated self-consistently with \eq{njlmasses3}, forming a set
of three coupled gap equations for the constituent masses. 
The irrelevant constant in \eq{njl3omega} is conveniently
chosen such that the minimal solution of $\Omega$ in vacuum is zero.

Once the self-consistent solutions are found, other thermodynamic
quantities can be derived in the standard way. In particular we can
calculate the pressure and the energy density,
\beq
    p \;=\; -\Omega~,\qquad
    \epsilon \;=\; \Omega + Ts + \sum_f \mu_f\,n_f~,
\eeq
where $s=\partial\Omega/\partial T$ is the entropy density and
$n_f=-\partial\Omega/\partial\mu_f$ are the number densities of the 
quarks of flavor $f$. The total quark number density and the baryon 
number density are given by $n = \sum_f n_f$ and $\rho_B = n/3$,
respectively.

Finally, in analogy to \eq{njlb}, we define the bag constant as 
\beq
     B \;=\; \Omega(T=0,\{\mu_f=0\};\{\phi_f = 0\})~. 
\label{njl3b}
\eeq

\subsection{Quark masses and effective bag constants at non-zero density}
\label{njl3beff}

\begin{figure}
\begin{center}
\epsfig{file=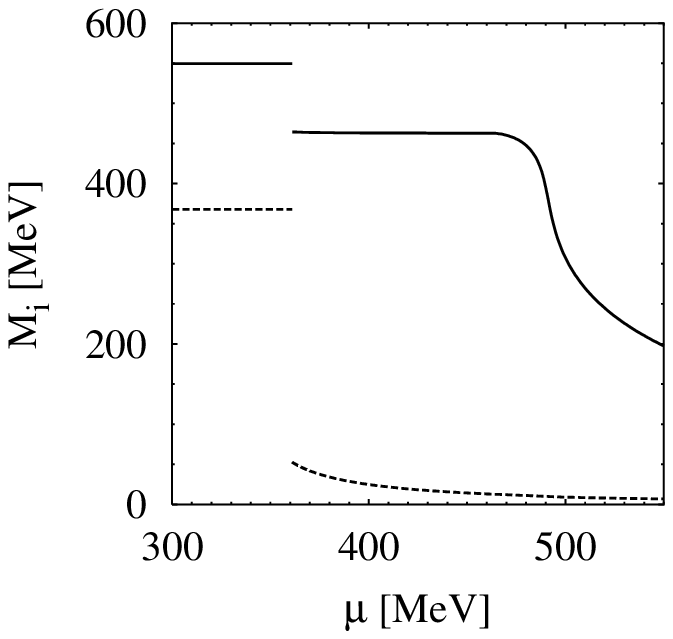,width=7.0cm}\hspace{0.5cm}
\epsfig{file=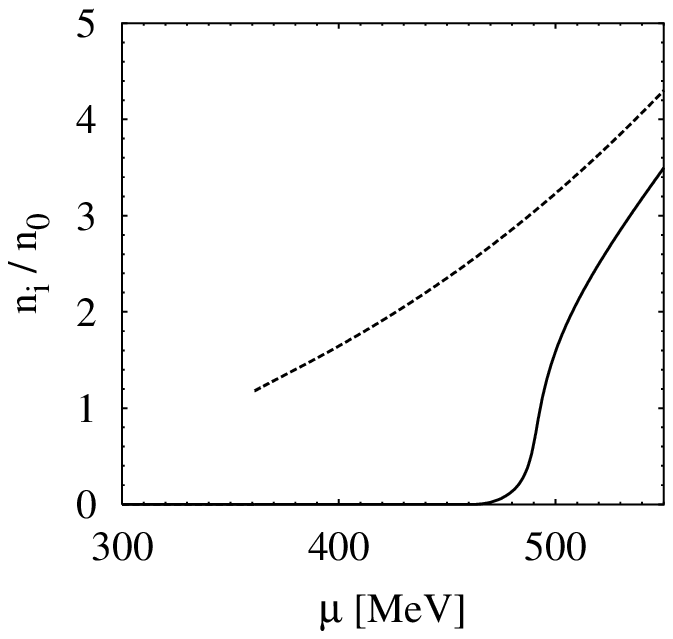,width=7.0cm}
\end{center}
\vspace{-0.5cm}
\caption{\small Quark matter properties as functions of a common quark
         number chemical potential $\mu$ (parameter set RKH~\cite{Rehberg}).
         Left: Constituent quark masses $M_u = M_d$ (dashed) and $M_s$
         (solid).
         Right: Number densities $n_u = n_d$ (dashed) and $n_s$ (solid)
         in units of $n_0 = 3\rho_0 = 0.51 {\rm fm}^{-3}$.}
\label{fignjl3rhomass}
\end{figure}

In the following we restrict ourselves to $T=0$.
In \fig{fignjl3rhomass} the constituent quark masses $M_i$ (left panel)
and the flavor densities $n_i$ (right panel) are plotted
as functions of a common quark number chemical potential $\mu$. 
Quantities related to the up and down quarks are indicated by dashed lines,
those related to the strange quarks by solid lines.
The results were obtained using the model parameters RKH~\cite{Rehberg}.   

At $\mu = \mu_c = 361$~MeV we find a first-order phase transition, 
where $M_u=M_d$ (dashed line) drops from 367.6~MeV to 52.5~MeV.
At this point the total baryon number density jumps from zero to
about $2.4 \rho_0$, equally carried by up and down quarks, while 
the density of strange quarks remains zero.  
Nevertheless, because of flavor-mixing, $M_s$ does not stay 
constant but drops at $\mu_c$ from 549.5 to 464.4~MeV.
Above $\mu_c$, the contributions of $\phi_u$ and $\phi_d$ to $M_s$ are
negligible and $M_s$ stays almost constant until $\mu$ exceeds $M_s$
and $n_s$ becomes non-zero as well. 
The behavior of the constituent masses shows some similarities with the
two-flavor model at finite isospin chemical potential with small flavor
mixing (central panel of \fig{figflamixmass}), although instead of a second
phase transition we now find a smooth cross-over above the strange quark
threshold. 
The main difference is that in the present example the flavor symmetry
is not broken by unequal chemical potentials, but be unequal masses.

It is again instructive to compare the NJL-model equation of state with
the bag model one. As discussed in \sect{njlbag}, both models behave
almost identically in the chiral limit, but already for bare quark masses
of a few MeV we found some differences, which could be expressed in
terms of density dependent effective quark masses and bag constants. 
In the three-flavor case, where chiral symmetry is broken much stronger
by the strange quark, we should expect much bigger effects.

Already in the two-flavor model, both, the constituent quark
masses and the effective bag constants, are in general not only functions 
of the total density, but also on the flavor composition, i.e., they
depend on each flavor density $n_i$ separately.
This point has not been discussed in \sect{njlbag} where we only considered 
isospin symmetric quark matter. 
However, in three-flavor systems, a restriction to equal flavor densities
is no longer a natural choice, as obvious from \fig{fignjl3rhomass}.
Therefore we generalize our previous definition of $B_\mi{eff}$,
\eq{Beff}, to include arbitrary flavor compositions, 
\beq
    \epsilon(n_u, n_d, n_s) \;=\;  \frac{N_c}{\pi^2} \sum_f 
    \int_0^{p_F^f} dp\,p^2\,\sqrt{p^2 + M_f^2(n_u, n_d, n_s)}
    \;+\; B_\mi{eff}(n_u, n_d, n_s)~,
\label{Beff3}
\eeq
where $p_F^f$ is the Fermi momentum of flavor $f$.

\begin{figure}
\begin{center}
\epsfig{file=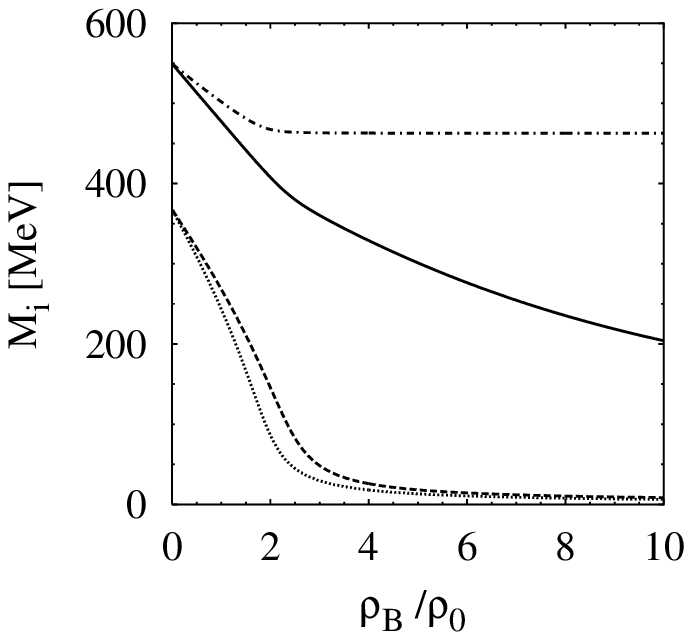,width=7.0cm}\hspace{0.5cm}
\epsfig{file=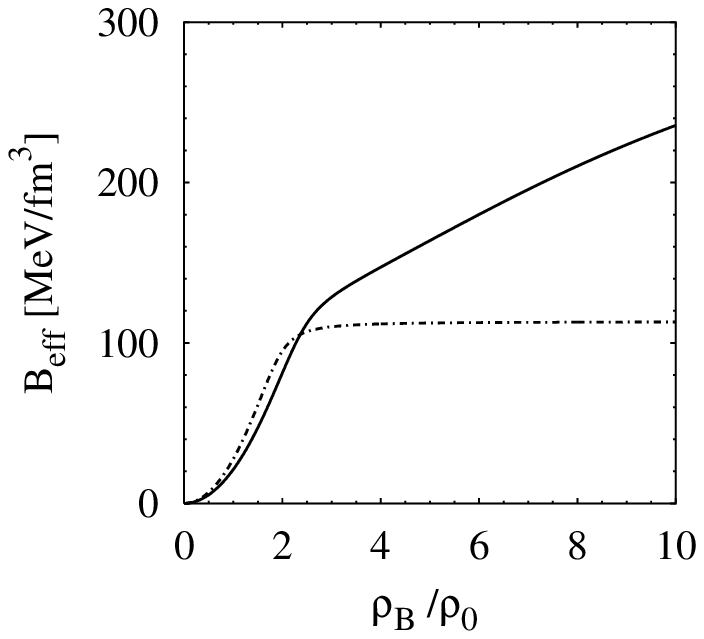,width=7.0cm}
\end{center}
\vspace{-0.5cm}
\caption{\small Various quantities for isospin symmetric matter, $n_u = n_d$,
         as functions of the total baryon number density $\rho_B$ 
         (parameter set RKH~\cite{Rehberg}).
         Left: Constituent quark masses for symmetric matter, 
         $n_u = n_d = n_s$: $M_u = M_d$ (dashed) and $M_s$ (solid),
         and without strangeness, $n_s=0$: $M_u = M_d$ (dotted) and 
         $M_s$ (dash-dotted).  
         Right: Effective bag constant $B_\mi{eff}$ for symmetric matter
         (solid) and for $n_s=0$ (dash-dotted).}
\label{fignjl3massbeffrho}
\end{figure}

To illustrate both, density dependence and dependence on flavor
composition, the constituent quark masses and the effective bag 
constant are displayed in \fig{fignjl3massbeffrho} for two different
cases. 
The first case corresponds to equal densities $n_u = n_d = n_s$.
The corresponding constituent quark masses as functions of $\rho_B$
are indicated by the dashed line ($M_u=M_d$) and by the solid line
($M_s$) in the left panel of the figure.  
As we have seen earlier, the constituent masses become equal to the 
current masses only at ``asymptotic'' densities, i.e., when the 
Fermi momentum reaches the cut-off. (In the present example this 
is $\rho_B = 17\rho_0$.) 
At this point all condensates vanish and
the effective bag constant $B_\mi{eff}$ (solid line in the right panel)
reaches the value of the bag constant $B = 291.7 {\rm MeV/fm}^3$. 
However, whereas in the non-strange sector chiral symmetry gets
quickly restored, at least on an absolute scale, this is not the
case for the strange sector, and neither $M_s$ nor $B_\mi{eff}$ can be
approximated by a constant above a certain density.

In the second case we consider equal non-strange densities $n_u = n_d$, 
but vanishing strangeness, $n_s = 0$. In this case the strange quark mass 
(dash-dotted line in the left panel) only drops through its mixing with 
the up and down masses (dashed line) and stays practically constant above 
$\rho_B \simeq 2 \rho_0$, where the latter are small. 
Obviously, the asymptotic value $M_s^\infty$ is just the vacuum mass
one would obtain for a vanishing six-point coupling. 
For the present example one finds $M_s^\infty = 462.8$~MeV, corresponding
to a strange quark condensate $\phi_s^\infty = (-251.6\,{\rm MeV})^3$.  
This means, chiral symmetry never gets restored in the strange sector,
and, as a consequence, $B_\mi{eff}$ (dash-dotted line in the right panel) 
does not approach the bag constant $B$, but the asymptotic value 
\beq
     B_\infty \;=\; 
    \Omega(T=0,\{\mu_f=0\};\phi_u=0,\phi_d=0,\phi_s=\phi_s^\infty )~. 
\label{njl3binf}
\eeq
This can be considerably smaller than $B$: In the present example
$B_\infty = 113.2~\MeV/\fm^3$, compared with $B =  291.7~\MeV/\fm^3$. 
Hence, at high enough densities, the system behaves like a two-flavor 
bag model with a bag constant $B_\infty$.

\begin{figure}
\begin{center}
\epsfig{file=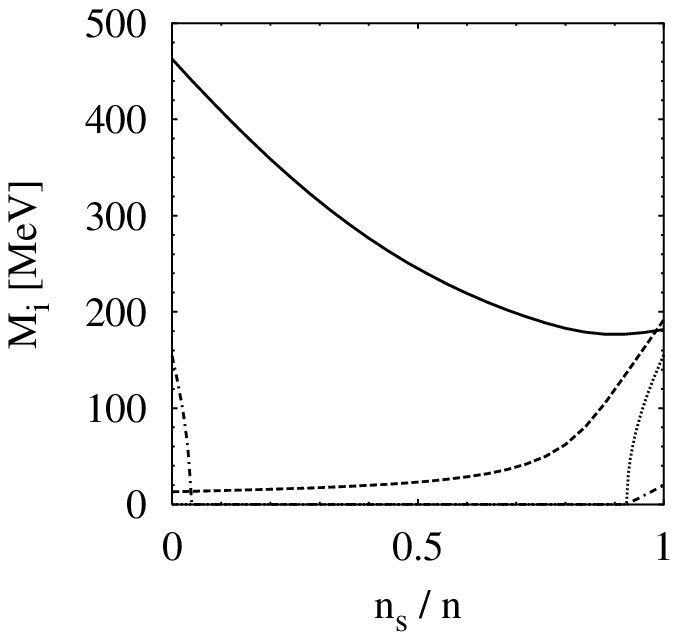,width=7.0cm}\hspace{0.5cm}
\epsfig{file=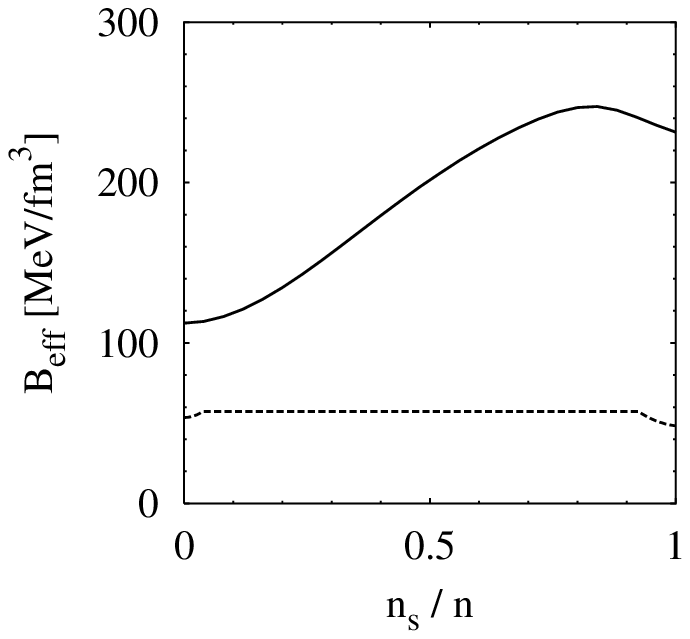,width=7.0cm}
\end{center}
\vspace{-0.5cm}
\caption{\small Various quantities as functions of the fraction of strange
         quarks, $n_s/n$, for isospin symmetric matter ($n_u=n_d$)
         at fixed total baryon number density $\rho_B = 5\rho_0$.
         The results were obtained using the model parameters
         RKH (see \tab{tabnjl3fit}) and the corresponding chiral limit
         ($m_u=m_d=m_s=0$).        
         Left: constituent quark masses $M_u=M_d$ (RKH: dashed, 
         chiral limit: dotted), $M_s$ (RKH: solid, chiral limit: dash-dotted). 
         Right: effective bag constant $B_\mi{eff}$ (RKH: solid, chiral 
         limit: dashed).}
\label{fignjl3massbeffrs}
\end{figure}

The two examples demonstrate, that the dependence of the effective quark
masses and bag constants on the flavor composition can be large. 
A complementary view on this point is given in \fig{fignjl3massbeffrs}
where the constituent masses and the effective bag constant
are displayed as functions of the strangeness fraction $n_s/n$
for $n_u=n_d$ at fixed baryon number density $\rho_B = 5\rho_0$.
At not too large values of $n_s/n$ we find a strong decrease of $M_s$
(left panel, solid line) and a weak increase of the non-strange masses 
(dashed line) with increasing $n_s/n$. This is easily understood from the 
fact that the density of strange quarks increases while the density of 
non-strange quarks is large, but decreases. 
The small increase of $M_s$ at very large $n_s/n$ is a flavor-mixing 
effect and related to the steeper increase of $M_u=M_d$ in this
regime. 
From the behavior of the constituent masses (and thus the condensates)
we can also qualitatively understand the behavior of the effective
bag constant (right panel, solid line), which is mostly rising and 
only turns around at large $n_s/n$. 

For comparison we have also plotted the results for the corresponding
chiral limit, ($m_u=m_d=m_s=0$). In this case one finds
a large regime ($0.04 < n_s/n < 0.92$) where chiral symmetry is exactly
restored, i.e., all constituent masses vanish. In this regime $B_\mi{eff}$
is equal to $B$ (= 57.3~MeV/fm$^3$ in the chiral limit) and therefore
constant. While this is in agreement with a bag model 
description, it is in sharp contrast to the NJL-model with realistic
quark masses where $B_\mi{eff}$  approximately doubles its value 
in the same regime. 

\section{Stability of strange quark matter}
\label{sqm}

\subsection{The strange quark matter hypothesis}
\label{sqmhypo}

The results of the previous sections may have interesting 
consequences for the existence of absolutely stable strange quark 
matter.

In 1984 Witten suggested that there could be a so-far unobserved
form of matter, ``strange quark matter'' (SQM), which is bound more 
strongly than ordinary nuclei and thus forms the true ground state of 
strongly interacting matter~\cite{Witten}.
In contrast to nuclei, where quarks are confined to individual color-less 
nucleons, SQM is supposed to be an extended or even macroscopic piece
of matter which is composed of {\it deconfined} up, down, and strange 
quarks. 
Witten's paper immediately attracted great attention and stimulated
a large number of further investigations,
although similar ideas had come up much earlier (Bodmer (1971)~\cite{Bodmer}).
 
Witten's original motivation for his conjecture was to give a possible
solution to the dark matter problem in terms of QCD effects.  
Later it was shown that most of the SQM possibly produced in the
early universe would quickly have converted into normal hadronic matter by 
evaporating nucleons~\cite{AlFa85}. Therefore SQM cannot be of cosmological
importance, even if it is the absolute ground state of matter.

Nevertheless, the existence of absolutely stable SQM -- besides
being interesting by itself -- could have other interesting consequences. 
For instance, there could be so-called ``strange stars'', i.e.,
compact stars entirely made of SQM. 
Being self-bound objects they could be arbitrarily small, in sharp contrast
to conventional neutron stars, which are bound by gravitation and therefore 
have a minimal radius of about 10~km (see, e.g., Ref.~\cite{Glendenning})
or even 12~km~\cite{LaPr00}.
Consequently the reported discovery of a compact star with a radius 
of 3.8-8.2~km~\cite{Drake02} received tremendous attention. 
However, the determination of compact star radii is of course very difficult
and a radius of 10-14~km was obtained in Ref.~\cite{WaLa02} for the same
object. In fact, so far all strange star candidates are highly controversial.

Another interesting scenario is the production of small lumps of SQM,
so-called ``strangelets'', in heavy-ion collisions.
For this it would be sufficient if SQM was stable with respect to
strong interactions, but not necessarily against weak decays.
On the other hand the production of positively charged absolutely stable
strangelets could in principle trigger a conversion of the earth into SQM.
This was one of the ``disaster scenarios'' discussed when RHIC was
commissioned~\cite{JBSW00}. Clearly, the most convincing argument against
this possibility is the existence of the Moon in spite of its long-term
exposure to high energetic cosmic rays~\cite{JBSW00}.

At first sight, the hypothesis of absolutely stable SQM seems to contradict 
the empirical stability of nuclei. 
In fact, we can immediately exclude the stability of non-strange quark
matter (NSQM) consisting of deconfined up and down quarks.
The essential point is that SQM may only be stable if it contains a large 
fraction of strange quarks, $n_s \approx n_u \approx n_d$. 
Since hypernuclei, i.e., nuclei which contain hyperons, have higher masses
than ordinary nuclei of the same mass number, this state cannot be reached 
via a series of
subsequent weak decay processes, but only via the simultaneous decay
of many quarks, associated with a very long lifetime\footnote{In addition 
it is possible that, because of surface effects, SQM is only absolutely 
stable for a very large number of particles, e.g., $N > 10^7$. Then
the decay of nuclei would not even be favored energetically.}. 

\begin{figure}
\begin{center}
\epsfig{file=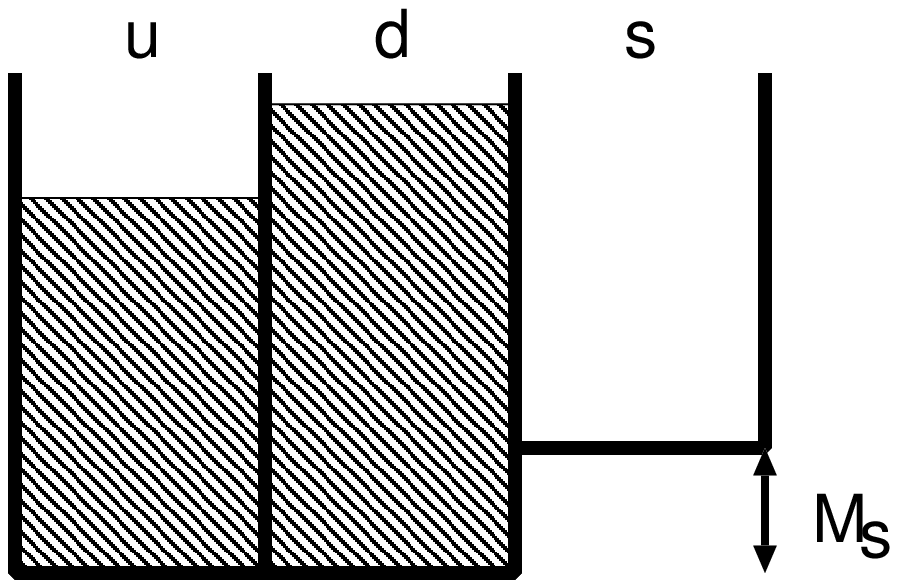,width=5.0cm}
\hspace{1.0cm}{$\longrightarrow$}\hspace{1.0cm}
\epsfig{file=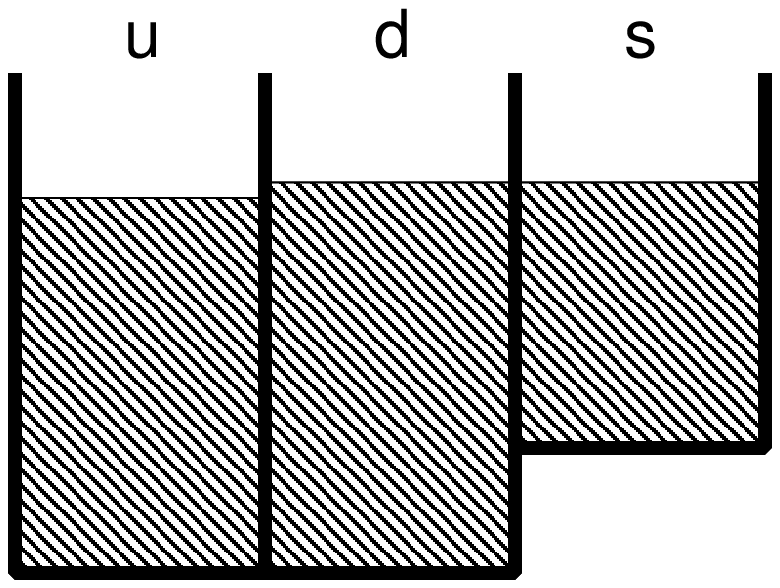,width=4.3cm}
\end{center}
\vspace{-0.5cm}
\caption{\small Basic principle of the SQM hypothesis (schematic):
For $\mu_d > M_s$, NSQM (left) can lower its energy by converting 
into SQM (right) via weak decay of $d$ quarks into $s$ quarks.}
\label{figsqmidea}
\end{figure}

A somewhat oversimplified picture of the basic idea is sketched in
\fig{figsqmidea}. Suppose we have a system of massless up and down 
quarks in a given volume. To be electrically neutral, the number of
down quarks should be twice as large as the number of up quarks and 
hence $\mu_d = 2^{1/3} \mu_u$ (left). Obviously, if this value is larger 
than the mass of the strange quark, the system could lower its
energy by transforming some of the down quarks into strange quarks
until they have equal Fermi energies (right).   
Thus, if the actual numbers are such that
\beq
    \Big(\frac{E}{A}\Big)_{SQM} \;<\;\Big(\frac{E}{A}\Big)_{nuclei}
    \;<\;\Big(\frac{E}{A}\Big)_{NSQM}
\label{sqmcondition}
\eeq
absolute stable SQM could exist without contradicting the empirical facts.
Of course, to be more realistic one should add electrons to the system
and consider neutral matter in weak equilibrium. Moreover, unlike 
\fig{figsqmidea} there is no fixed volume, but the system should be
bound by itself and the respective densities of SQM and NSQM could be
different. 

Right after Witten's paper, Farhi and Jaffe have performed an investigation
within the MIT bag model~\cite{Farhi}. To that end, they treated the bag 
constant, the strange quark mass and $\alpha_s$
as free parameters and searched for a window 
in parameter space, where \eq{sqmcondition} is fulfilled.
It turned out that there are indeed ``reasonable'' parameters for which
this could be achieved. For instance, for $\alpha_s = 0$ and 
$m_s = 150$~MeV, \eq{sqmcondition} would be fulfilled for 
$60~\MeV/\fm^3 \lesssim B \lesssim 80~\MeV/\fm^3$.
For larger values of $m_s$ the upper limit of $B$ becomes reduced, while
with increasing $\alpha_s$ both, upper and lower limit, are shifted to 
lower values.

Although none of the parameter fits listed in \tab{tabmitfit}
falls into Farhi and Jaffe's window, the authors pointed out that these 
parameters might not be applicable to describe dense quark matter. 
First, as we have seen
earlier, the hadron spectra strongly depend on parameters which become
irrelevant in infinite systems, like the zero-point energy or the treatment
of the center of momentum motion. 
But even those parameters which survive, could be effectively density
dependent. This is quite obvious for $\alpha_s$ which should become
smaller with increasing density.   
The relatively large values for $m_s$ (as compared with the particle data
book), point into the same direction. 
In this context it is certainly interesting to redo Farhi and Jaffe's 
analysis within the NJL model, where density dependent masses
emerge naturally. 
This has been done in Ref.~\cite{BuOe99}. In the following we discuss the
results.

\subsection{NJL-model analysis}
\label{sqmnjl}

\begin{figure}
\begin{center}
\epsfig{file=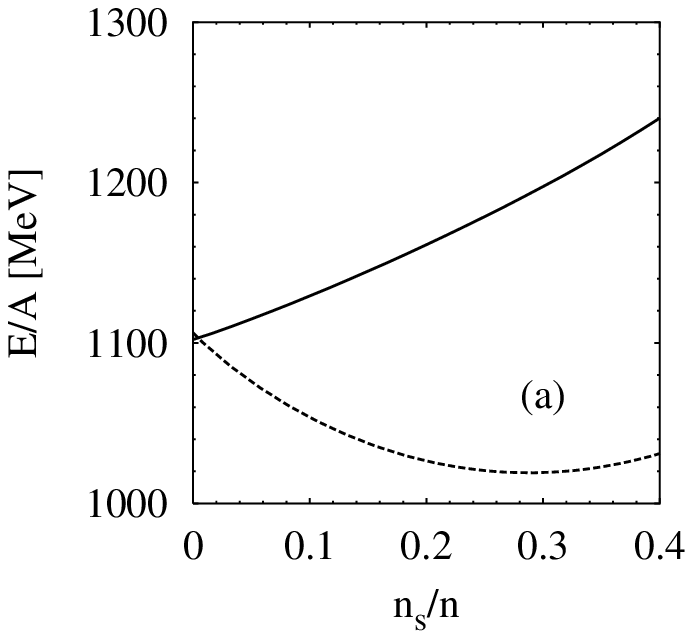,width=7.0cm}\hspace{0.5cm}
\epsfig{file=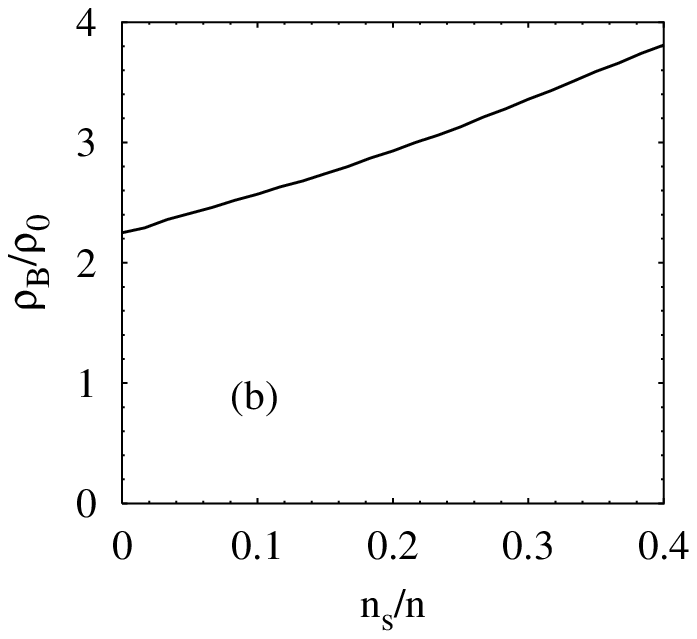,width=7.0cm}
\epsfig{file=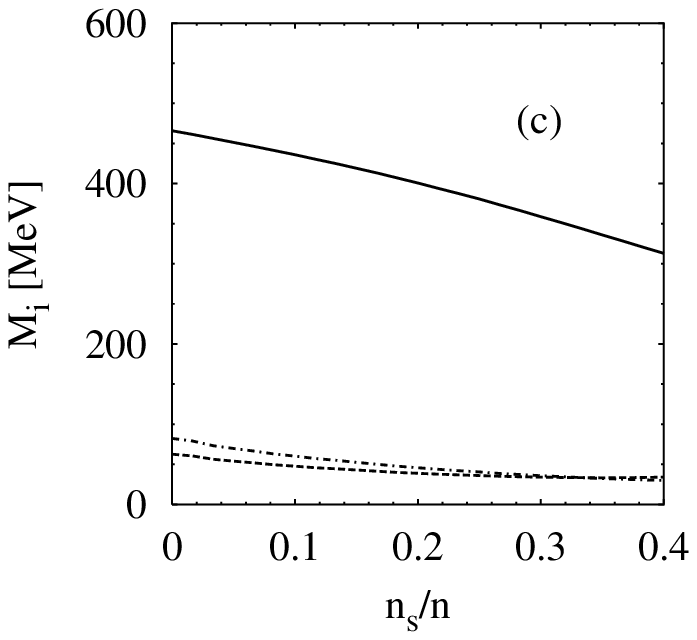,width=7.0cm}\hspace{0.5cm}
\epsfig{file=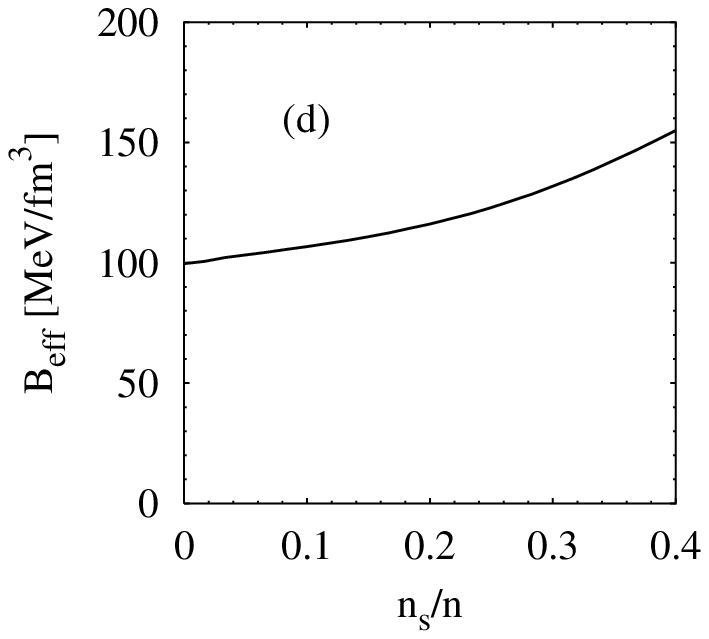,width=7.0cm}
\end{center}
\vspace{-0.5cm}
\caption{\small Properties of neutral NJL quark matter at zero pressure
         as functions of the fraction of strange quarks, $n_s/n$.
         (parameter set RKH~\cite{Rehberg}).
         (a) Energy per baryon number. Solid line:  NJL model.
          Dashed line:
          bag model with $m_u = m_d = 5.5$~MeV, $m_s = 140.7$~MeV,
          and $B = 113.2~\MeV/\fm^3$.
         (b) Total baryon number density.
         (c) Constituent quark masses $M_u$ (dash-dotted),
         $M_d$ (dashed), and $M_s$ (solid). 
         (d) Effective bag constant.}
\label{figmassbeffp0}
\end{figure}

When we compare the constituent masses and effective bag constants 
which we typically got in \sect{njl3beff} with Farhi and Jaffe's window, 
we can already anticipate that it will be hard to fulfill \eq{sqmcondition}.
This is confirmed by another pre-study which is displayed in 
\fig{figmassbeffp0}. In that figure various quantities are plotted 
as functions of the fraction of strange quarks $n_s/n$. However, unlike
in \fig{fignjl3massbeffrs}, matter is kept neutral by choosing 
$n_u/n = 1/3$ and $n_d/n = 2/3 - n_s/n$. 
Moreover, the results do not correspond to a fixed total density,
but to vanishing pressure.
The corresponding total baryon number density is shown in panel (b),
the constituent masses and the effective bag constant are given in 
panel (c) and (d), respectively.

The resulting values of $E/A$ are shown in panel (a). The NJL-model
result is indicated by the solid line. 
Since the pressure vanishes, each point corresponds to a minimum of
$E/A$ as a function of the total baryon density for fixed $n_s$.
However, the lowest value is reached at $n_s = 0$, i.e., it is not 
favorable for the non-strange matter to convert down quarks into
strange quarks.
The reason for this behavior is of course the large strange quark mass
$M_s = 466$~MeV. Therefore, since $\mu_d = 392$~MeV at this 
point\footnote{Note that $\mu_d > M_d^{vac}$. This means that
the neutral non-strange quark matter is unstable as well and could in 
principle reduce its energy by evaporating massive down quarks into the vacuum.
However, this is irrelevant for the present discussion.},
the conversion of a down quark into a strange quark costs 74~MeV.

For comparison we also show the result of a bag model calculation with
quark masses equal to the NJL current quark masses,
i.e., $m_u = m_d = 5.5$~MeV and $m_s = 140.7$~MeV. The bag constant
was taken to be equal to $B_\infty = 113.2~\MeV/\fm^3$, i.e., 
the asymptotic value of the dash-dotted line in the right panel of
\fig{fignjl3massbeffrho}.
The resulting $E/A$ is indicated by the dashed line in panel (a) of
\fig{figmassbeffp0}.
As one can see, for $n_s = 0$ the bag-model result agrees quite well
with the NJL result. However, because of the much smaller strange 
quark mass, in the bag model the conversion of down quarks to strange
quarks is energetically favored, and there is a minimum 
$E/A = 1019~\MeV/\fm^3$ at $n_s = 0.29$.
Note, however, that even this value is about $100~\MeV/\fm^3$ larger
than $E/A$ in atomic nuclei. This reflects the fact that the bag constant
is above Farhi and Jaffe's limit for $m_s \simeq 140$~MeV.

So far, we have only considered quarks. 
Of course, for a more realistic treatment of the problem, we have to
take into account weak decays, like
\beq
    d \;\leftrightarrow\; u \,+\, e \,+\, \bar\nu_e 
    \;\leftrightarrow\; s~,
\label{weak}
\eeq
and related processes. This implies that we have to include electrons and 
(in principle) neutrinos.
To large extent we can adopt the model of Farhi and Jaffe 
\cite{Farhi}, but replacing the MIT bag by the NJL mean field: 
Since we are interested in static properties of potentially stable
matter, we can safely assume that the neutrinos have enough time to
leave the system. 
The electrons are described by a non-interacting gas of massless
fermions,
\beq
    \Omega_e(T=0, \mu_e) \;=\; -\frac{\mu_e^4}{12\pi^2}~,
\label{Omegae}
\eeq
where $\mu_e$ is the electron chemical potential.
The total thermodynamic potential is then simply the sum of the quark 
part, \eq{njl3omega}, and $\Omega_e$, and consequently,  
\beq
   \epsilon_{tot} \;=\; \epsilon \;+\;\frac{\mu_e^4}{4\pi^2}
   \ , \qquad p_{tot} \;=\; p \;+\;\frac{\mu_e^4}{12\pi^2} \ ,
\label{epssqm} 
\eeq
where $\epsilon$ and $p$ refer to the quark contributions.
The electron density is given by $n_e = \mu_e^3/(3\pi^2)$.

Since the neutrinos can leave the system, lepton number is not conserved
and we effectively have two conserved charges, namely baryon number and
electric charge. 
Hence, in chemical equilibrium, only two of the four chemical potentials
which enter into the thermodynamic potential ($\mu_e$ and the three
quark chemical potentials $\mu_u$, $\mu_d$, and $\mu_s$) are independent
and could be expressed, e.g., in terms of a quark number chemical potential
$\mu$ and an electric charge chemical potential $\mu_Q$,
\beq 
    \mu_u \;=\; \mu \;+\; \frac{2}{3}\,\mu_Q~,\qquad
    \mu_d \;=\; \mu_s \;=\; \mu \;-\; \frac{1}{3}\,\mu_Q~,\qquad
    \mu_e \;=\; -\mu_Q~.
\label{njl3beta}
\eeq
Furthermore we demand charge neutrality,
\beq
   \frac{2}{3}\,n_u \;-\; \frac{1}{3}\,(n_d \,+\, n_s) 
                    \;-\; n_e \;=\; 0 \ .
\label{charge}
\eeq  
Thus, the system can be characterized by one independent variable, 
e.g., the baryon number density $\rho_B$.  

\begin{figure}
\begin{center}
\epsfig{file=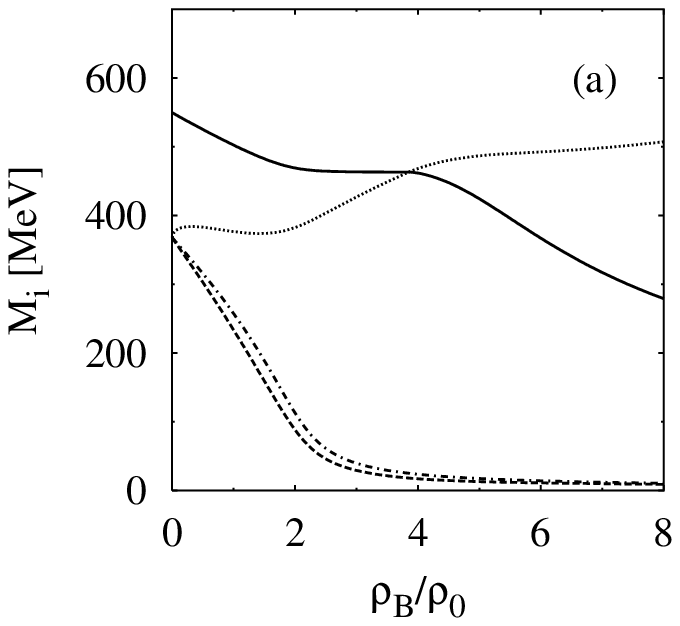,width=7.0cm}\hspace{0.5cm}
\epsfig{file=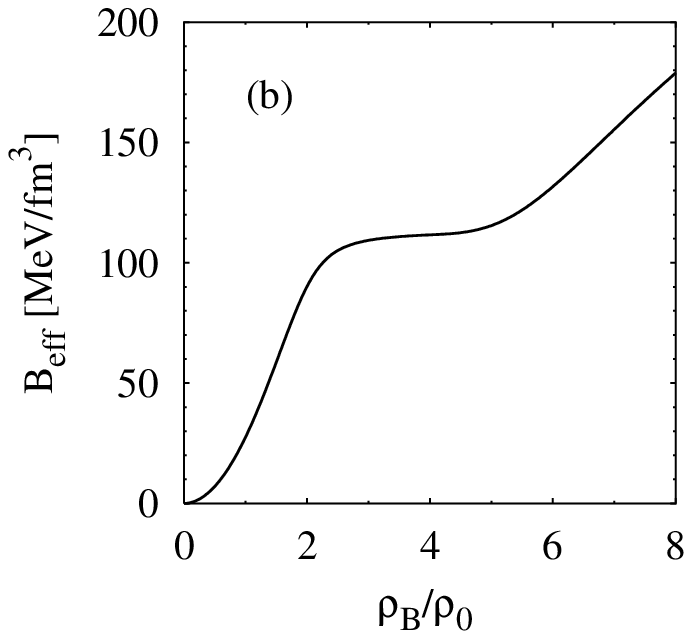,width=7.0cm}
\epsfig{file=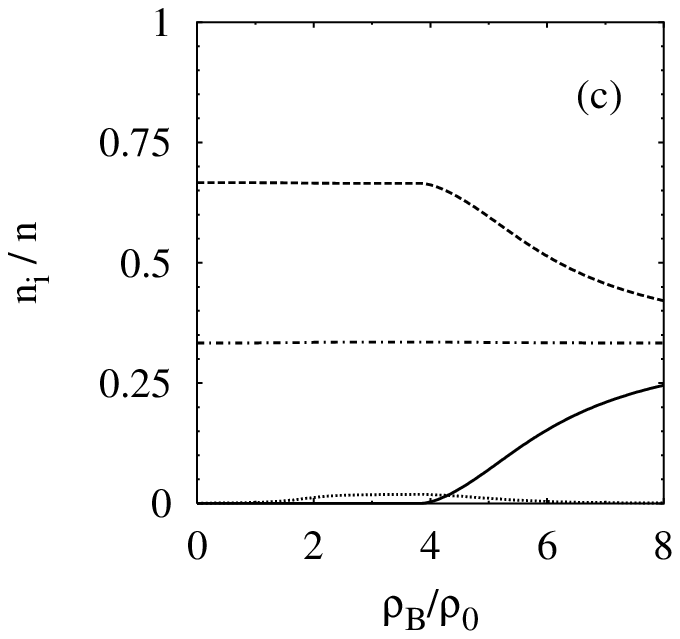,width=7.0cm}\hspace{0.5cm}
\epsfig{file=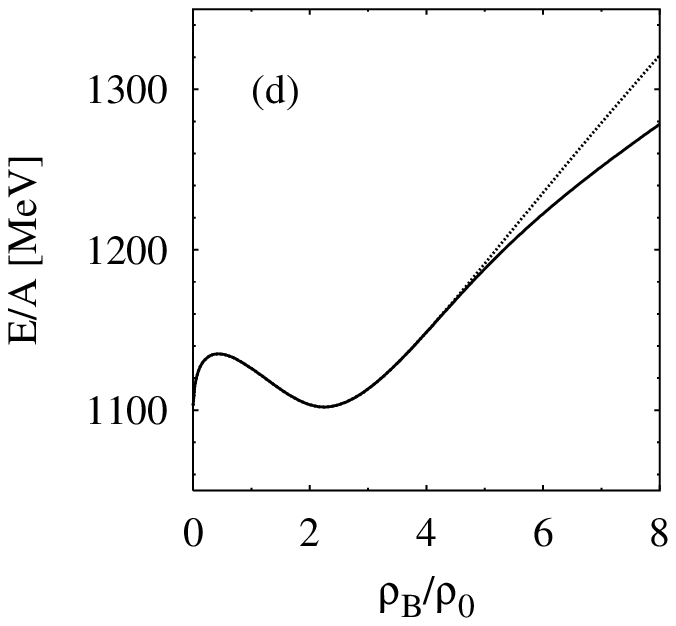,width=7.0cm}
\end{center}
\vspace{-0.5cm}
\caption{\small Properties of neutral matter of NJL model quarks 
         in beta equilibrium
         with electrons as functions of the total baryon number density
         $\rho_B$ (parameter set RKH~\cite{Rehberg}).
         (a) Constituent quark masses $M_u$ (dash-dotted),
         $M_d$ (dashed), and $M_s$ (solid). The dotted line indicates the
         chemical potential $\mu_d = \mu_s$. 
         (b) Effective bag constant.
         (c) Number densities $n_i$ divided by the total quark number density
         $n$: up quarks (dash-dotted), down quarks (dashed), strange quarks
         (solid), and electrons multiplied by 10 (dotted).
         (d) Energy per baryon number with (solid) and without (dotted)
         strangeness degrees of freedom.
         (a),(c), and (d) have been adapted from Ref.~\cite{BuOe99}.}
\label{fignjl3sqm}
\end{figure}

The results of our analysis are shown in \fig{fignjl3sqm}.
The energy per baryon number, $E/A = \epsilon_{tot}/\rho_B$,
is displayed in panel (d).
The solid line corresponds to the model described above, 
the dotted line to non-strange quark matter, where the strange quarks
have artificially been suppressed. Obviously the two curves only differ 
above $\sim~4\,\rho_0$. The reason for this is that below 
$\rho_B = 3.85~\rho_0$ the density of strange quarks vanishes.
This can be seen in panel (c) where the fractions $n_i/n$ of the various 
particles are plotted. Since electrons (dotted line) play 
practically no role (note that their fraction has been multiplied by a 
factor of 10 to be visible in the plot)
the fraction of up quarks (dash-dotted) is fixed by charge neutrality 
(\eq{charge}) to about 1/3. Thus, as long as $n_s=0$, the remaining
2/3 are mostly down quarks (dashed). 
For $\rho_B > 3.85\,\rho_0$, $n_s/n$ becomes non-zero (solid)
and $n_d/n$ drops accordingly\footnote{The small fraction of electrons 
can easily be understood in the following way: 
Suppose there were no electrons at all and no strange
quarks. Then, for charge neutrality, we must have $n_d = 2 n_u$ 
and thus $\mu_d \simeq 2^{1/3} \mu_u$, where we have neglected the masses of 
the up and down quarks. Employing beta equilibrium, \eq{njl3beta}, yields
$\mu_e \simeq (2^{1/3} -1) \mu_u$ and therefore 
$n_e/n\simeq N_c^{-1} (2^{1/3} -1)^3 n_u/n \simeq 0.002$.
This result justifies the neglect of $n_e$ in the first step, and one can
easily convince oneself that all neglected effects lead to a further
reduction of $n_e$. In fact, this estimate agrees well with with the maximum
value of $n_e/n$.}. 

The fact that there is no strangeness at lower densities is again due 
to the relatively large mass of the strange quark. The dynamical quark 
masses are displayed in panel (a) of \fig{fignjl3sqm}.
The strange quark mass is indicated by the solid line. 
For comparison we also show the chemical potential $\mu_s$ (dotted line).
For $\rho_B < 3.85~\rho_0$, $\mu_s$ is smaller than $M_s$, which is the
reason why $n_s$ vanishes in this regime.
The decrease of $M_s$ at small densities ($\rho_B \lesssim 2 \rho_0$)
is again a flavor-mixing effect and related to the drop of $M_u$ and
$M_d$. Above $2 \rho_0$, this contribution can be neglected and
$M_s$ stays almost constant until it starts decreasing again when
$n_s$ becomes non-zero\footnote{This behavior is rather different from 
typical  {\it parametrizations} which have been invented to obtain 
density dependent quark masses. Usually, these
parametrizations depend on the total baryon number density $\rho_B$ only
and show no plateau (see, e.g., \cite{Chak91,BeLu95,DBDRS99}).}.

A similar plateau structure is also seen in the effective bag constant 
which is shown in panel (b). At low densities, $\rho_B \lesssim 2 \rho_0$, 
and at high densities  $\rho_B \gtrsim 4 \rho_0$, $B_\mi{eff}$ rises due
to the symmetry restoration in the non-strange and in the strange
sector, respectively, while in the intermediate region $B_\mi{eff}$
is almost constant.
Note that the plateau values of $M_s$ and $B_\mi{eff}$ are roughly the
same as the asymptotic values $M_s^\infty = 462.8$~MeV and
$B_\infty = 113.2~\MeV/\fm^3$ of the dash-dotted lines in 
\fig{fignjl3massbeffrho}.

Let us come back to the energy per baryon (panel (d)) which is the main 
result of this analysis.
At those densities where strange quarks exist, they indeed lead to a 
reduction of the energy, as can be seen by comparing the solid curve 
with the dotted one. However, the minimum of $E/A$ does not belong to
this regime, but is located at a much lower density, $\rho_B = 2.25\,\rho_0$.
Here we find $E/A$ = 1102 MeV. Compared with the energy per baryon
in an iron nucleus, $E/A \simeq$ 930 MeV, this is still very large. 
In this sense our results are consistent with the empirical fact that
stable NSQM does not exist. 
However, since the energy of strange quark matter is even higher, our
calculation predicts that also SQM is not the absolute ground state of
strongly interacting matter. 

This result is very robust with respect to changes of the model 
parameters. 
Obviously stable quark matter with finite strangeness is 
only possible if the in-medium strange quark mass is much lower 
than the values we obtained above. The easiest way to reduce $M_s$
is to choose a lower value of the current mass $m_s$. If we leave all other 
parameters unchanged, $m_s$ must not be larger than 85~MeV if we require 
$n_s \neq 0$ at the minimum of $E/A$. With this value, we 
obtain much too low masses for $K$ and $\eta$ ($m_K \simeq$ 390 MeV, 
$m_\eta \simeq$ 420~MeV), while $E/A$ is still relatively large 
(1075 MeV). If we want to come down to $E/A \simeq$ 930 MeV we have 
to choose $m_s$ = 10 MeV or, alternatively, $m_s$ = 25 MeV and 
$m_u = m_d$ = 0. This is of course completely out of range.

We could also try to lower $M_s$ by choosing a smaller coupling 
constant $G$. (Since $\phi_u\phi_d$ is already very 
small at the densities of interest, $M_s$ is almost insensitive to
the coupling constant $K$). 
However, with a lower $G$ the vacuum masses of all quarks drop 
and correspondingly the bag constant $B$. This shifts the  
minimum of $E/A$ to lower densities and reduces the values of the
chemical potentials at the minimum.
In fact, $\mu_s$ at the minimum drops faster than $M_s$ at the minimum
and hence the density of strange quarks remains zero.  
To avoid this effect we could increase the coupling 
constant $K$ while decreasing $G$, e.g., in such a way that the vacuum 
masses of up and down quarks are kept constant. In order to get 
$n_s \neq 0$ at the minimum of $E/A$ we have to lower $G\Lambda^2$ 
to 1.5 and to increase $K\Lambda^5$ to 21.27, almost twice the value 
of parameter set RKH~\cite{Rehberg} and more than twice the values of
HK~\cite{hatsuda} or LKW~\cite{LKW92}. For these parameters the energy per 
baryon number is still 1077 MeV. On the other hand, if we
decrease the ratio $G/K$ further, this would flip the sign of the 
effective $q\bar q$ coupling in the pseudoscalar-flavor singlet 
channel, which is dominated by the combination 
$2G + \frac{2}{3} K (\ave{\bar u u}+\ave{\bar d d}+\ave{\bar s s})$. 
In that case there would be no solution for the $\eta'$-meson in 
vacuum. 

We could ask whether additional terms in the Lagrangian could help.
In particular vector interactions can be quite important at finite
density, as we have seen earlier.
However, since vector mean fields are repulsive\footnote{This depends
of course on the sign of the coupling constant $G_V$.
In a different context, the authors of Ref.~\cite{LRR97,LaRh99} have employed
a vector interaction with an unconventional sign, leading to attractive
mean-fields. Although this cannot be excluded in general, there are 
several arguments in favor of the ``conventional'' choice of sign:
(i) This sign emerges naturally, if the interaction is mediated via
a heavy boson in the s-channel. For instance, in the Walecka 
model~\cite{Walecka}, the $\omega$ mean-field is repulsive.
(ii) This sign can also be derived from a single gluon exchange via
Fierz transformation.
(iii) In order to describe vector mesons within the 
NJL model, the vector interaction must be attractive in the space-like 
components. It is then repulsive in the time-like components, which
are relevant in the mean field.
In particular, the fitted vector coupling of parameter set LKW~\cite{LKW92},
has the ``conventional'' sign and leads to a repulsive vector mean
field.}
the energy per baryon number will be even larger than before and SQM remains 
strongly disfavored compared with ordinary nuclei. 

We thus conclude that the NJL model does not support the idea of absolutely 
stable SQM if we want to keep the vacuum properties of the model at least
qualitatively unchanged.
The main reason is that the strange quark mass stays rather large
at densities where chiral symmetry is already approximately restored in 
the non-strange sector. As a consequence SQM tends to be disfavored
against NSQM.
But also the effective bag constant in that region is larger than the
upper limit of Farhi and Jaffe's window for any value of $m_s$.
This combination is hard to beat. 
Of course, we cannot exclude that the parameters which have been fixed in 
vacuum are not appropriate to describe high densities. 
However, in order to change our conclusions, rather drastic variations
would be required. 

The situation is not as clear for the possibility of strangelets to be
stable against strong but not against weak decays.
With the parameters used above (set RKH~\cite{Rehberg}) we find
$E/A = 1210.8$~MeV for quark matter consisting of equal numbers of
up, down, and strange quarks. Although this would still be unstable
against decay into $\Lambda$ or $\Sigma$ baryons ($M_\Lambda = 1116$~MeV,
$M_\Sigma = 1190$~MeV), the difference is smaller than before and 
could be more sensitive to the parameters 
(also see Refs.~\cite{MSSG01,Ratti}).

A quite important effect which has not been taken into account in
our analysis is the condensation of diquark pairs (``color
superconductivity''). 
As we will see, color superconductivity indeed provides an additional
binding mechanism which is more effective for strange matter
(``color-flavor locking'') than for non-strange matter. 
To include these effects, major extensions of the model are necessary. 
This  will be the central issue of the remaining part of this work.
In \sect{neutresultshom} we will come back to the SQM hypothesis and
investigate to what extent our conclusions change when diquark 
condensates are taken into account.

\chapter{Two-flavor color superconductors}
\label{tsc}

So far our discussion was restricted to quark-antiquark condensates,
$\ave{\bar q\,{\cal O}\,q}$, 
most importantly the ``quark condensate'' $\phi = \ave{\bar q q}$,
related to spontaneous chiral symmetry breaking.
At low temperatures and densities this led us to a 
non-trivial phase with $\phi \neq 0$, while the structure of the 
high-temperature or density regime was rather simple. 
In fact, if we neglect the current quark masses (chiral limit) and
possible vector interactions, the chirally restored phase is
completely trivial in the NJL mean-field, and the high-density effects 
discussed in the previous chapters were mainly based on the imperfect
chiral restoration in the presence of quark masses. 

It is known, however, that any Fermi system at sufficiently low 
temperatures is subject to a Cooper instability, as soon as an  
arbitrarily weak attraction is present (``Cooper theorem''~\cite{Cooper}).
The heuristic argument is very simple:
Consider an infinite system of non-interacting fermions. 
At $T=0$ they will form a Fermi sphere, with all states occupied up to the 
Fermi momentum $p_F$, and all other states empty.
Since the free energy $|E_{\vec p} - \mu|$
to create a particle or a hole with momentum $\vec p$ vanishes at the 
Fermi surface, one could create a pair of particles (or holes) directly at 
the Fermi surface without any free energy cost. 
If we now turn on a small attraction between the particles, this will
further lower the free energy, and thus the original Fermi sphere becomes
unstable. 

In BCS theory this problem is cured by the formation of a Cooper pair
condensate~\cite{BCS}. This leads to a gap in the excitation spectrum,
i.e., excitations with vanishing free energy do no longer exist.
In ordinary (metallic) superconductors, Cooper pairs are pairs
of electrons with opposite momentum and opposite spin.
In analogy we should expect that cold deconfined quark matter becomes a
``color superconductor'' where pairs of quarks condense (``diquark
condensate''). 
In fact, whereas the elementary interaction between electrons, i.e., 
photon exchange, is repulsive and the Cooper instability in metals
only comes about as a subtle effect of phonon exchange in the presence
of a screened photon field~\cite{Fro50} (see also Ref.~\cite{Po92}), 
the situation 
appears much more straight forward in QCD, where already the elementary 
interaction, i.e., gluon exchange is attractive in certain channels. 
Therefore the possibility of color superconductivity in high-density
QCD matter has already been suggested in 1975~\cite{CoPe75},  
only two years after the discovery of asymptotic freedom. 
However, in spite of further investigations in the 
70s~\cite{Ba77,Fr78}\footnote{Interestingly, the main intention of 
Barrois' paper, Ref.~\cite{Ba77}, was to argue in favor of a
six-quark condensate as an alternative to a BCS-like diquark pairing.}
and 80s~\cite{BaLo84}, until quite recently not much attention has been
payed to this possibility by a wider audience. 

This changed at the end of the 90s, after it had been discovered that in the 
region of interest, i.e., $\mu \sim 500$~MeV, the color superconducting 
gaps could be of the order of $\Delta\sim 100$~MeV~\cite{ARW98,RSSV98}, 
much larger than originally expected. 
Since in standard weak-coupling BCS theory the critical temperature 
for $J^P=0^+$-pairing is given 
by $T_c \simeq 0.57\,\Delta(T=0)$ \cite{FW71}, this would also imply
a sizeable extension of the color superconducting phases into the 
temperature direction \cite{PiRi99}.
Hence, color superconducting phases could be relevant
for neutron stars \cite{We99,BGS01} and -- if we are very lucky -- 
even for heavy-ion collisions \cite{PiRi00}. 

The calculations of Refs.~\cite{ARW98,RSSV98} have been performed within
NJL-type models with instanton-inspired interactions. 
Later, similar results (sometimes even larger gaps) have been obtained
within the instanton model~\cite{CaDi99,RSSV00}.
The large gaps have therefore first been attributed to non-perturbative 
effects which are effectively contained in these interactions,
whereas the old analyses were based on a single-gluon exchange. 
On the other hand, improved treatments of the gluon exchange, which
took into account that static magnetic gluons remain unscreened~\cite{Son99},
revealed gaps of similar size when the leading-order results are extrapolated 
from asymptotic densities down to chemical potentials below 
1~GeV~\cite{SchWi99}. Further improvements, however, 
seem to reduce the gap again by almost one order of magnitude~\cite{WaRi02}. 

Soon after the rediscovery of color superconductivity it was realized
that the consideration of diquark condensates opens the possibility for a
wealth of new phases in the QCD phase diagram (see \fig{figschemphase}
in the Introduction). 
For the exploration of these phases, often NJL-type models play a 
pioneering role. 
Since the models are relatively simple, they allow for the 
simultaneous consideration of several different condensates in 
order to investigate their competition and mutual influence. 
As we will discuss in the next chapter, this will again be most important
when strange quarks are involved.
We begin, however, with the two-flavor case. 
Besides being a somewhat simpler warm-up exercise, it turns out that
interesting results can already be obtained for this case.

\section{Diquark condensates}
\label{diquarks}

A diquark condensate is defined as an expectation value
\beq
   \ave{\,q^T {\cal O}\, q\,}~,
\eeq
where $q$ is a quark field with spin, flavor and color degrees of freedom,
and is $q^T$ the transposed (not adjoined!) field operator.   
${\cal O}$ denotes an operator acting in Dirac, flavor and color space,
\beq
    {\cal O} \;=\; {\cal O}_\mi{Dirac} \otimes {\cal O}_\mi{flavor} 
    \otimes {\cal O}_\mi{color}~.
\eeq
It can also contain derivatives, but we will not consider
this possibility here.

\subsection{Pauli principle}
\label{Pauli}

A priori, the only restriction to ${\cal O}$ is provided by the Pauli
principle. Since
\beq
    q^T {\cal O}\, q \;=\; {\cal O}_{ij} \, q_i \, q_j 
    \;=\; - {\cal O}_{ij} \, q_j \, q_i \;=\; -q^T {\cal O}^T q~,
\eeq
only totally antisymmetric operators ${\cal O}^T=-{\cal O}$ survive.

\begin{table}[t]
\begin{center}
\begin{tabular}{|c| c c |}
\hline
&&\\[-3mm]
&\quad antisymmetric\quad & \quad symmetric \quad
\\[1mm]
\hline
&&\\[-3mm]
        & $C\gamma_5$, \quad $C$, \quad $C\gamma^\mu\gamma_5$&
          $C\gamma^\mu$, \quad $C\sigma^{\mu\nu}$ \\[-1.5mm]
Dirac   & & \\[-1.5mm]
        & (S) \quad (P) \quad (V)\; & (A) \quad (T)\\[1mm]
\hline
&&\\[-3mm]
            &  $\tau_2$ &
            $\underbrace{\unity,\; \tau_1,\; \tau_3}$ \\[-2.5mm]
  $U(2)$    & & \\[-1.5mm]
            & singlet & triplet \\[1mm]
\hline
&&\\[-3mm]
            &  $\underbrace{\lambda_2,\; \lambda_5,\; \lambda_7}$ &
            $\underbrace{\unity,\; \lambda_1,\; \lambda_3,\; \lambda_4,\; 
             \lambda_6,\; \lambda_8}$ \\[-2.5mm]
  $U(3)$    & & \\[-1.5mm]
            & antitriplet & sextet \\[1mm]
\hline
\end{tabular}
\end{center}
\caption{\small Dirac operators and generators of $U(2)$ and $U(3)$,
         and their symmetries under transposition. 
         In this table $\tau_i$ denote Pauli matrices, and $\lambda_i$
         denote Gell-Mann matrices.  
         $C = i \gamma^2\gamma^0$ is the matrix of charge conjugation.
}
\label{tabpauli}
\end{table} 

The symmetry properties of various operators under transposition are given 
in \tab{tabpauli}.
In the first line operators in Dirac space are listed.
Here $C = i \gamma^2\gamma^0$ is the matrix of charge conjugation.
The five combinations correspond to definite properties of the bilinears
$q^T {\cal O}\, q$ under Lorentz transformations, as indicated below,
i.e., scalar, pseudoscalar, vector, axial vector, and tensor.
In the second and the third line of \tab{tabpauli} we have listed the 
generators of $U(2)$ and $U(3)$, respectively. The generators of $U(2)$ 
form a basis for the operator ${\cal O}_\mi{flavor}$ in the 
two-flavor case.  
The corresponding diquark bilinears transform as a singlet and a
triplet under isospin rotations, i.e., isospin 0 and isospin 1,
respectively. 
Finally, the generators of $U(3)$ form a basis for the operator 
${\cal O}_\mi{color}$ or for the operator ${\cal O}_\mi{flavor}$ in the 
three-flavor case. Here the diquark bilinears can be decomposed into
an antisymmetric antitriplet and a symmetric sextet.   

Since a totally antisymmetric operator ${\cal O}$ can be built as a product
of three antisymmetric operators ${\cal O}_\mi{Dirac}$, 
${\cal O}_\mi{flavor}$, and ${\cal O}_\mi{color}$, or of one antisymmetric 
and two symmetric operators, there are obviously many combinations which
are in principle permitted.
This highlights an important difference to ordinary superconductors
where color and flavor degrees of freedom do not exist. 
In which of these channels condensation takes place cannot be decided on
the basis of the Pauli principle alone, but depends on the details of
the interaction. 
(For a general overview about the classification of color superconducting
phases, see also Ref.~\cite{Rischkereview}.)

\subsection{Scalar color-antitriplet diquark condensate}
\label{sca}

The most important example is the diquark condensate
\beq
    s_{AA'} 
    \;=\; \ave{\,q^T \,C \gamma_5 \,\tau_A \,\lambda_{A'} \,q\,}~,
\label{saa}
\eeq
where $\tau_A$ and $\lambda_{A'}$ are the antisymmetric
generators of $U(N_f)$ and $U(N_c)$, acting in flavor space and in color 
space, respectively. In this work, we only consider
the physical number of colors, $N_c=3$. Then the $\lambda_{A'}$ denote the
three antisymmetric Gell-Mann matrices, $\lambda_2$, $\lambda_5$, and 
$\lambda_7$. 
Hence $s_{AA'}$ describes a diquark condensate in the scalar ($J^P = 0^+$)
color-antitriplet channel. 
This corresponds to the most attractive diquark channel for both, 
one-gluon exchange and instanton-mediated interactions
(see App.~\ref{fierzexamples}). 

In this chapter we discuss the case of two flavors ($N_f= 2$).
Then the flavor index in \eq{saa} is restricted to $A=2$, i.e.,
$s_{AA'}$ is a flavor singlet, 
describing the pairing of an up quark with a down quark.
The three condensates $s_{2A'}$, $A' = 2, 5, 7$, form a vector in 
color space.
Since this vector can always be rotated into the $A' = 2$-direction by a 
global $SU(3)$-color transformation, we may assume 
$s_{2A'} \,=\, \ct\,\delta_{A'2}$, without loss of generality. 
In the following, for convenience, we will denote $\ct$ by $\delta$,
\beq
     \delta \equiv \ct 
     \;=\; \ave{\,q^T \,C \gamma_5 \,\tau_2 \,\lambda_2 \,q\,}~.
\label{delta}
\eeq
Let us briefly summarize the main properties of a phase with 
non-vanishing $\delta$. (For further details, see Ref.~\cite{RaWi00} 
and references therein.)

As already mentioned, the vector $s_{2A'}$ transforms as an antitriplet 
under $SU(3)$-color. 
Denoting the three colors by ``red'', ``green'', and ``blue'',
the explicit color-flavor structure of $\delta$ reads 
\beq
     \delta  \;=\;
       -\; \ave{\,u_r^T \,C \gamma_5 \,d_g\,}
     \;+\; \ave{\,u_g^T \,C \gamma_5 \,d_r\,}
     \;+\; \ave{\,d_r^T \,C \gamma_5 \,u_g\,}
     \;-\; \ave{\,d_g^T \,C \gamma_5 \,u_r\,}~,
\label{deltacf}
\eeq
where $u_r$ corresponds to a red up quark, and so on.
This means, with this particular choice, only the red and green quarks
participate in the condensate, while the blue ones do not:
$SU(3)$-color is broken down to $SU(2)$\footnote{Strictly speaking,
gauge symmetries cannot be broken spontaneously~\cite{Elitzur} 
(see, however, Ref.~\cite{Splittorf} for possible caveats).
This also applies to the ``spontaneous breaking'' of the electromagnetic
$U(1)$ in ordinary superconductors or the breaking of the
$SU(2)_L\times U(1)_Y$ in electroweak theory.
In all these cases the spontaneous symmetry breaking can only be discussed
after gauge fixing.
Adopting Rajagopal and Wilczek's point of view, we therefore interprete the 
spontaneous breaking of a gauge symmetry as a ``convenient 
fiction''~\cite{RaWi00} in a fixed gauge.
The important point is that this ``fiction'' leads to correct 
predictions for ``real'', i.e., gauge invariant, observables.   
Whereas the vector $(s_{2A'})$, and thus $\delta$, is not a gauge 
invariant quantity, the gap in the quasiparticle spectrum only depends on 
its ``length'', $|\delta| = \sqrt{|s_{22}|^2 +|s_{25}|^2 + |s_{27}|^2}$, 
which is gauge invariant.
This may also be taken as some justification for studying color 
superconductivity within NJL-type models, although these models are only
symmetric under global $SU(3)$-color.}. 
Accordingly, five of the eight gluons receive a mass through the
Anderson-Higgs mechanism (``Meissner effect'').
The corresponding Meissner masses have been calculated in Ref.~\cite{Ri00a} 
(together with the Debye masses) for asymptotically high densities
and in Ref.~\cite{CaDi00} in the instanton liquid model.

Similarly, since the condensate carries a net electric charge, 
one might expect that also the photon acquires a mass,
giving rise to an ordinary (electromagnetic) Meissner effect.
In fact, $\delta$ is not invariant under a diagonal transformation
generated by the electric charge operator. One finds
\beq
q \rightarrow e^{\,i\alpha\,Q}\,q \qquad \Rightarrow \qquad
\delta \rightarrow e^{\,i\alpha/3}\,\delta~, \qquad \text{where} \quad   
Q = {\rm diag}_f \Big(\,\frac{2}{3},-\frac{1}{3}\,\Big)~. 
\label{qtrans2}
\eeq
However, since $\delta$ transforms in a similar way under the
color rotation
\beq
q \rightarrow e^{\,i\alpha'\,\lambda_8}\,q \qquad \Rightarrow \qquad
\delta \rightarrow e^{\,2i\alpha'/\sqrt{3}}\,\delta~,
\label{lambda8trans2}
\eeq
one can find a linear combination
\beq
    \tilde Q \;=\; Q \,-\, \frac{1}{2\sqrt{3}}\,\lambda_8 \;\equiv\;
    Q \,-\, {\rm diag}_{\,c}\Big(\,\frac{1}{6},\frac{1}{6},-\frac{1}{3}\,\Big)~, 
\label{qtilde2}    
\eeq
under which $\delta$ remains invariant. Indeed,
the pairs in \eq{deltacf} have vanishing $\tilde Q$ charge, i.e., 
$\delta$ is $\tilde Q$ neutral.
The physical relevance of $\tilde Q$ is related to the fact that 
the photon and the eighth gluon mix in the presence of $\delta$,
resulting in a state which becomes massive, while the orthogonal state
remains massless. 
This is quite analogous to the mixing which gives rise to the 
massive $Z$-boson and the massless photon in electroweak theory.
In the present case $\tilde Q$ is the charge the massless 
combination (the ``rotated photon'') is coupled to. 

Like all diquark condensates, $\delta$ breaks the $U(1)$ symmetry,
related to baryon number conservation, down to $Z_2$\footnote{
In finite systems, the total baryon number is of course conserved.
The correct interpretation is that there are long-range correlations 
$C(y,x) \sim \ave{q^\dagger(y) q^\dagger(y) q(x) q(x)}$, describing the 
superfluid transport of a fermion pair from a point $x$ to a distant point
$y$. In the grand canonical treatment of infinite systems
this is a result of the factorization
$C(y,x) \sim \ave{q^\dagger(y) q^\dagger(y)}\ave{q(x) q(x)} = |\ave{qq}|^2$.}. 
However, analogously to the case of electromagnetism, one can
construct a new unbroken global symmetry as a combination of 
$U(1)$ with the color rotation \eq{lambda8trans2}.
Thus, there is a conserved ``rotated'' baryon number.

As a flavor singlet, $\delta$ is invariant under isospin transformations
$SU(2)_V$. One can easily show that $\delta$ is also invariant under
the corresponding axial transformations $SU(2)_A$, i.e., $\delta$ leaves
chiral symmetry unbroken. 
Hence, there is no global symmetry broken by $\delta$ and, consequently,
there are no Goldstone bosons.  

We have already pointed out in the introduction to this chapter that
a non-vanishing diquark condensate leads to a gap in the quark excitation 
spectrum. Typically, the quasiparticle dispersion laws take the form
\beq
    \omega_{\mp}(\vec p) = \sqrt{(E_p\mp\mu)^2 + |\Delta|^2}~,
\label{E1}
\eeq
where $\omega_-$ corresponds to the free energy needed to create a particle 
above or a hole below the Fermi surface and $\omega_+$ is the corresponding 
antiparticle term. $E_p = \sqrt{\vec p^{\;2} + m^2}$ is again the on-shell 
energy of a non-interacting quark with mass $m$. 
Thus, there is a minimal free energy, $\delta F = 2|\Delta|$,
which is needed to excite a particle-hole pair from the ground state. 
In general, $\Delta$  is an energy and momentum dependent quantity
and can be different for quarks and antiquarks, but we
may ignore this in this more qualitative discussion.

Like in the case of chiral symmetry breaking, 
the most transparent, but also most elaborate way to derive $\Delta$
is via an explicit pairing ansatz for the ground state $\ket{g.s.}$ as 
a coherent state of red and green up and down quarks with zero total 
momentum~\cite{ARW98,RaWi00},
\begin{alignat}{1}
     \ket{g.s.} \;=\; \prod_{\vec p,s,c,c'} 
       &\Big[\cos{\theta_s^b(\vec p)}
       + \varepsilon_{3 c c'}\,e^{i\xi_s^b(\vec p)}\,\sin{\theta_s^b(\vec p)}\, 
       b^\dagger(\vec p,s,u,c)\,  
       b^\dagger(-\vec p,s,d,c') \Big]
\nonumber\\ 
       &\Big[\cos{\theta_s^d(\vec p)}
       + \varepsilon_{3 c c'}\,e^{i\xi_s^d(\vec p)}\,\sin{\theta_s^d(\vec p)}\, 
       d^\dagger(\vec p,s,u,c)\,  
       d^\dagger(-\vec p,s,d,c') \Big] 
       \ket{0}~.
       \label{bcsvac}
\end{alignat}
Here we have used the same notation as in \eq{njlvac}.
We have left out the part of the unpaired blue quarks, which 
decouples from the paired sector.
The upper line describes 
particle-particle and, if measured relative to a filled Fermi sea,
hole-hole pairing and basically corresponds to a standard BCS ansatz.
The lower line corresponds to antiparticle-antiparticle pairing. 
This term is of course missing in non-relativistic descriptions.

For given chemical potential $\mu$, the variational functions 
$\theta_s^{b,d}(\vec p)$ and $\xi_s^{b,d}(\vec p)$ are fixed by 
minimizing the free energy $\bra{g.s.} \hat H - \mu\,\hat N \ket{g.s.}$,
where $\hat H$ is the Hamiltonian and $\hat N$ is the quark number operator.
This minimization problem can be transformed into a self-consistency
problem for the gap parameter $\Delta$.
Details of this gap equation depend of course on the interaction.
For NJL-type models where the interaction is short ranged it is 
typically of the form 
\beq
    \Delta \;=\; 8\,H\,\Delta\,\int\dtp\,
    \Big\{\frac{1}{\omega_-(\vec p)} \,+\,\frac{1}{\omega_+(\vec p)}\Big\}~,
\label{bcsgap}
\eeq
where $H$ is the coupling constant in the scalar color-antitriplet 
channel. (This particular equation will be derived in \sect{2scformalism}
within a more generalized framework.)

Obviously, \eq{bcsgap} always has a trivial solution, $\Delta = 0$.
On the other hand, for $\mu > m$ and $\Delta \rightarrow 0$,
the integral on the r.h.s. becomes logarithmically divergent, due to a 
pole of $1/\omega_-$ at the Fermi surface.
Thus, even for arbitrarily small positive (attractive) values of $H$, there
is always a non-trivial solution which approximately behaves like
\beq
\Delta \;\propto\; \exp{(-\frac{\mi{const.}}{H})}~.
\eeq
The logarithmic divergence which guarantees the existence
of the non-trivial solution is the formal manifestation of the Cooper
instability. This feature is qualitatively different from the 
chiral gap equation, \eq{njlgap2}, where non-trivial solutions require
a certain minimal value of the coupling constant\footnote{Formally, 
in \eq{bcsgap} the same is true for $\mu < m$. Then the
trivial solution corresponds to zero density, i.e., there is no 
Fermi sphere which could become unstable. Nevertheless, there could be
a non-trivial solution of the gap equation, if $H$ is sufficiently 
strong~\cite{CaDi99,DFL96,VaJa00,RuRi03}.}.

\section{Interaction}
\label{interaction}

For the explicit computation of the color superconducting gaps and related
quantities, we have, of course, to specify the quark-quark interaction.
Unfortunately, this interaction in poorly known in the most interesting
regime of a few times nuclear matter density. 
Before we introduce the NJL-type interactions which will be the main basis
of our further investigations, we briefly discuss the microscopic 
(QCD based) approach at asymptotically large densities.

\subsection{Asymptotic densities}
\label{asymptotic}

Because of asymptotic freedom, the QCD coupling constant 
becomes small at large momentum scales and QCD can be treated perturbatively 
(see \eq{alpharun}).
Since at finite density the scale is set by the Fermi momentum,
it was realized rather early that QCD becomes relatively simple 
in the high-density phase~\cite{CoPe75}
and a description starting from first principles should be possible. 
This was also the basis of the early studies of color 
superconductivity~\cite{Ba77,Fr78,BaLo84}.
More recently, the large gaps which have been found within the more
phenomenological approaches of Refs.~\cite{ARW98,RSSV98} have also 
evoked renewed interest in asymptotic studies 
(see Refs.~\cite{Schaefer,Rischkereview} for recent reviews). 

\begin{figure}[t]
\begin{center}
\epsfig{file=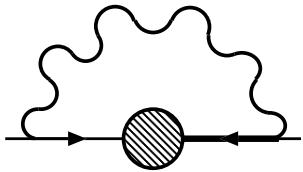,width = 4.0cm}
\caption{\small
         Anomalous quark self-energy diagram determining the diquark gap at 
         asymptotically large density. The wavy line symbolizes a 
         gluon. The shaded blob corresponds to the insertion of a diquark 
         condensate.}
\label{figbcsoge}
\end{center}
\end{figure}

The gap equation for the color superconducting phase can be derived 
within Dyson-Schwinger formalism~\cite{PiRi00,SchWi99,HMSW00}.
In the weak-coupling limit, the interaction between quarks is dominated
by single gluon exchange.
This amounts to evaluating the quark self-energy 
\beq
    \Sigma(p) \;=\; -ig^2 \int \dfk \Gamma_a^\mu\, S(k) \,  \Gamma_b^\nu
    \, D_{\mu \nu}^{ab}(k-p)~,
\label{sigmabcsoge}
\eeq
where $S(k)$ and $D_{\mu \nu}^{ab}(k-p)$ denote quark and gluon propagators,
respectively, $g$ is the QCD coupling constant,
and $\Gamma^\mu_a$ is the quark-gluon vertex. To leading order
it is basically the free vertex $\gamma^\mu \lambda_a$.  
The essential part of this diagram is depicted in \fig{figbcsoge}. 
Because of the non-conservation of (ordinary) baryon number, two quarks
can be absorbed or created by the condensate (shaded blob).
This gives rise to a so-called anomalous contribution to the self-energy
which is proportional to the gap. 
This will be discussed in more details in \sect{2SCHF}.

If the gluon in \eq{sigmabcsoge} is replaced by a point interaction
we arrive at a standard BCS-type gap equation, like in \eq{bcsgap}.
Originally it was concluded from this analogy that the gap should
behave like $\exp{(-\mi{const.}/g^2)}$ at weak coupling, 
leading to rather small gaps~\cite{BaLo84}.
However, it has long been known for ordinary superconductors that
the behavior can be rather different for long-range forces due to
retardation effects~\cite{Eli60}.

In Ref.~\cite{SchWi99}, the gluon propagator was taken to be of the
general form~\cite{Kapusta},
\beq
    D_{\mu \nu}^{ab}(q) \;=\; (\frac{P_{\mu \nu}^L}{q^2-F(q_0,|\vec q|)} \,+\, 
    \frac{P_{\mu \nu}^T}{q^2-G(q_0,|\vec q|)} 
    \,-\, \xi \frac{q_\mu q_\nu}{q^4})\,
    \delta_{ab}~,
\label{glueprop}
\eeq
where $P_{\mu \nu}^{L,T}$ are longitudinal and transverse projectors,
respectively, and $\xi$ is a gauge parameter. 
The function $F$ describes the Debye screening of the longitudinal
(electric) gluons. The transverse (magnetic) gluons are dynamically
screened due to Landau damping, but there is no screening of the
static ($q_0 = 0$) modes. 
To first approximation~\cite{SchWi99}
\beq
    F \;=\; 2N_f\,\frac{g^2\mu^2}{4\pi^2}~\quad
    \text{and}\quad    
    G \;=\; \frac{\pi}{4} \frac{q^0}{|\vec q|}\,F~.
\label{FGapp}
\eeq
Since typical frequencies are of the order of the gap, $q^0 \sim \Delta$,
the magnetic gluons cause a second logarithmic divergence, such that
the integral in \eq{sigmabcsoge}
diverges like $(\ln \Delta)^2$, rather than $\ln \Delta$. 
In this way the behavior of the gap is changed to 
\beq
    \frac{\Delta}{\mu} \;\propto\; \exp{(-\frac{3\pi^2}{\sqrt{2} g})}~,
\label{gapoge}
\eeq
which has first been found by Son~\cite{Son99} based on a renormalization
group study.
Afterwards this has been confirmed by several authors within the 
Dyson-Schwinger approach~\cite{PiRi99,PiRi00,SchWi99,HMSW00}. 
In particular, \eq{gapoge} remains valid if, instead of \eq{FGapp}, 
$F$ and $G$ are calculated within hard dense loop 
approximation~\cite{SchWi99}.

\eq{gapoge} has the striking consequence that asymptotically
${\Delta}$ grows to arbitrarily large values although the coupling
becomes weaker.Note that $g = \sqrt{4\pi\alpha_s}$ behaves like 
$1/\sqrt{\ln{\mu}}$, where we have assumed that the momentum scale $Q$
is proportional to $\mu$.
But also when the leading-order results are extrapolated downwards 
into the (astro-) physically more interesting regime below 1~GeV,
gaps of the order $\sim 100$~MeV are found~\cite{SchWi99}. 

However, although appreciated as nice support for the results based 
on phenomenological interactions it is clear that the perturbative
approach (i.e., one-gluon exchange) cannot be trusted at densities
of relevance. 
To estimate the range of validity of these calculations let us assume
(quite optimistically) that the perturbative regime begins at
$Q \approx 1.5$~GeV and identify this momentum scale with the
chemical potential. For two massless flavors this
corresponds to a baryon number density $\rho_B = 2/(3\pi^2) \mu^3 \approx
30$~fm$^{-3}$, which is about 175 times nuclear saturation density,
well beyond the maximum densities of about $10~\rho_0$ expected in 
the centers of compact stars.
On the other hand, at $Q \approx 1.5$~GeV the coupling is still rather
large, $\alpha_s \simeq 0.35$~\cite{alphas}, i.e., $g \simeq 2$.

In Ref.~\cite{RaSh00} Rajagopal and Shuster employed the gauge parameter
$\xi$ in \eq{glueprop} to study the relevance of higher-order terms.
Since (in principle) the gap in the excitation spectrum is an observable
and thus a gauge invariant quantity, it must not depend on $\xi$.
In Ref.~\cite{SchWi99} Sch\"afer and Wilczek showed, that the gap equation 
indeed becomes independent of $\xi$ at infinitely large chemical potential
and therefore they dropped the $\xi$-dependent terms in their calculations. 
In contrast, the authors of Ref.~\cite{RaSh00} kept these terms and 
investigated the $\xi$-dependence of the result at given chemical potential.
While they confirmed that these terms become negligible for 
$\mu\rightarrow\infty$, they found that they only begin to become small
for couplings $g \lesssim 0.8$.
Again identifying the momentum scale $Q$ with the chemical potential,
this corresponds to $\mu \gg 10^8$~MeV ($\rho_B \gg 5\cdot 10^{16}\rho_0$)!
Moreover, since the gap is dominated by almost collinear scattering,
$|\vec k - \vec p| \ll \mu$, it is likely that the relevant momentum scale 
is much lower than $\mu$, as also indicated by renormalization group 
studies~\cite{BeBe00}. This would push up the ``asymptotic regime''
even further.

Of course, the range of validity of the calculations can be
extended by taking into account higher-order corrections.
In the mean-time a complete analysis to subleading order in $g$
has been performed~\cite{WaRi02,BLR00}. 
Nevertheless, in view of our ``optimistic estimate'' above, it is
very unlikely that any calculation that is based on the use of
perturbation theory can be reliably extrapolated down to the physically 
interesting regime.

\subsection{NJL-type interactions}
\label{2scnjl}

Alternatively, color superconductivity can be studied within models
which are based on vacuum phenomenology,
like instanton models~\cite{CaDi99,RSSV00} 
or NJL-type models, e.g., Refs.~\cite{ARW98,RSSV98,BeRa99,SKP99}.
It is obvious that the use of models cannot overcome the problems
discussed at the end of the previous section:
While the gluon-exchange based studies become exact at asymptotic densities,
but cannot reliably be extrapolated down to densities at physical relevance,
NJL-type interactions are mostly constrained by fitting vacuum properties,
and it is not clear whether the parameters obtained in this way can still
be trusted in the deconfined phase.  

In principle, instanton models are in a somewhat better situation,
since they are based on a semiclassical description of the QCD vacuum and 
therefore less phenomenological.
In fact, the single instanton solutions and the corresponding zero modes
at finite chemical potential are known~\cite{Car80,Abr83}.
However, in the instanton liquid model there are additional 
parameters, namely the average instanton size $\bar\rho$ and the instanton 
density $1/\bar R^4$, which are well constrained in vacuum but
not yet fully under control at finite density.
In Ref.~\cite{CaDi99} these parameters have been kept constant.
On the other hand, at finite
density, instantons and anti-instantons could cluster to molecules, 
rather than being randomly distributed~\cite{RSSV00}.
In this way the characteristics of the effective quark-antiquark and
quark-quark interactions can change considerably since the strength
of the vertices related to isolated instantons (the 't Hooft interaction)
decreases, while new vertices with different quantum numbers 
emerge~\cite{SSV95}. 
In particular, in contrast to isolated instantons, molecules do not break 
the $U_A(1)$ symmetry.
As suggested in Ref.~\cite{Sch98}, instantons could also be
lined up in long diquark chains.
The competition between these possibilities has been investigated in 
Ref.~\cite{SSV95} within a simple statistical mechanics approach, 
but the issue is not yet completely settled.

It is clear that these uncertainties remain, if instanton effects are 
approximated by NJL-type interactions (e.g., neglecting the momentum 
dependence of the vertices),
and the situation is even worse for general NJL models, which are based
on symmetries and vacuum phenomenology only.
Another problem which arises in this case is the fact that 
the coupling constants in the quark-quark channel 
cannot unambiguously be related to the coupling 
constants in the quark-antiquark channel.
To see this, consider an NJL-type interaction of the form
\beq
    {\cal L}_\mi{int} \;=\; g_I (\bar q\, \hat \Gamma^{(I)} q)^2~,
\eeq  
where $\hat \Gamma^{(I)}$ is an operator corresponding to the 
quark-antiquark channel $I$. 
As detailed in App.~\ref{fierz}, we can perform a Fierz transformation
to describe the effect of the total (direct plus exchange) interaction by 
a Lagrangian
\beq
    {\cal L}_{q\bar q} \;=\; 
    \sum_{M} G_M\,(\bar q \, \hat \Gamma^{(M)} q)^2~,
\eeq  
where the sum runs over all quark-antiquark channels $M$ and
the effective coupling constants $G_M$ can be calculated from
the $g_I$. Of course, in order to avoid double counting, 
exchange diagrams should not be evaluated explicitly
when using this Lagrangian instead of ${\cal L}_\mi{int}$.

Analogously, we can perform a Fierz transformation into the 
particle-particle channel to derive an effective quark-quark interaction
\beq
    {\cal L}_{qq} \;=\; \sum_{D} H_D\,(\bar q\,\hat\Gamma^{(D)}\,C \,\bar q^T)
    (q^T\,C\,\hat\Gamma^{(D)} q)~,    
\eeq 
where $D$ corresponds to the various diquark channels.
Thus, if we know the underlying Lagrangian ${\cal L}_\mi{int}$, the
quark-antiquark coupling constants $G_M$ and the quark-quark coupling 
constants $H_D$ are uniquely fixed.

However, often we do not have an underlying theory.
In this case we may directly start from ${\cal L}_{q\bar q}$,
e.g., in order to describe the meson sector in vacuum,
and only impose constraints according to the symmetries. 
This is, what we have usually done in this work. 
In this case, if we know {\it all} $G_M$ (including those in the color 
octet channels), we still can calculate the $H_D$. 
However, usually only an incomplete subset of the coupling constants 
$G_M$ can be determined by fitting data. In this case, and without
underlying theory,
there is no unique solution for the quark-quark coupling 
constants $H_D$.

A popular example for ${\cal L}_\mi{int}$ is
the color current interaction
\beq
    {\cal L}_\mi{int} \;=\; -g\,(\bar q \gamma^\mu \lambda_a q)^2~, 
\eeq  
which can be thought of as abstracted from the QCD Lagrangian 
by converting the original $SU(N_c)$ gauge symmetry into a global
symmetry of the quark color currents.
For two flavors and three colors, the corresponding effective 
quark-antiquark Lagrangian reads (see App.~\ref{fierzexamples})
\beq
    {\cal L}_{q\bar q} \;=\quad G_S\,\Big[ 
     ({\bar q} q)^2 \,+\, ({\bar q}\,\vec\tau q)^2 
     \,+\, ({\bar q}\,i\gamma_5 q)^2 
     \,+\, ({\bar q}\,i\gamma_5\vec\tau q)^2\Big]
\quad + \quad \dots~,
\label{Lgqbq}
\eeq
with $G_S =\frac{8}{9}\,g$. 
The ellipsis stands for vector and axial-vector terms and for
color-octet terms. (Note that the terms we have written explicitly
are just the Lagrangian ${\cal L}_1$ defined in \eq{Lflamix1}.)

For the effective quark-quark Lagrangian one obtains
\beq
    {\cal L}_{qq} \;=\; H_S\,\sum_{A=2,5,7} \Big[ 
     (\bar q\,i\gamma_5 \tau_2 \lambda_A \,C \,\bar q^T)
    (q^T\,C\, i\gamma_5 \tau_2 \lambda_A q) \,+\,
     (\bar q\,\tau_2 \lambda_A \,C \,\bar q^T)
    (q^T\,C\,\tau_2 \lambda_A q) \Big]
    \quad + \quad \dots~,
\label{Lgqq}
\eeq 
with $H_S =\frac{2}{3}\,g$. Here the ellipsis comprises
vector and axial-vector terms, as well as color-sextet terms.

Comparing these results, we find that the coupling constant
$H_S$ in the scalar diquark channel is related 
to the coupling constant $G_S$ in the scalar quark-antiquark 
channel as $H_S : G_S = 3:4$.
Accidentally, the same relation follows if we start from a
two-flavor instanton induced interaction (see App.~\ref{fierzexamples}). 
Nevertheless, this relation is not universal. 
For instance, if we choose ${\cal L}_\mi{int} = {\cal L}_1$
as defined in \eq{Lflamix1}, we still obtain effective interactions
of the form of \eqs{Lgqbq} and (\ref{Lgqq}), but with $G_S = G_1$
and $H_S = 0$.
Thus, if we only know the value of $G_S$, e.g., from a fit to the 
pseudoscalar spectrum, we cannot infer the value of $H_S$ without
making assumptions about the underlying interaction.

If the quark-antiquark interaction has been constrained empirically,
the most natural solution to this problem would be to determine the
quark-quark coupling constants empirically, too. 
Unfortunately, the analog to the meson spectrum would be a diquark
spectrum, which of course does not exist in nature.
This means, one would have to fit the baryon spectrum by solving 
Fadeev equations, which is much more difficult. Fits which have been
performed so far seem to be consistent with the one-gluon relation
$H_S : G_S = 3:4$, but there are large 
uncertainties.
In this context it would also be interesting to look at 
the recently discovered $\Theta^+(1540)$ 
baryon~\cite{LEPS,DIANA,CLAS,SAPHIR,HERMES}
which is a candidate to be a $uudd\bar s$ pentaquark, being a member
of a flavor antidecuplet. This state and in particular its small width 
have been predicted some time ago by Praszalowicz~\cite{Pra87} and by
Diakonov, Petrov, and Polyakov~\cite{DPP97} within the chiral soliton 
model (but see Refs.~\cite{Cohen,Jaffe} for controversial opinions).
However, as recently suggested by Jaffe and Wilczek~\cite{JaWi03}, 
it could also
be understood as two highly correlated $ud$ pairs, forming a
scalar color and flavor antitriplet, bound to an $\bar s$ quark.
If true, this could provide interesting information about the 
interaction in the scalar diquark channel.
Unfortunately, there are many competing scenarios, like the model
of Shuryak and Zahed, where one of the scalar diquarks is replaced by a
tensor one~\cite{ShZa03}.

It is clear that NJL-type models cannot yield quantitative predictions, 
e.g., about the size of the color superconducting gap,
until more information about the parameters is available. Ultimately, 
this information must come from ``outside'', e.g., from Dyson-Schwinger
calculations, improved instanton models or, if possible, from lattice
calculations.   
The NJL coupling constants may then play the role of Fermi liquid
parameters, possibly derived using renormalization group 
techniques~\cite{EHS99,ScWi99a}, allowing for a simplified description 
of the quark matter in a given density 
regime\footnote{Non-Fermi liquid corrections to normal conducting
quark matter have been investigated in Refs.~\cite{BodV01,ScSc04,GIR04}.}.

At present, NJL models are basically used in this way although, of course,
on a much more speculative basis.
We have already pointed out that the simplicity of the NJL interactions
allows for the simultaneous investigation of several different condensates,
uncovering important interdependencies which are much harder
to explore in other approaches. 
This will be discussed in more technical details in the next section
and stays the most important theme of this work. 
Lacking better prescriptions, we will usually employ vacuum parameters
and simple relations, like $H_S : G_S = 3:4$, to fix the diquark coupling. 
With these parameters, we typically find a scalar diquark gap of the order 
$\sim 100$~MeV, in agreement with other calculations, e.g.~\cite{ARW98,RSSV98}.
Of course, since the latter were based on similar assumptions, this cannot be
taken as a proof for the correctness of these numbers.
However, we can study what happens to them if additional effects are taken 
into account. Our following investigations should mainly be interpreted
in this way.

When dealing with NJL-type models we should also keep in mind that
the Lagrangian is only {\it globally} symmetric under $SU(3)$-color. 
Hence, strictly speaking, there is no color superconductivity,
but only ``color superfluidity''. The spontaneous breaking of the
global $SU(3)$-color symmetry leads to false Goldstone modes
which are absent in QCD due to the Meissner effect. (The ``would-be"
Goldstone bosons are ``eaten'' by the gluons.)
Therefore the spectrum of bosonic excitations has to be interpreted
with great care. For the fermionic degrees of freedom the
replacement of a gauge symmetry by a global one should be less
problematic.

\section{Interplay with other condensates}
\label{interplay}

In Chap.~\ref{bag} we have discussed the properties of two-flavor
quark matter, mostly concentrating on the role of the quark (-antiquark) 
condensate 
\begin{equation}
     \phi \;=\;\ave{\bar q \,q}~,
\label{phi}
\end{equation}
which is related to chiral symmetry breaking and the non-trivial vacuum
structure. We have now seen that there are good arguments to believe
that two-flavor quark matter at high density is dominated by the 
scalar diquark condensate $\delta$ (\eq{delta}), which has been discussed
in some details in \sect{sca}.
In the chiral limit, and assuming that the chiral phase transition 
coincides with the deconfinement one, it is clear that $\delta$ and
$\phi$ characterize two disjoint regimes: In the hadronic phase we have
$\phi \neq 0$ but there are no free quarks which could condense, while 
in the deconfined phase $\phi = 0$ if chiral symmetry is restored. 
The situation is of course different when chiral symmetry is explicitly
broken by a non-vanishing current quark mass. 
In this case, as we have seen before,  $\phi$ cannot vanish completely
and coexists with the diquark condensate above $\mu_c$.
The question is then, whether this has sizeable effects or whether we 
can safely neglect the influence of $\phi$ in the deconfined phase. 

Obviously, NJL-type models which have been employed for studying both,
spontaneous chiral symmetry breaking and diquark condensation,
offer the nice possibility to study both condensates
and their interplay on the same footing. This has been done first
by Berges and Rajagopal~\cite{BeRa99}.
Their results are in qualitative agreement with our general considerations
above: In the chiral limit, 
they find a first-order phase transition from the vacuum
where $\phi \neq 0$ and $\delta = 0$ to a
high-density phase with $\delta \neq 0$ and $\phi = 0$.
After including a non-vanishing current quark mass, $\delta$ remains
zero below the phase transition, but $\phi$ does no longer drop to 
zero at the transition point. In fact, just above $\mu_c$, the gaps
related to $\delta$ and $\phi$ are of similar magnitude.

We can go further~\cite{BHO02}:
In a fully self-consistent treatment, one has to include all possible
condensates which are not protected by unbroken 
symmetries\footnote{This does not include equivalent condensates, i.e.,
condensates which can be obtained from the considered ones by one
of the spontaneously broken symmetry transformations.}.
First of all, at finite density, Lorentz invariance is broken and 
therefore the existence of Lorentz non-invariant expectation values
becomes possible. The most obvious example is of course the density itself,
\beq
    n \;=\; \ave{\bar q \,\gamma^0\,q} \;,
\label{n}
\end{equation}
which transforms like the time component of a 4-vector.
We have already seen in \sect{njlt0} that the influence of $n$ on the 
phase structure can be quite large if vector interactions are present.  

In a similar way, there could be a Lorentz non-invariant diquark 
condensate~\cite{BaLo84,LaRh99,ABR99},
\beq
     \delta_0 \;=\; \ave{q^T\,C\gamma^0\gamma_5\,\tau_2 \,\lambda_2 \,q}~,
\label{delta0}
\end{equation}
which also transforms like the time component of a 4-vector.

Moreover, since in the presence of $\delta$ or $\delta_0$  
color $SU(3)$ is broken, there is no reason to assume that all other
condensates are color-$SU(3)$ invariant in this state. For
instance, we should expect that the contributions of red and blue
quarks, $\phi_r$ and $\phi_b$,  to the quark condensate $\phi$
could be different, thus giving rise to a non-vanishing
expectation value \beq
     \phi_8 \;=\; \ave{\bar q \,\lambda_8\,q}
            \;=\; \frac{2}{\sqrt{3}}\,(\phi_r - \phi_b) \;.
\label{phi8}
\end{equation}
Here we have assumed that the condensate of green quarks and antiquarks,
$\phi_g$, is equal to $\phi_r$, because we do not want to break the
color-$SU(2)$ subgroup which was left unbroken by $\delta$ and $\delta_0$.

Similarly, the densities of red and
blue quarks will in general not be the same, i.e., in addition to
the total number density $n = 2n_r + n_b$ there could be
a non-vanishing expectation value 
\beq
     n_8 \;=\;\ave{\bar q \,\gamma^0\,\lambda_8\,q}
            \;=\; \frac{2}{\sqrt{3}}\,(n_r - n_b) \;.
\label{n8}
\end{equation}
Since these color-symmetry breaking expectation values,
induced by the presence of color-symmetry breaking diquark condensates,
could in turn influence the properties of the diquark condensates,
in principle, all condensates should be studied in a self-consistent way.

\begin{table}[t]
\begin{center}
\begin{tabular}{|l| c c c c c c c c c c c c c|}
\hline
&&&&&&&&&&&&&\\[-3mm]
& \quad & $n$   & \quad & $n_8$ & \quad & $\phi$ 
& \quad & $\phi_8$ & \quad & $\delta$ & \quad & $\delta_0$ & \quad
\\[1mm]
\hline
&&&&&&&&&&&&&\\[-3mm]
 Lorentz   && \ding{56} && \ding{56} && \ding{52} 
           && \ding{52} && \ding{52} && \ding{56} & \\
 $U(1) $   && \ding{52} && \ding{52} && \ding{52} 
           && \ding{52} && \ding{56} && \ding{56} & \\
 $U_A(1) $ && \ding{52} && \ding{52} && \ding{56} 
           && \ding{56} && \ding{56} && \ding{52} & \\
 $SU(2)_V$ && \ding{52} && \ding{52} && \ding{52} 
           && \ding{52} && \ding{52} && \ding{52} & \\
 $SU(2)_A$ && \ding{52} && \ding{52} && \ding{56} 
           && \ding{56} && \ding{52} && \ding{56} & \\
 $SU(3)_c$ && \ding{52} && \ding{56} && \ding{52} 
           && \ding{56} && \ding{56} && \ding{56} & \\
\hline
\end{tabular}
\end{center}
\caption{\small Symmetries conserved (\ding{52}) or not conserved 
(\ding{56}) by the condensates considered in this section.
}
\label{tab2scsym}
\end{table} 

The various condensates and their symmetry properties are listed in 
\tab{tab2scsym}.
Given that Lorentz invariance, color $SU(3)$, and chiral symmetry
are broken, this is the minimal set of condensates one has to take into
account in a fully self-consistent calculation.  
Of course, if favored by the interaction, other condensates which break
additional symmetries are possible. One example, a spin-1 condensate
which breaks the rotational invariance, will be discussed in \sect{aniso}.

\subsection{Hartree-Fock approach}
\label{2SCHF}

To illustrate the necessity of including the additional condensates
\eqs{n} to (\ref{n8}) in a ``fully self-consistent calculation''
and what we mean by this term we consider the generic NJL-type Lagrangian
\beq
    \hat{\cal L} \;=\; {\cal L} + \mu\,q^\dagger q \;=\;
    \bar q\,(i \delsl - m + \mu\gamma^0)\,q    
   + g_i \,(\bar q\,\Gamma_i \,q)^2~,
\label{LHF}
\eeq
where $\Gamma_i$ are arbitrary local operators in Dirac, flavor and 
color space, and $g_i$ are the corresponding coupling constants. (We 
implicitly assume a sum over $i$.) 
We have already added a chemical potential term $\mu\,q^\dagger q$,
which is formally equivalent to a Lorentz non-invariant energy term
in the Lagrangian. 

It is quite obvious that the fundamental Bogoliubov-Valatin approach
to derive the gap equation becomes extremely involved when more than
one condensate is present\footnote{A variational analysis, which includes
several color and flavor dependent diquark and quark-antiquark condensates 
has recently been presented in Ref.~\cite{MiMi03}.}. 
In this case it is most convenient to apply
Nambu-Gorkov formalism~\cite{NG} (see also \cite{FW71,Hos98}).
To that end, one introduces charge conjugated fields and operators,
\beq
    q^C(x) \;=\; C\,\bar q^T(x)~,\qquad \bar q^C(x) \;=\; q^T(x)\,C~,\qquad
    \Gamma_i^C \;=\;-\,C\,\Gamma^T\,C~,   
\eeq
and rewrites \eq{LHF} as
\beq
    \hat{\cal L} \;=\;
    \frac{1}{2}\,[\,\bar q \,(i \delsl - m + \mu\gamma^0)\,q \,+\,
    \bar q^C\,(-i \overleftarrow \delsl - m - \mu\gamma^0)\,q^C\,]    
    \,+\, \frac{1}{4}\,g_i \,[\,(\bar q\,\Gamma_i\,q)  
    + (\bar q^C \,\Gamma_i^C \,q^C)\,]^2~.
\eeq
Next, one formally doubles the number of degrees of freedom
by treating $q$ and $q^C$ as independent variables.
To this end one defines a bispinor field $\Psi$ and operators $\hat\Gamma_i$
in the corresponding bispinor space,
\beq
    \Psi(x) = \frac{1}{\sqrt{2}}\,\left(\begin{array}{c} \!\!q(x)\!\! \\
     \!\!q^C(x)\!\! \end{array}\right)~,\qquad
    \hat\Gamma_i = \left(\begin{array}{cc} \Gamma_i &  0 \\
    0 & \Gamma_i^C \end{array}\right)~.
\label{NGdef}
\eeq
In this way the Lagrangian can be written in a rather compact
form,
\beq
     \hat{\cal L} \;=\; \bar\Psi\,\tilde S_0^{-1}(x)\,\Psi 
     + g_i (\bar\Psi \,\hat\Gamma_i\, \Psi)^2~, 
\eeq
quite similar to the original Lagrangian, \eq{LHF}. 
Here $\tilde S_0^{-1}(x)$ plays the role of the inverse bare propagator
of the bispinor fields in coordinate space. Its Fourier transform in
momentum space is given by
\beq
    S_0^{-1}(p) \;=\; \left(\begin{array}{cc} \psl + \mu\gamma^0 -m &  0 \\
    0 & \psl - \mu\gamma^0 -m \end{array}\right)~.
\label{NGS0inv}
\eeq
Now we want to take into account self-energy contributions to the
propagator due to the interaction.
According to the quantum numbers of the six different condensates
discussed in the previous section we make the following ansatz:
\beq
    S^{-1}(p) \;=\;  S_0^{-1}(p) - \hat \Sigma(p)
\;=\; \left(\begin{array}{cc} \psl + \hat\mu\gamma^0 - \hat M 
    & (\Delta + \Delta_0\gamma^0)\,\gamma_5\tau_2\lambda_2
\\
      (-\Delta^* + \Delta_0^*\gamma^0)\,\gamma_5\tau_2\lambda_2
    & \psl - \hat\mu\gamma^0 - \hat M \end{array}\right)~,
\label{NGSinv}
\eeq
where
\beq
     \hat M = M_0 + M_8\,\lambda_8 \qquad \text{and} \qquad
     \hat \mu = \tilde\mu + \tilde\mu_8\,\lambda_8 
\eeq
are matrices in color space, describing color dependent constituent
quark masses and color dependent effective chemical potentials. 
Unlike the bare inverse propagator, \eq{NGSinv} also contains non-diagonal 
elements in Nambu-Gorkov space. These
connect quark fields with charge conjugated quark fields and thus
describe self-energy contributions due to diquark condensates.
Here the advantage of the bispinor notation becomes obvious.

We can now proceed analogously to \sect{njlmesons}:
We first invert \eq{NGSinv} to calculate the propagator $S(p)$.
The result is a rather lengthy expression which is given in App.~\ref{prop}.
Here we just mention that, due to the non-diagonal Nambu-Gorkov components
in \eq{NGSinv}, the propagator also gets non-diagonal components which
are the origin of the anomalous propagation mentioned earlier. 
Using this propagator we then calculate the quark self-energy in some given
approximation scheme. Identifying this with $\hat\Sigma$ in \eq{NGSinv}
we finally obtain a set of self-consistent gap equations for the 
parameters $M_0$, $M_8$, $\tilde\mu$, $\tilde\mu_8$, $\Delta$, and $\Delta_0$
of our ansatz. 

\begin{figure}[t]
\begin{center}
\epsfig{file=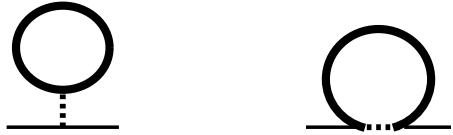,width = 6.0cm}
\caption{\small Hartree (left) and Fock (right) contribution to the quark
         self-energy. The bold solid lines indicate dressed
         Nambu-Gorkov propagators. The dotted lines symbolize the
         interaction.}
\label{fighf}
\end{center}
\end{figure}

In the following, we consider the quark self-energy in Hartree-Fock
approximation,
\beq
    \hat \Sigma \;=\; \hat \Sigma_H \;+\; \hat \Sigma_F
\eeq
which is illustrated in \fig{fighf}. The Hartree term (left) is given by
\beq
    \hat \Sigma_H \;=\; 2i\,g_i \, \int\dfp\;
    \frac{1}{2} \tr{\hat\Gamma_i\,S(p)} \; \hat\Gamma_i~. 
\label{SigmaH}
\eeq
Here the trace is to be taken over Dirac, flavor, color, and the two 
bispinor degrees of freedom. Note that in order to correct for the
artificial doubling of the number of
degrees of freedom we have to introduce
a factor $\frac{1}{2}$ in front of the trace\footnote{This can be
seen most easily, if we consider a scalar interaction 
$\Gamma_i = \Gamma_i^C = \unity$ and the simplified case $M_0 \neq 0$,
but $M_8 = \tilde\mu = \tilde\mu_8 = \Delta = \Delta_0 = 0$. In this case
$S^{-1}(p)$ is easily inverted and $S(p)$ is diagonal in Nambu-Gorkov
space with $S_{11} = S_{22}$ being ordinary fermion propagators with
mass $M_0$. Taking the trace gives thus twice the result without 
fermion doubling, which has to be corrected by the factor $\frac{1}{2}$.
A more general derivation of the factor $\frac{1}{2}$ can be found,
e.g., in Ref.~\cite{Ri00a}}.
It is immediately clear from \eq{SigmaH} that in
Hartree approximation only self-energy contributions proportional
to the operators $\hat\Gamma_i$ can arise.
In particular, since in our case all operators are diagonal in 
Nambu-Gorkov space, $\Sigma_H$ is also diagonal, and hence 
$\Delta = \Delta_0 = 0$.
In order to obtain non-vanishing diquark gaps we must therefore
consider the Fock contribution to the self-energy, which is given by
\beq
    \hat \Sigma_F \;=\; -2i\,g_i \, \int\dfp\;
       \hat\Gamma_i \,S(p)\, \hat\Gamma_i~.    
\label{SigmaF}
\eeq
Unlike the Hartree term its operator structure is not restricted to
terms proportional to the $\hat\Gamma_i$. 
In fact, because of the rather complicated form of the propagator, 
it is quite difficult to ``see'' without explicit calculation 
how the matrix $\hat\Gamma_i \,S(p)\, \hat\Gamma_i$ looks like.
Thus, if we started with some arbitrary inverse propagator $S^{-1}(p)$, 
it could happen that the Hartree-Fock self-energy calculated from 
the corresponding propagator contains operators which are not present 
in the original ansatz. For instance, if we started from \eq{NGSinv},
but without allowing for color dependent constituent quark masses
(i.e., terms proportional to $\lambda_8$ in the diagonal Nambu-Gorkov
components) it could turn out that the resulting Hartree-Fock self-energy
contains such terms, and the equations cannot be closed.  
If in such a case we simply ignored the extra terms,
this would be an example of a ``not fully self-consistent'' calculation.

As pointed out in the previous section, the only way to prevent this 
problem is to allow for all possible terms which are consistent with
the unbroken symmetries of the system. Here ``unbroken'' means unbroken,
including all other condensates. Hence, if we add a new condensate which 
breaks additional symmetries, this can induce further condensates 
which then have to be taken into account.
Since \eq{NGSinv} has been constructed in this way, we can be sure that 
this ansatz will lead to a closed set of gap equations.

\subsection{Thermodynamic potential}
\label{2scformalism}

As we have seen earlier, it is often advantageous to derive the gap 
equations from a thermodynamic potential, rather than directly from
a Dyson series. In the present case this even turns out to be simpler.
Therefore we do not pursue the programme outlined in \sect{2SCHF} in
detail, but start over again and first
calculate the thermodynamic potential of the system.

As before, the first step will be that we linearize the interaction 
terms of the Lagrangian in the presence of the condensates we want
to take into account. In this way, however, obviously only those 
condensates can contribute which correspond to an interaction channel
of the Lagrangian. In particular, when the Lagrangian contains only
quark-antiquark interactions, like \eq{LHF}, there will be no contribution
from diquark condensates to the linearized Lagrangian. 
This is because the linearization procedure corresponds to a Hartree
approximation. 
However, as briefly discussed in \sect{2scnjl}, particle-particle terms
can be included via Fierz transformation.
Starting from a given Lagrangian, one transforms the
interaction terms into the particle-particle channel (as well as into
the particle-antiparticle exchange channel) and adds the
resulting terms to the original Lagrangian. 
In this way one gets a new Lagrangian ${\cal L}_\mi{eff}$ which contains 
both, particle-antiparticle and particle-particle interactions.
By construction, this new Lagrangian is meant to be used in 
Hartree approximation only. A few examples for 
${\cal L}_\mi{eff}$ are given in App.~\ref{fierzexamples}.

At this point we stay rather general and consider the Lagrangian
\beq
    {\cal L}_\mi{eff} \;=\; \bar q (i \delsl - m) q
                      \;+\; {\cal L}_{q\bar q} \;+\; {\cal L}_{qq}
\label{Leff}
\end{equation}
with a quark-antiquark interaction of the form
\beq
     {\cal L}_{q\bar q} \;=\; G_s^{(0)}\,(\bar q q)^2
                        \;+\; G_s^{(8)}\,(\bar q \lambda_aq)^2
                        \;+\; G_v^{(0)}\,(\bar q \gamma^0q)^2
                        \;+\; G_v^{(8)}\,(\bar q \gamma^0 \lambda_aq)^2
                        \;+\; \dots~,
\label{Lqqbar2sc}
\end{equation}
and a quark-quark interaction
\beq
     {\cal L}_{qq} \;=\; H\,(\bar q \,i\gamma_5 C\,\tau_2 \lambda_A \,\bar q^T)
                           (q^T C\,i\gamma_5 \tau_2 \lambda_A \, q)
     \;+\;  H_0\,(\bar q \,\gamma^0\gamma_5 C\, \tau_2 \lambda_A \,\bar q^T)
              (q^T C \,\gamma^0\gamma_5 \tau_2 \lambda_A \, q)
     \;+\; \dots~.
\label{Lqq2sc}
\end{equation}
Here the dots indicate possible other channels, not related to
the condensates $\phi$, $\phi_8$, $\rho$, $\rho_8$, $\delta$, or $\delta_0$.
In \eq{Lqqbar2sc} $\lambda_a$, $a = 1, \dots, 8$, denotes the eight Gell-Mann 
matrices, while in \eq{Lqq2sc} $\lambda_A$, $A = 2, 5, 7$, denotes the 
antisymmetric  Gell-Mann matrices.
All color indices are understood to be summed over, i.e., the Lagrangian
is invariant under a global $SU(3)_c$.

Now we linearize this Lagrangian in the presence of the six condensates 
and then express the result in terms of bispinor fields, \eq{NGdef}.
In this way one finds
\beq
    {\cal L}_\mi{eff}^\mi{mean\,field} +  \mu\,q^\dagger q  \;=\;
     \bar\Psi\,\tilde S^{-1}\,\Psi \,-\, V~,
\label{2scLmf}
\eeq 
where $\tilde S^{-1}$ is the Fourier transform of the {\it dressed}
inverse Nambu-Gorkov propagator given in \eq{NGSinv}, and
\beq
    V \;=\; \frac{(M_0 - m)^2}{4G_s^{(0)}} \;+\; \frac{M_8^2}{4G_s^{(8)}}
      \;+\; \frac{(\tilde\mu - \mu)^2}{4G_v^{(0)}} \;+\; \frac{\tilde\mu_8^2}{4G_v^{(8)}}
      \;+\; \frac{|\Delta|^2}{4H} \;+\; \frac{|\Delta_0|^2}{4H_0} \;.
\label{2scpot}
\end{equation}
Here we have identified the constituent quark masses
\beq
    M_0 \;=\; m - 2G_s^{(0)}\phi~,\qquad
    M_8 \;=\; -2G_s^{(8)}\phi_8~,
\label{M08}
\end{equation}
the effective chemical potentials
\beq
    \tilde\mu \;=\; \mu + 2G_v^{(0)}n~,\qquad
    \tilde\mu_8 \;\;=\; 2G_v^{(8)}n_8~,
\label{mu08}
\end{equation}
and the diquark gaps
\beq
    \Delta \;=\; -2H\delta~,\qquad
    \Delta_0 \;=\; 2H_0 \delta_0~.
\label{Delta12}
\end{equation}

For later convenience, but also for the interpretation of the results,
it is useful to perform linear combinations to get red and blue quantities,
e.g., red and blue constituent quark masses
$M_r = M_0 + \frac{1}{\sqrt{3}}\,M_8$ and
$M_b = M_0 - \frac{2}{\sqrt{3}}\,M_8$.
We then find
\begin{alignat}{3}
  M_r \;=\;& m &\;-\;& &\frac{2}{3}(6G_s^{(0)} &+ 2G_s^{(8)})\phi_r
             \;-\; \frac{2}{3}(3G_s^{(0)} - 2G_s^{(8)})\phi_b\,,
\nonumber \\
  M_b \;=\;& m &\;-\;& &\frac{2}{3}(6G_s^{(0)} &- 4G_s^{(8)})\phi_r
             \;-\; \frac{2}{3}(3G_s^{(0)} + 4G_s^{(8)})\phi_b\,,
\nonumber \\
  \tilde\mu_r \;=\;& \mu &\;+\;& &\frac{2}{3}(6G_v^{(0)} &+ 2G_v^{(8)})n_r
                 \;+\; \frac{2}{3}(3G_v^{(0)} - 2G_v^{(8)})n_b\,,
\nonumber \\
  \tilde\mu_b \;=\;& \mu &\;+\;& &\frac{2}{3}(6G_v^{(0)} &- 4G_v^{(8)})n_r
                 \;+\; \frac{2}{3}(3G_v^{(0)} + 4G_v^{(8)})n_b\,.
\label{gapsr}
\end{alignat}

Since \eq{2scLmf} is bilinear in the bispinor fields $\Psi$
(+ the constant $V$), the thermodynamic potential at temperature $T$
and chemical potential $\mu$ is evaluated straight forwardly, 
\beq
    \Omega(T,\mu) \;=\; -T \sum_n \int \frac{d^3p}{(2\pi)^3} \;
    \frac{1}{2}\,{\rm Tr}\; \ln \Big(\frac{1}{T}\,S^{-1}(i\omega_n, \vec p)
    \Big)
    \;+\; V \;.
\label{2scOmega}
\end{equation}
Formally, the main difference to the earlier expression, \eqs{OmegaMF} and
(\ref{Omegatr}), is
that the trace has now been extended to the bispinor space, which
must again be corrected by a factor $\frac{1}{2}$ in front.
In practice, \eq{2scOmega} is more complicated because the inverse
propagator $S^{-1}$, \eq{NGSinv}, has not the form of a
free fermion inverse propagator, i.e., we can not simply copy a textbook 
result. 

Evaluating the trace and performing the Matsubara sum one obtains
\begin{alignat}{2}
    \Omega(T,\mu) \;=\; -4 \int \frac{d^3p}{(2\pi)^3} \;\Big\{\quad
    2\hspace{5mm} \Big(&\frac{\omega_- + \omega_+}{2} &
            &+\; T\,\ln(1 + e^{-\omega_-/T}) \;+\; T\,\ln(1 + e^{-\omega_+/T})
      \Big) \nonumber \\
           +\quad\Big(& \quad E_{p,b} &
            &+\; T\,\ln(1 + e^{-E_-/T})
              \;+\; T\,\ln(1 + e^{-E_+/T})
      \Big) \Big\} \nonumber \\
    \;+\; V\, \hspace{0.95cm} +\; const.~.&&&
\label{2scOmegaexp}
\end{alignat}
The factor 4 in front of the integral corresponds to the spin and flavor
degeneracy, while the factor 2 in the first line of the integrand
reflects the two paired colors.
The second line corresponds to the blue quarks
which do not participate in a diquark condensate. Their dispersion
laws are therefore the standard ones,
\beq
    E_\mp \;=\; E_{p,b} \mp \tilde\mu_b
                 \;=\; \sqrt{{\vec p}^{\;2}+ M_b^2} \mp \tilde\mu_b~.
\label{epspm}
\end{equation}
On the other hand the dispersion laws of the red and green quarks
which enter the first line of \eq{2scOmegaexp} are much more complicated,
\beq
    \omega_\mp \;=\; \sqrt{ {{\vec p}^{\;2} + M_r^2 + \tilde\mu_r^2 + |\Delta|^2
                    + |\Delta_0|^2 \mp 2s}}\,,
\label{2scepm}
\end{equation}
with
\beq
    s \;=\; \sqrt{(\tilde\mu_r^2 + |\Delta_0|^2){\vec p}^{\;2} + t^2}\,,\qquad
    t \;=\; M_r \tilde\mu_r - Re(\Delta\Delta_0^*) \,.
\label{stdef}
\end{equation}
These dispersion laws will be discussed in more detail in \sect{2scdisp}.

The self-consistent solutions of the condensates
correspond again to the stationary points of the thermodynamic potential, 
\beq
      \frac{\delta\Omega}{\delta M_0} \;=\; \frac{\delta\Omega}{\delta M_8}
   \;=\; \frac{\delta\Omega}{\delta\tilde\mu} \;=\; \frac{\delta\Omega}{\delta\tilde\mu_8}
   \;=\; \frac{\delta\Omega}{{\delta\Delta}^*}
   \;=\; \frac{\delta\Omega}{{\delta\Delta_0}^*} \;=\; 0 \,.
\label{stat}
\end{equation}
This leads to the following expressions for the various expectation values:
\begin{alignat}{2}
    \phi_r \;&=  & -4 &\int \frac{d^3p}{(2\pi)^3}\;\Big\{\; 
    \frac{M_r s - \tilde\mu_r t}{2s\,\omega_-}\;\tanh{(\frac{\omega_-}{2T})}
    \;+\;
    \frac{M_r s + \tilde\mu_r t}{2s\,\omega_+}\;\tanh{(\frac{\omega_+}{2T})} 
    \;\Big\}~,
\nonumber \\
    \phi_b \;&= & -4 &\int \frac{d^3p}{(2\pi)^3}\; \frac{M_b}{E_{p,b}} \;
    \Big(1 - n_{p,b}(T,\tilde\mu_b) - \bar n_{p,b}(T,\tilde\mu_b)\Big)~,
\nonumber \\
    n_r \;&= & 4 &\int \frac{d^3p}{(2\pi)^3}\;\Big\{\; 
    \frac{\tilde\mu_r(s-{\vec p}^{\;2}) - M_r t}{2s\,\omega_-}\;
    \tanh{(\frac{\omega_-}{2T})}
    \;+\;
    \frac{\tilde\mu_r(s+{\vec p}^{\;2}) + M_r t}{2s\,\omega_+}\;
    \tanh{(\frac{\omega_+}{2T})}
    \;\Big\}~,
\nonumber \\
    n_b \;&= & 4 &\int \frac{d^3p}{(2\pi)^3}\; 
    \Big(n_{p,b}(T,\tilde\mu_b) - \bar n_{p,b}(T,\tilde\mu_b)\Big)~,
\nonumber \\
    \delta \;&= & -8 &\int \frac{d^3p}{(2\pi)^3}\;\Big\{\; 
    \frac{\Delta s + \Delta_0 t}{2s\,\omega_-}\;\tanh{(\frac{\omega_-}{2T})}
    \;+\;
    \frac{\Delta s - \Delta_0 t}{2s\,\omega_+}\;\tanh{(\frac{\omega_+}{2T})}
    \;\Big\}~,
\nonumber \\
    \delta_0 \;&= & 8 &\int \frac{d^3p}{(2\pi)^3}\;\Big\{\; 
    \frac{\Delta_0 (s-{\vec p}^{\;2}) + \Delta t}{2s\,\omega_-}\;
    \tanh{(\frac{\omega_-}{2T})}
    \;+\;
    \frac{\Delta_0 (s+{\vec p}^{\;2}) - \Delta t}{2s\,\omega_+}\; 
    \tanh{(\frac{\omega_+}{2T})}
    \;\Big\}~,
\label{2scallgaps}
\end{alignat}
where $n_{p,b}$ and $\bar n_{p,b}$ are the usual Fermi occupation functions
for the blue quarks and antiquarks. Note that the thermal factors 
$\tanh{(\omega_\pm/2T)}$ which arise in the expressions related
to the paired quarks go to 1 for $T \rightarrow 0$.

Together with \eqs{Delta12} and (\ref{gapsr}) the above equations
form a set of six
coupled gap equations for $M_r$, $M_b$, $\tilde\mu_r$, $\tilde\mu_b$,
$\Delta$, and $\Delta_0$. Although the expressions for the blue
expectation values $\phi_b$ and $n_b$ formally look like the
corresponding formulae for free particles (cf. \eq{phif}),
they depend on the effective quantities $M_b$ and $\tilde\mu_b$, 
which are influenced by the red quarks via \eq{gapsr}. 

The above equations illustrate nicely the emergence of induced
condensates which are not protected by symmetries.
One immediately sees that the red and blue quark condensates and
and densities are in general different from each other, i.e.,
$\phi_8$ and $n_8$ do not vanish.
Also, in general $\delta$ and $\delta_0$ cannot vanish separately.
This means, the standard scalar diquark condensate 
$\delta$ is in general accompanied by an induced non-vanishing 
expectation value $\delta_0$, even when the coupling in the 
$\delta_0$-channel is repulsive~\cite{ABR99}. 
This is not the case, however, for $m=0$. In this case
there are solutions with
$\phi_r = \phi_b = \delta_0 = 0$, even for $\delta \neq 0$, 
corresponding to unbroken chiral symmetry,
but also solutions with $\phi_r = \phi_b = \delta = 0$ and $\delta_0 \neq 0$
which correspond to an unbroken $U_A(1)$ symmetry
(see \tab{tab2scsym}).
Also, there is always a solution $\delta = \delta_0 = 0$,
corresponding to unbroken $U(1)$ symmetry.
In this case the expressions for $\phi_r$ and $n_r$ get the same
structure as the analogous expressions for the blue quarks
and become equal to them, unless the interaction favors the breaking
of $SU(3)_c$, even without diquark condensates.

Note that this discussion
could only be performed a posteriori: After we have included, e.g.,
$\delta_0$ we see that the resulting gap equations do in general 
not allow for $\delta_0 = 0$.
However, if we had not taken into account $\delta_0$ we would not
immediately have noticed an inconsistency. 
This is different from the Hartree-Fock scheme discussed in the
previous section. 

On the other hand, usually not all condensates which are there
are relevant. For instance for $H_0=0$, which is the case, e.g.,
for instanton mediated interactions (see App.~\ref{fierzexamples}), 
we have $\Delta_0 = 0$.
Thus, although $\delta_0$ does in general not vanish and could be calculated
from \eq{2scallgaps}, it does not influence the other quantities.
For instance the dispersion law for the red and green quarks 
and the gap equation for $\Delta$ reduce 
to the ``standard forms'' (cf. \eqs{E1} and (\ref{bcsgap})),
\beq
\omega_{\mp}(\vec p) = \sqrt{(\sqrt{\vec p^2 + M_{r}^2} \mp \tilde\mu_{r})^2 +
|\Delta|^2}
\label{epmstandard}
\eeq
and 
\beq
    \Delta \;=\; 8\,H\,\Delta\,\int\dtp\,
    \Big\{\frac{1}{\omega_-}\;\tanh{(\frac{\omega_-}{2T})} 
    \;+\;
    \frac{1}{\omega_+}\;\tanh{(\frac{\omega_+}{2T})}\Big\}~,
\label{gapstandard}
\eeq
which are obviously much simpler than the general expression given above.
(They are, however, less ``standard'' than they appear, because
in general $\tilde\mu_r \neq \mu$ and $M_r \neq m$.)

But even when $H_0\neq 0$, $\Delta_0$ (or other induced gap parameters)
might be small and have only a small impact on other quantities.
In this case it makes certainly more sense to neglect these condensates 
and include others, imposed by physics arguments, 
than dealing with a fully self-consistent set, which misses
important channels.
For the remainder of \sect{interplay}, however, we will keep all condensates
discussed above to get some flavor of their possible importance.

\subsection{Dispersion laws and gapless solutions}
\label{2scdisp}

We have just seen that the dispersion laws \eq{2scepm} reduce to the
standard form, \eq{epmstandard}, when $\Delta_0 = 0$.
In this case, there is a gap of $2\Delta$ in the particle-hole
excitation spectrum of the red and green components.
Among others, this has the consequence that their specific heat at 
$T \ll \Delta$ is exponentially suppressed, like in
ordinary superconductors. (The specific heat of color superconducting
quark matter will be discussed in more details in \sect{anisot}.)

In this section we want to discuss the modification of the dispersion law,
assuming that the interaction gives rise to a non-negligible gap parameter
$\Delta_0$.
To that end we first notice that the general expression for $\omega_-$,
\eq{2scepm}, can be written in a way, analogous to \eq{epmstandard},
\begin{equation}
\omega_-(\vec p) \;=\; \sqrt{(\sqrt{\vec p^{\;2} + M_\mi{eff}^2} 
- \mu_\mi{eff})^2 +
|\Delta_\mi{eff}|^2}\,,
\label{2seeff}
\end{equation}
where
\begin{equation}
\mu_\mi{eff} \;=\; \sqrt{\tilde\mu_{r}^2 + |\Delta_0|^2}~, \qquad
M_\mi{eff}^2 \;=\; \frac{(M_{r} \tilde\mu_{r} - Re (\Delta\Delta_0^*))^2}
                      {\mu_\mi{eff}^2}\,,
\label{2scmueff}
\end{equation}
and
\begin{equation}
|\Delta_\mi{eff}|^2 \;=\; |\Delta|^2 + M_{r}^2 - M_\mi{eff}^2~. 
\label{2scDeltaeff}
\end{equation}
Thus, the gap in the spectrum is now given by $\Delta_\mi{eff}$
not by $\Delta$ or $\Delta_0$. 
Formally, this implies the interesting possibility that, 
even for non-vanishing  $\Delta$ or $\Delta_0$, the gap vanishes if 
\beq
    M_r \Delta_0 = -\tilde\mu_r \Delta~.
\label{condition}
\eeq
In this case $\omega_-$ would have a node at
\beq
    \vec p_\mi{node}^{\,2} \;=\; \mu_\mi{eff}^2 - M_\mi{eff}^2~,
\eeq
provided the r.h.s.\@ is positive.
For $\tilde\mu_r \neq 0$ \eq{condition} can be used to eliminate
$\Delta$ in \eq{2scmueff},
\beq
    \mu_\mi{eff}^2 = \tilde\mu_r^2 \, (1 + \frac{|\Delta_0|^2}{\tilde\mu_r^2})
    \,, \quad
    M_\mi{eff}^2 = M_r^2 \, (1 + \frac{|\Delta_0|^2}{\tilde\mu_r^2}) \,,
\end{equation}
and hence
\beq
    \vec p_\mi{node}^{\,2} 
    = (\tilde\mu_r^2 - M_r^2)\, (1 + \frac{|\Delta_0|^2}{\tilde\mu_r^2})~.
\end{equation}
This means, $\tilde\mu_r^2$ must be greater or equal to $M_r^2$ and it
immediately follows from \eq{condition} that a node in the quasiparticle
spectrum is only possible if $|\Delta_0| \geq |\Delta|$.

In the vicinity of the node the quasiparticle takes the form of
a non-relativistic fermion,
\begin{equation}
\omega_-(\vec p) \approx \frac{\vec p^{\;2}}{2\mu_\mi{eff}} - 
\frac{\mu_\mi{eff}^2 - M_\mi{eff}^2}{2\mu_\mi{eff}}  \equiv
\frac{\vec p^{\;2}}{2m^*} - \mu^*~.
\end{equation}
Therefore, in spite of non-vanishing diquark condensates, the specific heat 
of such a system would be linear in $T$. 

In the above discussion it is tacitly assumed that there is an
interaction which yields solutions of the coupled gap equations
for which \eq{condition} holds. How realistic is this assumption? 
Since all quantities which enter \eq{condition} are in general
$\mu$ dependent, it is obvious, that the gapless solutions can 
exist at most at certain values of $\mu$. The only exception
would be that both sides of \eq{condition} vanish. Apart from
the trivial case $\Delta = \Delta_0 = 0$ this could be realized
in the form $M_r = \Delta = 0$. 
It is instructive to study this possibility in a simple toy model
with $m = G_i^{(k)} = H = 0$ and only $H_0 \neq 0$.
If we regularize the divergent integrals by a sharp 3-momentum cut-off
$\Lambda$ and restrict ourselves to $T=0$, the thermodynamic potential
of this schematic model is readily calculated:
\beq
    \Omega_\mi{schem}(T=0,\mu;\Delta_0) \;=\;
    -\frac{1}{6\pi^2} \Big( \;2(\mu^2 + |\Delta_0|^2)^2 + \mu^4 \; \Big)
    \;+\; \frac{|\Delta_0|^2}{4H_0} \quad +\; const.~.
\end{equation}
As a function of $\Delta_0$, $\Omega_{schem}$ is not bounded from below.
However, as discussed earlier, only the self-consistent solutions are
physically meaningful, i.e., $\Delta_0$ is constrained to the stationary
points, $\delta\Omega_{schem} /\delta \Delta_0 = 0$.
There is always a trivial solution with
$\Delta_0 = 0$. For $0 < H_0 < \frac{3\pi^2}{8\mu^2}$ there are
also non-trivial solutions with $|\Delta_0|^2 = \frac{3\pi^2}{8H_0}
- \mu^2$. However, whenever these non-trivial solutions exist, they
correspond to maxima of $\Omega_{schem}$, while at same time the
trivial solution is a local minimum with a lower value of
$\Omega_{schem}$. This means, the non-trivial gapless solution is
unstable. We will come back to this point in the end of
the next section.

\subsection{Numerical results}
\label{results}

In the schematic example discussed in the end of the previous
section the problem was reduced to a single condensate. In this
section we want to present the results of a numerical study of the
full coupled set of gap equations derived in 
Sec.~\ref{2scformalism}~\cite{BHO02}.
As an example we consider color current interaction 
\beq
     {\cal L}_\mi{col} \;=\; \bar q (i \delsl - m) q
              \;-\; g_E \,(\bar q \gamma^0 \lambda_a q)^2
              \;+\; g_M \,(\bar q \vec\gamma \lambda_a q)^2~,
\label{Lhge}
\end{equation}
because it allows to study the interplay of all condensates discussed
above. (As mentioned earlier, for instanton-type interactions 
$\Delta_0$ would vanish.) 
The effective Lagrangian at finite densities
does not need to be Lorentz invariant. We underline this
by explicitly allowing for different ``electric'' and
``magnetic'' coupling constants, $g_E$ and $g_M$.
In the following we will study two cases, namely a Lorentz-invariant
interaction, $g_E = g_M$, and a purely magnetic interaction, $g_E=0$.
The latter might be motivated by the fact that at high densities
electric gluons are Debye screened, whereas, as mentioned earlier,
magnetic gluons are 
only dynamically screened and therefore dominate the interaction.

The effective quark-antiquark interaction ${\cal L}_{q \bar q}$ and the
effective quark-quark interaction ${\cal L}_{qq}$ as given in \eqs{Lqqbar2sc}
and (\ref{Lqq2sc}) can be derived from ${\cal L}_{col}$ by performing the
appropriate Fierz transformations, see App.~\ref{fierzexamples}.
The resulting coupling constants are
\begin{alignat}{3} 
    G_s^{(0)} &= \frac{2}{9}\,(g_E+3g_M)~,\quad &
    G_s^{(8)} &= \phantom{-g_E}-\frac{1}{24}\,(g_E+3g_M)~,\quad&
    H &= \frac{1}{6}\,\,(g_E+3g_M)~, \nonumber \\
    G_v^{(0)} &= \frac{2}{9}\,(g_E-3g_M)~,\quad &
    G_v^{(8)} &= -g_E-\frac{1}{24}\,(g_E-3g_M)~,\quad&
    H_0 &= \frac{1}{6}\,\,(g_E-3g_M)~. \nonumber \\
\label{geff}
\end{alignat}

We begin our discussion with the case of a Lorentz-invariant interaction,
\beq
    g \;:=\; g_E \;=\; g_M\,.
\label{lii}
\end{equation}
Of course, this does not mean that there are only
Lorentz-invariant condensates, since Lorentz-invariance is still broken
by the chemical potential.

If we insert \eq{lii} into \eq{geff} we find that for $g > 0$
the interaction is attractive in the scalar
quark-antiquark channel and in the scalar diquark channel and repulsive
in all other channels of interest.
Of course, non-vanishing expectation values in the repulsive channels
do not develop spontaneously, but only as a result of an external
source, like the chemical potential, or induced by non-vanishing
expectation values in attractive channels.
In fact, as we have already seen in \sect{njlthermo}
in the context of the repulsive vector interaction,
the solutions of the gap equations correspond to maxima of the
thermodynamic potential with respect to variations in the repulsive channels,
whereas they can be maxima or minima with respect to variations in the
attractive channels.

In our numerical calculations we restrict ourselves again to $T$~=~0
and take a sharp 3-momentum cut-off $\Lambda$ to regularize the integrals.
In the following we take $\Lambda$~=~600~MeV, $g\Lambda^2$~=~2.75,
and a bare quark mass $m$~=~5~MeV~\cite{BHO02}. With these parameter values
we obtain a vacuum constituent quark mass $M_r = M_b$~=~407.7~MeV.
This corresponds to a quark condensate $\phi = -2(245.7 {\rm MeV})^3$,
while $\phi_8$, $n$, $n_8$, $\delta$ and $\delta_0$ vanish in vacuum.

\begin{figure}
\epsfig{file=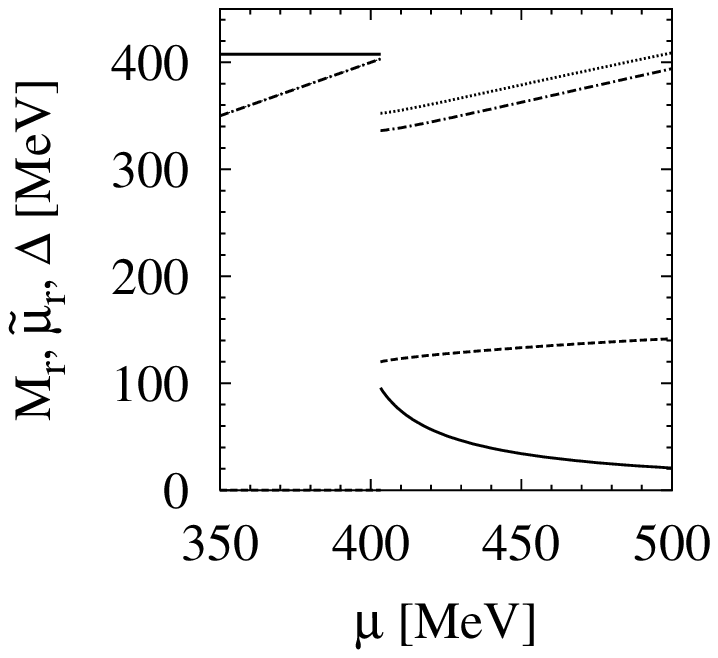,width=7.4cm}
\hfill
\epsfig{file=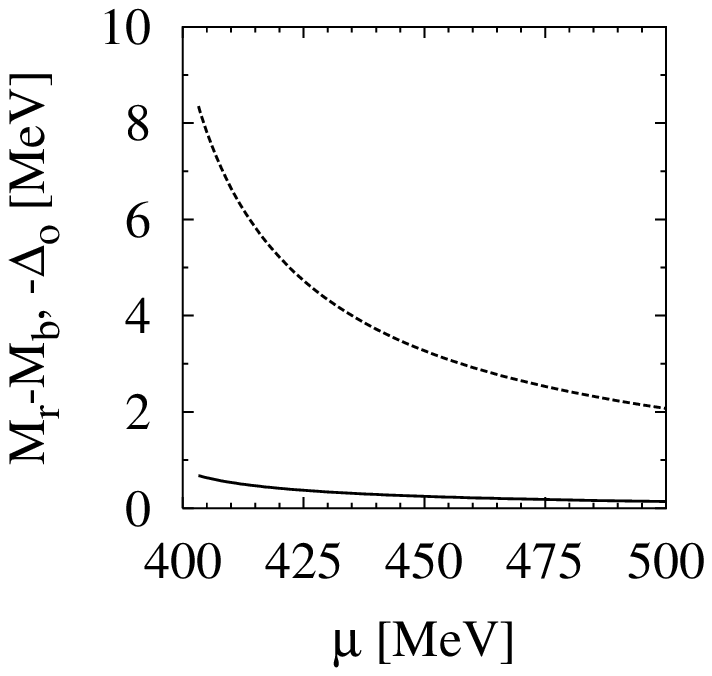,width=7.4cm}
\caption{\small Various quantities obtained with the Lorentz invariant
                interaction $g_E = g_M = 2.75/\Lambda^2$
                as functions of the quark chemical potential $\mu$.
                Left panel: $M_r$ (solid), $\Delta$ (dashed), $\tilde\mu_r$
                (dash-dotted), $\tilde\mu_b$ (dotted).
                Right panel: $M_r-M_b$ (solid), $-\Delta_0$ (dashed).
                Adapted from Ref.~\cite{BHO02}.}
\label{fig2scgaps}
\end{figure}

When we increase the quark chemical potential nothing happens up
to a critical value $\mu_{crit} =$ 403.3~MeV. At this point a
first-order phase transition takes place (type (a) in the classification
of \sect{njlt0}) and all expectation
values under consideration receive non-vanishing values.
This can be inferred from \fig{fig2scgaps} where various
quantities are displayed as functions of $\mu$. In the left panel
we show the constituent quark mass $M_r$, the diquark gap
$\Delta$, and the effective chemical potentials $\tilde\mu_r$ and
$\tilde\mu_b$. In the right panel the mass difference $M_r-M_b$ and the
diquark gap $-\Delta_0$ are plotted\footnote{The gap equations fix $\Delta$
and $\Delta_0$ only up to a common phase. Here we choose $\Delta$ to be 
real and positive. It then follows that $\Delta_0$ is real and negative.}. 
At $\mu=\mu_{crit}$ the
constituent quark masses drop by more than 300~MeV and are no
longer identical. The difference, however, is small,
$M_r$~=~95.7~MeV and $M_b$~=~95.0~MeV. With increasing chemical
potential, both, the masses and their difference, decrease further.
In the diquark channel we find $\Delta$~=~120.0~MeV
at $\mu=\mu_{crit}$, whereas -- similar to what has been found in
Ref.~\cite{ABR99} -- the second diquark gap parameter is more than one 
order of magnitude smaller, $\Delta_0$~=~-8.4~MeV. 
Like the constituent masses, it decreases
with increasing $\mu$, while $\Delta$ is slightly growing in
the regime shown in \fig{fig2scgaps}.

\begin{figure}
\begin{center}
\epsfig{file=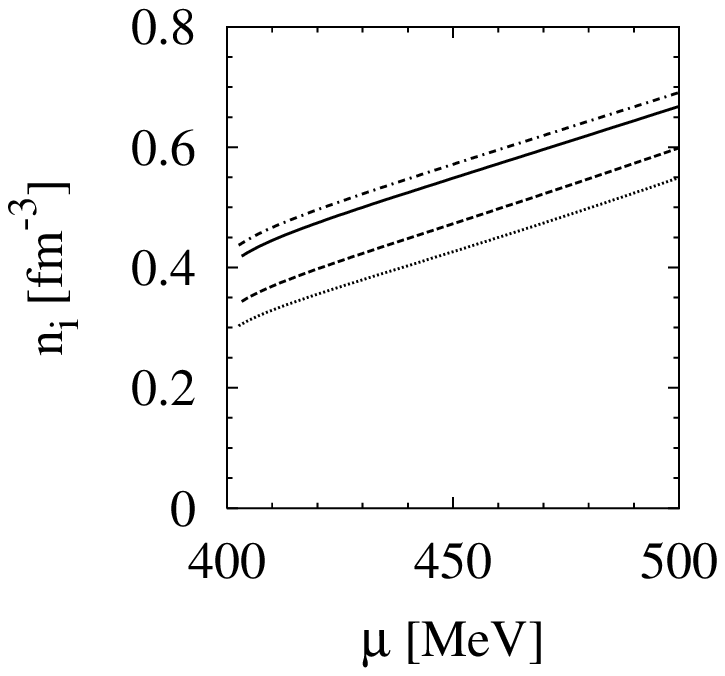,width=7.4cm}
\hfill
\epsfig{file=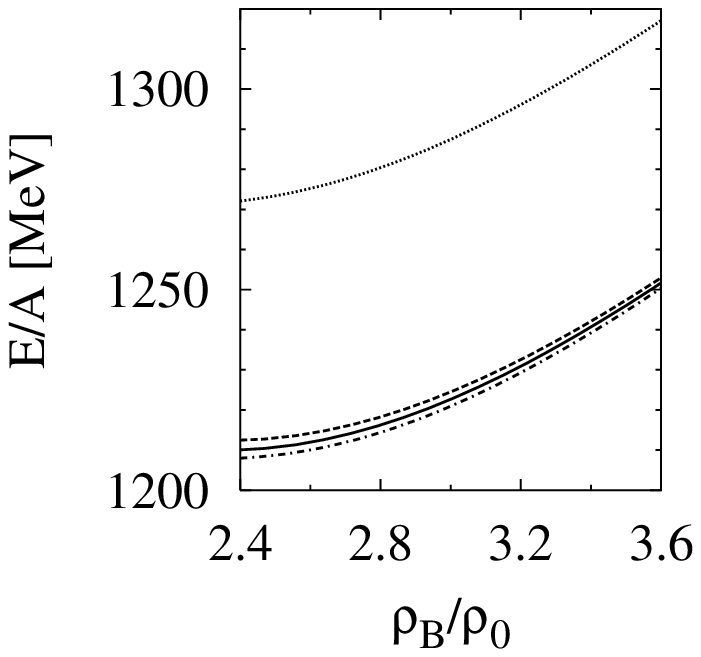,width=7.4cm}
\end{center}
\vspace{-0.5cm}
\caption{\small Left: Quark number densities as functions of the 
         quark chemical potential $\mu$,
         obtained with the Lorentz-invariant interaction 
         $g_E = g_M = 2.75/\Lambda^2$: $n_r = n_g$ (solid), $n_b$ (dashed).
         The two other lines indicate the results without the contribution
         of the Hartree term (``gluon tadpoles''): 
         $n_r = n_g$ (dash-dotted), $n_b$ (dotted).
         Right: Energy per baryon number as function of the total baryon
         number density for a color superconducting system with equal 
         densities of gapped and ungapped colors (dashed) 
         and equal chemical potentials, $\mu_8=0$ (solid). 
         The dash-dotted line corresponds 
         to the calculation without ``gluon tadpoles'', the dotted line to 
         a calculation without diquark pairing.}
\label{fig2scdens}
\end{figure}

Below the phase transition the densities are zero and, thus,
the effective chemical potentials
$\tilde\mu_r$ and $\tilde\mu_b$ are equal to the external chemical potential
$\mu$. At the phase transition point, $\tilde\mu_r$ and $\tilde\mu_b$ drop by
67~MeV and 51~MeV, respectively and then grow again as functions
of $\mu$. The corresponding  number densities of red and blue
quarks are shown in \fig{fig2scdens}. At $\mu=\mu_{crit}$,
$n_r$ (solid line) jumps from zero to $0.42$~fm$^{-3}$, 
whereas the density of blue quarks (dashed line)
reaches only $n_b = 0.34$~fm$^{-3}$ at this point. Both densities grow of 
course with increasing chemical potential, but their difference remains 
nearly constant.

The unequal densities of red and blue quarks in this state
can be understood as follows~\cite{DiRi04}:
For the (unpaired) blue quarks the occupation number is of course a 
step function, whereas for the red and green quarks it is smoothened
by the gap, leading to
a depletion below and to an enhancement above the nominal Fermi surface.
However, because of phase space, i.e., the integral measure $p^2 dp$,
this leads to an overall enhancement of the number of red and green
quarks at fixed Fermi momentum. 

On the other hand, it was argued in a recent paper that in QCD the two-flavor
color superconducting phase is automatically color neutral~\cite{GeRe03}. 
The arguments are based on gauge invariance and therefore they do not apply
directly to NJL-type models. 
However, a key role in neutralizing color was attributed to the so-called 
``gluon tadpole'' which is basically the Hartree diagram shown in \fig{fighf} 
with the dotted line identified with a gluon.
It is therefore interesting to analyze the effect of this diagram 
for our color current interaction. 

In our calculations the contribution of the Hartree term is hidden in
the ``$-g_E$'' in the expression for the effective coupling constant 
$G_v^{(8)}$ in \eq{geff}, whereas all other couplings correspond to the
Fock diagram (see App.~\ref{fierzexamples}). First of all, this means 
that the ``gluon tadpole'' is already contained in our calculations
but, obviously, this is not sufficient to get a color neutral
result. Nevertheless, it does lead to a reduction of the color charge.
To demonstrate this, we have switched off the Hartree contribution to
$G_v^{(8)}$ and kept only the Fock terms. The resulting red and blue
quark number densities are also displayed in \fig{fig2scdens} (dash-dotted
and dotted lines in the left panel). As one can see, the effect of the
``gluon tadpole'' is to reduce the difference by about 50\%.
Qualitatively, this is due to the fact that the Hartree contribution
to $G_v^{(8)}$ is repulsive, i.e., $n_8$ is reduced by this term.
(Note that, due to this term, $\tilde\mu_r < \tilde\mu_b$.)
We should stress again that our results do not contradict the claim
Ref.~\cite{GeRe03} which is based on gauge invariance. 
It would be nice to see, however, how this comes about in QCD in terms
of explicit diagrams. 
In fact, more recently it has been shown that the color neutralization
is provided by constant gluon field~\cite{DiRi04} which is of course missed 
in an NJL-model description. Similar results have also been obtained
for three-flavor QCD~\cite{Kry03}.

But even though color neutrality is not guaranteed to be automatically
fulfilled in our model, the total number of quarks of each color is a 
conserved quantity because the Lagrangian ${\cal L}_{col}$ is symmetric 
under global color $SU(3)$ transformations. One could therefore ask what
happens if we start with a large but finite system with equal
numbers of red, green and blue quarks at low densities and then
compress it. Obviously, we cannot get one single color superconducting 
phase with the properties discussed above, i.e., with unequal numbers
of red and blue quarks.
A possible scenario could be that several domains
emerge in which the symmetry is broken into different color directions,
such that the total number of red, green and blue quarks remains
unchanged. Clearly, in a realistic system, large colored domains would 
be disfavored because of the color-electric energy, related to 
long-range color forces, which are not included in our NJL mean-field
description. For small domains, on the other hand, surface effects 
should be taken into account. 

Alternatively, we can construct a homogeneous superconducting state with 
equal densities for the gapped and ungapped quarks.
To that end we have to introduce different {\it external} chemical potentials
for red and blue quarks, or, equivalently, an additional external
chemical potential $\mu_8 = (2/\sqrt{3})\,(\mu_r-\mu_b)$.
Then the second equation in \eq{mu08} becomes
\beq
    \tilde\mu_8 \;\;=\; \mu_8 + 2G_v^{(8)}n_8~,
\label{mu8tilde}
\end{equation}
and in \eq{2scpot} we should replace $\tilde\mu_8^2$ by $(\tilde\mu_8 - \mu_8)^2$.
With this additional external parameter we can enforce the densities of all
colors to be equal, even in the superconducting state.
Obviously this is the case if $\tilde\mu_8 = \mu_8$.

Within our mean field approximation such a state would be
energetically less favored than a state with the same total density, but
$\mu_8$~=~0. This is shown in the right panel of \fig{fig2scdens}.
The energy density $\epsilon$ of the system is given by
\beq
    \epsilon(T=0,n,n_8)
    \;=\; \Omega(T=0,\mu,\tilde\mu_8) \;+\; \mu n \;+\; \mu_8n_8 \,.
\end{equation}
As usual, we take the pressure (and thus the energy density) 
of the (non-trivial) vacuum to be zero, $\Omega(0,0,0)=0$. 
In the right panel of \fig{fig2scdens} the energy per baryon,
$E/A$, is displayed as a function of the total baryon
number density $\rho_B$. The solid line is the result for
$\mu_8$~=~0, i.e., it corresponds to the unequal red and
blue quark densities as shown in the left panel. The dashed line
corresponds to the result for equal red and blue quark densities.
As one can see in the figure, the energy of this solution is
always higher than the energy of the solution with unequal
densities. This means, according to this result, a large
homogeneous system of equally many red, green and blue quarks is
unstable against decay into several domains in which the density
of the gapped quarks is larger than the density of the
ungapped quarks. On the other hand, the energy difference is not
very large, less than 3~MeV per baryon. Therefore it seems likely, 
that the homogeneous solution with equal densities would
be favored, once color-electric and surface energies are taken into
account.

For comparison we also show the result for the calculation without
the Hartree contribution (dash-dotted line). 
In this case $E/A$ is slightly lower. Thus, the ``gluon tadpole''
acts like an external color chemical potential, as obvious from
\eq{mu8tilde}. In particular, if the ``gluon tadpole'' had indeed been
strong enough to neutralize the color charge it would have shifted
the solid line on top of the dashed one.   

Finally, we show the energy per baryon without diquark pairing,
i.e., $\Delta = \Delta_0 = 0$ (dotted line). As one can see the pairing
energy is about 60~MeV per baryon. From this point of view, the energy
needed to neutralize color is a minor effect. 
We will come back to this in more details in Chap.~\ref{neut}, 
where we construct
color and {\it electrically} neutral matter for applications to
neutron stars. 

The small values for $\Delta_0$ we found in \fig{fig2scdens}
seem to justify the common practice to neglect this
condensate completely. 
To get some idea about to what extent this result depends
on the coupling $H_0$ in the $\delta_0$ channel we now turn to 
a purely magnetic interaction, $g_E = 0$. 
In order to keep the vacuum properties fixed we chose 
$g_M = 4/3 g = 3.67 \Lambda^{-2}$, where $g$ is the common electric and 
magnetic coupling constant used before. With this choice the effective 
coupling constant $G_s^{(0)}$, but also $G_s^{(8)}$ and the scalar diquark 
coupling constant $H$ remain unchanged.
On the other hand the $\delta_0$ channel becomes more repulsive:
We now obtain $H_0 = -H$, whereas we had $H_0 = -H/2$ for the
Lorentz-invariant interaction. Similarly, we get a strong vector repulsion,
$G_v^{(0)} = -G_s^{(0)}$ instead of $G_v^{(0)} = -G_s^{(0)}/2$ which we had
before.  Finally, $G_v^{(8)}$ becomes attractive but is strongly reduced.

\begin{figure}
\epsfig{file=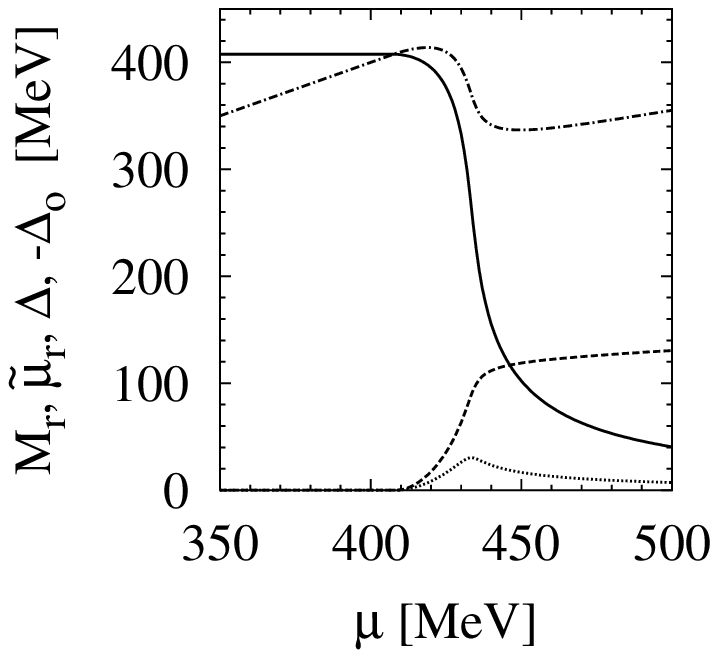,width=7.4cm}
\hfill
\epsfig{file=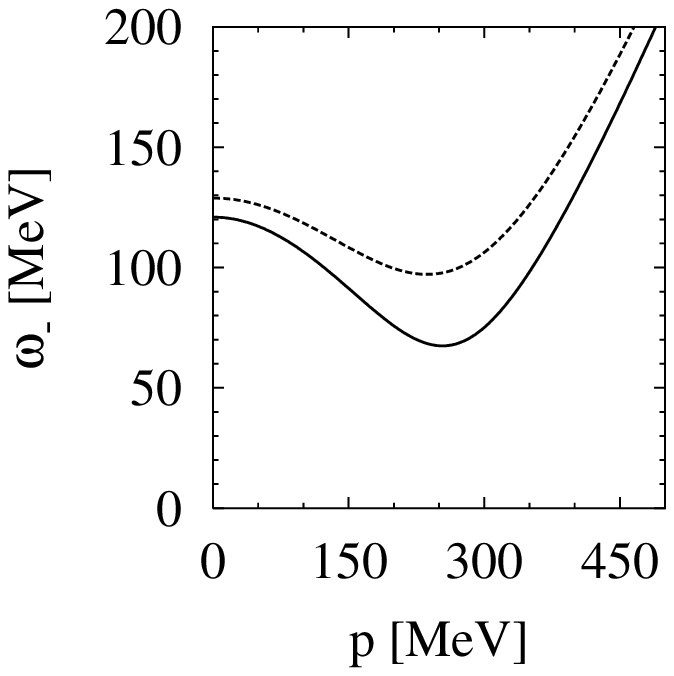,width=7.4cm}
\caption{\small Various quantities obtained with the purely magnetic
                interaction, $g_E = 0$ and
                $g_M = 3.67/\Lambda^2$.
                Left panel: $M_r$ (solid), $\Delta$ (dashed),
                $-\Delta_0$ dotted, and $\tilde\mu_r$ (dash-dotted)
                as functions of the quark chemical potential $\mu$.
                Right panel: 
                Dispersion law $\omega_-(\vec p)$ at $\mu$~=~433.4~MeV.
                The dashed line was calculated neglecting $\Delta_0$ in the
                gap equations, whereas the solid line corresponds to the
                exact solution.
                A similar figure is shown in Ref.~\cite{BHO02} for
                slightly different parameters.}
\label{figgapsge0}
\end{figure}

Our results are displayed in \fig{figgapsge0}. In the left
panel we show the behavior of $M_r$, $\Delta$, $-\Delta_0$, and
$\tilde\mu_r$ as functions of $\mu$. The most striking difference to our
previous example (\fig{fig2scgaps}) is the fact that we now find
a smooth crossover instead of a first-order phase transition.
In the chiral limit the phase transition becomes second order.
This is obviously due to the strong vector repulsion, similar to
the examples we have discussed in \sect{njlt0}.
The new aspect of the present result is that it was obtained including
diquark condensates. The fact, that there are two competing condensates,
$\phi$ and $\delta$, 
was one of the main arguments supporting the 
belief that the chiral phase transition at zero temperature and large
$\mu$ should be first order~\cite{RaWi00}.
Here we found a counter example to this argument\footnote{
It is clear, however, that this particular result
should not be taken literally, because it would imply the existence
of a color superconducting quark gas at arbitrarily low densities.
The same words of caution are in order 
for the so-called ``coexistence phase'' which has 
recently been discussed in Refs.~\cite{KKKN,BVY03}.
Based on NJL model calculations, it was claimed in these references
that even in the chiral limit, $\phi$ and $\delta$ could coexist in 
one phase. (Such a phase has also been observed in Ref.~\cite{RSSV00}.)
The reason for this behavior is that in these examples the chiral
phase transition takes place at a chemical potential which is larger
than the vacuum quark mass (i.e., it does not correspond to ``case (a)'', 
according to the classification of \sect{njlt0}).
Therefore, there is a regime, $M_{vac} < \mu < \mu_c$ which corresponds 
to a phase with $\phi \neq 0$, but with a non-zero density of quarks.
Then, according to the Cooper theorem, $\delta$ must be non-zero, too,
if there is an attractive interaction in the scalar diquark channel. 
From this point of view, the existence of a color superconducting
low-density constituent quark gas in the model is not very surprising,
although unphysical.}.

Because of the larger value of the coupling constant $|H_0|$, the
absolute value of the induced diquark gap $|\Delta_0|$ is now larger 
than in the previous example, although it remains well below the value
of $\Delta$.
This can be understood from the fact that a large $\delta_0$ requires
both, color and chiral symmetry, to be broken strongly (see \tab{tab2scsym}).
i.e., large values of $\Delta$ and large constituent quark masses at the
same time. 
This also explains why $|\Delta_0|$ has a maximum in the transition
region, where both, $\Delta$ and $M$, are not small.
 
At the maximum, located at $\mu = 433.4$~MeV, 
we find $\Delta_0 = -30.4$~MeV, while $\Delta = 89.4$~MeV,
$M_r$~=~264.3~MeV, and $\tilde\mu_r$~=~369.9~MeV. In terms of
\eqs{2scmueff} and (\ref{2scDeltaeff}) this corresponds to 
$\mu_\mi{eff}$~=~371.1~MeV, $M_\mi{eff}$~=~270.7~MeV, and
$|\Delta_\mi{eff}|$~=~67.5~MeV. The resulting dispersion law 
$\omega_-(\vec p)$ is shown in the right panel of \fig{figgapsge0} (solid
line). At $|\vec p| = \sqrt{\mu_\mi{eff}^2 - M_\mi{eff}^2}$~=~253.8~MeV it
has a minimum with $\omega_- = |\Delta_\mi{eff}|$~=~67.5~MeV.
On the other hand, if we neglect $\Delta_0$ in
the gap equation, we get $M_\mi{eff} = M_r$~=~286.5~MeV, $\mu_\mi{eff} =
\tilde\mu_r$~=~372.5~MeV, and $\Delta_\mi{eff} = \Delta$~=~97.6~MeV.
Consequently, the minimum value of $\omega_-$ is now 97.6~MeV, almost 
50\% more as without neglecting $\Delta_0$.
The corresponding dispersion law is indicated by the dashed line in the
right panel of \fig{figgapsge0}. As one can see, the entire
function $\omega_-(\vec p)$ is shifted to higher energies as compared
with the solid curve and the minimum is more shallow.

Finally, we would like to come back to the question of possible
gapless solutions. Obviously, none of our numerical
examples presented so far came close to condition
(\ref{condition}). For instance, if we take $M_r$~=~264.3~MeV,
$\tilde\mu_r$~=~369.9~MeV, and $\Delta$~=~89.4~MeV, as found 
for the purely magnetic interaction at $\mu$~=~433.4~MeV, one would need 
$\Delta_0 =$~-125.1~MeV, about four times as large as the actual value. 
The situation was even worse for the Lorentz-invariant interaction
where the discrepancy was about a factor 50 at $\mu = \mu_{crit}$ and became
larger with increasing chemical potential. In fact, none of our
numerical examples fulfilled $|\Delta_0| \geq |\Delta|$, which
we identified as a necessary condition for gapless color
superconducting states.

To get some insight on what an interaction which yields such a state 
could look like, we invert the problem and employ the gap
equations to calculate the effective coupling constants which are
consistent with a given set of gap parameters. For instance,
\eq{condition} is obviously fulfilled if we choose $M_r =
\Delta$~=~100~MeV and $\tilde\mu_r = -\Delta_0$~=~350~MeV. For
simplicity we might assume $M_b = M_r$ and $\tilde\mu_b = \tilde\mu_r$. 
Except of $\Delta_0$ this is within the typical range of these quantities
in the earlier examples. If we now take $m$~=~5~MeV and a cut-off
$\Lambda$~=~600~MeV, as before, and $\mu$~=~450~MeV, the gap
equations yield $G_s^{(0)}\Lambda^2 = 3.36$, $G_v^{(0)}\Lambda^2 =
-1.41$, $H\Lambda^2 = 6.80$, $H_0\Lambda^2 = 6.18$, and
$G_s^{(8)}\Lambda^2 = G_v^{(8)}\Lambda^2 = 0$. Here the essential
difference to our earlier examples is the need of an {\it attractive}
interaction in the $H_0$ channel. Furthermore, the interaction is
relatively strong in both diquark channels. However, for these
parameters there is another solution with $M_r = M_b$~=~58.1~MeV,
$\tilde\mu_r = \tilde\mu_b$~=~362.8~MeV, $\Delta$~=~966.6~MeV, and
$\Delta_0$~=~16.1~MeV. It turns out that for the gapless
solution the value of the thermodynamic potential $\Omega$ is about 
900 MeV/fm$^3$ higher than for the other solution. 

Hence, similar to what 
we found in the schematic example of Sec.~\ref{2scdisp}, the gapless state 
does not correspond to a stable solution. In fact, we did not succeed to
construct any stable gapless color superconducting solution. 
A similar observation was made in
Ref.~\cite{ABeR00} for gapless states in the color-flavor locked
phase. This suggests that gapless color superconductors might
in general be unstable. Although a rigorous proof is still missing, 
this could be understood as follows:
For a single gap parameter, e.g., $\phi$ or $\delta$, 
it is easy to see, that the effect of the gap is to lower the kinetic part
of the thermodynamic potential, i.e., to make the first term in
\eq{2scOmega} more negative to the expense of a positive condensation
energy $V$, see \eq{2scpot}. However, in the case of a gapless color 
superconductor the various gap parameters conspire in such a way,
that the advantage in the kinetic term gets lost (at least partially), 
but we still have to pay the price of a positive $V$. 
The instability of the gapless solutions is therefore quite 
reasonable.

More recently, it has been shown that the standard scalar 
diquark condensate $\delta$ can have a gapless excitation spectrum
if the chemical potentials for up and down quarks are very 
different~\cite{ShHu03,HuSh03}. (Similar case for three flavors have been 
discussed in Ref.~\cite{GLW03,AKR04}.)
In this case, the mechanism is rather different from ours, since the
gapless solutions do not come about through the interplay of several 
condensates, but as a consequence of the stress imposed by the unequal
chemical potentials (or masses).
In the sense of the above discussion, these gapless solutions are unstable 
as well. However, as shown in Ref.~\cite{ShHu03,HuSh03},
the decay of these solutions could be forbidden by additional constraints,
like charge neutrality. We will come back to this scenario in \sect{neutdisc}.

\section{Spin-1 pairing of the ``blue'' quarks}
\label{aniso}

The diquark condensates we have discussed so far, i.e., the scalar 
condensates $\delta$ and $\delta_0$, only involved two colors
(``red'' and ``green'') while the quarks of the third color (``blue'') 
were left unpaired.
Of course, these quarks will also be subject to a Cooper instability 
if there is an attractive channel in which they can pair.
Since only a single color is involved, the pairing must take place in
a channel which is symmetric in color. Assuming $s$-wave condensation
in an isospin-singlet channel, a possible candidate is a spin-1
condensate. This had already been suggested in Ref.~\cite{ARW98}.
Although in that reference the size of the corresponding gap was estimated 
to be much smaller than the gap in the scalar channel, its existence could
have interesting astrophysical consequences~\cite{RaWi00}:

Suppose a new-born neutron star contains a quark core consisting of up
and down quarks. Within the first few minutes the temperature of the
star drops below 1~MeV~\cite{BuLa86} and hence well below the critical 
temperature $T_c \simeq 0.57\,\Delta(T=0)$ for forming a scalar condensate,
if $\Delta$ is of the order of several tens of MeV, as estimated above. 
This means, practically all red and green quarks are gapped
and the specific heat of the quark core is completely determined by the
blue quarks. As long as these remain unpaired, they can radiate neutrinos
via the direct URCA process and dominantly contribute to the cooling
of the star. A possible pairing of the blue quarks could thus change the
cooling behavior dramatically, once the temperature drops below
the corresponding critical temperature. 
Another interesting point is the emergence of an electromagnetic Meissner
effect, which would of course strongly affect the magnetic field of 
the neutron star. 

In this context a more detailed knowledge about the properties of
a possible spin-1 condensate, in particular its size, and its thermal
properties would be desirable. 
This will be the subject of the following discussion.

\subsection{Condensation pattern and symmetries}
\label{anisosym}

We consider the complex vector order parameter
\beq
    \zeta_n  
    \;=\; \langle\,q^T \;C\,\sigma^{0n}\;\tau_2\;\hat P_3^{(c)}\;q\, 
  \rangle~, 
\label{zetan}
\eeq
where $\sigma^{\mu\nu} = i/2\,[\gamma^\mu,\gamma^\nu]$ and
$\hat P_3^{(c)} = 1/3 - 1/\sqrt{3}\,\lambda_8$ is the projector on color 3,
i.e., the blue quarks.
$\zeta_n$ describes the spin-1 pairing of two quarks in a relative 
s-state. (Other forms of spin-1 condensates are discussed, e.g., 
in Refs.~\cite{BaLo84,Rischkereview,Sch00,ABCC02,Schmitt}.)

An interesting feature of $\zeta_n$ is that it is
not neutral with respect to the ``rotated'' electric charge $\tilde Q$,
defined in \eq{qtilde2}.
For the transformations given in \eqs{qtrans2} and (\ref{lambda8trans2})
we find
\begin{alignat}{1}
q \rightarrow e^{\,i\alpha\,Q}\,q \qquad &\Rightarrow \qquad
\zeta_n \rightarrow e^{\,i\alpha/3}\,\zeta_n 
\nonumber\\
q \rightarrow e^{\,i\alpha'\,\lambda_8}\,q \qquad &\Rightarrow \qquad
\zeta_n \rightarrow e^{\,-4i\alpha'/\sqrt{3}}\,\zeta_n~,
\end{alignat}
and hence
\beq
q \rightarrow e^{\,i\alpha\,\tilde Q}\,q \qquad \Rightarrow \qquad
\zeta_n \rightarrow e^{\,i\alpha}\,\zeta_n~. 
\eeq
We can find a different linear combination,
\beq
    \tilde Q' = Q + \frac{1}{4\sqrt{3}}\,\lambda_8~,     
\eeq
under which $\zeta_n$ remains invariant.
However, there is no generalized electric charge for which both,
$\delta$ and $\zeta_n$, are neutral. This means, if both, $\delta$ and 
$\zeta_n$, are present in a neutron star, there will be an electromagnetic 
Meissner effect, which would strongly influence the magnetic field.
Recently, similar effects have been discussed in 
Refs.~\cite{Schmitt,SWR03,SWR04}.
The detailed evaluation of the Meissner masses for our case (two flavors,
one color) remains to be done.

It is obvious, that a non-vanishing vector $\vec \zeta$, pointing in
some direction in space, also breaks the $O(3)$ rotational symmetry of 
the system spontaneously.
There are well-known examples for spin-1 pairing in condensed matter physics,
e.g., superfluid $^3\mi{He}$, 
where some phases are also anisotropic~\cite{He3}.  
In relativistic systems this is  certainly not a very frequent phenomenon.
It is possible only at finite chemical potential, which itself breaks 
Lorentz invariance explicitly. 
Since $O(3)$ is a global symmetry of QCD, 
there should be collective Nambu-Goldstone excitations in the spectrum.
However, in Lorentz non-invariant systems, there are subtleties which
can spoil the standard proof of the Goldstone theorem, leading to
peculiarities, like excitations with quadratic dispersion laws or an
unusual number of Goldstone 
bosons~\cite{NiCh76,Leut94,SSSTV,SaSch,Ho98,OM98,MiSh01}.

In Refs.~\cite{Ho98,OM98},
this problem has been analyzed within an effective Ginzburg-Landau type
potential for the complex order parameter $\vec\zeta$\footnote{This has been
done in the context of Bose-Einstein condensation in ultra-cold alkali atoms.}
The potential consists of a mass term with the ``wrong sign'' in order
to get a non-trivial solution and two different fourth-order terms,
\beq
    V(\vec\zeta) \;=\; -a^2\, \zeta_n^\dagger\zeta_n 
    \;+\; \frac{1}{2}\,\lambda_1\, (\zeta_n^\dagger\zeta_n)^2  
    \;+\; \frac{1}{2}\,\lambda_2\, 
    \zeta_m^\dagger\zeta_m^\dagger\zeta_n\zeta_n~,
\label{anisopot}
\eeq
with $\lambda_1+\lambda_2 > 0$ for stability.
In fact, the $\lambda_2$-term explicitly breaks the $O(3)$ invariance,
but its introduction is conceptionally useful because it lifts the 
degeneracy between the $M_J = 0$ and the $M_J = \pm 1$ states: 

For $\lambda_2 < 0$ the potential is minimized by a $(J=1,M_J=0)$-state,
\beq
    \vec\zeta(M_J=0) = \zeta \, 
      \left(\begin{array}{c} 0 \\ 0 \\ 1 \end{array}\right)~,
\label{TM0}
\eeq    
with $|\zeta| = a/\sqrt{\lambda_1+\lambda_2}$. 
This solution corresponds to an anti-ferromagnet. 
The spectrum of small oscillations above this ground state consists
of 1+2 Nambu-Goldstone bosons, all with linear dispersion law: one zero-sound
phonon and two spin waves \cite{Ho98,OM98}. Implying a finite
Landau critical velocity, this fact is crucial for a macroscopic
superfluid behavior of the system \cite{MiSh01}.

On the other hand, for $\lambda_2 > 0$ the potential is minimized by
$(J=1,M_J=\pm 1)$-states,
\beq
    \vec\zeta(M_J=\pm 1) = \mp \frac{\zeta}{\sqrt2}\, 
      \left(\begin{array}{c} 1 \\ \pm i \\ 0 \end{array}\right)~,
\label{TM1}
\eeq
with $|\zeta| = a/\sqrt{\lambda_1}$. This corresponds to ferromagnetic
solutions.
In this case the Nambu-Goldstone spectrum above this ground state 
consists of one phonon with linear dispersion law
and one spin wave whose energy tends to zero with momentum
squared~\cite{Ho98,OM98}. 

As mentioned above, 
if the underlying Hamiltonian is exactly rotational invariant,
the last term in \eq{anisopot} must vanish, i.e., $\lambda_2=0$. 
The $M_J=0$-solutions are then degenerate with the 
$M_J=\pm 1$-solutions~\cite{ABCC02}. In real systems, it will therefore
depend on details of the surrounding which solution is realized (if any).

In the following, we will mostly concentrate on the $M_J=0$ case
which has been discussed in Ref.~\cite{BHO03}. 
The case $M_J=\pm 1$ is more complicated and has so far only been investigated
at zero temperature~\cite{ABCC02}. This will briefly be discussed
afterwards.

\subsection{Gap equation}
\label{anisogap}

In order to analyze the spin-1 condensate more quantitatively, 
we employ again an NJL-type model. As mentioned above, we will
first restrict ourselves to the case $M_J=0$, \eq{TM0},
i.e., we consider a non-vanishing expectation value
\beq
    \zeta  
    \;=\; \langle\,q^T \;C\,\sigma^{03}\;\tau_2\;\hat P_3^{(c)}\;q\, 
  \rangle~. 
\label{zeta}
\eeq
We also keep the dominant condensation channels at high and
at low density, i.e., the scalar color-antitriplet diquark condensate 
$\delta$, \eq{delta}, and the quark-antiquark condensate 
$\phi$, \eq{phi}. However, for simplicity, we neglect the 
other condensates discussed in \sect{interplay} and all other condensates
possibly induced by a non-vanishing $\zeta$. 

To allow for condensation in the
$\zeta$-channel in Hartree approximation, we need an attractive 
quark-quark interaction with the quantum numbers of a Lorentz-tensor,
flavor-singlet and color-sextet\footnote{Again, we may in principle
allow for Lorentz non-invariant Lagrangians, where the time-space components
(the $\sigma^{0i}$-terms) and the space-space components (the 
$\sigma^{ij}$-terms) of the interaction enter with different coupling
constants. However, since only the time-space components are relevant
for our condensation pattern, this would not make any difference in the
results.}.  
Guided by the structure of instanton-induced interactions 
(see App.~\ref{fierzexamples}) we consider a quark-antiquark term
\begin{equation}
{\cal L}_{q\bar q} = G\,\Big\{ (\bar qq)^2 - (\bar q\vec\tau q)^2
   - (\bar q \,i\gamma_5q)^2 + (\bar q \,i\gamma_5\vec\tau q)^2 \Big\}
\label{Lqqbar}
\end{equation}
and a quark-quark term
\begin{alignat}{2}
{\cal L}_{qq}\;=\quad
      &H_s\,\Big\{&(\bar q\, i\gamma_5\, C\tau_2\lambda_A\,\bar q^T)
                  &(q^T\, C\, i\gamma_5 \tau_2\lambda_A\, q)
                -\;(\bar q\, C\tau_2\lambda_A\,\bar q^T)
                   (q^T\, C \tau_2\lambda_A\, q)\,\Big\}
\nonumber \\
    - &H_t        &(\bar q\, \sigma^{\mu\nu} C\tau_2\lambda_S\,\bar q^T)
                  &(q^T\, C \sigma_{\mu\nu} \tau_2\lambda_S \,q)~,
\label{Lqq}
\end{alignat}
where $\lambda_A$ and $\lambda_S$ are again the antisymmetric and symmetric
color generators, respectively. For instanton induced interactions
the coupling constants fulfill the relation
$G : H_s : H_t =  1 : 3/4 : 3/16$,
but for the moment we will treat them as arbitrary parameters.
As long as they stay positive, the interaction is attractive in the
channels giving rise to $\delta$, $\zeta$, and $\phi$.

Applying the same techniques as in \sect{2scformalism} it is straight 
forward to 
calculate the mean-field thermodynamic potential $\Omega(T,\mu)$ in the 
presence of these condensates. One finds
\begin{alignat}{1}
    \Omega(T,\mu) = &-4 \sum_{i=1}^3 \int\!\!\frac{d^3p}{(2\pi)^3}
                        \Big[\frac{\omega_i^- + \omega_i^+}{2} 
    + T\ln\Big(1+e^{-\omega_i^-/T}\Big) 
    + T\ln\Big(1+e^{-\omega_i^+/T}\Big) \Big]
\nonumber\\
&+\frac{1}{4G}(M-m)^2 +\frac{1}{4H_s}|\Delta|^2
+\frac{1}{16H_t}|\Delta'|^2~,
\label{Omegaaniso}
\end{alignat}
where again $m$ is the bare quark mass, $M = m -2G\,\phi$, 
$\Delta= -2H_s\,\delta$, and
\beq
    \Delta' = 4H_t\,\zeta.
\eeq
The dispersion laws for the red and green quarks are the ``standard'' ones,
i.e.,
\beq
\omega_{1,2}^{\mp}(\vec p) = \sqrt{(\sqrt{\vec p^2 + M^2} \mp \mu)^2 +
|\Delta|^2}~,
\eeq
whereas the dispersion laws for the blue quarks now read
\begin{equation}
\omega_3^\mp(\vec p) \;=\; \sqrt{ (\sqrt{M_{\mathit{eff}}^2 
+ \vec p^{\,2}} \mp \mu_{\mathit{eff}})^2 + |\Delta_{\mathit{eff}}'|^2 }~,
\label{E3}
\end{equation}  
where the effective chemical potential, the effective mass, and the
effective gap are angle dependent quantities,
\beq
    \mu_{\mathit{eff}}^2 = \mu^2 + |\Delta'|^2 \sin^2{\theta}~,\qquad
    M_{\mathit{eff}} = M \mu/\mu_{\mathit{eff}}~, \qquad
|\Delta_{\mathit{eff}}'|^2 = |\Delta'|^2\,(\cos^2{\theta} + \frac{M^2}{\mu_{\mathit{eff}}^2}\,\sin^2{\theta})~. 
\label{Dpeff}
\eeq
\begin{figure}[t]
\begin{center}
\epsfig{file=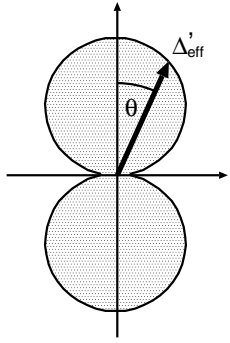, width = 3.0cm}
\end{center}
\caption{\small Schematic illustration of the angle dependence of 
         $\Delta_{\mathit{eff}}'$.}
\label{figanisodeff}
\end{figure}
Here $\cos{\theta} = p_3/|\vec p|$. 
Thus, as expected, for $\Delta'\neq 0$, $\omega_3^\mp(\vec p)$ is an 
anisotropic function of $\vec p$, reflecting the spontaneous breakdown of 
rotational invariance. 
The dependence of the effective gap $\Delta_{\mathit{eff}}'$ 
on the polar angle $\theta$ is illustrated in \fig{figanisodeff}. 
For $M = 0$, it vanishes at $\theta = \pi/2$. 
In general, its minimal value is given by 
\begin{equation}
    \Delta'_\mi{min} \;=\; \Delta_{\mathit{eff}}'(\theta = \frac{\pi}{2}) \;=\;
\frac{M |\Delta'|}{\sqrt{\mu^2 + |\Delta'|^2}}~.
\label{M0}
\end{equation}

For our later analysis of the specific heat we will need the density
of states,
\beq
    N(E) \;=\; \int\dtp\,\delta(\omega_3^-(\vec p) - E)
\label{NEdef}
\eeq
for the low-lying quasiparticle spectrum.
To that end we expand $\omega_3^-$ about its minimum. One obtains 
\begin{equation}
  \omega_3^-(\vec p)\approx \sqrt{ \Delta_\mi{min}^{\prime 2} 
+ v_\perp^2 (p_\perp - p_0)^2
                                  + v_3^2 p_3^2}~,  
\end{equation}
where $p_\perp^2 = p_1^2 + p_2^2$, and
\begin{equation}
  v_\perp = \sqrt{1-(\frac{\mu M}{\mu^2 + |\Delta'|^2})^2},\qquad
  v_3 = \frac{\Delta'_\mi{min}}{M},\qquad 
  p_0 = \frac{v_\perp}{v_3}|\Delta'|. 
\end{equation}
Inserting this into \eq{NEdef}, we find that the density of states
is linear in energy, 
\begin{equation}
  N(E) = \frac{1}{2\pi} \frac{\mu^2 + |\Delta'|^2}{|\Delta'|}\;E\;
  \theta(E-\Delta'_\mi{min})~.
\label{NE}
\end{equation}
This linear dependence is typical for effectively 2-dimensional systems:
The angular structure of the gap restricts the low-lying excitations
to stay in the ``equator plane'', i.e., $\theta=\pi/2$. 

The gap equations for $\Delta$, $\Delta'$ and $M$ are again derived 
by minimizing $\Omega$ with respect to these
variables. Imposing $\partial\Omega/\partial{\Delta'}^* = 0$,
we get
\begin{equation}
\Delta' = 16\,H_t\,\Delta' \sum_{-+}\int \frac{d^3 p}{(2\pi)^3} \, 
(1\mp\frac{{\vec p}_\perp^{\,2}}
{\mu_{\mathit{eff}}\,\sqrt{\vec p^{\,2} + M_{\mathit{eff}}^2}})\,
\frac{1}{\omega_3^\mp}\,\tanh{\frac{\omega_3^\mp}{2T}}~. 
\label{Deltapgap}
\end{equation}  
Note that $\Delta$ does not explicitly enter this equation.
In turn, the gap equation for $\Delta$, resulting from
$\partial\Omega/\partial\Delta^* = 0$, takes the standard form
(\eq{gapstandard} with $H\rightarrow H_s$ and 
$\omega_\mp \rightarrow \omega_1^\mp$)
and does not explicitly depend on $\Delta'$. 
On the other hand,
both, $\Delta$ and $\Delta'$, enter the gap equation for $M$, 
\begin{equation}
M \;=\; m \;+\,4\,G\,M \sum_{-+}\int \dtp \,\Big\{
2\, \frac{E_p\mp\mu}{E_p \omega_1^\mp}\,\tanh{\frac{\omega_1^\mp}{2T}}
\;+\;(1\mp\frac{\mu^2}
{\mu_{\mathit{eff}}\,\sqrt{\vec p^{\,2} + M_{\mathit{eff}}^2}})\,
\frac{1}{\omega_3^\mp}\,\tanh{\frac{\omega_3^\mp}{2T}} \Big\}~, 
\label{Mgap}
\end{equation}  
which follows from the requirement $\partial\Omega/\partial M = 0$.
We thus have a set of three equations, where 
the equations for $\Delta$ and $\Delta'$ are not directly 
coupled, but only through their dependence on $M$.
In particular, for $M=0$ they decouple.

\begin{figure}[t]
\begin{center}
\epsfig{file=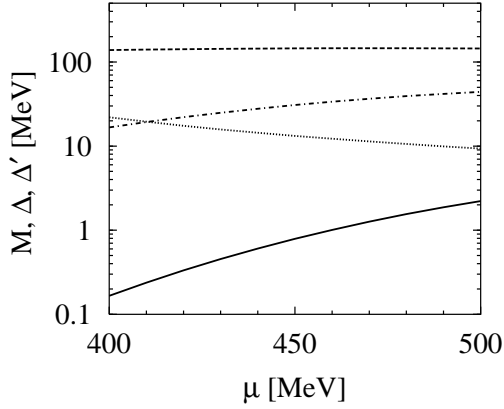, width = 7cm}
\caption{\small 
         $M$ (dotted), $\Delta$ (dashed), and $\Delta'$ (solid) at $T$~=~0
         as functions of the quark chemical potential $\mu$ using 
         $\Lambda = 600$~MeV, $m = 5$~MeV, and $G\Lambda^2 = 2.4$.
         The coupling constants $H_s$ and $H_t$ are fixed by
         the instanton relation $G : H_s : H_t =  1 : 3/4  : 3/16$.
         The dash-dotted line indicates the result for $\Delta'$
         if the value of $H_t$ is doubled.
         Adapted from Ref.~\cite{BHO03}.}
\label{figanisocond}
\end{center}
\end{figure}

Numerical solutions of the gap equations are presented in
\fig{figanisocond}. We have chosen a sharp 3-momentum cut-off 
$\Lambda = 600$~MeV, a current quark mass $m =$~5~MeV, and 
$G\Lambda^2 = 2.4$ for the coupling constant in the quark-antiquark 
part~\cite{BHO03}. Obviously, these
parameters are close to the region fixed by fitting $f_\pi$ and $m_\pi$
(see \tab{tabnjl2fit}) as well as those employed in \sect{results}
and lead to reasonable vacuum properties ($M = 393$~MeV, $f_\pi = 93.6$~MeV, 
$m_\pi = 129$~MeV, $\langle\bar u u\rangle = (-244\rm{MeV})^3$).
In order to fix the coupling constants $H_s$ and $H_t$ we have employed
the instanton relation, $G : H_s : H_t =  1 : 3/4  : 3/16$. 
The resulting values of $M$, $\Delta$, and
$\Delta'$ as functions of $\mu$ at $T$~=~0 are displayed in
the figure. The chemical potentials correspond to baryon
densities of about 4 - 7 times nuclear matter density.  In agreement
with the earlier expectations of Ref.~\cite{ARW98} $\Delta'$ is 
relatively small, about two to three orders of magnitude smaller than $\Delta$
in this regime. However, it is strongly
$\mu$-dependent and rises  by more than a factor of 10 between 
$\mu = 400$~MeV and $\mu = 500$~MeV.
Being a solution of a self-consistency problem, $\Delta'$ is also 
extremely sensitive to the coupling constant $H_t$.
If we double the value of $H_t$, we arrive at the dash-dotted line for 
$\Delta'$, which is then almost comparable to $\Delta$
(see also \fig{figanisox}).
We also find that $\Delta'$ is very sensitive to the cut-off.
This can be traced back to the factor
$(1-{\vec p}_\perp^{\,2}/s)$ in the gap equation
(\ref{Deltapgap}) which can become negative for large momenta. 
It is quite obvious then, that also the form of the regularization,
i.e., sharp cut-off, form factor, etc., will have a strong impact on the
results.

\subsection{Thermal behavior}
\label{anisot}

With increasing
temperature both condensates, $\delta$ and $\zeta$, are reduced
and eventually vanish in second-order phase transitions at critical
temperatures $T_c$ and $T_c'$, respectively. It has been shown 
\cite{PiRi99} that $T_c$ is approximately given by the 
well-known BCS relation 
\beq
T_c \approx 0.57\, \Delta(T = 0)~.
\label{Tcapp}
\end{equation}
In order to derive a similar relation for 
$T_c'$ we inspect the gap equation (\ref{Deltapgap}) at $T$~=~0 and in the limit $T \rightarrow T_c'$. Neglecting
$M$ (since $M\ll \mu$ this is valid up to higher
orders in $M^2/\mu^2$) and antiparticle contributions one gets
\beq
\int \frac{d^3 p}{(2\pi)^3}
\Big\{\Big[(1-\frac{{\vec p}_\perp^{\,2}}{s})\frac{1}{\omega_3^-(\vec p)}
\Big]_{\Delta'(T=0)}
   \,-\,(1-\frac{{\vec p}_\perp^{\,2}}{\mu\,|\vec p|})
     \frac{1}{|\mu-|\vec p||} \tanh{\frac{|\mu-|\vec p||}{2T_c'}}\Big\} 
\;\approx\; 0~.
\label{Tcest}
\eeq  
Since the integrand is strongly peaked near the Fermi surface,
the $|\vec p|$-integrand must approximately vanish at  $|\vec p| = \mu$,
after the angular integration has been performed. 
From this condition one finds to lowest order in $\Delta'/\mu$
\begin{equation}
  T_c' \;\approx\; \frac{1}{3}\;\Delta'(T=0)~.
\end{equation}
For the scalar condensate, the analogous steps
would lead to $T_c/\Delta$($T$=0) $\approx \frac{1}{2}$ instead of \eq{Tcapp}.
This gives a rough idea about the quality of the approximation.    
Note that there are other examples of diquark condensates, where
$T_c \neq 0.57 \Delta$($T$=0)~\cite{SWR02}. This is also the case for
crystalline superconductors~\cite{BKRS01}.

\begin{figure}[t]
\begin{center}
\epsfig{file=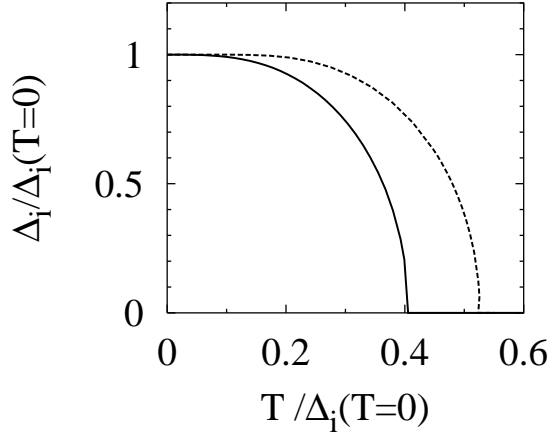,width = 7.5cm}
\caption{\small $\Delta_i/\Delta_i$($T$=0) as function of $T/\Delta_i$($T$=0).
Dashed: $\Delta_i=\Delta$. Solid: $\Delta_i=\Delta'$. The calculations
have been performed at $\mu=450$~MeV for the same parameters as 
employed in \fig{figanisocond}. Adapted from Ref.~\cite{BHO03}.}
\label{figanisocondt}
\end{center}
\end{figure}

Numerical results for $\Delta(T)$ and 
$\Delta'(T)$ are shown in \fig{figanisocondt}. The quantities have been 
rescaled in order to facilitate a comparison with the above relations
for $T_c$ and $T_c'$. The results
are in reasonable agreement with our estimates. These findings turn
out to be quite insensitive to the actual choice of parameters.

The specific heat is given by\footnote{Although standard,
this formula is not quite correct (I thank Igor Shovkovy for having pointed 
out this problem to me.):
Strictly, $c_v$ is defined as the temperature
dependence of the internal energy $\epsilon$ at fixed volume and at
fixed {\it particle number}, i.e.,
$c_v = (\frac{\partial\epsilon}{\partial T})|_{V,N} =  
\frac{T}{V} (\frac{\partial S}{\partial T})|_{V,N}$.
If we were allowed to evaluate the derivative at fixed chemical potential
we would get \eq{cvdef}. Keeping the particle number fixed, 
the correct expression is
\beq
    c_v \;=\; -T \,\Big\{\,\frac{\partial^2 \Omega}{\partial T^2} \,-\,
             \Big(\frac{\partial^2 \Omega}{\partial \mu^2}\Big)^{-1}
             \Big(\frac{\partial^2 \Omega}{\partial T\partial\mu}\Big)^2 
             \,\Big\}~.
\eeq
However, we have checked numerically that the correction term is 
negligible (see also Ref.~\cite{FW71}).
}
\beq
    c_v \;=\; -T \,\frac{\partial^2 \Omega}{\partial T^2}~.
\label{cvdef}
\eeq
For $T \ll T_c$ it is completely dominated by the blue quarks,
since the contribution of the red and green ones
is suppressed by a factor $\rm{e}^{-\Delta/T}$.  
Thus, keeping the $\omega_3^-$-part only and neglecting the
$T$-dependence of $M$ and $\Delta'$, one gets from \eq{Omegaaniso}
\beq
     c_v \;\approx\;\int\dtp\,\Big(\frac{\omega_3^-}{T}\Big)^2\,
     \frac{1}{\cosh^2{(\frac{\omega_3^-}{2T})}}~.
\eeq
At low temperatures we can replace the $\cosh$ by an exponential.
Employing the density of states, Eq.~(\ref{NE}), the integral is then
readily turned out. 
One finds
\begin{alignat}{1}
  c_v \;\approx\; &\frac{12}{\pi} \frac{\mu^2 +
    |\Delta'|^2}{|\Delta'|}\,T^2\, e^{-\frac{\Delta'_\mi{min}}{T}}\sum_{n = 0}^3
    \frac{1}{n!}\left(\frac{\Delta'_\mi{min}}{T}\right)^n~.
\label{cvapp}
\end{alignat}
According to the approximations made, this expression
should be valid for $T \ll T_c'$. 
In this regime $c_v$ depends quadratically on $T$ for $T \gtrsim \Delta'_\mi{min}$,  
and is exponentially suppressed at lower temperatures.
\begin{figure}[t]
\begin{center}
\epsfig{file=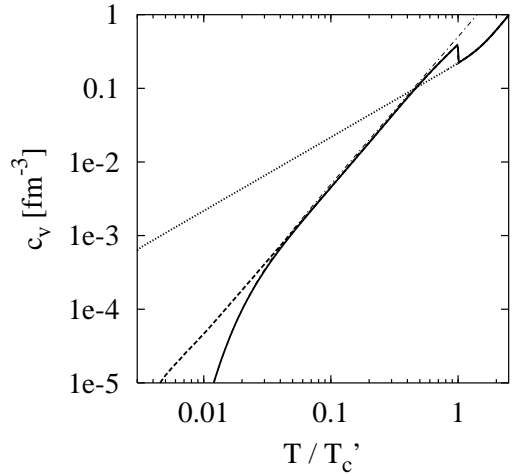,width = 7.0cm}
\caption{\small Specific heat as function of $T/T_c'$. The calculations
have been performed at $\mu=450$~MeV for the same parameters as 
employed in \fig{figanisocond}, however, for numerical convenience
with the doubled value of $H_t$.
Solid: full calculation,
dashed: result for $M = 0$,
dotted: without spin-1 condensate. The dash-dotted line indicates the
result of Eq.~(\ref{cvapp}). Adapted from Ref.~\cite{BHO03}.}
\label{figanisocv}
\end{center}
\end{figure}

To test this relation we evaluate $c_v(T)$ numerically using
Eq.~(\ref{Omegaaniso}) and (\ref{cvdef}).
The results for fixed $\mu$~=450~MeV are displayed
in \fig{figanisocv}. For numerical convenience we have doubled the
value of $H_t$,
leading to a relatively large $\Delta'$($T$=0)~=~30.8~MeV. The
critical temperature is $T_c'\simeq$~0.40~$\Delta'$($T$=0). For the
energy gap we find $\Delta'_\mi{min}$~=~0.074~$T_c'$.  It turns out that
Eq.~(\ref{cvapp}), evaluated with constant values of $\Delta'$ and
$M$, (dash-dotted line) is in almost perfect agreement with the
numerical result (solid line) up to $T \approx T_c'/2$.  The phase
transition, causing the discontinuity of $c_v$ at $T=T_c'$, is of
course outside the range of validity of Eq.~(\ref{cvapp}).
We also display $c_v$ for $M=0$ (dashed line). 
Since $\Delta'_\mi{min}$ vanishes in this case, there is no
exponential suppression, and $c_v$ is proportional to $T^2$ down to
arbitrarily low temperatures.  However, even when $M$ is included, the
exponential suppression is partially canceled by the sum on the r.h.s.
of Eq.~(\ref{cvapp}).  
For comparison, we also show $c_v$ for a system with $\Delta'$~=~0,
which exhibits a linear $T$ dependence at low 
temperatures (dotted line).

Our results show that, even though the magnitude of the gap parameter
$\Delta'$ is strongly model dependent, 
its relations to the critical temperature and the specific heat are 
quite robust. Thus, if we had empirical data, e.g., for the specific heat
of dense quark matter, they could be used to extract information about the 
existence and the size of $\Delta'$.
In this context neutron stars and their cooling properties are the 
natural candidates to look at.

As already mentioned, in Ref.~\cite{RaWi00} it was suggested that 
the spin-1 pairing of the blue quarks might have observable
consequences for the cooling of a neutron star. 
(The relevance of $c_v$ and the possible effect of diquark condensates
on neutron star cooling was also discussed in 
Refs.~\cite{neutrino1,neutrino2,neutrino3}.)
According to the original idea, 
quite soon after the temperature has dropped
below the critical temperature for the spin-1 pairing, the contribution
of the blue quarks to the specific heat will be exponentially suppressed.
Obviously, this argument has to be somewhat refined since, as seen above, 
$c_v(T)$ first behaves as $T^2$ and the exponential suppression 
sets in only at $T < \Delta'_\mi{min}$.

On the other hand, we should admit that there are good reasons to doubt 
that a spin-1 isospin-singlet condensate is stable in a neutron star,
where we have to impose neutrality constraints. 
In order to pair, the up and down quarks should 
have similar Fermi momenta, whereas for charge neutrality one needs
roughly twice as many down quarks as up quarks.  
In fact, it has been argued that these constraints
could completely prohibit the existence of two-flavor color superconducting 
matter in neutron stars~\cite{AR02}. 
In that case, there is the possibility that up and down quarks are
separately paired in single-{\it flavor} spin-one 
condensates~\cite{Rischkereview,Sch00,ABCC02,Schmitt}.)
However, as we will see in Chap.~\ref{neut}, it is possible that the
standard spin-0 condensate $\delta$ is not destroyed by the 
neutrality conditions. For $\zeta$, 
this question has not yet been investigated, but, given that $\Delta'$ is 
presumably small, $\zeta$ should be much more fragile and will probably 
not survive. In this case, the fate of the blue quarks is rather unclear.

\subsection{Spin-1 pairing in the red-green sector}

As we have discussed earlier, the coupling strengths in the various
channels are poorly known and the choice of the instanton relation,
to fix their ratios is not at all stringent. 
At least in principle, this implies the possibility that the red
and green quarks are also paired in a spin-1 state if the ratio
$H_t:H_s$ is large enough.
To investigate this scenario, we extend the formalism of \sect{anisogap}
to include a condensate of the form
\beq
    \zeta_{rg}  
    \;=\; \langle\,q^T \;C\,\sigma^{03}\;\tau_2\;\hat P_{12}^{(c)}\;q\, 
  \rangle~. 
\label{zetarg}
\eeq
together with the other condensates. 
$\hat P_{12}^{(c)} = 2/3 + 1/\sqrt{3}\,\lambda_8$ is the projector on 
the first two colors, i.e., the red and the green quarks.
Denoting the corresponding gap parameter by $\Delta_{rg}'$, the 
thermodynamic potential, \eq{Omegaaniso} gets an additional term
$|\Delta_{rg}'|^2/(8H_t)$, and the dispersion laws for the red and
green quarks are now given by an expression analogous to \eqs{E3} and
(\ref{Dpeff}), but with an effective gap
\beq
|\Delta_{\mathit{eff}}^{rg}|^2 \;=\; |\Delta|^2 \;+\;
|\Delta_{rg}'|^2\,(\cos^2{\theta} + 
\frac{M^2}{\mu_{\mathit{eff}}^{rg\;2}}\,\sin^2{\theta})~. 
\eeq
 
\begin{figure}[t]
\begin{center}
\epsfig{file=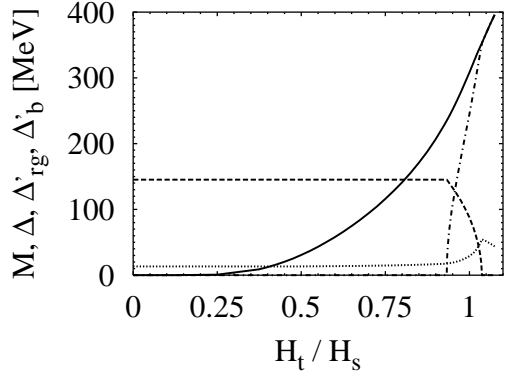,width = 7.5cm}
\caption{\small Gap parameters as functions of the ratio 
between tensor and scalar coupling constant $H_t$ and $H_s$: 
constituent quark mass $M$ (dotted), scalar diquark gap
$\Delta$ (dashed), and spin-1 gap parameters $\Delta_{rg}'$ 
(dash-dotted) and $\Delta_{b}'$ (solid). The calculations
have been performed at $\mu=450$~MeV. Except of $H_t$, which is
varied, the parameters are the same as in \fig{figanisocond}.}
\label{figanisox}
\end{center}
\end{figure}

Numerical results are presented in \fig{figanisox} where the various
gap parameters are displayed as functions of the tensor coupling 
$H_t$ at fixed chemical potential $\mu$. For the other parameters we
have taken the same values as in Secs.~\ref{anisogap} and \ref{anisot}.
For the sake of clarity we now call the spin-1 gap parameter of
the blue quarks $\Delta'_b$, instead of $\Delta'$.

We find three different regimes:
For $H_t < 0.93\,H_s$, we have the ``standard situation'' that
the red and green quarks are paired in a scalar
condensate, i.e., $\Delta \neq 0$ (dashed line), whereas  
$\Delta_{rg}' = 0$. 
Then, in some intermediate regime, $0.93 < H_t/H_s < 1.04$, both,
$\Delta$ and $\Delta_{rg}'$, are non-zero.
Here one can nicely see that both condensates compete for the same
quarks and, hence, $\Delta$ decreases while $\Delta_{rg}'$ (dash-dotted
line) rises steeply.
Finally, for $H_t > 1.04 H_s$, $\Delta = 0$ and all quarks are paired
in a spin-1 condensate. As a consequence, $\Delta_{rg}'$ is equal to
$\Delta_b'$ (solid line) in this regime. 
It is quite remarkable that the transition from
spin-0 to spin-1 pairing in the red/green sector is smooth, i.e., there
are two second-order phase transitions, instead of a single first-order
transition.
This demonstrates again that the presence of two competing condensates
does not automatically guarantee a first-order phase transition
(see \sect{results}).

The properties of the two ``new'' phases have not yet been investigated
in detail. In fact, it might be difficult to motivate such an analysis,  
since ``typical'' interactions, like instantons 
($H_t:H_s = 1:4$) or single gluon exchange ($H_t = 0$), do by far not 
provide enough strength in the tensor channel.

\subsection{$M_J = \pm 1$}
\label{ferro}

As discussed in \sect{anisosym}, the condensation could also take place
in the $(J=1, M_J = \pm 1)$-channel.
According to \eq{TM1}, this means that we have a condensate of the 
form
\beq
    \mp\langle\,q^T \;C\,\sigma^{01}\;\tau_2\;\hat P_3^{(c)}\;q\,\rangle \;=\;
    i\langle\,q^T \;C\,\sigma^{02}\;\tau_2\;\hat P_3^{(c)}\;q\,\rangle \;=\;
    \frac{\zeta}{\sqrt{2}}~,
\label{zetapm1}
\eeq
where we have assumed that only the blue quarks condense in this channel.
With this ansatz we can again apply the techniques of \sect{2scformalism} 
to calculate the thermodynamic potential.
In practice, however, this turns out to be more difficult than for
$M_J=0$.  The dispersion laws $\omega(\vec p)$ of the quasiquark excitations 
generally correspond to the eigenvalues of the inverse Nambu-Gorkov propagator.
In the present case, this leads to the following secular equation
in the blue quark sector
\begin{alignat}{1}
    f(\omega^2) \;=\; 
    \Big[(\omega^2 - E_-^2)(\omega^2 - E_+^2)
    - |\Delta'|^2\,(\omega^2 + \vec p^{\;2}\cos^2(\theta) - \mu^2 - M^2) 
    \Big]^2&
\nonumber\\
    \;-\; 4|\Delta'|^4\,\Big((\omega^2 - M^2)\, \vec p^{\;2}\cos^2(\theta)
    + M^2\mu^2\Big)&
    \;=\; 0~, 
\label{ferrodispeq}
\end{alignat}
where $E_\mp = \sqrt{\vec p^{\;2} + M^2} \mp \mu$, as before,
while the gap parameter is given by $\Delta' = 4\sqrt{2}\,H_t\,\zeta$. 
This equation is of fourth order in $\omega^2$. 
Hence, instead of two, there are four different dispersion laws
for the blue quasiquarks, and their explicit expressions are in general 
rather complicated.  In this context it is instructive to inspect the 
case $\theta = 0$ (or $\theta = \pi$), where the solutions take relatively 
simple forms. One finds:
\begin{alignat}{1}
\omega_{-,1}^2(|\vec p|,\theta=0) \;=\; E_-^2~, \qquad & 
\omega_{-,2}^2(|\vec p|,\theta=0) \;=\; E_p^2 + \mu^2 + |\Delta'|^2
- 2\sqrt{E_p^2\mu^2 + \vec p^{\;2}|\Delta'|^2}~,
\nonumber \\
\omega_{+,1}^2(|\vec p|,\theta=0) \;=\; E_+^2~, \qquad & 
\omega_{+,2}^2(|\vec p|,\theta=0) \;=\; E_p^2 + \mu^2 + |\Delta'|^2
+ 2\sqrt{E_p^2\mu^2 + \vec p^{\;2}|\Delta'|^2}~.
\label{ferrodisp0}
\end{alignat}
Hence, only two of the solutions are affected by the gap, while the other
two are not. In fact, from a non-relativistic point of view one would expect
that only quarks with spin up (down) can participate in an $M_J = +1$
($M_J = -1$) condensate and, thus, the spin down (up) quarks are ``blind'' to
this condensate. Relativistically, this remains true for quarks moving 
parallel or antiparallel to the $z$-direction, because
\beq
     C(\mp \sigma^{01} + i \sigma^{02}) \;=\; 2i
     \left(\begin{array}{cc} 
     \unity \pm \sigma_3 & 0 \\ 0 &  \unity \pm \sigma_3
     \end{array}\right)~.
\eeq
Here $\sigma_3$ is a Pauli matrix in spin space.
This observation trivially implies that there is always one gapless mode,
$\omega_{-,1}^2 = 0$ at $\theta = 0$ and $\vec p^{\;2} = \mu^2 - M^2$. 

The two other solutions at $\theta = 0$
turn out to be identical to the $M_J=0$-dispersion laws at $\theta = \pi/2$,
\beq
    \omega_{\mp,2}(|\vec p|,\theta=0) \;=\; 
    \omega_3^\mp\Big(|\vec p|,\theta=\frac{\pi}{2}\Big)~
\eeq
with $\omega_3^\mp$ as given in \eq{E3}.
Hence, for massless quarks, there is a second gapless mode at $\theta = 0$,
which corresponds to the $\omega_{-,2}$ branch at
$\vec p^{\;2} = \mu^2 + |\Delta'|^2$. 
In general, $\omega_{-,2}$ has a minimal value which is given by
\eq{M0}.

For arbitrary polar angles, the solutions of \eq{ferrodispeq} are quite
complicated.
Therefore Alford et al.~\cite{ABCC02} have determined the
dispersion laws numerically. 
In addition they have also derived an approximate expression, valid for
$M=0$ and $\omega$, $||\vec p| - \mu| \ll \mu$. 
In this case, they find that there are not only gapless modes at
the ``poles'', i.e., $\theta = 0,\pi$, but in a finite regime of order
 $\Delta'/\mu$ around them. When they introduce finite quark masses 
$M \gtrsim \Delta'$, this effect goes away and there remains a single 
gapless solution at the poles. 
It would be interesting to analyze the corresponding
thermal and transport properties of the system, but this has not yet
been done.

As pointed out in \sect{anisosym}, the ground state energy of the system
must be the same for $M_J=0$  and for $M_J=\pm 1$ 
if rotational invariance is not broken explicitly, i.e.,
if the Hamiltonian commutes with the total angular momentum $\vec J$.
This has been confirmed in Ref.~\cite{ABCC02} by explicit calculation. 
This result, although expected on general grounds, 
is quite remarkable in view of the rather different properties of the
dispersion laws.

\chapter{Three-flavor color superconductors}
\label{3sc}

In this chapter our analysis is extended to three quark flavors.
The additional flavor degree of freedom allows for  new condensation
patterns, most important the ``color-flavor locking'' (CFL) 
where both color and flavor $SU(3)$ are broken, leaving a residual
unbroken $SU(3)$ symmetry under a certain combination of color and flavor
rotations~\cite{ARW99}. 
The main features of this phase are briefly summarized in \sect{CFLideal} 
for the idealized case of three degenerate flavors. 
After that we discuss the influence of explicit symmetry breaking
due to a realistic strange quark mass. We will show that the 
self-consistent treatment of the quark masses has important effects.

\section{Three degenerate flavors}
\label{CFLideal}

\subsection{Condensation patterns}
\label{patterns}

In \eq{saa} we have introduced the scalar diquark condensate,
\beq
    s_{AA'} 
    \;=\; \ave{\,q^T \,C \gamma_5 \,\tau_A \,\lambda_{A'} \,q\,}~,
\label{saa3}
\eeq
where $\tau_A$ and $\lambda_{A'}$ are the antisymmetric
generators of $SU(N_f)$ and $SU(N_c)$, acting in flavor space and in color 
space, respectively. Hence, in general $s_{AA'}$ is a matrix with a 
flavor index $A$ and a color index $A'$. As before, we restrict ourselves to
the physical number of colors, $N_c=3$. Then $\lambda_{A'}$ denotes the
three antisymmetric Gell-Mann matrices, $\lambda_2$, $\lambda_5$, and 
$\lambda_7$. 

In the two-flavor case, which we have discussed in the previous chapter,
there is only one anti-symmetric generator in flavor space, $\tau_A=\tau_2$,
and the condensates $s_{2A'}$ form a vector with three color components 
$A'$.
Since we can always rotate this vector into the $A'=2$-direction by a
global $SU(3)$ color transformation, it was sufficient to restrict the
discussion of two-flavor color superconductors to a non-vanishing $s_{22}$.
(The only exception was in \sect{results} where we remarked that {\it global}
color neutrality could be achieved by the formation
of domains where the vector $s_{2A'}$ points into different directions.)

For three flavors, the flavor operators $\tau_A$ also denote the
three antisymmetric Gell-Mann matrices, i.e., $s\equiv (s_{AA'})$ is a 
$3\times3$ matrix, $A,A' \in \{2,5,7\}$.  
Since the rows and columns of this matrix are in general three linearly 
independent vectors in color or flavor space, respectively,
it is obvious that it usually cannot be reduced to a matrix with
a single non-vanishing element by color or flavor rotations. 
In general, the best we can do is to bring the 
matrix in triangular form, e.g., 
\beq
    s \;=\;\left(\begin{array}{ccc} \ct & 0 & 0 \\ 
                                      s_{52} & \cf & 0 \\
                                      s_{72} & s_{75} & \cs
                          \end{array}\right)~, 
\eeq
where five of the six non-vanishing components can be chosen to be real.
To this end we only need to perform general $SU(3)$ transformations in 
color space and diagonal $U(3)$ transformations in flavor space, i.e.,
a transformation to this form is still possible if the flavor $SU(3)$ is
explicitly broken by unequal masses.

In the following we assume that we have three degenerate flavors, i.e.,
the Lagrangian is symmetric under $SU(3)$ transformations, both, in color
and in flavor space. In this case we can perform a Ginzburg-Landau 
analysis. Similar to \eqs{anisopot} to (\ref{TM1})
the effective potential reads
\beq
    V(s) \;=\; -a^2\, {\rm tr}\, s^\dagger s  
    \;+\; \frac{1}{2}\lambda_1\,({\rm tr}\, s^\dagger s)^2 
    \;+\; \frac{1}{2}\lambda_2\,{\rm tr} (s^\dagger s)^2~, 
\label{LGpot3}
\eeq
with two invariant quartic terms.
For $\lambda_2 < 0$ the ground state is of the form~\cite{Sch00b,PiRi99b,Pat81}
\beq
    \ct \neq 0  \quad \text{and}\quad
    s_{AA'} = 0 \quad \text{if}\quad (A,A') \neq (2,2) \,.
\label{2sc}
\end{equation}
Obviously, this is identical to the two-flavor condensate, 
embedded into the enlarged flavor space.
A phase with this characteristics is therefore called ``2SC phase''
(i.e., ``two-flavor color superconducting phase'').
Most properties of this phase are of course identical to the
real two-flavor case and do not need to be discussed again. 
The main difference is that in the enlarged flavor group 
the condensate is no longer a singlet but transforms as an antitriplet. 
As a consequence also chiral $SU(3)_L\times SU(3)_R$ is broken down to
$SU(2)_L\times SU(2)_R$.  

For $\lambda_2 > 0$ the ground state takes the form of a unit 
matrix~\cite{Sch00b,PiRi99b,Pat81},
\beq
    \ct \;=\; \cf \;=\; \cs \;\neq\;0 \quad \text{and}\quad
    s_{AA'} = 0 \quad \text{if}\quad A \neq A'~.
\label{cfl}
\end{equation}
This is the so-called ``color-flavor locked'' (CFL) phase because the
color index is locked to the flavor index (a better explanation of this
name will be given in \sect{cflprop}). 
A similar condensation pattern is well known from the 
B-phase of liquid $^3$He where the components of the spin
and the orbital angular momentum of the pair are locked in the same way 
(see, e.g., Ref.~\cite{He3}).
In the context of color superconductors it was suggested first in 
Ref.~\cite{ARW99}. These authors considered a slightly more general ansatz
with a totally symmetric color-flavor wave function. Besides combining
a color antitriplet with a flavor antitriplet, as in \eq{saa3}, this can
also be achieved with two sextets (see \tab{tabpauli}).
In fact, in the CFL phase, the antitriplet terms are in general 
accompanied by induced sextet terms, which, in an obvious generalization of 
our notation, are of the form 
$s_{00} = s_{11} = s_{33} = s_{44}= s_{66} 
= s_{88}$~\cite{ARW99,Sch00b,PiRiwhy,Sho99}.
However, at least for interactions with the quantum numbers of a single
gluon exchange, where the color-sextet channel is repulsive, 
these terms have been found to be small~\cite{ARW99,Sho99}.
Therefore we will neglect them in the following.

For three-flavor QCD at asymptotic densities it can be shown that
the CFL phase is the correct ground state~\cite{Sch00b,EHHS00}.
The same is true for NJL-type models with three degenerate flavors. 
The main features of this condensation pattern will be summarized 
below (also see Ref.~\cite{RaWi00}).

\subsection{Properties of the CFL phase}
\label{cflprop}

The three non-vanishing diquark condensates which form the CFL phase
(\eq{cfl}) are listed in \tab{tabcflcond}.
Separately, each of them looks like a two-flavor color superconductor, 
being a color and flavor antitriplet in the scalar channel.
The first condensate is identical to the ``standard'' two-flavor
scalar condensate and consists of paired red and green $u$ and $d$ 
quarks (cf. \eq{deltacf}). The two other condensates have the same
structure but rotated in color and flavor space, i.e.,
$(r,g,b) \rightarrow (g,b,r) \rightarrow (b,r,g)$ and 
$(u,d,s) \rightarrow (d,s,u) \rightarrow (s,u,d)$.
Hence, all of the nine color-flavor combinations participate in
a condensate\footnote{As one can see in \tab{tabcflcond}, 
six species are paired with a fixed partner, whereas the remaining
three ($u_r$, $d_g$, $s_b$) form a ``triangle''.
Formally, this is related to the fact that the $9\times 9$ matrix
$a\,\tau_2\lambda_2 + b\,\tau_5\lambda_5+ c\,\tau_7\lambda_7$
can be decomposed into three $2\times2$ blocks and one $3\times 3$ 
block~\cite{ABR99}.
This will become important later on.},
and therefore all fermionic excitations are ``gapped''.
In general, one finds an octet with 
$\Delta_\mi{oct} = \Delta_{\bar 3} - \Delta_{6}$  and a singlet with  
with $\Delta_\mi{sing} = 2\Delta_{\bar 3} + 4\Delta_{6}$, 
where ${\bar 3}$ and $6$ refer to pairing in the color-antitriplet and 
color-sextet channel, respectively \cite{ARW99,Sch00,Sho99}. 
Thus, $\Delta_\mi{sing} = 2\Delta_\mi{oct}$ if the
color-sextet contribution is neglected.
Moreover, for QCD at asymptotic densities it can be shown that
$\Delta_\mi{oct} = 2^{-1/3} \Delta_\mi{2SC}$, where $\Delta_\mi{2SC}$ is 
the gap of the corresponding 2SC solution~\cite{Sch00b}.  
For NJL-type interactions, further details of the dispersion laws and 
the gap equations will be presented in \sect{cfulformalism} within a 
more general framework.

\begin{table}[t]
\begin{center}
\begin{tabular}{|l|c|c|c|}
\hline
&&&\\[-3mm]
condensate 
& $\ct =  \ave{\,q^T \,C \gamma_5 \,\tau_2 \,\lambda_2 \,q\,}$
& $\cf =  \ave{\,q^T \,C \gamma_5 \,\tau_5 \,\lambda_5 \,q\,}$
& $\cs =  \ave{\,q^T \,C \gamma_5 \,\tau_7 \,\lambda_7 \,q\,}$
\\[1mm]
\hline
&&&\\[-3mm]
diquark pairs
& $(u_r,d_g)$, $(u_g,d_r)$
& $(d_g,s_b)$, $(d_b,s_g)$
& $(s_b,u_r)$, $(s_r,u_b)$
\\
\hline
\end{tabular}
\end{center}
\caption{\small Color-flavor structure of the diquark pairs involved 
in the CFL phase.}
\label{tabcflcond}
\end{table} 

In the CFL phase, unlike the 2SC phase, color and flavor $SU(3)$ as well 
as chiral symmetry are broken completely: The $SU(2)$ subgroups which are
left unbroken, e.g., by $\ct$ are broken by $\cf$ and $\cs$.
However, the CFL ground state is invariant under certain
combinations of color and flavor rotations. This is more or less
evident from the color-flavor structure of the condensates (see 
\tab{tabcflcond}). For instance, if we 
declare the up quarks to be down quarks and vice versa, and at the 
same time interchange the meaning of red and green,
the pairing pattern remains unchanged. Formally, one can easily check
that 
\beq
    q^T C\gamma_5\,(\tau_2\lambda_2+\tau_5\lambda_5+\tau_7\lambda_7)\,q
    \quad\text{is invariant under}\quad
    q \;\rightarrow e^{i\,\theta_a(\tau_a - \lambda_a^T)}\,q~,
\eeq
where $a$ runs from 1 to 8: Color and flavor transformations are 
locked to a common $SU(3)_{c+V}$ which explains the name
``color-flavor locking''. Note that this mechanism is not new. 
We already mentioned that an analogous pairing pattern and thus
similar symmetry properties exist in liquid $^3$He
where the orbital angular momentum of the pair is locked to the spin.
But also in the case of chiral symmetry breaking in vacuum the
left- and right-handed $SU(N_f)$ transformations are locked to a common
$SU(N_f)_V$ (see also Ref.~\cite{PiRiwhy}).
In the CFL phase, both left- and right-handed flavor rotations, are 
locked to the color rotations and thereby indirectly to each other. 

As before, the standard $U(1)$ symmetry, related to baryon number
conservation, is broken down to $Z_2$. However, in contrast to the two-flavor 
case (cf. \sect{sca}) it is not possible to define an unbroken ``rotated''
baryon number. Similarly,
if we assume that the non-superconducting state is symmetric under
$U_A(1)$ (which should be the case at very high densities), this is
also broken down to $Z_2$.  
Thus, the pattern of symmetry breaking in the CFL phase reads
\beq
    SU(3)_c \times SU(3)_L \times SU(3)_R \times U(1)\,(\times U_A(1))
\;\longrightarrow\;
    SU(3)_{c+V}\times Z_2\,(\times {Z_2}_A)~.
\label{cflsymmetries}
\eeq

As a consequence of the complete breaking of $SU(3)_c$, all eight 
gluons acquire a mass in the CFL phase. The corresponding Meissner masses at
asymptotic densities have been calculated in Ref.~\cite{RischkeSU3}.
In contrast to the two-flavor case, \eq{cflsymmetries} contains also
spontaneously broken {\it global} symmetries of QCD. Hence, there are Goldstone
bosons, namely a pseudoscalar octet related to the broken chiral symmetry,
and a scalar singlet related to the breaking of $U(1)$. In addition 
there could be a massless pseudoscalar singlet which corresponds to the
breaking of $U_A(1)$.

Gauging a diagonal subgroup of the unbroken vector symmetry we can 
again define a ``rotated'' electromagnetic charge,
\beq
    \tilde Q \;=\; Q \,-\, \frac{1}{2}\,\lambda_3 \,-\,    \frac{1}{2\sqrt{3}}\,\lambda_8 \;\equiv\;
    Q \,-\, {\rm diag}_{\,c}\Big(\,\frac{2}{3},-\frac{1}{3},-\frac{1}{3}\,
    \Big)~. 
\label{qtilde3}    
\eeq
This definition differs from that in the two-flavor case, \eq{qtilde2}, 
by the $\lambda_3$-term. Since the 2SC ground state is invariant under 
$\exp{(i\alpha\lambda_3)}$, we could have added this term in that case, too,
but there was no need to do so.
One can easily check that all diquark pairs listed in \tab{tabcflcond}
have vanishing net $\tilde Q$-charge, i.e., the CFL ground state is a
perfect insulator for $\tilde Q$-photons. 

Another interesting result is that the $\tilde Q$-charges of all excitations,
including the quarks, are integers (in units of the rotated charge of 
the electron). This is exactly what we expect from a confining theory!
This observation has led Sch\"afer and Wilczek to the hypothesis of
``quark-hadron continuity''~\cite{ScWi99}:
It turns out that the CFL phase has the same symmetries and the same
low-lying excitation spectrum as confined hypernuclear matter at 
low densities, i.e., there is a one-to-one correspondence between the 
quarks, gluons, and Goldstone bosons in the CFL phase, and the baryons,
vector mesons, and Goldstone bosons in superfluid hypernuclear matter.
This would imply that no phase transition is needed between the low-density
and the high-density regime, and it is just a matter of convenience to
describe the former in the language of hadronic degrees of freedom and the 
latter referring to quarks and gluons\footnote{There are interesting,
although controversial,
attempts to explain the structure of the vacuum in an analogous way,
assuming that the vacuum is a Higgs phase with color-flavor locked
octets of quark-antiquark condensates, 
$\ave{\bar q \tau_a^T \lambda_a q}$~\cite{Wetterich}
(also see Ref.~\cite{Sch01} how this could be tested on the lattice). 
Basically, this would lead to the same spectrum, i.e., integer charged quarks,
massive gluons and massless pseudoscalar Goldstone bosons.    
Possible consequences of CFL diquark condensates at zero density have 
been discussed much earlier~\cite{SrSu81}.}. 

Of course, all this only holds in the idealized case of an exact flavor
$SU(3)$ symmetry. In reality we know that the ground state at low 
densities is ``ordinary'', i.e., non-strange, nuclear matter (unless the 
strange quark matter hypothesis turns out to be correct). Still there 
could be some modified quark-hadron continuity if the nuclear matter phase
is followed by (almost symmetric) hypernuclear matter. 
On the other hand, coming from high densities,
the CFL phase could be followed by a 2SC phase at 
lower densities, before matter turns into the hadronic phase
(cf. upper right panel of \fig{figschemphase} in the Introduction). In this
case the transition would not be continuous. This so-called 
``color-flavor unlocking'' transition from the CFL to the 2SC phase in 
the presence of realistic strange quark masses will be studied in great
detail in the next section.

\section{Realistic strange quark masses}
\label{CFUL}

The situations discussed so far are idealizations of the real world, where
the strange quark mass $M_s$ is neither infinite, such that strange quarks 
can be neglected completely (as we did in Chap.~\ref{tsc}), nor degenerate 
with the masses of the up and down quarks (as assumed in \sect{CFLideal}).
This leads to the question, whether quark matter at moderate densities
behaves more like a two-flavor color superconductor or like
a color-flavor locked state or like something else.    

Since standard BCS pairing involves fermions with opposite momenta,
$\vec p_a = - \vec p_b$, near their respective Fermi surface,
pairing is only possible, if the two Fermi momenta are not too far 
apart, or, equivalently, if the attraction is sufficiently strong.
(Note that for unequal Fermi momenta a Cooper instability is no longer
guaranteed for arbitrarily weak attractions, since in the non-interacting
case the creation of a BCS pair would always enhance the free energy.)
A rough estimate for the pairing condition is given by
\beq
    | p_F^a - p_F^b | \;=\; 
    \;\lesssim\; \sqrt{2}\,\Delta_{ab}~,
\label{stablegap}
\eeq
where $p_F^i = \theta(\mu_i-M_i) \sqrt{\mu_i^2 - M_i^2}$ are the 
nominal Fermi momenta and $\Delta_{ab}$ is the corresponding pairing 
gap\footnote{This approximate relation has been derived in 
Ref.~\cite{RaWi01} for a simplified model with two quark species,
requiring that the paired state is more favored than an unpaired
state. For quark matter with three colors and three flavors 
the authors of Ref.~\cite{ARRW01} find a similar relation with 
$\sqrt{2}$ on the r.h.s. replaced by $2$ for CFL pairing being 
more favored than no pairing at all and by a number not less than 
$\sqrt{3}$ for CFL pairing being more favored than 2SC pairing. 
For most of our purposes these details do not really matter. 
In particular we will never employ \eq{stablegap} to determine
the stability of a given phase, but mostly for qualitative arguments.
We should also mention that in certain physical situations the formally
stable solutions are forbidden by additional constraints and the
``unstable'' solution is the only allowed one. In these cases \eq{stablegap}
can be violated strongly (see \sect{neutdisc}).
\label{foot}}. 
In this section we will assume exact isospin symmetry ($m_u=m_d$) and that
the chemical potentials for all quarks are equal.
(Unequal chemical potentials will be discussed in Chap.~\ref{neut}.)
The above criterion is then always fulfilled for the pairing of up and down 
quarks.

Obviously, for sufficiently large quark chemical potentials $\mu \gg M_s$, 
the strange quark mass (and thus its difference to the light quark masses)
is negligible in \eq{stablegap} and $us$- and $ds$-pairing becomes 
possible as well. Hence, we expect a phase similar to the CFL phase, i.e.,
with non-vanishing values for $\ct$, $\cf$, and $\cs$. 
However, as a consequence of the mass difference, $M_s > M_u = M_d$, 
this phase will be less symmetric than in the idealized case discussed
above. In particular we expect the condensates which contain one strange and 
one non-strange quark, i.e., $\cf$ and $\cs$, to be smaller than $\ct$
which contains only non-strange quarks. This corresponds to an 
$SU(2)_{color + V}$ subgroup  of the original symmetry, where isospin 
rotations are locked to certain $SU(2)$ rotations in color space.
Later, when we also consider isospin breaking by non-equal chemical 
potentials (Chap.~\ref{neut}), all condensates will in general be different
from each other. Following common practice, we will nevertheless call 
this phase ``CFL phase'' whenever $\ct$, $\cf$, and $\cs$ do not vanish.

At small chemical potentials, the strange quark mass cannot
be neglected against $\mu$ and a phase with $\cf = \cs \neq 0$ is no longer
favored. From this we would conclude, that quark matter at low chemical
potentials, $\mu \simeq M_s$, is a two-flavor color superconductor (``2SC''),
where only up and down quarks participate in a diquark condensate.
In addition, there also could be unpaired strange quarks\footnote{Similar to 
the blue up and down quarks (see \sect{aniso}), the strange quarks 
could pair in a spin-1~\cite{ABCC02,Sch00} or in a
color-sextet~\cite{jiri} channel.}. In this case, the phase is often
called ``2SC+s'' phase, but we will usually not make this distinction.

On the other hand, below a certain value of $\mu$, one 
should finally reach the hadronic phase. 
This brings us back to the discussion in the end of the previous section.
If we start in the CFL phase and keep lowering the chemical potential,
the question is whether we first observe a transition to a 2SC phase 
followed by a transition to the hadronic phase at lower $\mu$ or whether
the CFL phase is directly connected to the hadronic phase without an 
intermediate 2SC phase. The latter would be particularly interesting, 
since it would again imply the possibility of a quark-hadron continuity 
scenario. In this case, however, there must be at least one phase 
transition, e.g., from ordinary (non-strange) nuclear matter to superfluid 
hypernuclear matter which then can continously evolve to the CFL phase. 
(For a systematic discussion of the symmetry properties of the various
quark and hadron phases, see Ref.~\cite{ScWi99b}.)

It is obvious from the discussion above
that the answer to this question depends on the strange 
quark mass. If $M_s$ is large, the CFL phase is disfavored already
at large values of $\mu$, possibly above the critical $\mu$ for the transition
to the hadronic phase, whereas small values of $M_s$ would favor a
quark-hadron continuity scenario. 
More quantitative investigations have been performed first in
Refs.~\cite{RSSV00,ABR99,ScWi99b}.
The authors of Ref.~\cite{ABR99} 
have studied the color-flavor unlocking phase transition in a
model calculation with different values of $M_s$. 
Assuming that the region below $\mu \simeq$~400~MeV belongs to the hadronic
phase, these authors came to the conclusion that a 2SC-phase exists if
$M_s \gtrsim$~250~MeV. Here $M_s$ is the constituent mass of the strange 
quark, which was treated as a free parameter in Ref.~\cite{ABR99}. 
A similar analysis has been performed in Ref.~\cite{RSSV00} 
employing an instanton mediated interaction. 

We have already seen, however, that constituent quark masses, if treated
self-consistently, are $T$- and $\mu$-dependent quantities, which in general
depend on the presence of quark-antiquark and diquark condensates and which
can even change discontinuously at a first-order phase boundary.
This means, not only the phase structure depends on the effective quark 
mass, but also the quark mass depends on the phase. 
This interdependence has not been taken into account in 
Refs.~\cite{RSSV00,ABR99}.
In this situation the natural next step is to generalize the analysis of
the previous chapter to three flavors and to study the interplay between
diquark condensates and quark-antiquark condensates within an NJL-type model. 
In fact, the early analysis of Ref.~\cite{ScWi99b} went already in this
direction, although the authors made some simplifying assumptions
about the quark dispersion laws. 
Below, we present a detailed discussion of these issues, mainly following
Refs.~\cite{BuOe02,OeBu02}.

\subsection{Formalism}
\label{cfulformalism}

To get started, we supply the three-flavor NJL-type Lagrangian
discussed in Chap.~\ref{njl3} with a quark-quark interaction term, i.e.,
\beq
    {\cal L}_\mi{eff} \;=\; {\bar q} (i \delsl - \hat{m}) q
                      \;+\; {\cal L}_{q\bar q} \;+\; {\cal L}_{qq}~,
\label{LCFUL}
\end{equation}
where
\beq
    {\cal L}_{q\bar q} \;=\; G\, \sum_{a=0}^8 \Big[({\bar q} \tau_a q)^2
    \;+\; ({\bar q} i\gamma_5 \tau_a q)^2\Big] 
    \;-\; K\,\Big[{\rm det}_f\Big({\bar q}(1+\gamma_5)q\Big) \,+\
                   {\rm det}_f\Big({\bar q}(1-\gamma_5)q\Big)\Big]
\label{LqbarqCFUL}
\end{equation}
as before, and 
\beq
    {\cal L}_{qq} \;=\;
    H\sum_{A = 2,5,7} \sum_{A' = 2,5,7}
    ({\bar q} \,i\gamma_5 \tau_A \lambda_{A'} \,C{\bar q}^T)
    (q^T C \,i\gamma_5 \tau_A \lambda_{A'} \, q)~.
\label{LqqCFUL}
\end{equation}
Again, these effective interactions might arise via Fierz rearrangement
from some underlying more 
microscopic theory and are understood to be used at mean-field level 
in Hartree approximation.

Note that ${\cal L}_\mi{eff}$ is only the simplest Lagrangian which
combines the phenomenologically constrained quark-antiquark interaction
of Chap.~\ref{njl3} 
with a term which allows for diquark condensation in the scalar 
color-antitriplet channels. 
For instance, we neglect the interesting
possibility of a combined quark-quark and quark-antiquark six-point
interaction which naturally arises from a Fierz transformation
of the instanton interaction~\cite{RSSV00}. 
We also neglect further condensates, like induced condensates
or possible spin-1 pairing of so-far unpaired species.

Starting from ${\cal L}_\mi{eff}$ we proceed in the usual way.
In order to calculate the mean-field thermodynamic potential at temperature
$T$ and quark chemical potential $\mu$, we linearize the interaction
in the presence of the diquark condensates
$s_{AA}$ and the quark-antiquark condensates $\phi_i$.
Introducing the constituent quark masses as in \eq{njlmasses3},
\beq
    M_i \;=\; m_i \,-\, 4G\phi_i \,+\, 2 K \phi_j \phi_k~, 
    \qquad \text{$(i,j,k)$ = any permutation of $(u,d,s)$}~, 
\label{MCFUL}
\end{equation}
and the diquark gaps
\begin{equation}
\Delta_A \;=\; -2 H \,s_{AA} ~,
\end{equation}
and employing Nambu-Gorkov formalism one gets
\begin{alignat}{1}
    \Omega(T,\mu) 
     \;=\; &-T \sum_n \int \frac{d^3p}{(2\pi)^3} \;
    \frac{1}{2}\,{\rm Tr}\;\ln\, \Big(\frac{1}{T}\,S^{-1}(i\omega_n, \vec p)
    \Big)
    \nonumber\\
    &+\, 2G\,(\phi_u^2 \,+\, \phi_d^2 \,+\, \phi_s^2) 
     \;-\; 4K \phi_u\,\phi_d\,\phi_s 
    \;+\; H\,(|\ct|^2 \,+\, |\cf|^2 \,+\, |\cs|^2)~. 
\label{OmegaCFUL}
\end{alignat}
Here
\beq
    S^{-1}(p) \;=\; \left(\begin{array}{cc}
    \psl - \hat{M} + \mu\gamma^0 &
    \sum_A \Delta_A \, \gamma_5\tau_A\lambda_A \\
    -\sum_A \Delta_A^* \, \gamma_5\tau_a\lambda_A &
    \psl - \hat{M} - \mu\gamma^0
    \end{array}\right)
\label{SinvCFUL}
\end{equation}
is the inverse fermion propagator, where $\hat{M} = diag(M_u,M_d,M_s)$.
Taking into account the Dirac structure, color, flavor, and the 
Nambu-Gorkov components, $S^{-1}$ is a $72\times72$ matrix, and the
trace in \eq{OmegaCFUL} has to be evaluated in this 72 dimensional space. 

Since we are only dealing with one common chemical potential, 
there is some simplification due to isospin symmetry, 
$m_u = m_d$, which implies
\beq
    \phi_u = \phi_d \qquad \text{and} \qquad \cf = \cs~, 
\label{cfuliso}
\eeq
and thus $M_u = M_d$ and $\Delta_5 = \Delta_7$.
In this case a tedious but straight-forward calculation yields
\begin{alignat}{1}
\frac{1}{2}\,{\rm Tr}\;\ln\, \Big(\frac{1}{T}\,S^{-1}(&i\omega_n,\vec p)\Big)
\nonumber\\
\;=\; 3\ln\Big(\frac{1}{T^4}(&\xup\xum +2\dt^2\yu + \dt^4) \Big)
\nonumber\\
\;+\; 2\ln\Big(\frac{1}{T^4}(&\xup\xsm +2\df^2\yus + \df^4) \Big)
\;+\; 2\ln\Big(\frac{1}{T^4}(\xsp\xum +2\df^2\yus + \df^4) \Big)
\nonumber\\
\;+\;\phantom{2} \ln\Big(\frac{1}{T^8}(&\xup\xum\xsp\xsm + 2\dt^2 \xsp\xsm\yu
       + 4\df^2(\xup\xsm+\xsp\xum)\yus 
\nonumber\\ 
      &+ \dt^4\xsp\xsm + 4\dt^2\df^2(\xsp\xusm + \xusp\xsm) 
\nonumber\\ 
      &+ 4\df^4(\xup\xsm+\xsp\xum + 4\yus^2) \phantom{\Big)} 
\nonumber\\ 
      &+ 8\dt^2\df^4\ys + 32\df^6\yus \;+\; 16\df^8)\; \Big) \;,
\label{trlog}
\end{alignat}
where we have introduced the abbreviations
\beq
    x_{ff'}^\pm \;=\; (\omega_n\pm i\mu)^2 + \vec p^{\,2} + M_f M_{f'}
    \quad\text{and}\quad
    y_{ff'} = \omega_n^2 + \mu^2 + \vec p^{\,2} + M_f M_{f'}\;.
\end{equation}
With these definitions one finds for the argument of the first
logarithm on the r.h.s. of \eq{trlog}
\beq
    \xup\xum +2\dt^2\yu + \dt^4 \;=\; 
    (\omega_n^2 + {\omega_u^-}^2)(\omega_n^2 + {\omega_u^+}^2)
\end{equation}
with
\beq
    \omega_u^\mp \;=\; 
    \sqrt{(\sqrt{\vec p^{\,2} \;+\; M_u^2} \mp \mu)^2 + \dt^2}\;.
\end{equation}
Obviously, these are exactly the dispersion laws of the paired quarks in a 
two-flavor color superconductor, \eq{epmstandard}. 
The corresponding Matsubara sums are readily turned out using \eq{Matsu}.

The other terms in \eq{trlog} are in general more complicated.
There are, however, two simplifying limits.
The first one corresponds to a two-flavor color superconductor, together with
unpaired strange quarks. In this case $\Delta_5$ vanishes and 
\eq{trlog} becomes
\begin{alignat}{1}
\frac{1}{2}\,{\rm Tr}\,&\ln\, \Big(\frac{1}{T}\,S^{-1}(i\omega_n,\vec p)
\Big)\Big|_{\Delta_5 = 0} 
\nonumber\\
=\quad 
&4\Big[\ln(\frac{\omega_n^2 + {\omega_u^-}^2}{T^2}) \,+\,
       \ln(\frac{\omega_n^2 + {\omega_u^+}^2}{T^2})\Big]
\;+\; 2\Big[\ln(\frac{\omega_n^2 + {E_u^-}^2}{T^2}) \,+\,
\ln(\frac{\omega_n^2 + {E_u^+}^2}{T^2}) \Big]
\nonumber \\
\;+\; &3\Big[\ln(\frac{\omega_n^2 + {E_s^-}^2}{T^2}) \,+\,
\ln(\frac{\omega_n^2 + {E_s^+}^2}{T^2}) \Big]~,
\label{trlog2cs}
\end{alignat}
with $E_f^\mp = \sqrt{\vec{p}^{\,2} + M_f^2} \mp \mu$.
Here we recover the fact, that only four of the six light quarks (two colors)
participate in the 2SC condensate, while the two remaining ones 
and all strange quarks fulfill the dispersion laws of free 
particles with effective masses $M_f$.

We can also reproduce the structure of the dispersion laws of the idealized
three-flavor symmetric CFL-state. 
To this end we evaluate \eq{trlog} for 
$M_u=M_s$ and $\Delta_2 = \Delta_5$. One finds
\begin{alignat}{1}
\frac{1}{2}\,{\rm Tr}\,&\ln\, \Big(\frac{1}{T}\,S^{-1}(i\omega_n,\vec p)
\Big)\Big|_{M_u=M_s,\,\Delta_2=\Delta_5} 
\nonumber\\ 
&\;=\; 8 \Big[\ln(\frac{\omega_n^2 + {\omega_\mi{oct}^-}^2}{T^2}) \,+\,
           \ln(\frac{\omega_n^2 + {\omega_\mi{oct}^+}^2}{T^2}) \Big]
\;+\;  \Big[\ln(\frac{\omega_n^2 + {\omega_\mi{sing}^-}^2}{T^2}) \,+\,
         \ln(\frac{\omega_n^2 + {\omega_\mi{sing}^+}^2}{T^2}) \Big]
\label{trlog3cfl}
\end{alignat}
with $\omega_\mi{oct}^\mp = \omega_u^\mp$ and
$\omega_\mi{sing}^\mp = \sqrt{(\sqrt{\vec{p}^{\,2} \;+\; M_u^2} \mp \mu)^2 + 
|2\Delta_2|^2}$. Thus $\Delta_\mi{sing} = 2\Delta_\mi{oct}$, as already 
mentioned in \sect{cflprop}.

The Matsubara sums over \eqs{trlog2cs} and (\ref{trlog3cfl}) can again 
be turned out with the help of \eq{Matsu}.
In general, i.e.,  for \eq{trlog} with arbitrary values of the 
condensates, this cannot be done so easily. 
If one combines the second with the third logarithm on the r.h.s., the
argument becomes a polynomial of fourth order in $\omega_n^2$. 
The same is true for the argument of the fourth logarithm. 
The corresponding dispersion laws are related to the zeros of these
polynomials. Although, in principle, the zeros of a polynomial of fourth 
order can be determined analytically, the resulting expressions
are usually difficult to handle. Therefore, in practice one has to determine
the dispersion laws numerically. After that, one can again employ \eq{Matsu} 
to calculate the Matsubara sum.  
Alternatively, one can turn out the Matsubara sum numerically without
previous determination of the dispersion laws. 
To that end, in order to get a convergent result, one should subtract and add 
a properly chosen term,
$
    \sum_n A_n \;=\; \sum_n\, (A_n - B_n) \;+\; \sum_n B_n~,
$
where $A_n$ stands for \eq{trlog}, $\sum_n B_n$ can be turned out 
analytically and $A_n - B_n$ is well-behaved. This is the method we have
used\footnote{Some authors use simplified dispersion laws to 
circumvent this problem. The authors of Ref.~\cite{ScWi99b} assume that,
even in the presence of a symmetry breaking dynamical strange quark mass,
the dispersion laws in the CFL phase are of the standard form,
\eq{epmstandard}, with an octet gap attributed to the six non-strange
and two strange quarks, and a singlet gap attributed to the third
strange quark. 
The authors of Ref.~\cite{GNA02}
make the assumption that the particle part and the antiparticle part of
the Hamiltonian separate. This enables them to derive an analytical
expression for the quasiparticle energies. It turns out that the numerical 
results obtained in this approximate way are very similar to our exact 
solutions.}.

Because of the isospin relations, \eq{cfuliso}, the thermodynamic potential
depends on four different condensates, $\phi_u$, $\phi_s$, $\ct$, and $\cf$. 
The self-consistent solutions are again given by the stationary points of the
potential,
\beq
    \frac{\delta \Omega}{\delta \phi_u} \;=\; 
    \frac{\delta \Omega}{\delta \phi_s} \;=\;
    \frac{\delta \Omega}{\delta \ct^*} \;=\;
    \frac{\delta \Omega}{\delta \cf^*} \;=\; 0~.
\end{equation}
The explicit evaluation of these derivatives is trivial, but 
leads to rather lengthy and not very illuminating expressions, which we 
do not present. 
However, since the thermodynamic potential is a function of the squared
diquark gaps, $|\Delta_A|^2$, it is obvious that there are always
trivial solutions $\Delta_A = 0$, independent of the values of the other
gap parameters. 
For the non-trivial solutions the four equations are coupled and have to be 
solved simultaneously. The stable solution is again the 
one which corresponds to the lowest value of $\Omega$.

\subsection{Numerical results without 't Hooft interaction}
\label{resultsk0}   

To fix the parameters for the numerical analysis we begin again
with a color current interaction (cf.~\eq{Lhge}),
\beq
    {\cal L}_{int} \;=\; -g\,\sum_{a=1}^8 ({\bar q} \gamma^\mu \lambda_a q)^2~.
\label{Linthge}
\end{equation}         
This interaction was also the starting point of the model calculations 
in Refs.~\cite{ARW99,ABR99,ScWi99b}. 
Performing Fierz transformations we find that the effective coupling 
constants which enter \eqs{LqbarqCFUL} and (\ref{LqqCFUL}) are related
to each other as (see App.~\ref{fierzexamples})
\beq
     G \;:\; K \;:\; H \;=\; 1 \;:\; 0 \;:\; \frac{3}{4}~. 
\label{CFULFierz}
\end{equation}
In addition, there are again various other channels,
which in principle should be taken into account to be fully 
self-consistent\footnote{As discussed in \sect{interplay}, already for two 
flavors, a self-consistent treatment requires the simultaneous 
consideration of possible expectation values in six different channels.
Here we should have at least twice as many, because of the broken
flavor symmetry. Moreover, as mentioned earlier, even in the idealized
case of three massless flavors, there is an induced
color-sextet diquark condensates in the CFL-phase \cite{ARW99,Sch00b,Sho99}.}.
This is, however, not the goal of the present calculation. 
At this point, \eq{Linthge} should only be viewed as a ``typical''
interaction, used to relate the coupling constant in the diquark channel
to the quark-antiquark coupling constant. 

Perhaps the most severe limitation of this choice is the fact that the
six-point interaction completely vanishes, i.e., $K=0$. 
Similar to what we have discussed in \sect{flamix}, this means that there is
no flavor mixing in the quark-antiquark channel. There is of
course flavor mixing in the (flavor-antitriplet) diquark condensates, 
where always quarks of two different flavors are paired. Hence, the strange 
quarks decouple from the non-strange quarks in all but the CFL 
phase. The consequences of this limitation will be investigated in 
\sect{resultsk} where we introduce non-vanishing values of $K$.

Employing \eq{CFULFierz}, there remain four parameters: the coupling
constant $G$, the cut-off $\Lambda$ , and the two current quark masses, 
$m_u$ and $m_d$.
We take $\Lambda = 602.3$~MeV and $m_u = 5.5$~MeV as in parameter set
RKH~\cite{Rehberg} of \tab{tabnjl3fit},
and tune the two remaining parameters, $G$ and $m_s$, to reproduce the 
vacuum constituent quark masses $M_u = 367.6$~MeV and $M_s = 549.5$~MeV
of that set. In this way we find $G\Lambda^2 = 2.319$ and
$m_s = 112.0$~MeV.

We begin with the discussion of the results at zero temperature.  
\begin{figure}
\begin{center}
\epsfig{file=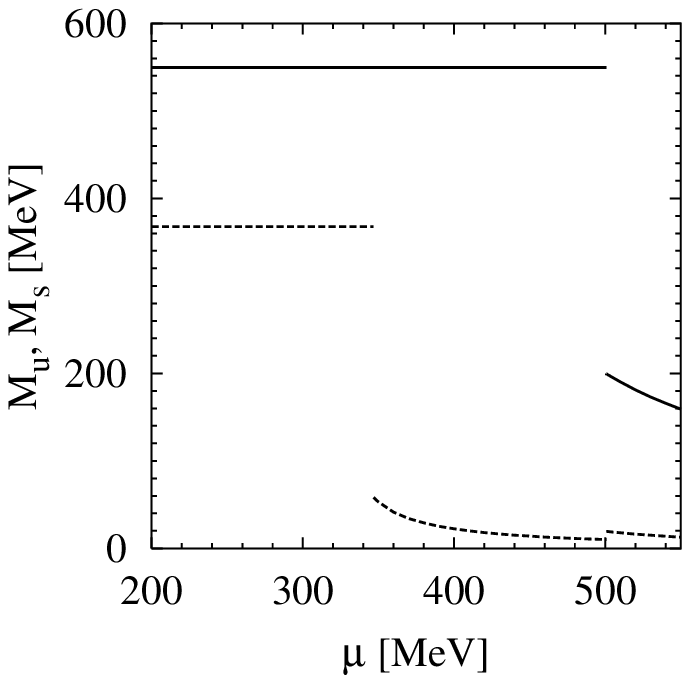,width=7.4cm}
\hfill
\epsfig{file=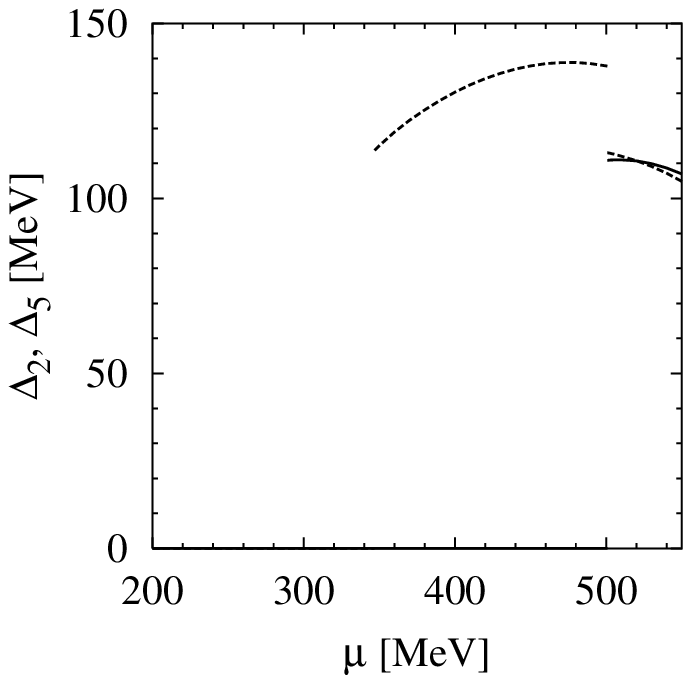,width=7.4cm}
\end{center}
\vspace{-0.5cm}
\caption{\small Gap parameters at $T=0$ as functions of the quark chemical 
          potential $\mu$ for $K=0$.
          Left: Constituent masses of up and down quarks (dashed), 
          and of strange quarks (solid).
          Right: Diquark gaps $\Delta_2$ (dashed) and $\Delta_5$ (solid).
          Adapted from Ref.~\cite{BuOe02} with slightly different parameters.}
\label{figk0t0}
\end{figure}
The behavior of the four gap parameters as functions of the quark chemical
potential $\mu$ is displayed in \fig{figk0t0}. In the left panel we
show the constituent quark masses $M_u$ and $M_s$, in the right panel the
diquark gaps $\Delta_2$ and $\Delta_5$.
One can clearly distinguish between three phases. 
At low chemical potentials $\mu < \mu_1 =346.9$~MeV, the system remains 
in the vacuum phase, i.e., the diquark gaps 
vanish and the constituent quark masses stay at their vacuum values.
For the condensates this means $\ct = \cf = \cs = 0$, while
$\phi_u$ and $\phi_s$ are large. Hence, in a schematic sense, 
we can identify this phase with the ``hadronic phase'',
keeping in mind the limitations of this picture we have discussed earlier.

At $\mu = \mu_1$ a first-order phase transition takes place and the 
system becomes a two-flavor color superconductor:
The diquark condensate $\ct$ has now a non-vanishing expectation value,
related to a non-vanishing diquark gap $\Delta_2$, whereas $\Delta_5$ 
remains zero.
Just above the phase boundary we find $\Delta_2 = 113.7$~MeV. At the
same time the mass of the up quark drops from the vacuum value to 
$M_u = 58.3$~MeV. With increasing $\mu$, $M_u$ decreases further, while
$\Delta_2$ increases until it reaches a maximum at $\mu \simeq 475$~MeV. 
Just below the next phase boundary at $\mu = \mu_2 = 500.8$~MeV we find 
$\Delta_2 = 137.8$~MeV and $M_u = 10.0$~MeV. 

In the 2SC phase the baryon number density is of course no longer
zero and increases from about 2.5 times nuclear matter density at 
$\mu = \mu_1$ to about 6.5 times nuclear matter density at $\mu = \mu_2$. 
The density of strange quarks remains zero up to $\mu = \mu_2$. 

At $\mu = \mu_2$ the system undergoes a second phase transition, now
from the 2SC phase into the CFL phase, which is characterized by a 
non-vanishing diquark gap $\Delta_5$ (together with a non-vanishing
$\Delta_2$). The phase transition is again of first order: At the
transition point $\Delta_5$ jumps from zero to 110.9~MeV, while $M_s$
drops from 549.5~MeV to 199.6~MeV. The non-strange quantities are also
discontinuous and change in the opposite direction: 
$M_u$ jumps from 10.9~MeV to 19.4~MeV, and
$\Delta_2$ drops from 137.8~MeV to 113.1~MeV, 
in rather good agreement with the asymptotic relation 
$\Delta_\mi{oct} = 2^{-1/3} \Delta_\mi{2SC}$~\cite{Sch00b}.  
The density jumps from 6.5 to 9.3 times nuclear matter density. 

As anticipated, unlike the ideal $SU(3)$-flavor symmetric case, the
sizes of the gaps $\Delta_2$ and $\Delta_5$ are no longer equal.
In fact, as we are quite far away from an exact $SU(3)$ symmetry, 
it is remarkable that the diquark gaps $\Delta_2$ and $\Delta_5$ are so
similar at the transition point. 
At least partially, this may be attributed to the cut-off.
Obviously, when the Fermi momentum comes close to $\Lambda$
important parts of the gap equation are cut off, and the gap becomes
smaller with increasing chemical potentials. For the up and down quarks
this situation is reached somewhat earlier than for the strange quarks
because of their higher Fermi momenta, related to their lower mass.  
This also explains why we find $\Delta_5 > \Delta_2$ above 
$\mu \simeq 520$~MeV. Here we certainly approach the limits of the 
model. 

The most important point of our analysis is the discontinuous behavior 
of the strange quark mass at the phase boundary.
As we have argued in the introductory part of
this section, we expect the CFL phase to be stable if $\mu$ is
considerably larger than $M_s$ and that it becomes unstable when
$\mu \lesssim M_s$. Clearly, this kind of reasoning can only be used
to estimate the critical $\mu$ if $M_s$ is more or less constant.
Obviously this is not the case. Just above the phase boundary
we have $\Delta_5^{crit} := |p_F^u - p_F^s|/\sqrt{2} = 29$~MeV
$< \Delta_5 = 111$~MeV, which means that the approximate
stability condition, \eq{stablegap}, is well satisfied and far away 
from its limit. 
In fact, the reason for the phase transition at $\mu = \mu_2$
is not that $u$ and $s$ quarks can no longer pair below that value. 
We still find a CFL solution down to much lower values of $\mu$.
However, for $\mu < \mu_2$, this solution is
only metastable, and there is a more favored solution with a much higher 
strange quark mass which cannot support a pairing of $u$ and $s$ quarks.
In this sense we may say that the 2SC-CFL phase transition is driven
by the chiral phase transition in the strange sector, although of 
course all condensates mutually influence each other.

Relations like \eq{stablegap} have been used to argue, that the
color-flavor-unlocking transition at $T=0$ must be first order, 
because the mismatch of the Fermi surfaces of strange and non-strange
quarks prevents the gap parameter $\Delta_5$ from becoming arbitrarily
small~\cite{ABR99,RaWi01,ARRW01}. Qualitatively, our results support
these arguments, even though the quantitative values for the minimal gap 
derived in these references cannot be applied to cases with density dependent 
masses: We find that there is indeed a non-vanishing lowest possible value 
of $\Delta_5$ for metastable CFL solutions and the minimal value of 
$\Delta_5$ in an absolutely stable CFL phase is even larger. 

\begin{figure}
\begin{center}
\epsfig{file=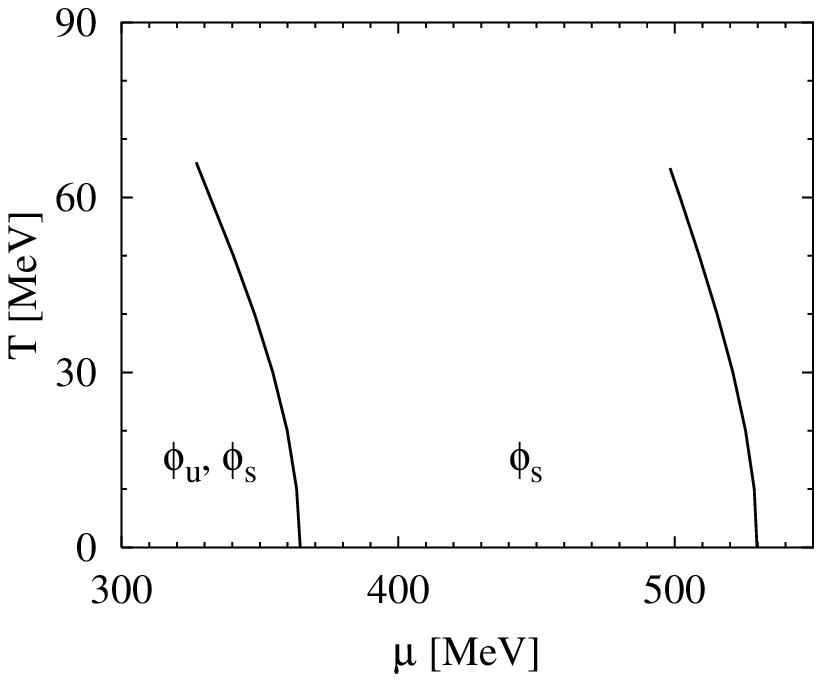,width=7.4cm}
\hfill
\epsfig{file=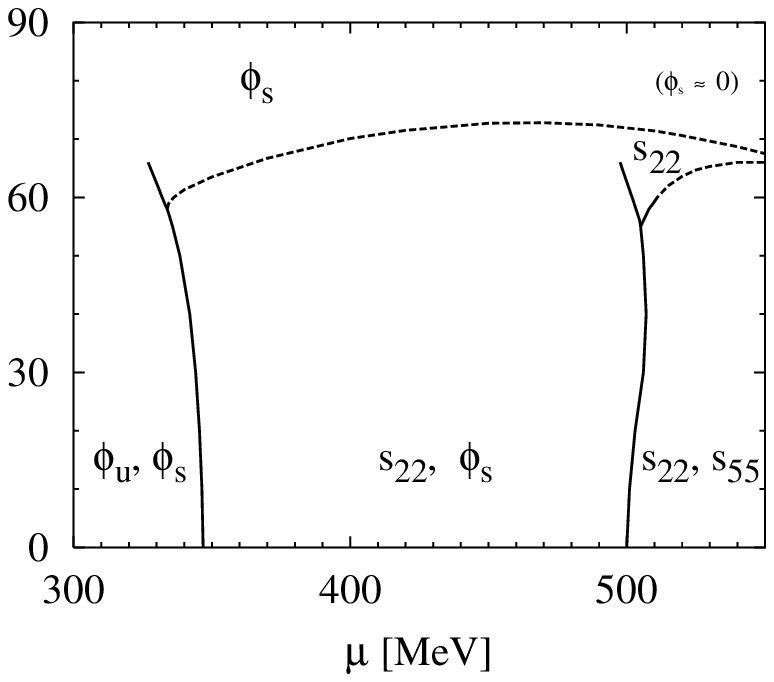,width=7.4cm}
\end{center}
\vspace{-0.5cm}
\caption{\small Phase diagram for $K=0$ in the $\mu-T$ plane without
         (left) and with (right) diquark condensates taken into account. 
         The solid lines indicate phase boundaries with a first-order 
         phase transition, the dashed lines correspond to second-order 
         phase transitions. The different phases can be distinguished by 
         different values for the various condensates. Within the figure 
         we have indicated only those condensates which are significantly 
         different from zero. (Note that because of isospin symmetry, 
         $\phi_d = \phi_u$ and $s_{77} = s_{55}$.)
         The right figure has been adapted from Ref.~\cite{BuOe02} for
         slightly different parameters.}
\label{figphasek0}
\end{figure}

We now extend our analysis to non-vanishing temperatures.
For an easier interpretation, let us first neglect the diquark condensates.
The resulting phase diagram in the $\mu-T$ plane is shown in the left panel
of \fig{figphasek0}. Since for $K=0$  the different quark flavors decouple, 
there are two separate phase boundaries, corresponding to the chiral 
phase transition of the non-strange quarks at lower chemical potentials and
of the strange quarks at higher chemical potentials.
Thus, at low temperatures, we have three different regimes. 
When the temperature is increased, both phase boundaries end in a 
second-order endpoint.  

When diquark condensates are included (right panel) these first-order phase
boundaries are partially shifted to lower chemical potentials, but remain
qualitatively unchanged. 
As discussed for zero temperature, at low temperatures the disappearance of
one type of quark-antiquark condensate is always accompanied
by the appearance of a new diquark condensate. 
We have the ``hadronic'' (better: normally conducting) phase with
$\phi_u,\phi_s \neq 0$ and vanishing diquark condensates,
the 2SC phase with $\phi_s, \ct \neq 0$, but $\phi_u \approx 0$
and $\cf \neq 0$, and the CFL phase with $\ct, \cf \neq 0$, but 
$\phi_u, \phi_s \approx 0$. 
On the other hand, at high temperatures all condensates vanish. 
This leads to two additional phase boundaries.   
The first one corresponds to the melting of the diquark condensate in the
2SC phase. As we have seen before, this is a second-order phase transition
with a transition temperature approximately given by the standard
BCS relation, \eq{Tcapp}.
For instance, at $\mu = 400$~MeV we find $\Delta_2(T = 0 ) = 130.3$~MeV,
and $T_c = 70.1$~MeV, whereas from \eq{Tcapp} 
we would expect $T_c = 74.3 $~MeV.

\begin{figure}
\begin{center}
\epsfig{file=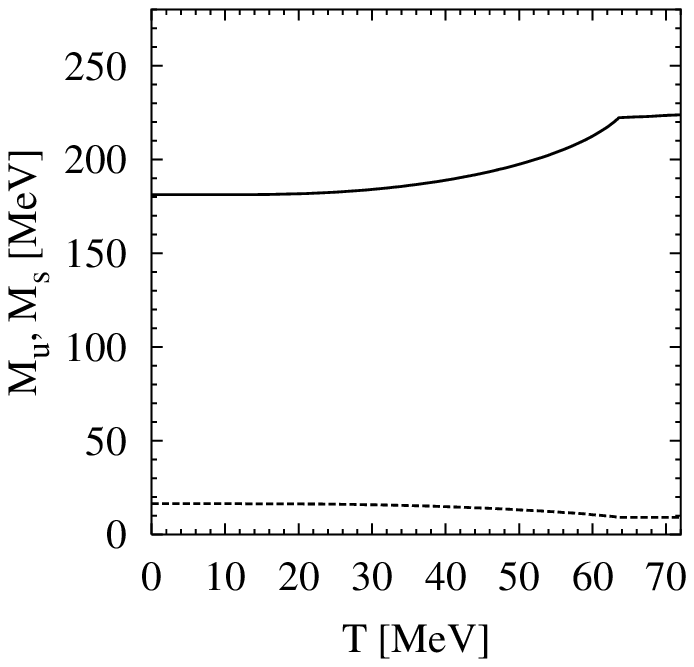,width=7.4cm}
\hfill
\epsfig{file=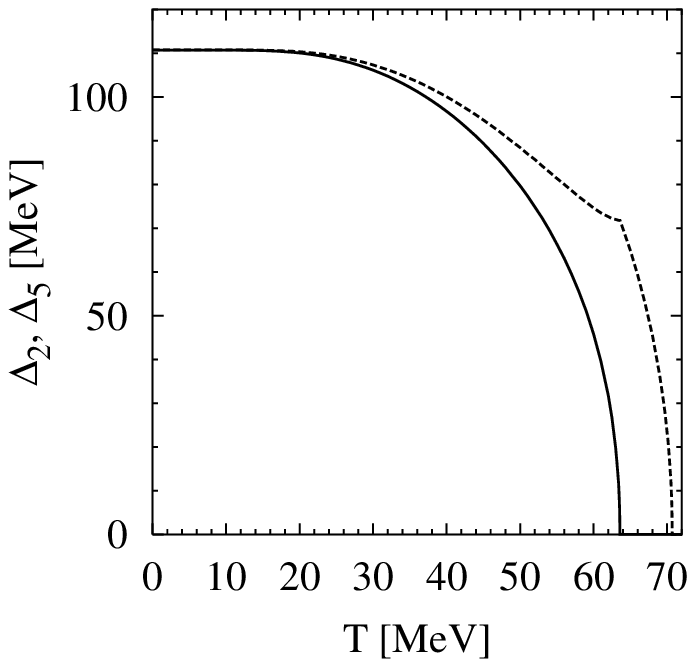,width=7.4cm}
\end{center}
\vspace{-0.5cm}
\caption{\small Gap parameters at $\mu = 520$~MeV as functions of the 
          temperature.
          Left: Constituent masses of up and down quarks (dashed), 
          and of strange quarks (solid).
          Right: Diquark gaps $\Delta_2$ (dashed) and $\Delta_5$ (solid).
          Adapted from Ref.~\cite{BuOe02} for slightly different parameters.
}
\label{figcfulmu=520}
\end{figure}
 
Starting from the CFL phase and increasing the temperature, one first 
observes a melting of the diquark gap $\Delta_5$, before at
somewhat higher temperatures $\Delta_2$ vanishes as well. The intermediate
2SC phase ``above'' the CFL-phase is partially separated from the 
2SC regime ``left'' to the CFL-phase by the upper part of the chiral
phase boundary of the strange quark. Note that this part of the phase
boundary is not affected at all by the diquark condensate, since the strange 
quarks are completely decoupled from the non-strange sector.
In particular the critical endpoint is at the same place as in the left 
figure. (The same is of course true for the endpoint of the non-strange
chiral phase boundary which is located in a regime where the diquark
condensates vanish.)

Based on the assumption that $\Delta_5$ is always smaller than $\Delta_2$
(which is not really true at $T=0$, as we have seen), the earlier 
disappearance of $\Delta_5$ and hence the existence of a 2SC phase ``above''
the CFL phase was already anticipated in Ref.~\cite{RaWi00}. 
Applying similar arguments as at zero temperature 
(cf.~Refs.~\cite{ABR99,RaWi01,ARRW01}) it was also predicted in that 
reference, that the corresponding color-flavor unlocking 
transition should stay first order. This was corroborated by a second 
argument, claiming that the phase transition
corresponds to a finite temperature chiral restoration phase transition 
in a three-flavor theory. In this case the universality arguments of 
Ref.~\cite{PiWi84} should apply, stating that the phase transition 
should be of first order. 

Indeed, following the phase boundary from the left, 
we find that the transition continues to be first order. 
However, above a critical point at $\mu \simeq 511$ MeV and 
$T \simeq 60$ MeV the phase transition becomes second order.
This is illustrated in \fig{figcfulmu=520}, where the
constituent masses (left panel) and the diquark gaps (right panel)
are displayed as functions of the temperature for fixed $\mu = 520$~MeV.  
As one can see, both condensates smoothly go to zero at 
$T=63.5$~MeV and $T=70.7$~MeV, respectively.
Of course it is possible that the second-order phase transitions
are artifacts of the mean-field approximation and become first order
if fluctuations are included~\cite{Vosk03,GHRR04}.
In any case, the arguments of Refs.~\cite{ABR99,RaWi01,ARRW01} in 
favor of a first-order color-flavor unlocking
phase transition are much less stringent at finite
temperature, where even for vanishing condensates the Fermi surfaces are 
smeared due to thermal effects. The applicability of the universality
argument is also questionable in the present situation because
the 2SC phase is not a three-flavor chirally
restored phase, but only $SU(2)\times SU(2)$ symmetric. 
Recall that even $SU(3)_V$ is broken spontaneously by the flavor-antitriplet 
diquark condensates, and explicitly by the unequal current quark masses. 
Therefore a rigorous prediction of the true order of the
phase transition is rather difficult.  

Finally, we should repeat that our results at $\mu \gtrsim$~500~MeV 
are sensitive to the cut-off. This is somewhat disturbing,
since the entire CFL phase belongs to this region. As we will see below,
the situation improves when a flavor mixing interaction
is included, which shifts the 2SC-CFL phase boundary to lower chemical
potentials.

\subsection{Influence of the 't Hooft interaction}
\label{resultsk}   

We now study the effect of a non-vanishing 't Hooft interaction
on the results of the previous section~\cite{OeBu02}. 
To this end we replace the color current interaction, \eqs{Linthge}
and (\ref{CFULFierz}), by a set of parameters with $K \neq 0$.
For a better comparison with our previous results we leave the cut-off
$\Lambda$ and the light current mass $m_u$ unchanged, but
vary the coupling constant $G$  and the strange current mass $m_s$
in such a way that the vacuum masses $M_u$ and $M_s$ remain constant. 
We also keep the quark-quark coupling constant $H$ at a fixed value.
This means that instead of $H:G = 3:4$ we now have $H:G_\mi{eff} = 3:4$,
where $G_\mi{eff} = G - \frac{1}{2}K\phi_s^{vac}$ is the effective
4-point coupling which determines the light quark constituent mass
in vacuum (cf. \eq{alphaeff}). 

\begin{figure}
\begin{center}
\epsfig{file=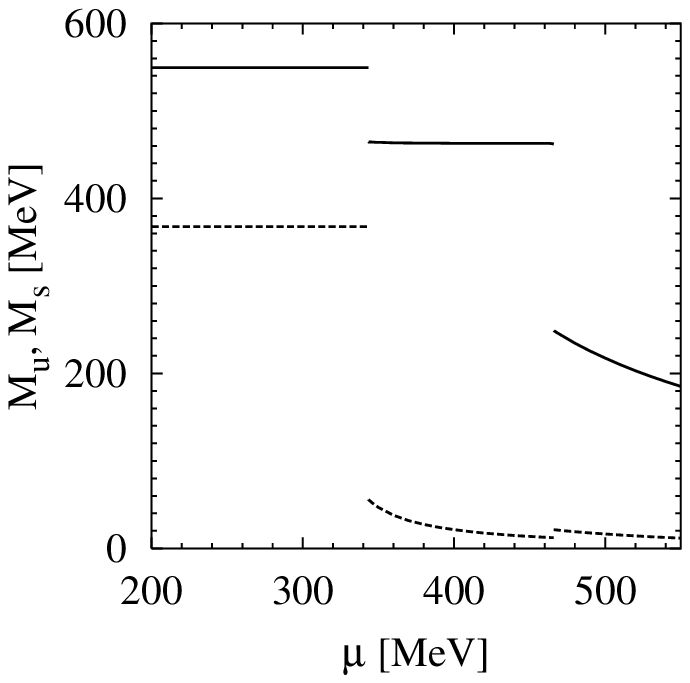,width=7.4cm}
\hfill
\epsfig{file=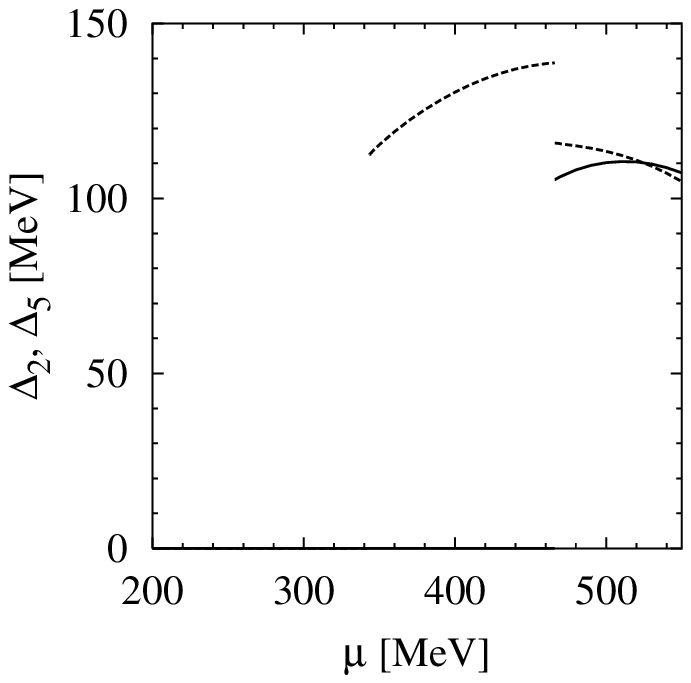,width=7.4cm}
\end{center}
\vspace{-0.5cm}
\caption{\small The same as \fig{figk0t0}, but for $K\Lambda^5 = 12.36$
(\tab{tabnjl3fitk}, parameter set III).}
\label{figk12.36t0}
\end{figure}

Four sets of parameters obtained in this way
(including the $K=0$ parameters of the previous section) are listed 
in \tab{tabnjl3fitk}.
\begin{table}[b]
\begin{center}
\begin{tabular}{| c | c c c c c c | c |}
\hline
&&&&&&&\\[-3mm]
set & $\Lambda$ [MeV] & $m_u$ [MeV]  & $m_s$ [MeV] & $G\Lambda^2$ 
  & $K\Lambda^5$ & $H\Lambda^2$ &  $H : G$  
\\[1mm]
\hline
&&&&&&&\\[-3mm]
I   & 602.3 & 5.5 & 112.0 & 2.319 & {\phantom 1}0.00 & 1.739 & 0.75 \\
II  & 602.3 & 5.5 & 123.6 & 2.123 & {\phantom 1}5.00 & 1.739 & 0.82 \\
III & 602.3 & 5.5 & 140.7 & 1.835 &            12.36 & 1.739 & 0.95 \\
IV  & 602.3 & 5.5 & 158.5 & 1.536 &            20.00 & 1.739 & 1.13 \\
\hline
\end{tabular}
\end{center}
\caption{\small Model parameters employed in the numerical studies 
of this section.
}
\label{tabnjl3fitk}
\end{table}  
Set III is of particular interest because, apart from the quark-quark
coupling constant, it is identical to parameter set RKH of \tab{tabnjl3fit}
(i.e., the empirical fit of Ref.~\cite{Rehberg}).
This was our standard parameter set for the numerical studies
in Chap.~\ref{njl3}, e.g., for investigating the strange matter hypothesis
in \sect{sqm}. Therefore let us begin with these parameters.

In \fig{figk12.36t0}, the constituent masses and the diquark
gaps at $T=0$ are displayed
as functions of the chemical potential. When we compare
this figure with \fig{figk0t0}, the analogous figure for $K=0$,
we see that the essential features remain unchanged: There are 
three distinct phases, i.e., vacuum with spontaneously broken chiral symmetry,
2SC, and CFL, separated by first-order phase boundaries at which the
constituent quark masses change discontinuously.
Of course, as pointed out several times before, due to the flavor mixing
at $K\neq 0$ the strange quark mass stays no longer constant across the 
first phase boundary, such that $M_s < M_s^{vac}$ in the 2SC phase.
This has the consequence that the 2SC phase is less stable against transitions
to the CFL phase and thus the critical chemical potential $\mu_2$ for the 
2SC-CFL phase transition is shifted to lower values. (The critical chemical
potential $\mu_1$ corresponding to the vacuum-2SC phase transition is only
slightly reduced.) In turn, since the CFL phase starts at lower chemical
potentials, the value of $M_s$ just above $\mu_2$ is larger than for
$K=0$ and this causes a larger difference between the diquark condensates
$\Delta_2$ and $\Delta_5$ in this regime. Roughly speaking, we may say that
the CFL solutions are not very sensitive to $K$, but solutions which have 
only been metastable for $K = 0$ become absolutely stable down to lower
values of $\mu$ when $K$ is increased. 

For $K=0$ we found that the density of strange quarks is zero in the entire
2SC phase. One could have expected that this is changed by 
a flavor mixing interaction which reduces $M_s$ in the 2SC phase and in
this way the threshold for populating strange quark states. We see,
however,
that the situation is more complicated because at the same time the 
2SC-CFL phase boundary is lowered as well. Nevertheless, for parameter 
set III there is indeed a small regime (about 3~MeV below $\mu_2$) where 
the density of strange quarks is non-zero. 

\begin{figure}
\begin{center}
\epsfig{file=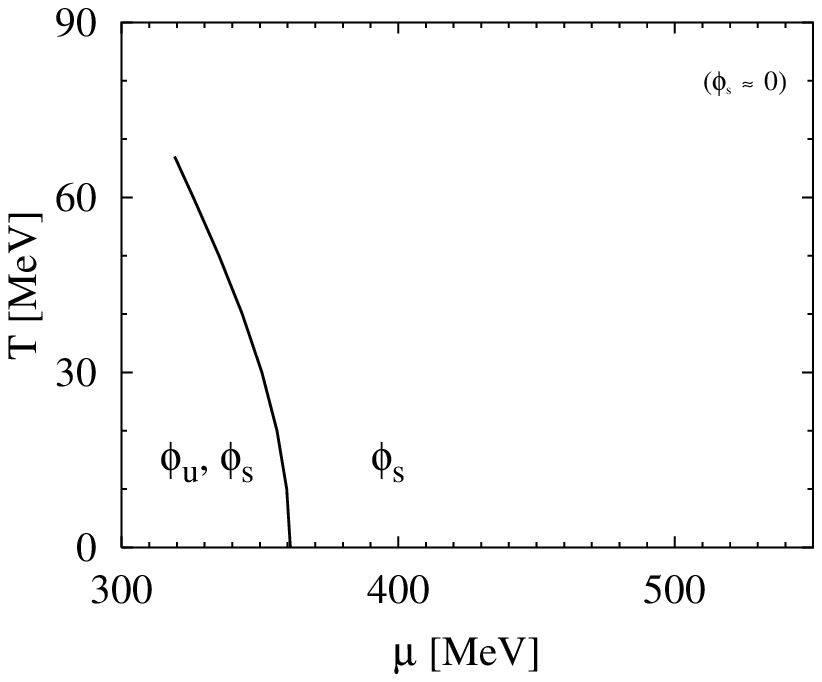,width=7.4cm}
\hfill
\epsfig{file=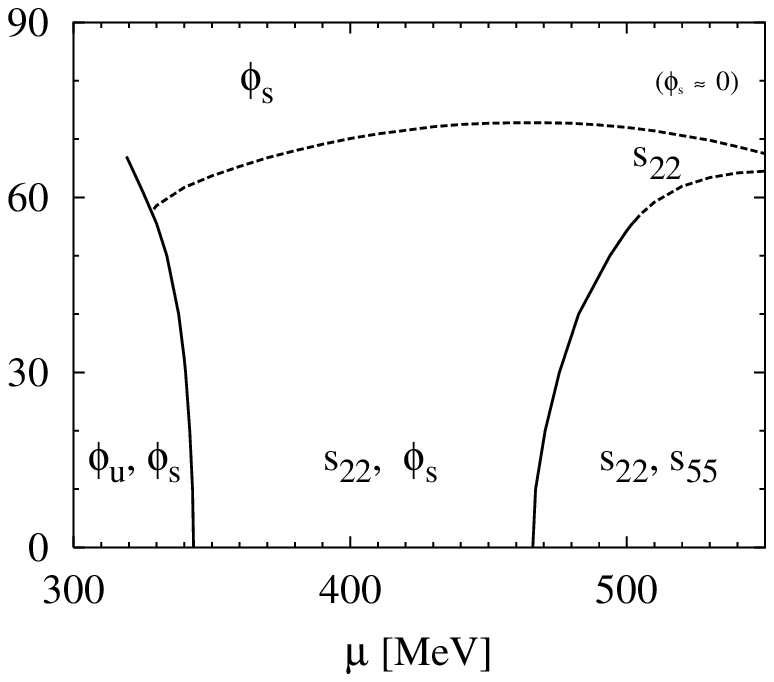,width=7.4cm}
\end{center}
\vspace{-0.5cm}
\caption{\small The same as \fig{figphasek0}, but for $K\Lambda^5 = 12.36$
(\tab{tabnjl3fitk}, parameter set III).}
\label{figphasek12.36}
\end{figure}

We now turn to non-vanishing temperature.
The $\mu-T$ phase diagram for parameter set III is displayed in the
right panel of \fig{figphasek12.36}.
In order to understand the main differences to the $K=0$ case 
(\fig{figphasek0}) it is again instructive, first to neglect the
influence of the diquark condensates. The corresponding phase diagrams
are displayed in the left panels of both figures. 
As one can see, the main effect of the flavor mixing is that the
second phase boundary, i.e., the one which for $K=0$ was related to the 
chiral phase transition of the strange quark has completely disappeared
for $K\Lambda^5 = 12.36$.
This difference has important consequences for the phase diagrams with diquark
condensates included (right panels): 
For $K=0$ (\fig{figphasek0}) the phase boundary between 2SC and CFL phase
at low temperatures has two origins: the chiral phase transition of the 
strange quark and the color-flavor locking transition. Whereas the former
tends to turn the phase boundary to the left (the strange quark condensate
$\phi_s$ is more stable at lower $\mu$) the latter tends to turn the
boundary to the right (the diquark condensates $\cf$ and $\cs$ are more
stable at higher $\mu$). As a consequence of this competition the
phase boundary goes up more or less vertically and finally splits into 
two branches.
This is quite different for $K\Lambda^5 = 12.36$ (\fig{figphasek12.36}).
Since without diquark condensates (left) there is no second phase
transition, the behavior of the phase boundary is basically dictated
by the diquark condensates. Consequently, it turns to the right and there
is no splitting into two branches.
As for $K=0$, the phase transition is first order at lower $T$ but becomes
second order above $T \simeq 56$~MeV. (We repeat that it must be first
order at low temperature because of the mismatch of the strange and the
non-strange Fermi surfaces.)  

\begin{figure}
\begin{center}
\epsfig{file=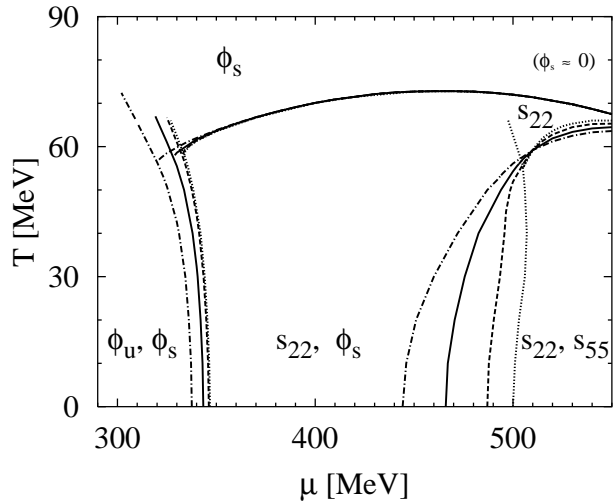,width=8cm}
\end{center}
\vspace{-0.5cm}
\caption{\small Phase diagrams in the $\mu$-T plane for parameter set I
($K=0$, dotted), II ($K\Lambda^5 = 5$, dashed), III ($K\Lambda^5 = 12.36$, 
solid), and IV ($K\Lambda^5 = 20$, dash-dotted).}
\label{figphaseallk}
\end{figure}

A more general overview about the $K$ dependence of the phase structure
is given in \fig{figphaseallk} where the phase boundaries are displayed
for the four parameter sets of \tab{tabnjl3fitk}. For the sake of
clarity, we do not distinguish between first-order and second-order phase
transitions in this diagram. Obviously, except of the 2SC-CFL boundary,
all other phase boundaries are rather insensitive to $K$, once the vacuum
masses and the value of the quark-quark coupling constant $H$ are fixed.
As discussed above, the 2SC-CFL phase boundary is shifted to lower chemical 
potentials when $K$ is increased. However, even for the relatively
large value  $K\Lambda^5 = 20$, there is still a large region where the
2SC phase is the most dominant quark phase.

\begin{figure}
\begin{center}
\epsfig{file=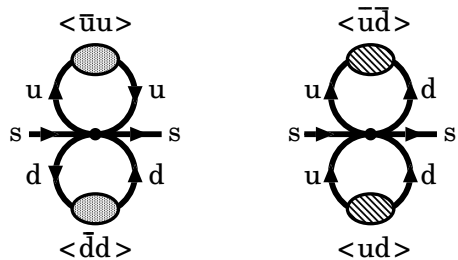,width=6cm}
\end{center}
\vspace{-0.5cm}
\caption{\small Two contributions to the constituent mass of the strange
quark. Left: self-energy proportional to $\phi_u\phi_d$. 
Right: self-energy proportional to $|\ct|^2$.          
}
\label{figsmass}
\end{figure}

In this context we should note that the above analysis of flavor 
mixing effects is not complete,   
since some terms which could arise from the instanton interaction
are missed
in our starting point, \eqs{LCFUL} to (\ref{LqqCFUL}). 
To see this, consider a six-point vertex as shown in \fig{figsixv}.
We have seen earlier that this vertex can contribute to the constituent
quark mass if we close two quark loops (\fig{fignjl3gap}).
Thus the self-energy of the strange quark (for instance) contains 
a term proportional to the non-strange condensates $\phi_u$ and $\phi_d$,
as depicted in the left diagram of \fig{figsmass}.
This term has been taken into account in our model (see \eq{MCFUL}). 
However, in the presence of diquark condensates there is also a term
proportional to $|\ct|^2$ due to the fact, that a diquark pair can be
created or destroyed in the condensate (right diagram of \fig{figsmass}). 
This contribution has not been taken into account in our calculations. 
Similarly, if one closes a single quark loop and Fierz transforms the 
resulting four-point vertex one obtains flavor and density dependent
contributions to the quark-quark interaction which are not contained in 
\eq{LqqCFUL}. 
Instanton effects of this type have partially been discussed in 
Ref.~\cite{RSSV00}, but not self-consistently including constituent
quark masses away from the chiral limit.
An extention of our model in this direction would certainly be very
interesting. 

Our calculations suggest that the 2SC phase is favored in a relatively 
large regime, even for rather large values of $K$.
However, before drawing early conclusions we should recall that our model 
is lacking a realistic description of the hadronic phase. 
Hence, it is still possible that the 2SC phase is excluded if the hadron-quark 
phase transition takes place at a relatively large chemical potential. 
Examples for this scenario will be discussed in \sect{hqphasetr}.
We should also mention that the stress imposed to the Cooper pairs by
unequal masses could result in new phases, like crystalline color
superconductors~\cite{ABR01,BR02}
or a CFL phase with condensed kaons~\cite{Sch00c,BS02,KR02}.
These phases will briefly be discussed in \sect{neutdisc}.
Before, we will introduce another source of stress which comes about
by the requirement of electric and color neutrality.
As a consequence, the chemical potentials and, thus, the Fermi momenta
of up and down quarks become unequal, thereby disfavoring the
2SC phase. This will be investigated in mored detail in the next
chapter.

\chapter{Neutral quark matter}
\label{neut}

In the previous chapter we have restricted ourselves to the thermodynamics
of quark matter depending on temperature and one common quark chemical
potential. 
However, for the description of a possible quark core of a neutron star 
we must consider neutral quark matter in beta equilibrium. 
As in our discussion of strange quark matter in \sect{sqm}, this means that
we have to deal with more than one independent chemical potential.
In the present case the situation is even more involved. In addition 
to electric neutrality, we also have to impose color neutrality, which is 
not automatically realized in color superconducting phases with color 
independent chemical potentials (see \sect{results})\footnote{This applies
at least to NJL-type models. As mentioned earlier, there are arguments that
color superconducting phases in QCD are automatically color 
neutral~\cite{DiRi04,GeRe03,Kry03}.
This would be one more reason to correct the NJL model for color neutrality.}.
In fact, strictly speaking, the matter has to be in a color singlet state.
However, as shown in Ref.~\cite{ABMW01}, the energy related
to projecting color neutral systems onto color singlets becomes
negligible in the thermodynamic limit. It is thus sufficient to 
require electric and color neutrality.

It has been pointed out by Alford and Rajagopal that these
constraints strongly disfavor the 2SC phase, such that this phase
might not be present in compact stars~\cite{AR02}.
The basic arguments can be understood in the following way.
Suppose we have a system of up, down, and strange quarks together with
leptons.
As we have seen in \sect{results}, imposing color neutrality in 2SC matter
does not cost much energy and we may concentrate on electric neutrality.
Moreover, as discussed in \sect{sqm}, the lepton fraction is very small in 
beta equilibrated matter, such that their contribution to the electric
charge can be neglected. The quark densities must then satisfy the relation
(cf. \fig{fignjl3sqm})
\beq
    2\,n_u \;-\; n_d \;-\; n_s \;\approx\; 0~.
\label{elneutapp}
\eeq
Obviously, for equal quark masses this relation is realized by equal 
densities $n_u = n_d = n_s$ and quark matter is in the CFL phase.
The arguments of Alford and Rajagopal are based on an expansion in the
strange quark mass, which they assumed to be small compared with the
chemical potential. 
In this case the densities and thus the Fermi momenta are still similar 
to each other and we may write $p_F^i = \bar p_F + \delta p_F^i$
with $\delta p_F^i \ll \bar p_F$.
Hence
\beq
    n_i \;=\; \frac{1}{\pi^2}\,{p_F^i}^3 \;\approx\; 
              \frac{1}{\pi^2}\,({\bar p_F}^3 + 3 {\bar p_F}^2\,\delta p_F^i)~.
\eeq
Inserting this into \eq{elneutapp} we find
\beq
    p_F^d \;-\; p_F^u \;\approx\;  p_F^u \;-\; p_F^s~, 
\eeq
i.e., the Fermi momenta of up and down quarks differ approximately by the
same amount as the Fermi momenta of up and strange quarks. 
This implies that $us$ pairing is as likely as $ud$ pairing. 
Indeed, within their expansion scheme Alford and Rajagopal found that, 
whenever $ud$ pairing is more favored than no pairing at all, the CFL phase 
is even more favored, and there is thus no room left for the 2SC phase. 
Assuming that the strange quark mass is smaller in the CFL phase than in
the 2SC phase, as motivated by our previous results, the argument 
becomes even stronger~\cite{AR02}.

On the other hand the assumption of a small strange  quark mass which
was the basis of the expansion in Ref.~\cite{AR02} might not be justified 
in the 2SC phase. Instead, as we have seen in \sect{CFUL}, it might be
more realistic to assume that there are no strange quarks at all. 
In this case \eq{elneutapp} implies $p_F^d \approx 2^{1/3} p_F^u$. 
For instance, for $p_F^u = 400$~MeV this means that the Fermi momenta 
of $u$ and $d$ differ by about 100~MeV and hence, according to 
\eq{stablegap}, the gap should be larger than about 70~MeV for the
2SC phase to be stable. From this point of view a stable neutral 2SC
phase seems not to be excluded. 

In this situation it is obviously worthwhile to extend the NJL-model
analysis to include unequal chemical potentials. This has been done first 
by Steiner, Reddy, and Prakash~\cite{SRP02}, and shortly afterwards by 
Neumann, Buballa, and Oertel~\cite{NBO03}, focusing on somewhat different 
issues. The crucial point is again, that NJL model calculations allow to
study the effects of density and phase dependent quark 
masses, which have not been included self-consistently in the estimates 
of Ref.~\cite{AR02}.
In this chapter we discuss the results of these investigations.

\section{Formalism}
\label{formalism}

\subsection{Conserved charges and chemical potentials}
\label{charges}

We consider a system of quarks and leptons described by the 
thermodynamic potential
\beq
    \Omega(T,\{\mu_{f,c}\},\{\mu_{\ell_i}\}) \;=\;
    \Omega_q(T,\{\mu_{f,c}\}) \;+\;
    \Omega_\ell(T,\{\mu_{\ell_i}\})~,
\eeq
where $\mu_{f,c}$ is the chemical potential of a quark with flavor $f$
and color $c$, while $\mu_{\ell_i}$ refers to a lepton of type 
$\ell_i \in \{e, \mu, \dots\}$.
For $\Omega_\ell$ we simply take a gas of non-interacting leptons,
while the quark part is derived from the NJL model and will be discussed
in more detail in \sect{neutnjl}.

The various particle densities can be derived from $\Omega$ in the 
standard way,
\beq
    n_{f,c} \;=\; -\frac{\partial \Omega}{\partial\mu_{f,c}}~,\qquad 
    n_{\ell_i} \;=\; -\frac{\partial \Omega}{\partial\mu_{\ell_i}}~. 
\eeq
The total flavor and color densities are then given by
\beq
n_f = \sum_c n_{f,c}~,\qquad n_c = \sum_f n_{f,c}~.
\eeq
We are mainly interested in describing
the conditions present in compact stars older than a few minutes, 
when neutrinos can freely leave the system. 
In this case lepton number is not conserved and we have four independent
conserved charges, namely the total electric charge $n_Q$ 
and the three color charges.
Neglecting the $\tau$-lepton, which is too heavy to play a role in neutron
stars, the total electric charge is given by
\beq
n_Q =   \frac{2}{3} n_u - \frac{1}{3} n_d  -\frac{1}{3}n_s - n_e -n_\mu~. 
\eeq
For the color charges, instead of $n_r$, $n_g$, and $n_b$, we will often
use the linear combinations
\beq
n = n_r + n_g + n_b~,\qquad 
n_3 = n_r - n_g~,\qquad 
n_8 = \frac{1}{\sqrt{3}}(n_r + n_g - 2 n_b)~.
\eeq
Here $n$ corresponds to the total quark number density. Since $n=3\rho_B$, 
it is also related to the conserved baryon number. 
$n_3$ and $n_8$ describe color asymmetries. 
Note that in contrast to \eq{n8} we do not assume $n_r=n_g$.

The four conserved charges $\{n_i\}= \{n,n_3,n_8,n_Q\}$
are related to four independent chemical potentials
$\{\mu_i\}= \{\mu,\mu_3,\mu_8,\mu_Q\}$, such that
\beq
n_i = - \frac{\partial\Omega}{\partial\mu_i}~.
\label{densities}
\eeq
The individual quark chemical potentials $\mu_{f,c}$ are then given by
\beq
    \mu_{f,c} = \mu \;+\; \mu_Q\; 
\Big(\frac{1}{2} (\tau_3)_{ff} + \frac{1}{2 \sqrt{3}} (\tau_8)_{ff}\Big) 
\;+\; \mu_3\; (\lambda_3)_{cc} \;+\; \mu_8\; (\lambda_8)_{cc}~.
\label{mus}
\eeq
Here, as before, $\tau_i$ and $\lambda_j$ are Gell-Mann matrices corresponding
to flavor and color, respectively.
The electron and muon chemical potentials are simply 
\beq
   \mu_e =\mu_\mu = -\mu_Q~.
\label{muemu}
\eeq 
\eqs{mus} and (\ref{muemu}) imply
\beq
    \mu_{d,c} = \mu_{s,c} = \mu_{u,c} + \mu_e \quad \text{for all \it{c}}~,
\label{beta}
\eeq
which is usually referred to as beta equilibrium.

We are mostly interested in electrically and 
color neutral matter, which is characterized by
\beq
n_Q = n_3 = n_8 = 0~.
\label{ecneutral}
\eeq
Since we have four conserved charges and three neutrality conditions
the neutral solutions can be characterized by one
independent variable, namely the quark number density $n$.  
In the four-dimensional space spanned by the chemical potentials 
$\{\mu_i\}$ these solutions form one or several one-dimensional lines. 
This is a straight forward generalization of the situation in \sect{sqm}.
Since for normal conducting quark matter $n_3$ and $n_8$ automatically
vanish for $\mu_3=\mu_8=0$, these chemical potentials did not play any
role in that context and effectively we only had to deal with
two chemical potentials and one neutrality condition. 
(Similarly, many cases to be discussed below are restricted to
$\mu_3=0$.)

On the other hand, the above situation can be generalized further.
For instance, in a proto-neutron star a few seconds after the collapse
of the progenitor, neutrinos are trapped, and we get an additional chemical 
potential related to the conserved lepton number.
At the same time, however, we get another constraint from the fact that
the lepton fraction, i.e., the total lepton number divided by the total
baryon number, is fixed to the value present in the progenitor star. 
Although we will mostly refer to the above case of four conserved charges,
the formalism we develop in this part is straight forwardly generalized to 
other cases.

\subsection{Thermodynamic potential for non-uniform quark chemical potentials}
\label{neutnjl}

In order to attack the problems discussed above we need to extend the
NJL model of \sect{cfulformalism} to non-uniform chemical potentials
for different flavors and colors. This amounts to replacing 
the chemical potential $\mu$ in the inverse fermion propagator
\eq{SinvCFUL} by a diagonal matrix $\hat\mu$ with flavor and color
dependent components $\mu_{f,c}$. 
Although not all of these components are independent 
if we impose beta equilibrium (see \eq{mus})
we keep them as arbitrary inputs at this point.
The essential difference to the situation in \sect{cfulformalism} is the 
fact that unequal chemical potentials for up and down quarks
violate isospin symmetry. Hence \eq{cfuliso} does no longer hold,
and we have to deal with six different condensates, i.e., the three 
constituent masses $M_u$, $M_d$ and $M_s$, and the three diquark condensates
$\ct$, $\cf$ and $\cs$.
Because of this lower degree of symmetry the explicit evaluation of the 
integrand in \eq{OmegaCFUL} becomes of course much more involved.

As we have discussed in \sect{cflprop}, even though quarks of all colors
and flavors participate in a condensate in the CFL phase, not every
quark species is paired with all others, but there are certain combinations
(see \tab{tabcflcond}). In particular, six of the nine color-flavor
species have only one fixed partner species, while the remaining three
form a triangle. 
As a consequence,  $S^{-1}$ can be decomposed into several independent 
blocks~\cite{ABR99,AR02,SRP02},
and ${\rm Tr}\;\ln\, S^{-1}(p)$ can be written as a sum, 
\begin{alignat}{1}
    {\rm Tr}\;\ln\, S^{-1}(p) \;=\quad
      ({\rm Tr}\;\ln\,{\cal M}_{u g, d r} \,&+\, 
       {\rm Tr}\;\ln\,{\cal M}_{d r, u g}) \;+\; 
      ({\rm Tr}\;\ln\,{\cal M}_{u b, s r} \,+\, 
       {\rm Tr}\;\ln\,{\cal M}_{s r, u b})
\nonumber \\ 
 +\; ({\rm Tr}\;\ln\,{\cal M}_{d b, s g}\,&+\, 
       {\rm Tr}\;\ln\,{\cal M}_{s g, d b}) \quad+\; 
      {\rm Tr}\;\ln\,{\cal M}_{ur, dg, sb}~,
\label{blocks}
\end{alignat}
where the matrices ${\cal M}$ correspond to the different independent
blocks of $S^{-1}$.
Six of them have a $2\times 2$ structure in the 18-dimensional space
spanned by color, flavor and Nambu-Gorkov degrees of freedom,
\beq
{\cal M}_{f_1 c_1, f_2 c_2} = 
\left(\begin{array}{cc} \psl^+_{f_1,c_1} -M_{f_1} &
\Delta_{f_1,f_2} \,\gamma_5 \\ 
- \Delta_{f_1,f_2}^* \,\gamma_5 & \psl^-_{f_2,c_2} -
M_{f_2}\end{array}\right)~,
\label{block2}
\eeq
where $\psl^\pm_{f,c} = \psl \pm \mu_{f,c} \gamma_0$.
These blocks describe the pairing of two species of quarks with flavors
$f_1$ and $f_2$ and colors $c_1$ and $c_2$, respectively.
$\Delta_{f_1,f_2}$ is the corresponding diquark gap, i.e.,
$\Delta_{ud} \equiv \Delta_{du} \equiv \Delta_2$, 
$\Delta_{us} \equiv \Delta_{su} \equiv \Delta_5$, and 
$\Delta_{ds} \equiv \Delta_{sd} \equiv \Delta_7$.
 
The remaining block involves three quark species
and is a $6\times 6$ matrix in color--flavor--Nambu-Gorkov space,
\beq
{\cal M}_{ur,dg,sb} = \left(
\begin{array}{cccccc}\psl^+_{u,r}-M_u&0&0&0&-\Delta_2\gamma_5
&-\Delta_5\gamma_5 \\ 
0&\psl^+_{d,g}-M_d&0&-\Delta_2\gamma_5&0&
-\Delta_7\gamma_5 \\ 
0&0&\psl^+_{s,b}-M_s&-\Delta_5\gamma_5&
-\Delta_7\gamma_5&0 \\
0&\Delta_2^*\gamma_5&\Delta_5^*\gamma_5&\psl^-_{u,r}-M_u&0&0\\ 
\Delta_2^*\gamma_5&0&\Delta_7^*\gamma_5&0&\psl^-_{d,g}-M_d&0\\ 
\Delta_5^*\gamma_5&\Delta_7^*\gamma_5&0&0&0&\psl^-_{s,b}-M_s
\end{array} \right)~.
\label{block6}
\eeq
(Including the Dirac components, this is of course a $24\times 24$ matrix,
while \eq{block2} describes $8\times 8$ matrices.)

This block structure has interesting consequences:
It is known from ordinary superconductors, where electrons with spin up
are paired with electrons with spin down, that the respective number densities,
$n_\uparrow$ and $n_\downarrow$, are always equal to each other. This is
true even in the presence of a magnetic field, as long as the superconducting
state remains intact~\cite{equalninsupercond}. 
Some time ago it has been shown that the analogous statement holds for
color superconductors, if one considers two species with unequal chemical 
potentials~\cite{Bedaque02} or masses~\cite{RaWi01}\footnote{In 
Ref.~\cite{Bedaque02} it was shown for a system of two massless flavors
that $n_u = n_d$ for $|\mu_u-\mu_d| < 2\Delta$, but the densities become
unequal for $|\mu_u-\mu_d| > 2\Delta$. Without further constraints, the
latter belongs to the regime where the diquark condensate is unstable,
confirming the statement that the densities in the stable regime are
equal. However,
under certain conditions there could be stable color superconducting 
solutions with $|\mu_u-\mu_d| > 2\Delta$ if their decay is prohibited 
by the requirement of local charge neutrality~\cite{ShHu03,HuSh03}.
This situation is never realized in the numerical examples to be discussed 
below. In principle, however, this possibility should be kept in mind
and will briefly be discussed in \sect{neutdisc}.}.
From this it was originally concluded that in the CFL phase
where all quarks participate in a condensate, the densities $n_{f,c}$ are
equal for all flavors and colors. This 
would mean that the CFL phase is always electrically and color neutral,
even for unequal quark masses~\cite{RaWi01}.
However, as pointed out by Steiner, Reddy, and Prakash~\cite{SRP02}, 
only those quarks which are paired in the same  $2\times 2$ block have the 
same density, whereas the densities could differ for different blocks.
Furthermore the argument does not apply to the $6\times 6$ block.
According to the color-flavor structure discussed above this means 
for the CFL phase
\beq
n_{u,g} = n_{d,r}~, \qquad n_{u,b} = n_{s,r}~, \qquad n_{d,b} = n_{s,g}
\qquad \text{(CFL)}~,
\label{denscfl1}
\eeq
leading to the remarkable identities
\beq
n_{u} = n_{r}~, \qquad n_{d} = n_{g}~,\qquad  
n_{s} = n_{b}\qquad \text{(CFL)}~.
\label{denscfl}
\eeq
These relations guarantee neutrality of CFL matter under the rotated
electromagnetism $\tilde Q$, \eq{qtilde3},
but in general they do
not preclude the presence of ordinary electric or color charges~\cite{SRP02}.
Note, however, that color neutral CFL matter is automatically electrically 
neutral as long as no leptons are present.

In the 2SC phase, where $\Delta_5 = \Delta_7 = 0$, \eq{block6} can be
decomposed further, and we obtain a new $2\times 2$ block involving
red $u$-quarks and green $d$-quarks. Together with the other  
$2\times 2$ block which contains $\Delta_2$ this leads to the relations
\beq 
n_{u,r} = n_{d,g}~,\qquad n_{u,g} = n_{d,r} \qquad \text{(2SC)}~.
\label{dens2sc}
\eeq 
The corresponding relations for other possible phases, e.g., 
with two non-vanishing
diquark condensates, can be obtained analogously.

The further elaboration of the thermodynamic potential contains only
straight forward manipulations, but the result is extremely lengthy
and will not be presented here.
The self-consistent solutions for the condensates 
$\phi_f$ and $s_{AA}$, i.e., the stationary points of 
$\Omega_q$ are determined numerically~\cite{fredo}.

\section{Numerical results}
\label{neutresults}

In this section we first explore the phase structure 
at $T=0$ in the space of the different chemical potentials $\mu_i$
and determine the electric and color charge densities in the various 
regimes~\cite{NBO03}. 
Based on these results, we then construct solutions of homogeneous 
electrically and color neutral quark matter and analyze the corresponding
equation of state~\cite{SRP02}.

\subsection{Equal chemical potentials}
\label{neutresult0}

To have a well-defined starting point we begin with the 
``standard case'' of a uniform, color and flavor independent,
chemical potential for all quarks, i.e., 
$\mu_3 = \mu_8 = \mu_Q = 0$. This implies that no leptons are present.
As model parameters we adopt again parameter set RKH of 
\tab{tabnjl3fit}~\cite{Rehberg},
for the bare quark masses, the cut-off, and the coupling constants 
$G$ and $K$. For the diquark coupling we take $G = H$.   
This is the value we have chosen in Ref.~\cite{NBO03}, following the choice
of parameters in the first version of Ref.~\cite{SRP02}\footnote{In their
final version these authors have taken $H:G = 3:4$.}.
Obviously this is very similar to parameter set III in \tab{tabnjl3fitk},
where we had $H = 0.95~G$. 

\begin{figure}
\begin{center}
\epsfig{file=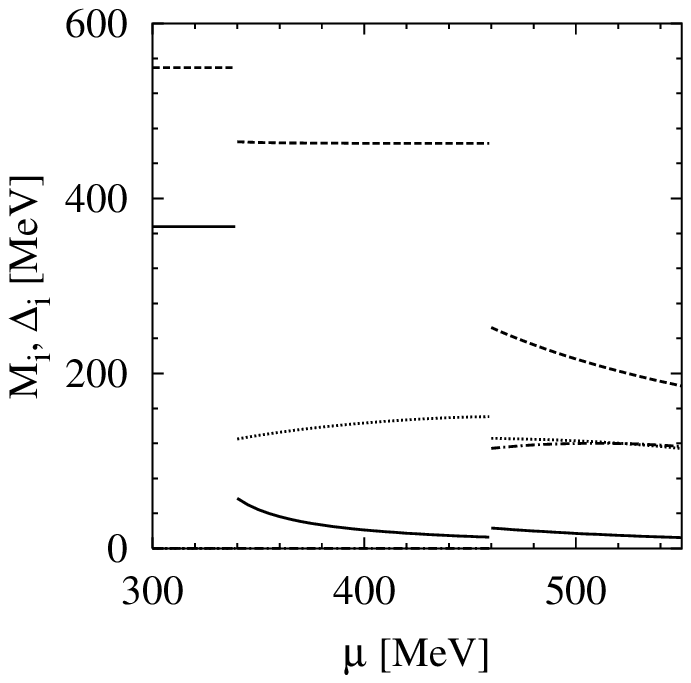,width=7.cm}
\epsfig{file=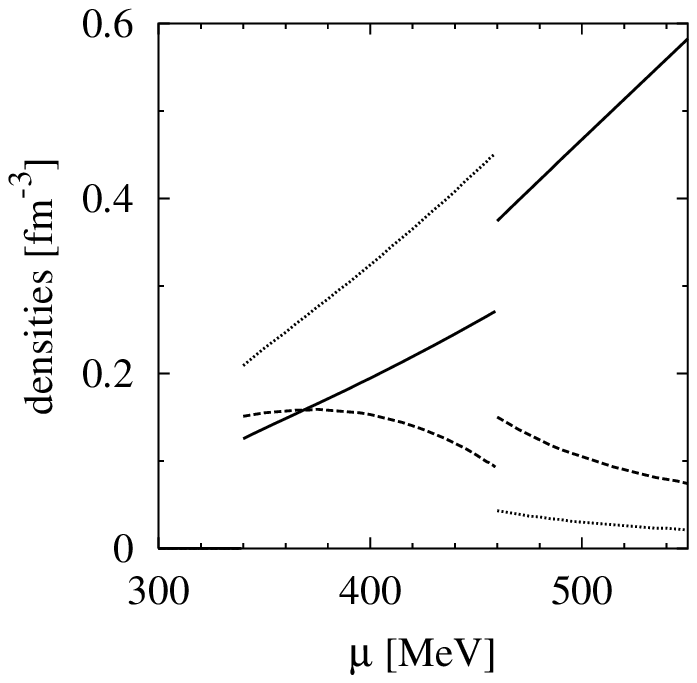,width=7.cm}
\end{center}
\vspace{-0.5cm}
\caption{\small Gap parameters and densities at $T =\mu_3 = \mu_8 = \mu_Q = 0$
as functions of $\mu$. Left: $M_u=M_d$ (solid), $M_s$ (dashed), $\Delta_2$
(dotted), and $\Delta_5=\Delta_7$ (dash-dotted). Right: $n/10$ (solid),
$n_8$ (dashed), and $n_Q$ (dotted). Adapted from Ref.~\cite{NBO03}.
}
\label{figneut0}
\end{figure}

With these parameters we obtain the results which are displayed in 
\fig{figneut0}. In the left panel we show once again the
constituent masses and diquark gaps as functions of $\mu$.
Since isospin symmetry is still preserved, we still have $M_u = M_d$
and $\Delta_5 = \Delta_7$, as in \sect{CFUL}.
Since the parameters are almost the same, the results  
are of course very similar to those presented in \fig{figk12.36t0}:
We find three phases:
a normal conducting phase at low chemical potentials, followed by a
2SC phase and, finally, a CFL phase. 
At the first-order phase boundaries we observe again strong discontinuities 
in the quark masses.

In the right panel we show the corresponding densities. Note that the
quark number density $n$ (solid line) has been divided by 10 to fit to the
scale. 
The dotted line corresponds to the electric charge density $n_Q$, the 
dashed line to the color density $n_8$. The color density $n_3$ is identically
zero.
We find again that all densities vanish in the ``normal phase'', i.e., this
phase corresponds to the vacuum. As discussed earlier
it has to be like this: As soon as up and down quarks are present, 
their Fermi surfaces are subject to a Cooper instability leading to the 
formation of the diquark condensate $\ct$. This argument will no longer go 
through, once we have switched on one of the other chemical potentials which 
lift the degeneracy of the Fermi surfaces of all up and down quarks.

The two other phases carry both, electric and color charges. The electric
charge of the 2SC phase is easily understood. Since $\mu_Q = 0$, there are
no leptons and the densities of up and down quarks are equal. Moreover,
in this example
there are no strange quarks, which are too heavy to be populated in this
regime. Hence the total electric charge density is given by $n_Q = n/6$. 
The non-vanishing color density $n_8$ is the same effect we have already
encountered in \sect{results} and reflects the fact that for equal
chemical potentials the densities of the paired (red and green) quarks are 
larger than the density of the unpaired (blue) quarks.
Numerically, we find $(n_r-n_b)/n = 10\%$ at the lower boundary and
$(n_r-n_b)/n = 3\%$ at the upper boundary of the 2SC phase.

Just above the transition to the CFL phase this ratio does not change
very much, whereas the electric charge density drops significantly due to a
strong increase of the density of strange quarks.
To a large extent, this is caused by a sudden drop of the strange quark mass,
but this is only part of the story. 
For instance, at $\mu = 500$~MeV we have $M_u = M_d = 17.2$~MeV and 
$M_s = 216.5$~MeV.
Using these numbers in a free gas approximation we would expect
$n_Q = 0.049$~fm$^{-3}$, whereas numerically we find $n_Q = 0.030$~fm$^{-3}$.
This difference is caused by the diquark pairing, which links the 
flavor densities in the CFL phase directly to the color densities, as 
discussed in \eq{denscfl}. For $n_3 = 0$ one finds 
$n_Q = 1/(2\sqrt{3})\,n_8$, in agreement with our numerical results.

\subsection{Phase structure}
\label{neutresultsphases}

Aiming the construction of electrically and color neutral quark matter
we have in general to introduce non-vanishing chemical potentials $\mu_8$ 
and $\mu_Q$ to remove the charge densities $n_8$ and $n_Q$ we have found
above.
In order to see how these additional chemical potentials influence
the phase boundaries, we first study them separately. 

\begin{figure}
\begin{center}
\epsfig{file=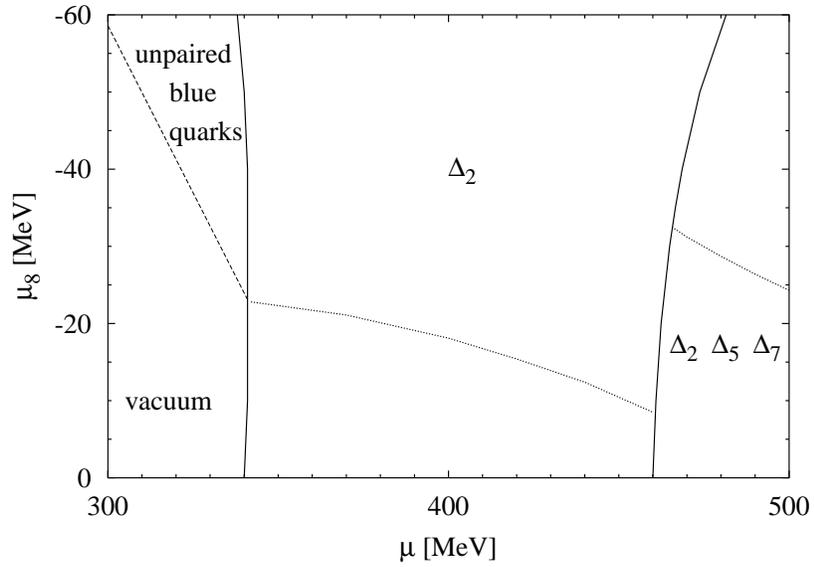,width=11.cm}
\end{center}
\vspace{-0.5cm}
\caption{\small Phase diagram in the $\mu-\mu_8$ plane for 
$T = \mu_3 = \mu_Q = 0$. 
The various phases separated by the solid lines are  characterized by 
different non-vanishing diquark gaps $\Delta_i$ as indicated in the figure.
In the non-superconducting phase quarks are present only above the dashed 
line. The dotted lines are the lines of color neutral matter. In the 
CFL phase this also corresponds to electrically neutral matter.
Taken from Ref.~\cite{NBO03}.
}
\label{figneutphasemu8}
\end{figure}

In \fig{figneutphasemu8} we show the phase diagram in 
the $\mu-\mu_8$-plane for $\mu_Q = \mu_3 = 0$. The (first-order) phase
boundaries are indicated by solid lines. We find again the three
phases discussed before, i.e., the normal phase, the 2SC phase, and the CFL 
phase. For $\mu_8 = 0$ we have seen that the ``normal phase'' actually 
corresponds to the vacuum. However, when $\mu_b = \mu - 2/\sqrt{3} \mu_8$
becomes larger than the vacuum masses of the light quarks (the region above 
the dashed line), blue up and down quark states can be populated forming 
a gas of unpaired blue quarks. Here we have neglected that in principle
these quarks could pair in a different channel as discussed in \sect{aniso}.
Moreover, we should repeatedly note that our model is not suited for a
realistic description of the low-density regime, where confinement and 
hadronic degrees of freedom have to be taken into account.

In the color superconducting phases we have indicated the lines of 
color neutral matter (dotted). In the CFL phase, as discussed below 
\eq{denscfl}, color neutral quark matter is automatically electrically 
neutral as well, i.e., in the CFL phase the dotted line already corresponds 
to a neutral matter solution, we are looking for.
It meets the phase boundary to the 2SC phase at
$\mu = 465.7$~MeV and $\mu_8 = -32.5$~MeV. The 2SC matter which is in 
chemical and mechanical equilibrium with the neutral CFL matter at this
point carries both, electric and color charge, $n_Q = 0.464$~fm$^{-3}$ and
$n_8 = -0.329$~fm$^{-3}$. In \sect{neutmix}, this point 
will be our starting point to construct neutral mixed phases.
Unlike color neutral CFL matter, color neutral 2SC matter is not
electrically neutral but positively charged. In fact, a non-vanishing 
$\mu_8$ does not change the ratio of up and down quarks and hence,
as long as no strange quarks are present, $n_Q/n = 1/6$, as before.

\begin{figure}
\begin{center}
\epsfig{file=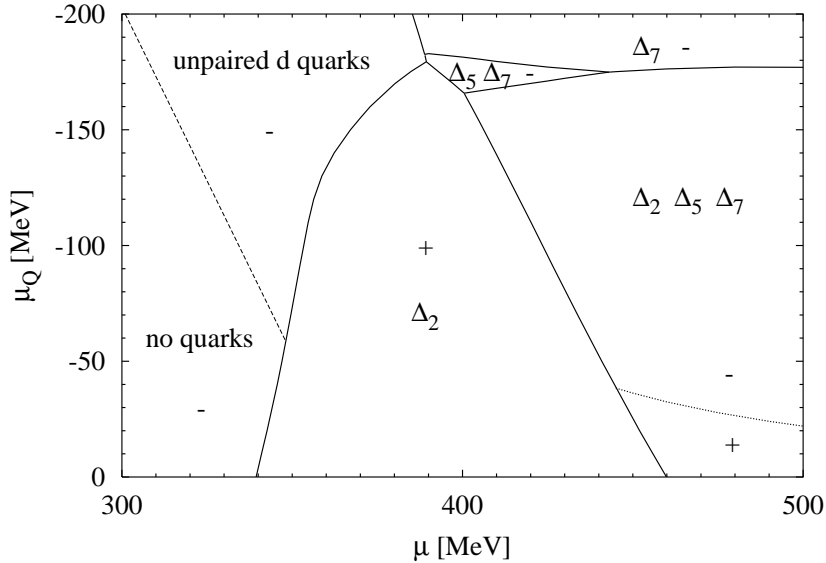,width=11.cm}
\end{center}
\vspace{-0.5cm}
\caption{\small Phase diagram in the $\mu-\mu_Q$ plane for 
$T = \mu_3 = \mu_8 = 0$. 
The various phases separated by the solid lines are  characterized by 
different non-vanishing diquark gaps $\Delta_i$ as indicated in the figure.
In the non-superconducting phase quarks are present only above the dashed 
line.
The ``$+$'' and ``$-$'' signs indicate the sign of the electric charge
density in the corresponding region. 
The dotted line corresponds to electrically (but not color)
neutral matter in the CFL phase. Taken from Ref.~\cite{NBO03}.
}
\label{figneutphasemuq}
\end{figure}

In \fig{figneutphasemuq} we show the phase diagram in 
the $\mu-\mu_Q$-plane for $\mu_8 = \mu_3 = 0$. 
Since we are interested in neutralizing the electrically positive
2SC phase, we choose $\mu_Q$ to be negative.
As long as this is not too large, we find again the normal phase at lower
values of $\mu$, the 2SC phase in the intermediate region and the CFL phase
for large $\mu$. This changes dramatically around $\mu_Q \simeq -180$~MeV
where both, the 2SC phase and the CFL phase disappear and a new phase 
emerges. This phase is analogous to the 2SC phase but with $ds$ pairing, 
instead of $ud$ pairing (``2SC$_{ds}$''). In a small intermediate regime
there is yet another phase which contains $us$ and $ds$ but no $ud$
pairs (``SC$_{us+ds}$''). 

Qualitatively, the existence of these phases is quite plausible: 
At low values of $|\mu_Q|$, the Fermi momenta of the up and down quarks 
are relatively similar to each other, whereas the strange quarks are suppressed
because of their larger mass. With increasing negative $\mu_Q$, however,
the up quarks become more and more disfavored and eventually the 
Fermi momenta are ordered as $p_F^u < p_F^s < p_F^d$. It is then 
easy to imagine that only $ds$ pairing or -- in some intermediate regime --  
only $us$ and $ds$ pairing is possible. 

\begin{figure}
\begin{center}
\epsfig{file=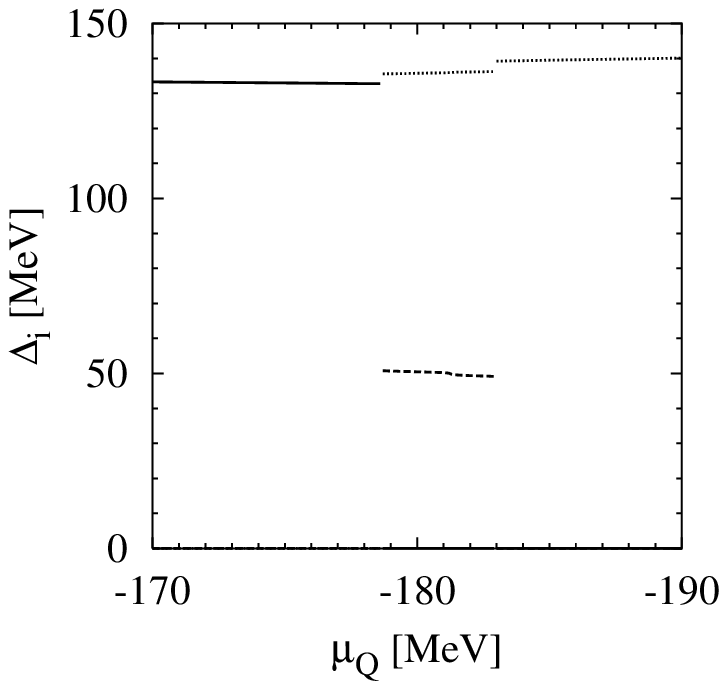,width=7.cm}
\epsfig{file=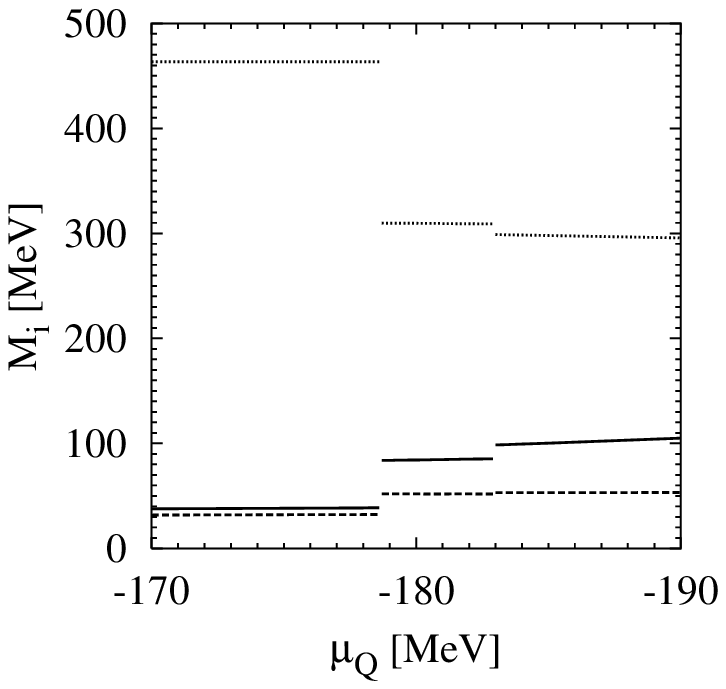,width=7.cm}
\end{center}
\vspace{-0.5cm}
\caption{\small Diquark gaps and quark masses for
$T =\mu_3 = \mu_8 = 0$, and $\mu = 390$~MeV as functions of $\mu_Q$. 
Left: $\Delta_2$ (solid), $\Delta_5$ (dashed), and $\Delta_7$ (dotted).
Right: $M_u$ (solid), $M_d$ (dashed), and $M_s$ (dotted).
Adapted from Ref.~\cite{NBO03}.
}
\label{figgapsq}
\end{figure}

Following this argument, one might expect that there is always a value
of $\mu_Q$, where the Fermi momenta of up and strange quarks are equal 
and hence the 2SC phase should either be followed by the CFL phase or by
a phase with $us$-pairing only. However, this is not the case
because of the discontinuous behavior of the quark masses.
This is illustrated in \fig{figgapsq} where the
diquark gaps and constituent quark masses are shown as functions of
$\mu_Q$ for fixed $\mu = 390$~MeV and  $\mu_3 = \mu_8 = 0$.
The 2SC--SC$_{us+ds}$ phase transition takes place at $\mu_Q = -178.6$~MeV,
corresponding to $\mu_u = \mu + 2/3 \mu_Q \simeq 270$~MeV and
$\mu_d = \mu_s = \mu - 1/3 \mu_Q \simeq 450$~MeV. 
Below the transition point the strange quark mass is larger than 
460~MeV and, consequently, no strange quarks are present.
At the transition point the strange quark mass drops to 310~MeV and 
the nominal Fermi momentum $p_F^s = \sqrt{\mu_s^2-M_s^2}$
is immediately larger than $p_F^u$.

It turns out that the stability of the various condensates is rather well 
described by the criterion given in \eq{stablegap}. 
In the 2SC phase, just below the phase boundary, we have
$\Delta_2 = 132.8$~MeV, slightly larger than 
$\Delta_2^\mi{crit} := |p_F^d - p_F^u|/\sqrt{2} =127.4$~MeV. 
At the phase boundary the latter rises to 133.6~MeV due to a sudden
increase of the up quark mass by more than 40~MeV. Taking the earlier
value of $\Delta_2$, the above criterion is no longer fulfilled, which
is consistent with our numerical result that the $ud$-pairs break up.
This level of agreement is certainly better than one should
expect (see footnote~\ref{foot} on page \pageref{foot}).
In fact, in the SC$_{us+ds}$ phase we find $\Delta_5$ continuously
decreasing from 50.8~MeV to 49.1~MeV whereas 
$\Delta_5^\mi{crit} = |p_F^s - p_F^u|/\sqrt{2}$ 
increases from 48.2~MeV to 52.6~MeV, slightly violating \eq{stablegap}. 
Nevertheless, qualitatively, one can understand the break-up of the $us$ 
pairs, which occurs at $\mu_Q = -183.0$~MeV, from the fact that at this point 
$\Delta_5^\mi{crit}$ jumps to 62.6~MeV due to a further increase of 
$M_u$ and a further decrease of $M_s$. Moreover, the fact that we always find
$\Delta_5 \approx |p_F^s - p_F^u|/\sqrt{2}$, at least in this example, 
indicates that the SC$_{us+ds}$ phase is rather fragile and might disappear 
upon small variations of the model parameters.

In the phase diagram, \fig{figneutphasemuq}, we also indicate the
sign of the electric charge density for the various regions,
and the line of electrically neutral matter in the CFL phase (dotted line). 
Note that there is no other electrically neutral regime in this diagram
(apart from the vacuum at small $\mu$ and $\mu_Q = 0$). In the normal phase, 
there are again no quarks below the dashed line, corresponding to the line
$\mu - 1/3 \mu_Q = M_d$. This region is nevertheless negatively charged
due to the leptons which are present for any $\mu_Q < 0$. Above the dashed 
line there are also down quarks rendering the matter even more negative.
(In the right corner of this phase there is also a tiny fraction
of up quarks.)

\begin{figure}
\begin{center}
\epsfig{file=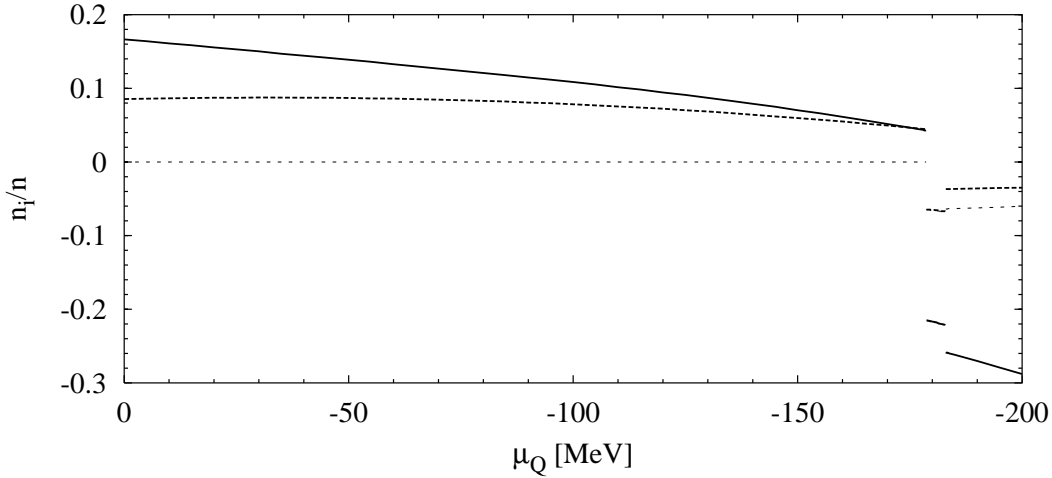,width=14.cm}
\end{center}
\vspace{-0.5cm}
\caption{\small Relative densities for
$T =\mu_3 = \mu_8 = 0$, and $\mu = 390$~MeV as functions of $\mu_Q$: 
$n_Q/n$ (solid), $n_8/n$ (dashed), and $n_3/n$ (dotted). 
Adapted from Ref.~\cite{NBO03}.
}
\label{figneutdensq}
\end{figure}

The ``new'' phases, 2SC$_{ds}$ and SC$_{us+ds}$, are negatively charged
as well.
On the contrary, the entire 2SC phase is positively charged, even at the
largest values of $|\mu_Q|$. This is illustrated in \fig{figneutdensq}
where the various charge densities $n_i$ divided by the total quark
number density $n$ are plotted as functions of $\mu_Q$, again for 
fixed $\mu = 390$~MeV and $\mu_3 = \mu_8 = 0$. As expected, $n_Q/n$ 
(solid line) decreases with increasing negative $\mu_Q$. However, in
the 2SC phase ($0 \ge \mu_Q > -178.6$~MeV) it stays positive and
before the point of neutrality is reached the phase transition to the 
SC$_{us+ds}$ phase takes place. 

\begin{table}[b]
\begin{center}
\begin{tabular}{|c|c|c|c|c|c|c|c|c|}
\hline
phase        &  N  & 2SC & 2SC$_{us}$ & 2SC$_{ds}$  & SC$_{ud+us}$
             & SC$_{ud+ds}$ & SC$_{us+ds}$ & CFL  \\ \hline
diquark gaps & --- & $\Delta_2$ & $\Delta_5$ & $\Delta_7$ 
             & $\Delta_2$, $\Delta_5$ & $\Delta_2$, $\Delta_7$ 
             & $\Delta_5$, $\Delta_7$ 
             & $\Delta_2$, $\Delta_5$, $\Delta_7$
\\ \hline
\end{tabular}
\end{center}
\caption{\small Phases and corresponding non-vanishing diquark gaps. 
}
\label{tablephases}
\end{table}  
The difficulty to obtain electrically neutral 2SC matter can be traced
back to the fact that, according to \eq{dens2sc}, the sum of red and
green $u$ quarks is equal to the sum of red and green $d$ quarks. As
long as no strange quarks are present, the related positive net charge
can only be compensated by the blue quarks and the leptons. This
requires a very large negative $\mu_Q$. However, before this
point is reached it becomes more favored to form a different phase
with a relatively large fraction of strange quarks which then also
participate in a diquark condensate\footnote{This is very similar to
the arguments of Alford and Rajagopal~\cite{AR02} discussed in the 
introductory part to the present chapter. The main difference
is that we do not compare different {\it neutral} phases with each
other, but phases in chemical equilibrium.}.  Again, the
self-consistent treatment, which leads to a sudden drop of the strange
quark mass and hence to a sudden increase of the strange Fermi
momentum, is crucial in this context.

So far we have not considered the effect of a non-vanishing
chemical potential $\mu_8$ on top of a non-vanishing $\mu_Q$.
Since the blue quarks are the main carriers of negative 
electric charge in the 2SC phase, one could hope that increasing the 
number of blue quarks, as necessary for color neutrality, 
could also help to electrically neutralize 2SC matter.  
It turns out, however, that the rather small values of $\mu_8$ which are
needed for color neutrality (see \fig{figneutphasemu8})
do not change the above results qualitatively.

At this point we should note that the non-existence of {\it stable}
neutral 2SC solutions in the present grand canonical treatment 
does not mean that neutral 2SC matter does not exist at all.
At large negative values of $\mu_Q$  there are neutral 2SC solutions
which are metastable. This means, there are other solutions, e.g., in the
$2SC_{ds}$ phase, which have a larger pressure for the same chemical 
potentials. However, these other solutions are not neutral but in general
colored and negatively charged. Therefore a finite piece of neutral
2SC matter cannot simply decay into a different phase. Instead, there are
basically two possibilities. The first one is a phase separation,
leading to a globally neutral mixed phase of two 
or more charged components in chemical and mechanical equilibrium.
This scenario will be discussed in more detail in \sect{neutmix}.
On the other hand, if the corresponding Coulomb and surface energies
are too large, this phase separation is not favored. In this case 
the ``metastable'' solutions could be stable, if there is no other
homogeneous neutral solution with a higher pressure at the same
value of $\mu$. These solutions, which have been constructed first
in Ref.~\cite{SRP02} will be discussed in \sect{neutresultshom}.

To conclude this section, we consider an example where $\mu_3\neq 0$.
At first sight, there seems to be no motivation for this.
In fact, there is no need to vary $\mu_3$, as long as we are only interested 
in finding electrically and color neutral solutions of homogeneous normal, 
2SC, or CFL matter. 
For normal and 2SC matter, this follows from the fact that both phases
have an unbroken $SU(2)_c$ symmetry, and thus $n_3=0$ for $\mu_3 = 0$.
The situation is more complicated for the CFL phase, but we have seen
already that CFL matter can be neutralized by applying a non-vanishing
chemical potential $\mu_8$ and $\mu_3=\mu_Q=0$.
Nevertheless, as we will see in \sect{neutmixresults}, the construction
of neutral mixed phases requires also non-vanishing 
values of $\mu_3$. In this context we will encounter another phase,
which is not present in Figs.~\ref{figneutphasemu8} and \ref{figneutphasemuq}.
For illustration we consider a plane in the four-dimensional 
$\{\mu_i\}$-space where $\mu$ and $\mu_Q$ are taken as independent 
variables and $\mu_3$ and $\mu_8$ are given by
$\mu_3 = -\mu_Q/2$ and $\mu_8 = -\mu_Q/7 -30$~MeV.
The relevance of this particular choice will become more clear in 
\sect{neutmixresults}.
Here we just note that $\mu_3 = -\mu_Q/2$ means that
$\mu_{u,r} = \mu_{d,g}$ .
Also the sum $\mu_{s,r}+\mu_{u,b}$, corresponding to the chemical potential
related to a pair of a red strange quark and a blue up quark, equals the
sum $\mu_{s,g}+\mu_{d,b}$, corresponding to the chemical potential
related to a pair of green strange quarks and blue down quarks.
Together with the relations given in \eq{denscfl1} and the isospin
symmetry of the original Lagrangian this implies for the CFL phase
that $n_u = n_d$ or, according to \eq{denscfl}, 
$n_r = n_g$ and thus $n_3 = 0$.

\begin{figure}
\begin{center}
\epsfig{file=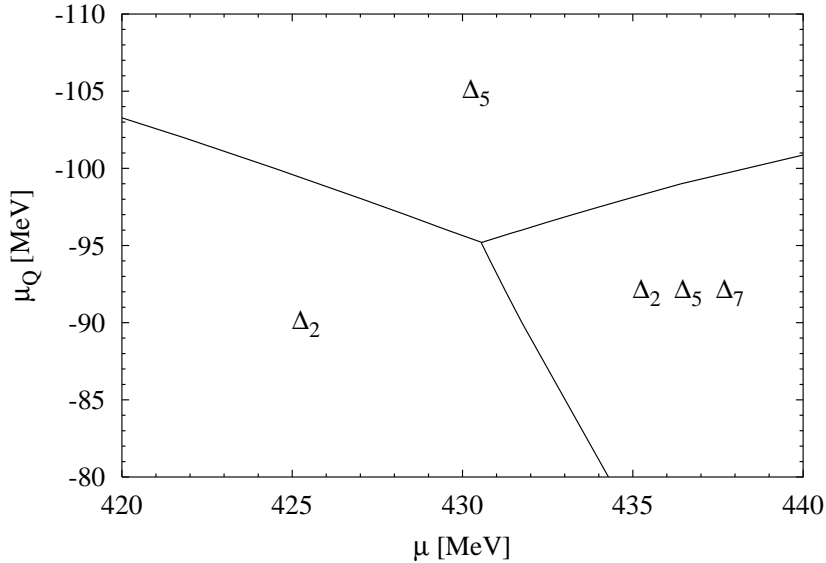,width=11.cm}
\end{center}
\vspace{-0.5cm}
\caption{\small Phase diagram in a plane defined by $\mu$ and $\mu_Q$ as 
independent variables and $\mu_3 = -\mu_Q/2$ and $\mu_8 = -\mu_Q/7 -30$~MeV.
The various phases separated by the solid lines are  characterized by 
different non-vanishing diquark gaps $\Delta_i$ as indicated in the figure.
Taken from Ref.~\cite{NBO03}.
}
\label{figneutphase3}
\end{figure}

In \fig{figneutphase3} we show a small part of the resulting phase diagram.
Here, in addition to the standard 2SC and CFL phases, we find a phase
where only $u$ and $s$ quarks are paired (``2SC$_{us}$'').
We have thus found examples for all three  
$\Delta_A$, $A=2,5,7$, being the only non-vanishing scalar diquark gaps
in some regime. Taking all possible combinations of no, one, two, or three
of these condensates (see \tab{tablephases}),
the phases SC$_{ud+us}$ and SC$_{ud+ds}$, i.e., 
the combinations $\Delta_2+\Delta_5$ and $\Delta_2+\Delta_7$ are 
the only ones we have not encountered. These phases might exist as well,
but we have not searched for them systematically..

\subsection{Homogeneous neutral solutions}
\label{neutresultshom}

\begin{figure}
\begin{center}
\epsfig{file=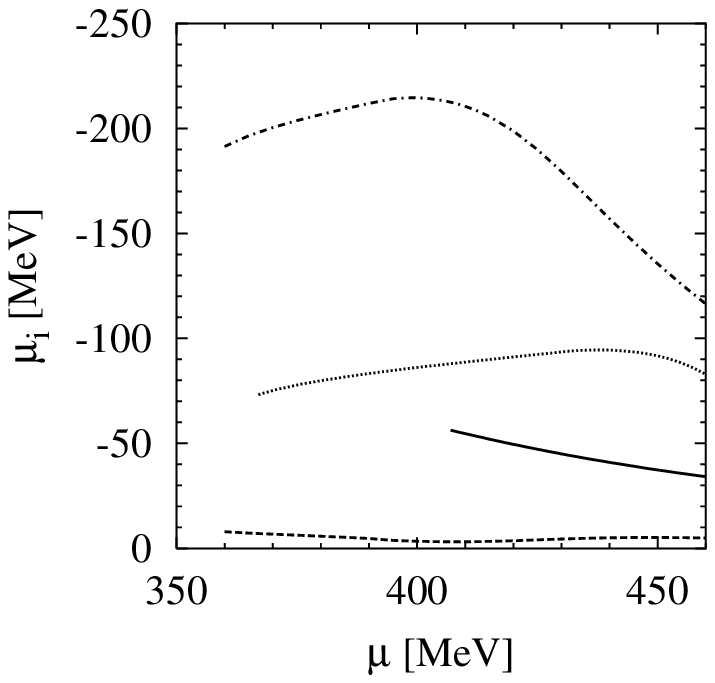,width=7.cm}
\epsfig{file=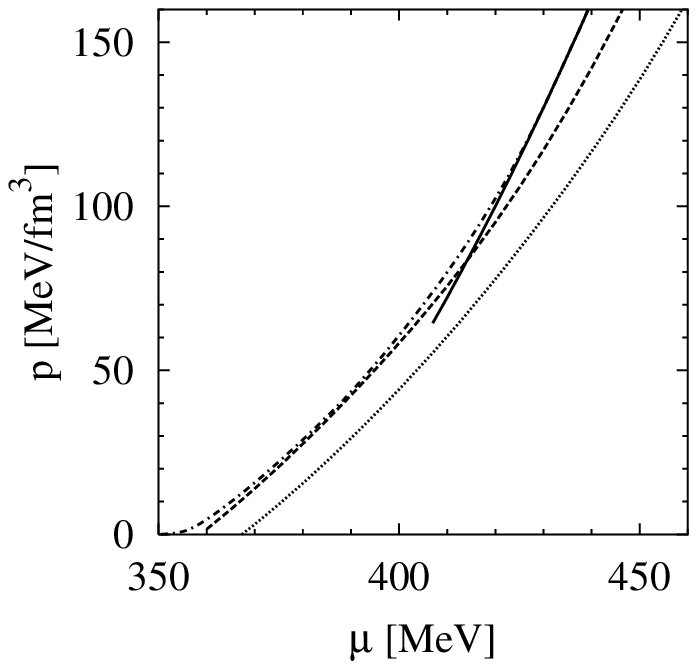,width=7.cm}
\end{center}
\vspace{-0.5cm}
\caption{\small Quantities related to electrically and color neutral
         homogeneous CFL, 2SC, and normal quark matter as functions of the
         quark number chemical potential $\mu$. 
         Left: chemical potentials $\mu_8^{(CFL)}$ (solid), $\mu_8^{(2SC)}$ 
         (dashed), $\mu_Q^{(2SC)}$ (dash-dotted), $\mu_Q^{(normal)}$ 
         (dotted). Note that $\mu_Q^{(CFL)} = \mu_8^{(normal)} = 0$.
         Right: pressure in the CFL phase (solid), 2SC (dashed), and normal
         quark matter (dotted). Also shown is the pressure of the mixed
         phase solution constructed in \sect{neutmixresults} (dash-dotted).}
\label{figmuphom}
\end{figure}

Solutions of homogeneous electrically and color neutral quark matter 
obtained within our model are presented in \fig{figmuphom}. 
In the left panel we show the chemical potentials $\mu_8$ and
$\mu_Q$, needed to neutralize the matter in a given phase.
As pointed out above, the fact that we did not find regions of absolutely
stable solutions of neutral normal or 2SC quark matter in the phase
diagram does not exclude the existence of metastable solutions.
These might finally turn out to be stable if mixed
phases are suppressed due to surface and Coulomb effects and if
there is no other neutral solution with a higher pressure at the same 
value of $\mu$.

\begin{figure}
\begin{center}
\epsfig{file=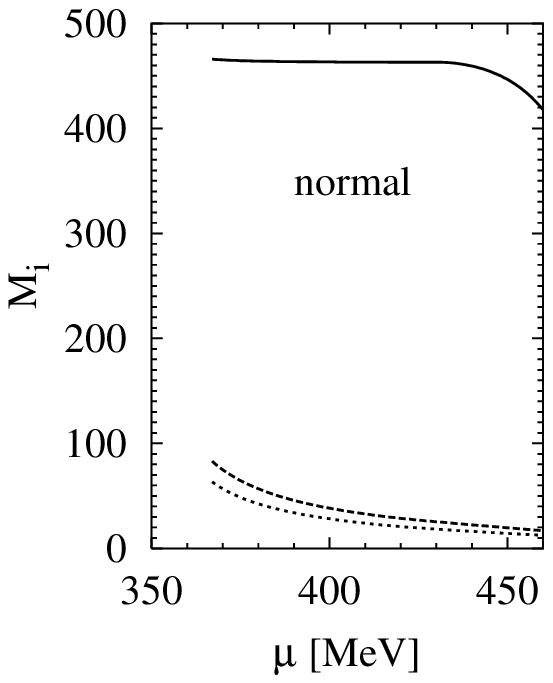,width=4.8cm}
\epsfig{file=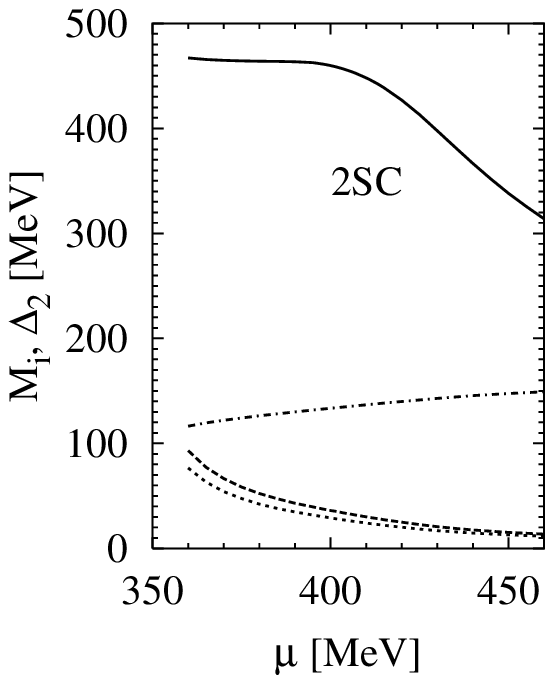,width=4.8cm}
\epsfig{file=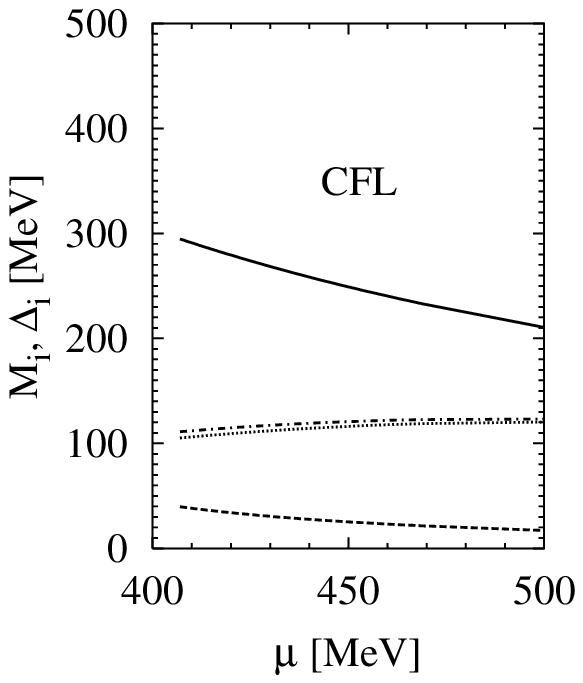,width=4.8cm}
\end{center}
\vspace{-0.5cm}
\caption{\small Constituent masses and diquark gaps for homogeneous
         electrically and color neutral quark matter. Left: $M_u$ (dashed),
         $M_d$ (dotted), and $M_s$ (solid) in the normal phase.
         Center: $M_u$ (dashed), $M_d$ (dotted), $M_s$ (solid), and
         $\Delta_2$ (dash-dotted) in the 2SC phase.
         Right: $M_u=M_d$ (dashed), $M_s$ (solid), 
         $\Delta_2$ (dash-dotted), and $\Delta_5=\Delta_7$ in the CFL
         phase.}
\label{figcondhom}
\end{figure}

The latter is analyzed in the right panel of \fig{figmuphom}.
We find that the CFL phase (solid line) is the phase with the highest 
pressure for $\mu > 414$~MeV. Below this point the 2SC phase (dashed)
is most favored, whereas the pressure of normal quark matter (dotted) 
is always lower. This means, we do not confirm the predictions of Alford 
and Rajagopal according to which the 2SC phase is always disfavored against 
CFL or normal quark matter~\cite{AR02}. The reason for this is the fact 
that the arguments of Alford and Rajagopal are based on an expansion in the
strange quark mass, which fails if $M_s$ is large. 
As one can see in \fig{figcondhom}, where the constituent masses and
diquark gaps are displayed for the various solutions, this is obviously
the case. In fact, the fraction of strange quarks in the 2SC phase is
only 5\% at $\mu = 414$~MeV and vanishes below $\mu = 395$~MeV.   

In this context it is interesting to revisit the arguments for the 
limit of large strange quark masses which we gave in the 
introduction to this chapter. There we estimated the  
difference of the up and down Fermi momenta for neutral matter in the 
normal phase to be about 100~MeV, which agrees well with the value we  
find for $|\mu_Q|$ in this phase (see \fig{figmuphom}). 
From this value, applying \eq{stablegap} we concluded that there
will be a stable 2SC solution if $\Delta_2 \gtrsim 70$~MeV.
Since we find gaps of more than 100~MeV this is consistent with our 
results. 
(Note, that $|\mu_Q|$ in the 2SC phase, which is greater than 200~MeV
at $\mu \simeq 400$~MeV, is not the relevant quantity to compare
with, because we first have to prepare {\it neutral} quark matter in the 
normal phase and than check whether the mismatch of the corresponding
Fermi surfaces can be overcome by the gap.)

Our results are in qualitative agreement with Ref.~\cite{SRP02}
where a similar NJL model calculation has been presented first. 
The main difference is that the authors of Ref.~\cite{SRP02} used a 
smaller diquark coupling ($H = 0.75\; G$ instead of $H = G$)
and therefore find smaller gaps. This could be the reason why
the neutral 2SC solutions presented in that reference cease to exist
for $\mu \lesssim 440$~MeV. Therefore, in contrast to our results,
neutral normal quark matter is favored below this value, simply 
because there is no other solution. On the other hand, whenever 
there is a neutral 2SC solution, it is more favored than neutral
normal quark matter, in agreement with our findings.    

We should recall that we have only considered the case of
zero temperature and non-conserved lepton number, appropriate 
for neutron stars older than a few minutes.
Steiner, Reddy, and Prakash~\cite{SRP02} have also analyzed
the case of finite temperature and conserved lepton number, which is
relevant during the evolution from a proto-neutron star, where neutrinos
are trapped, to a cold compact star, where the neutrinos can freely leave
the system. The authors showed that in this case the 2SC phase is favored 
because neutral CFL matter excludes electrons (at least at $T = 0$ and still
disfavors them at $T > = 0$)
and can therefore not easily accommodate a finite lepton number. 

\begin{figure}
\begin{center}
\epsfig{file=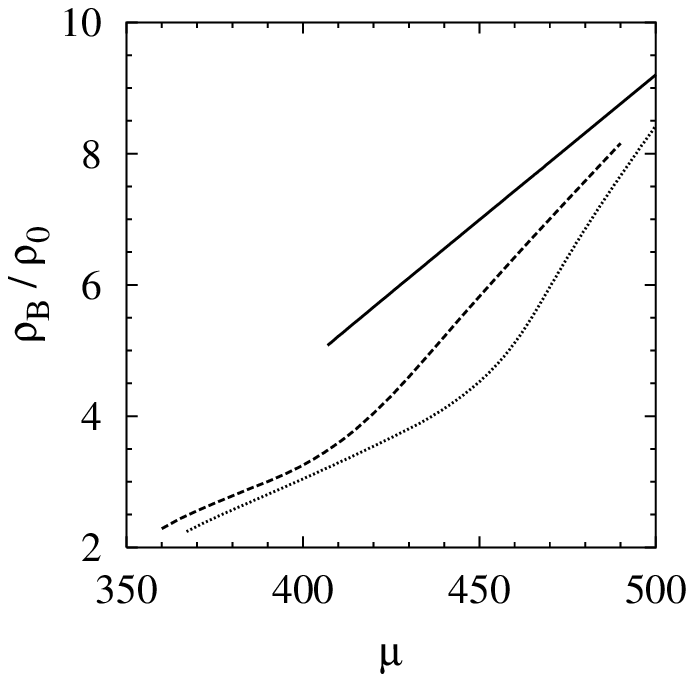,width=7.cm}
\epsfig{file=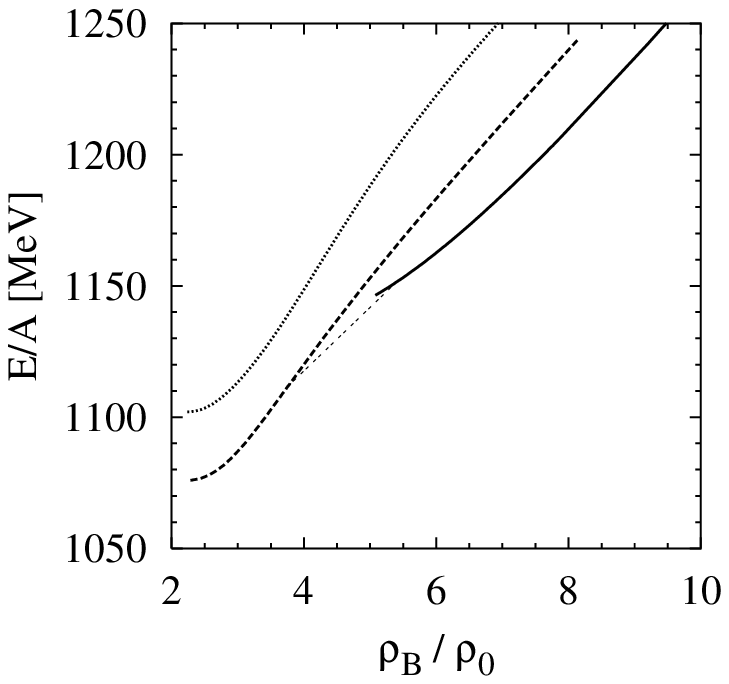,width=7.cm}
\end{center}
\vspace{-0.5cm}
\caption{\small Baryon number density $\rho_B$ as function of the quark number
         chemical potential $\mu$ (left), and energy per baryon number
         $E/A$ as function of $\rho_B$ (right) for homogeneous electrically 
         and color neutral quark matter: CFL (solid), 2SC (dashed), and
         normal quark matter (dotted). The thin double-dashed line indicates
         $E/A$ in a 2SC-CFL mixed phase obtained in a Maxwell construction
         with two separately neutral components.} 
\label{figdenseahom}
\end{figure}

Finally, we come back to the hypothesis of absolutely stable
strange quark matter~\cite{Bodmer,Witten}. In our earlier analysis in
\sect{sqm} the effects of color superconductivity have not been taken 
into account.
Since the formation of diquark condensates gives rise to extra binding 
energy, it is in principle possible that SQM which would be unstable in the 
normal quark phase becomes absolutely stable in the CFL 
phase~\cite{Mad01,LuHo02}.
To analyze this question, we calculate the energy per baryon 
number $E/A$ as function of the baryon number density $\rho_B$ for our 
solutions of neutral quark matter. The results are displayed in the
right panel of \fig{figdenseahom}. For clarity, we also show the relation
between $\rho_B$ and $\mu$ (left panel). In both figures the dotted line
corresponds to normal quark matter, i.e., to the equation of state used in
\sect{sqm}. The solid and the dashed lines correspond to CFL and 2SC matter,
respectively. 

We see that the diquark condensates indeed lead to an appreciable reduction
of $E/A$. Although this effect is stronger in the CFL phase than in the 
2SC phase, it is not strong enough to produce a minimum in the CFL phase. 
In fact, this could have been anticipated from \fig{figmuphom} since we
did not find a CFL solution with zero pressure\footnote{In our numerical
analysis, we did not succeed to find a neutral CFL solution for 
$\mu \lesssim 407$~MeV. Since the neutral 2SC solution is favored already 
for $\mu \leq 414$~MeV, this is of no relevance for the present discussion.
Nevertheless, we should note that the fate of the CFL solution below
$\mu \simeq 407$~MeV is not clear. Since solutions of the gap equation
correspond to stationary points of the thermodynamic potential
they cannot simply ``end'' at some point, but only ``turn around''.
Hence, if there is no neutral CFL solution below a certain value of
$\mu$, there must be two solutions above this point. Of course, this
second solution would be unstable.}. 
Thus if we prepare a large but finite piece of
CFL matter and then slowly decrease the external pressure, the matter 
will expand. Eventually, when the pressure reaches the critical value
(corresponding to the crossing of the 2SC and CFL lines in \fig{figmuphom}),
the system will follow the thin double-dashed line in \fig{figdenseahom} and
make a transition into the 2SC phase. 

We also see that the minimal values of $E/A$, both in the CFL phase and in
the 2SC phase are still considerably larger as 930~MeV, the value in 
$^{56}Fe$. Thus, although a more systematic study of the parameter dependence
remains to be done, our results suggest, that including diquark condensates
will not alter the main conclusions drawn in \sect{sqm}.   

\section{Mixed phases}
\label{neutmix}

As one can see in \fig{figmuphom},
the values of the chemical potentials $\mu_8$ and $\mu_3$ needed to neutralize
matter for a given value of $\mu$ depend on the phase. 
Hence, neutral matter in one phase is never in chemical
equilibrium with neutral matter in a different phase, even if their quark
number chemical potentials $\mu$ are the same. In particular the
points of equal pressure do not fulfill the Gibbs condition for a 
phase transition, stating that the pressure and {\it all} chemical
potentials should be the same in coexisting phases.

To see what this means, suppose two volumes of neutral quark matter  
in different phases, but with with equal pressure 
and equal quark number chemical potential, are brought in contact with 
each other. According to the above, the two phases must differ in at
least one chemical potential $\mu_i$ . Hence the total free energy of the 
system can be lowered by transferring a part of the corresponding charge 
$Q_i$ from the phase with the higher $\mu_i$ to the phase with the lower
$\mu_i$, keeping the two volumes and the total number of quarks in each 
phase constant. If not prevented by the emerging Coulomb forces, this
process will go on until the chemical potentials in both phases are 
equal. Obviously, in this state the two phases are no longer separately
neutral but oppositely charged. 

These considerations suggest that instead of requiring locally neutral
matter it might be energetically more favored to form mixed phases of several
components in chemical equilibrium which are only neutral in total.  
This scenario has been pushed forward by Glendenning about a decade 
ago~\cite{G92}. Although formulated in a quite general way for systems
with more than one conserved charge, Glendenning mainly applied his
results to the quark-hadron phase transition in neutron stars where he
only had to care about electric neutrality.
In this case a neutral mixed phase can obviously be constructed in those
regions of the phase boundary where the charge densities of the two
components have opposite signs.
Below we generalize this procedure to the more complex case of
constructing electrically and color neutral mixed phases.

\subsection{Formalism}
\label{neutmixform}

We consider a mixed phase consisting of two components, 1 and 2,
in thermal, chemical, and mechanical equilibrium. 
In terms of the thermodynamic potential the phase equilibrium can be
expressed as the equality
\beq
    \Omega(T,\{\mu_i\};\chi^{(1)}) \;=\; \Omega(T,\{\mu_i\};\chi^{(2)})~, 
\label{phaseboundary}
\eeq
where $\chi^{(1)}$ and $\chi^{(2)}$ are two different sets of 
condensates which solve the coupled gap equations at temperature $T$
and chemical potentials $\{\mu_i\} = \{\mu,\mu_3,\mu_8,\mu_Q\}$.
At fixed temperature, \eq{phaseboundary} defines a 3-dimensional 
first-order phase boundary in the 4-dimensional space spanned by the 
chemical potentials $\{\mu_i\}$.

Since $\chi^{(1)} \neq \chi^{(2)}$, the densities
\beq
n_i^{(\alpha)} = - \frac{\partial\Omega(T,\{\mu_i\};\chi^{(\alpha)})}
{\partial\mu_i}~.
\label{densitiesalpha}
\eeq
are in general different in the two coexisting phases. 
In particular, the neutrality condition, \eq{ecneutral}, 
is in general not fulfilled simultaneously for both components.
However, as indicated above it is sufficient to demand that 
the {\it average} charge and color densities of the mixed phase 
vanish~\cite{G92}. If the two components occupy
the volume fractions $x^{(1)}$ and $x^{(2)} = 1-x^{(1)}$, respectively,
the average densities are given by
\beq
    n_i \;=\; x^{(1)}\,n_i^{(1)} \;+\;  (1-x^{(1)})\,n_i^{(2)}~.
\label{xidef}
\eeq
This is zero for
\beq
    x^{(1)} = \frac{n_i^{(2)}}{n_i^{(2)}-n_i^{(1)}}~.
\label{fraction}
\eeq
To be meaningful the solution must be in the interval $0<x^{(1)}<1$. 
This is fulfilled when the charge densities
$n_i^{(1)}$ and $n_i^{(2)}$ have opposite signs, which is an obvious
prerequisite for a charge neutral mixture.
For a single charge, e.g., $n_Q$, it is the only one.
However, in order to get simultaneous neutrality for three charges,
\eq{ecneutral}, we have to require that the result of \eq{fraction}
is the same for $i=Q$, 3, and 8. This is the case when
\beq
    n_Q^{(1)} \,:\, n_Q^{(2)} \;=\;
    n_3^{(1)} \,:\, n_3^{(2)} \;=\;
    n_8^{(1)} \,:\, n_8^{(2)}~.
\label{mix}
\eeq

In our numerical calculations we will restrict ourselves to $T=0$. Then,
as already mentioned,  the phase boundaries, \eq{phaseboundary}, 
are three-dimensional surfaces in the four-dimensional space of chemical
potentials. Since \eq{mix} imposes two additional constraints,
electrically and color neutral mixed phases can be constructed along a
one-dimensional line. In the simplest case
this line starts at a point where the 
neutrality line of phase 1 ($n_Q^{(1)} = n_3^{(1)} = n_8^{(1)} = 0$), 
meets the phase boundary and it ends where the neutrality line of phase 2 
meets the phase boundary. Between these two points $x^{(1)}$ changes 
continuously from 1 to 0. 

However, in our system there are also two-dimensional manifolds with
three coexisting phases
\beq
    \Omega(T,\{\mu_i\};\chi_1) \;=\; 
    \Omega(T,\{\mu_i\};\chi_2) \;=\; 
    \Omega(T,\{\mu_i\};\chi_3)~. 
\label{phaseboundary3}
\eeq
It is thus possible that the neutrality line in a two-component mixed phase 
meets a third phase boundary before the fraction of one of the two components
has become zero. In this case we can construct a neutral mixed phase
consisting of three components. The corresponding neutrality condition reads
\beq
    \hat N\,\vec x \;\equiv\; \left (
    \begin{array}{c c c}
     n_Q^{(1)} & n_Q^{(2)} & n_Q^{(3)}\\
     n_3^{(1)} & n_3^{(2)} & n_3^{(3)}\\
     n_8^{(1)} & n_8^{(2)} & n_8^{(3)}
    \end{array}
    \right )  \left (
    \begin{array}{c}
     x^{(1)} \\ x^{(2)}\\ x^{(3)}
    \end{array}
    \right ) \;=\; 0~.
\eeq
In order to find a non-trivial solution for $\vec x$, we must have 
$\det \hat N = 0$. Together with \eq{phaseboundary3}, this again restricts 
the possible solutions to a one-dimensional subspace. Moreover, since the
fractions $x^{(\alpha)}$ should be non-negative, for each $i=Q,3,8$
the densities $n_i^{(\alpha)}$ must not have the same sign for all 
$\alpha=1,2,3$.
(The correct normalization $\sum_\alpha x^{(\alpha)} = 1$ can always be
achieved and does not lead to further constraints.) 

Finally, there even could be a mixed phase, consisting of four components.
The corresponding phase boundary is one-dimensional and again the region of
possible neutral mixed phases is further restricted by the requirement
that the various fractions $x_i$ should not be negative.

\subsection{Numerical results}
\label{neutmixresults}

\begin{figure}
\begin{center}
\epsfig{file=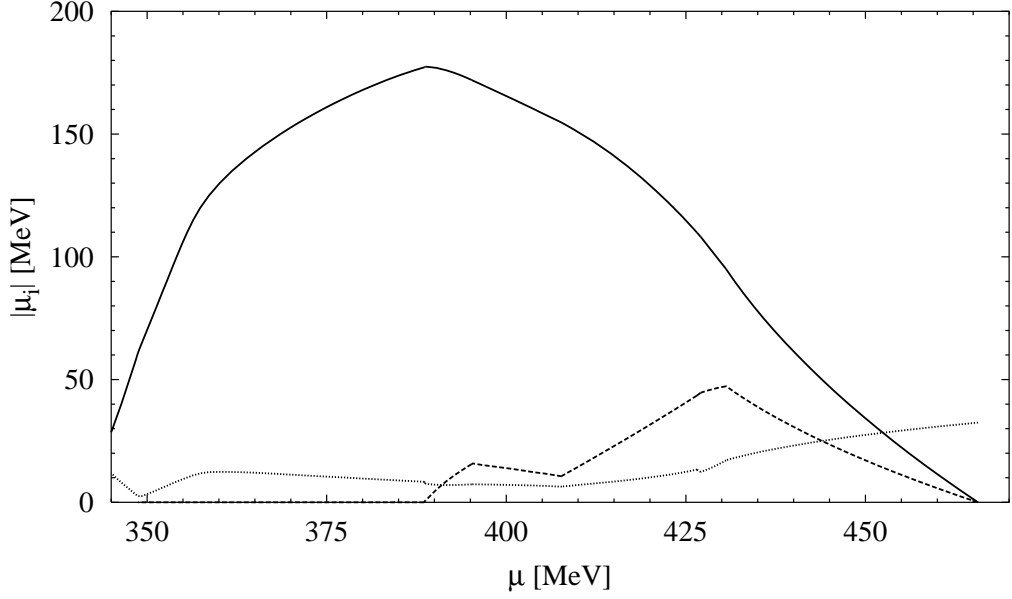,width=14.cm}
\end{center}
\vspace{-0.5cm}
\caption{\small Chemical potentials $\mu_i$ corresponding to the 
                electrically and color neutral mixed phases, listed in
                \tab{tablemix}: 
                $-\mu_Q$ (solid), $\mu_3$ (dashed), and $-\mu_8$ (dotted).
                Adapted from Ref.~\cite{NBO03}.
}
\label{figmixmutot}
\end{figure}

Following Ref.~\cite{NBO03},
we now apply the formalism developed above to the model of \sect{neutresults}
to construct electrically and color neutral mixed phases. 
Starting point is $\mu = 465.7$~MeV, $\mu_8
= -32.5$~MeV, and $\mu_3 = \mu_Q = 0$ where the line of neutral CFL
matter meets the boundary to the 2SC phase (see \fig{figneutphasemu8}).
At lower values of $\mu$,
mixed phases become possible and are energetically favored as long as
Coulomb and surface effects are neglected. Altogether we find nine 
mixed phase regimes characterized by
different compositions of coexisting phases (see \tab{tablemix}).
The corresponding chemical potentials $\mu_i$ as functions of $\mu$
are displayed in \fig{figmixmutot}.
In \tab{tablemix} we also list the corresponding minimal and 
maximal quark number densities, averaged over the components of the
respective mixed phase.

\begin{table}[b!]
\begin{center}
\begin{tabular}{|l|c|c|}
\hline
components                        & $\mu$ [MeV]   & $\rho_B / \rho_0$ \\ 
\hline
N, 2SC                            & 340.9 - 388.6 & 0.00 - 2.94 \\ \hline
N, 2SC, SC$_{us+ds}$              & 388.6 - 388.7 & 2.94 - 2.94 \\ \hline
N, 2SC, SC$_{us+ds}$, 2SC$_{us}$  & 388.7 - 388.8 & 2.94 - 3.06 \\ \hline
2SC, SC$_{us+ds}$, 2SC$_{us}$     & 388.8 - 395.4 & 3.06 - 3.40 \\ \hline
2SC, SC$_{us+ds}$                 & 395.4 - 407.7 & 3.40 - 3.86 \\ \hline
2SC, SC$_{us+ds}$, CFL            & 407.7 - 426.5 & 3.86 - 5.69 \\ \hline
2SC, SC$_{us+ds}$, CFL, 2SC$_{us}$& 426.5 - 427.1 & 5.69 - 5.75 \\ \hline
2SC, CFL, 2SC$_{us}$              & 427.1 - 430.6 & 5.75 - 6.10 \\ \hline
2SC, CFL                          & 430.6 - 465.7 & 6.10 - 7.69
\\ \hline
\end{tabular}
\end{center}
\caption{\small Composition of electrically and color neutral mixed phases,
                corresponding quark number chemical potentials and average
                baryon number densities. The various components are defined 
                in \tab{tablephases}.}
\label{tablemix}
\end{table}  

\begin{figure}
\begin{center}
\epsfig{file=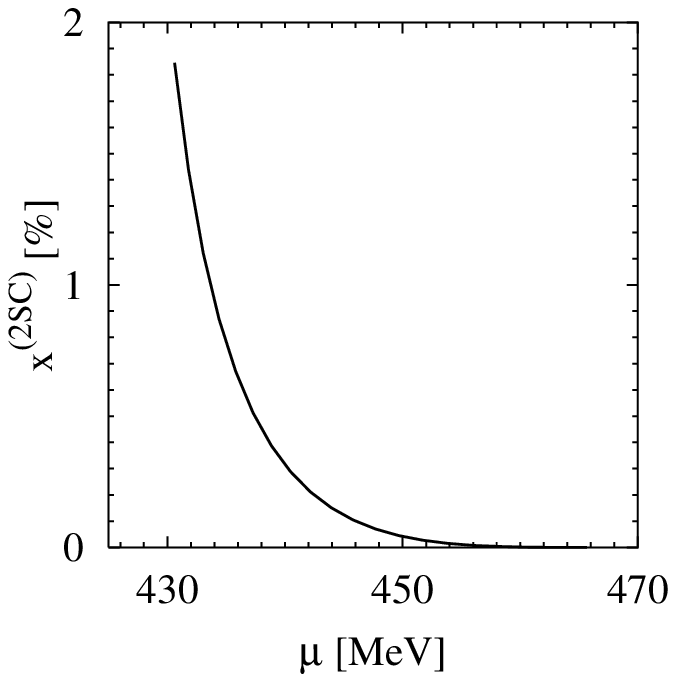,width=7.cm}\qquad
\epsfig{file=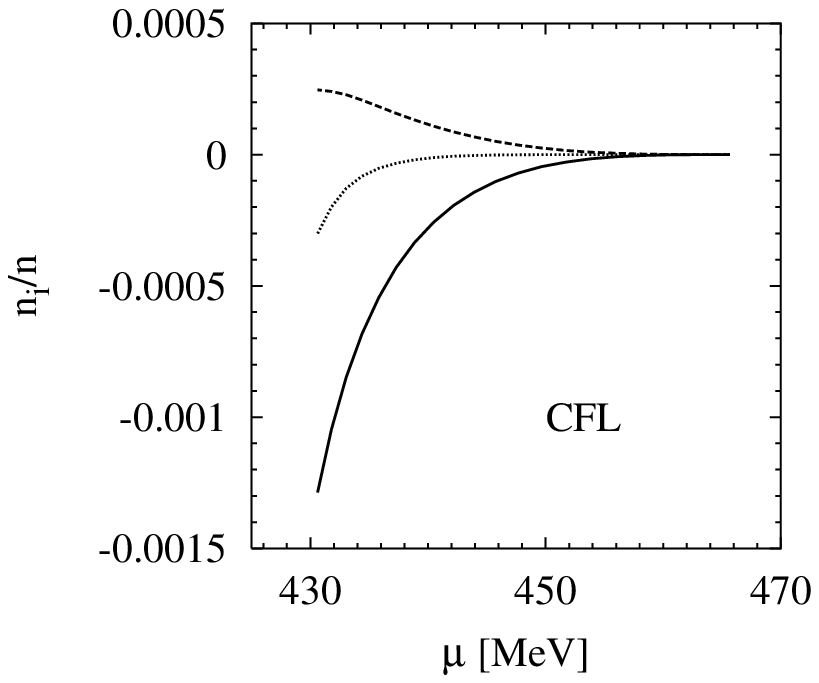,width=7.cm}
\end{center}
\vspace{-0.5cm}
\caption{\small Quantities related to the neutral 2SC-CFL mixed phase as 
functions of the quark number chemical potential $\mu$. 
Left: Volume fraction $x^{(2SC)}$ of the 2SC component.
Right: Relative densities in the CFL component:
$n_Q/n$ (solid), $n_8/n$ (dashed), and $n_3/n$ (dotted). 
Adapted from Ref.~\cite{NBO03}.
}
\label{figmixmux}
\end{figure}

In the regime closest to the region of homogeneous neutral CFL matter 
(430.6~MeV~$< \mu <$ 465.7~MeV), we find a mixed phase consisting of
a CFL component and a 2SC component. The volume fraction $x^{(2SC)}$
of the 2SC component is displayed in the  left panel of \fig{figmixmux}.
In the higher-$\mu$ part of this region it is completely negligible, but
even at the lower end it remains below 2\%. 
Consequently, the CFL component must stay 
almost neutral by itself. Indeed, the relative charge densities 
$n_i/n$, $i=3,8,Q$, (right panel of \fig{figmixmux}) are very small.
As we have discussed in \sect{neutresultsphases}, $n_3$-neutrality of the 
CFL phase 
is maintained by the relation $\mu_3 = -\mu_Q/2$. For the actual values of
$\mu_3$ and $\mu_Q/2$ in the 2SC-CFL mixed phase we find a deviation 
of less than 1\% from this relation, while $\mu_8$ can approximately be
fitted by $\mu_8= -\mu_Q/7 - 30$~MeV. This is the reason why we have
calculated the phase diagram
shown in \fig{figneutphase3} with these constraints.

In that figure it can be seen that the 2SC-CFL phase boundary meets the 
boundary to the 2SC$_{us}$ phase at $\mu = 430.6$~MeV. Below that point
we get a three-component
neutral mixed phase, consisting of 2SC, CFL and 2SC$_{us}$.  Then,
on a short interval in $\mu$, we even find a four-component
neutral mixed phase (2SC, CFL, 2SC$_{us}$, and SC$_{us+ds}$) before
upon further decreasing $\mu$ the system goes over into a neutral
2SC-CFL-SC$_{us+ds}$ mixed phase.

\begin{figure}
\begin{center}
\epsfig{file=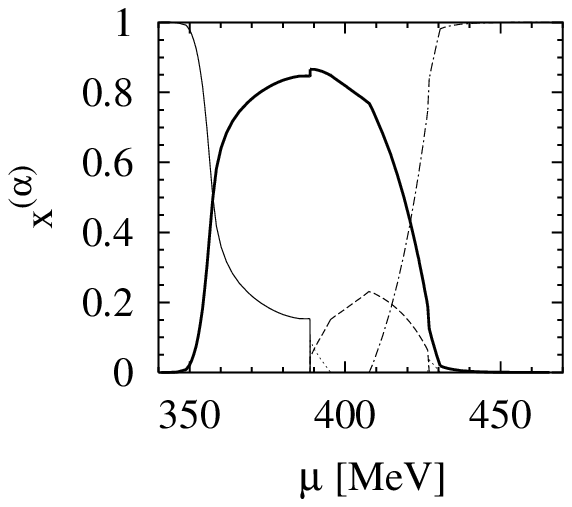,width=7.4cm}
\epsfig{file=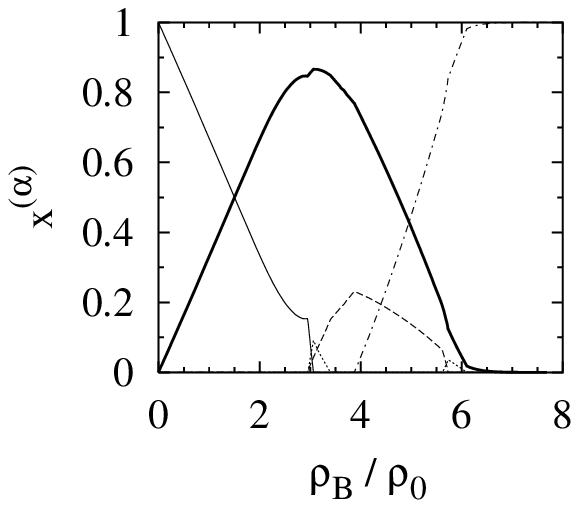,width=7.4cm}
\end{center}
\vspace{-0.5cm}
\caption{\small Volume fractions $x^{(\alpha)}$ of the various components 
         $\alpha$ in the 
         mixed phase region as functions of the quark number chemical
         potential $\mu$ (left) and as functions of the averaged
         baryon number density $\rho_B = n/3$ in units of 
         $\rho_0 = 0.17$~fm$^{-3}$ (right): normal (thin solid),
         2SC (bold solid), CFL (dash-dotted), SC$_{us+ds}$ (dashed),
         2SC$_{us}$ (dotted). Left figure adapted from Ref.~\cite{NBO03}.
}
\label{figmixxtot}
\end{figure}

In the left panel of \fig{figmixxtot} the volume
fractions of the various components of the mixed phases are plotted as
functions of $\mu$. 
To get some idea about the corresponding densities, we also show 
the volume fractions as functions of the average baryon number density
$\rho_B$ (right panel).
Whereas the 2SC-CFL-mixed phase (\fig{figmixmux}) is completely
dominated by the CFL component, thereafter the CFL fraction becomes
quickly smaller with decreasing $\mu$, while in particular the 2SC
component, becomes more and more important.
Below $\mu = 407.7$~MeV ($\rho_B \simeq 3.9~\rho_0$)
the CFL component disappears completely.

An admixture of normal quark matter is found below $\mu =
388.8$~MeV ($\rho_B \simeq 3.1~\rho_0$). The fractions of the superconducting 
phases other than the
2SC phase then rapidly become smaller and vanish at $\mu = 388.6$~MeV
($\rho_B \simeq 3.0~\rho_0$),
while the fraction of normal matter strongly increases.\footnote{Note,
however, that by construction all volume fractions $x^{(\alpha)}$ are 
continuous functions of both, $\mu$ and $\rho_B$ in the entire mixed phase 
region.
Therefore the pretended steps in the figure only correspond to very
rapid changes, but not to discontinuities.} 
Nevertheless the 2SC phase stays the dominant component for
$\mu\gtrsim 360$~MeV ($\rho_B \approx 1.5~\rho_0$). 

As discussed in \sect{neutresultsphases}, apart from the vacuum there is 
no solution of stable neutral non-superconducting quark matter
in our model. Therefore the
normal-2SC mixed phase cannot end in normal homogeneous quark matter 
but only in the vacuum. Hence, the chemical potential $\mu_Q$ must 
finally go to zero. Eventually, at $\mu = 348.6$~MeV 
($\rho_B \simeq 0.02~\rho_0$), $|\mu_Q|$ drops below
60~MeV and we enter the regime where the normal phase only consists of
electrons without quarks (cf. \fig{figneutphasemuq}). This means, the 
corresponding mixed phase consists of (electrically positive) 2SC-droplets 
surrounded by regions without quarks and neutralized by a homogeneous 
background of electrons\footnote{In the spirit of the schematic picture 
discussed in \sect{njlbag} where the solutions of dense quark 
matter in equilibrium with the vacuum were identified with ``nucleons'', 
one might be tempted to view this mixed phase as a gas or plasma of 
``atoms''. 
However, even on a very schematic level, this would not make sense.
Clearly, there is no 2SC phase inside nucleons or nuclei.
Thus, although it is interesting to see how our procedure brings itself
to an end by finally reaching the vacuum in a consistent way,
the results should not be trusted in the low-density regime.}. 
Since the electrons are color neutral, the
2SC component must be color neutral by itself. This is maintained
by an increase of $|\mu_8|$ in this regime. 
At $\mu = 340.9$~MeV we finally reach the vacuum.

\subsection{Surface and Coulomb effects}
\label{neutmixsurface}

\begin{figure}
\begin{center}
\epsfig{file=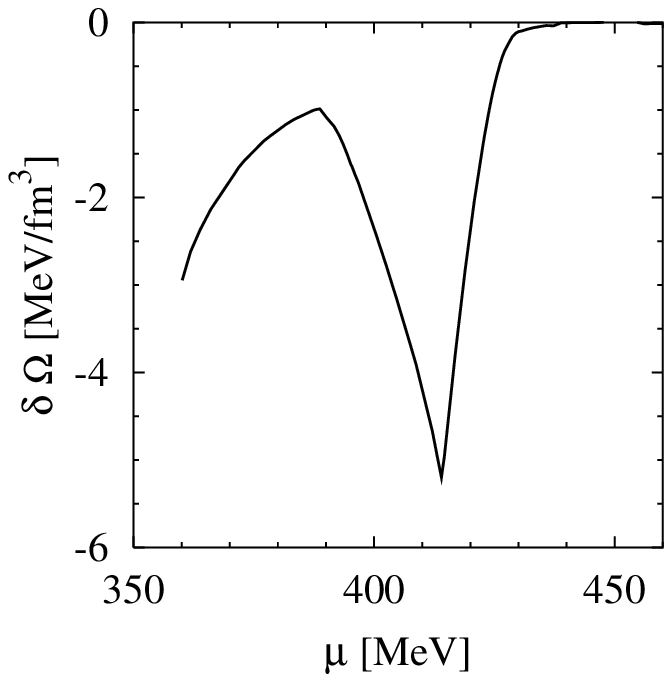,width=7.4cm}
\epsfig{file=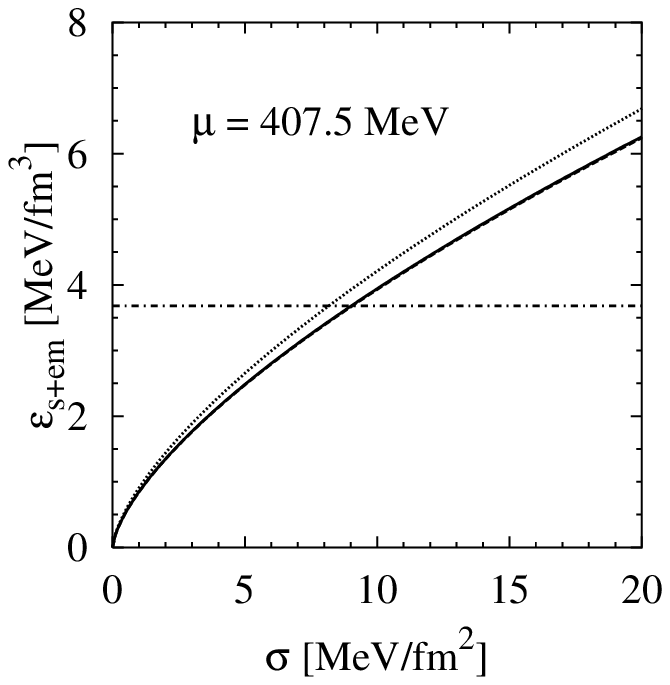,width=7.4cm}
\end{center}
\vspace{-0.5cm}
\caption{\small Left: Difference in bulk free energy, 
$\delta\Omega = \Omega^{(mix)} - \Omega^{(hom)}$ between the mixed phase
solution and the most favored homogeneous neutral solution (CFL matter 
for $\mu > 414$~MeV, 2SC matter for $\mu < 414$~MeV, see \fig{figmuphom}). 
Right: Surface and Coulomb energy per volume for the $2SC+SC_{us+ds}$
mixed phase at $\mu = 407.5$~MeV as a function of the surface tension
$\sigma$: slabs (dotted), 
$SC_{us+ds}$ rods (dashed), and $SC_{us+ds}$ drops (solid).  
(The results for rods and drops are almost indistinguishable.)
The dash-dotted line indicates the value of $-\delta\Omega$ at this
point. Note that the color-electric Coulomb energy has not been included.  
}
\label{figbulk}
\end{figure}

When we compare the mixed phase results with the results obtained in
\sect{neutresultshom} for homogeneous phases we see that the regions
where the mixed phases are dominated by the CFL phase or by the 2SC phase, 
roughly correspond to those regions 
where we found homogeneous CFL or 2SC matter, respectively, to be favored.
There are of course differences in details.
In order to decide which of the two scenarios is the more realistic one
we should recall that so far we have 
neglected the surface energy and the energies of the electric and 
color-electric fields in the mixed phases.
Without these terms the mixed phases would be favored, which follows 
already from the general arguments given in the beginning of \sect{neutmix}.
Quantitatively this is shown in the left panel of \fig{figbulk} where the 
difference in bulk free energy, 
$\delta\Omega = \Omega^{(mix)} - \Omega^{(hom)}$ between the mixed phase
solution and the most favored homogeneous neutral solution is plotted.
$\delta\Omega$ is negative, as expected, and has a minimum of about 
$-5$~MeV/fm$^3$
at the chemical potential which corresponds to the 2SC-CFL phase transition
point in the homogeneous case. (The second rise of $|\delta\Omega|$ below 
$\mu \simeq 490$~MeV is do to the appearance of a normal quark matter
component in the mixed phase.)

Of course, the mixed phases are only stable if this 
gain in bulk free energy is larger than the neglected surface and
(electric and color-electric) Coulomb contributions.
For the two-component mixed phases these can in principle be estimated
adapting the techniques which have been developed by Ravenhall, Pethick, and 
Wilson~\cite{RPW83} in the context of nuclear matter at sub-saturation
densities and which have been applied, among others, in 
Refs.~\cite{ARRW01,HePeSt93,GlePei95} to analyze possible quark-hadron 
mixed phases in neutron stars. 
According to these references, the surface and the electromagnetic
Coulomb contribution to the energy density are given by
\beq
    \epsilon_s \;=\; \frac{d\,x\,\sigma}{r_0}~,\qquad
    \epsilon_{em} \;=\; 2\pi\,\alpha_{em}\,f_d(x) (\delta n_Q)^2\,r_0^2~,
\label{epssc}
\eeq
where $\sigma$ is the surface tension, $x$ the volume fraction of the
rarer phase, and $\delta n_Q$ the difference of the electric charge
density in the two phases. 
The result depends on the geometry which is controlled by the dimension
$d$, corresponding to slabs (``lasagna phase'', $d=1$), 
rods (``spaghetti phase'', $d=2$), and drops ($d=3$),
and the parameter $r_0$, which denotes the radius of the rods or drops, 
or the half-thickness of the slabs of the rarer phase.
The geometrical factor $f_d(x)$ is given by
\beq
    f_d(x) \;=\; \frac{1}{d+2}\,\Big(\frac{2-d\, x^{1-2/d}}{d-2} + x \Big)~.
\eeq
For $d=2$ one must interprete $f_d$ as a continuous function of $d$ and
take the appropriate limit.

In principle, the surface plus Coulomb contribution to the energy is now 
easily obtained, minimizing the sum of $\epsilon_s$ and $\epsilon_{em}$
with respect to $r_0$. This gives
\beq
    \epsilon_{s+em} \;=\; \frac{3}{2}\,
    \Big(4\pi\,\alpha_{em}\,d^2\,f_d(x)\,x^2\,(\delta n_Q)^2\sigma^2
    \Big)^{1/3}~.
\label{epsscmin}
\eeq
In practice, however, one generally has to deal with the problem that the 
surface tension is poorly known. For the quark-hadron case, typical estimates 
range from 10 to 100~MeV~fm$^{-2}$~\cite{HePeSt93} or even 
300~MeV~fm$^{-2}$~\cite{ARRW01}, whereas until very recently
practically nothing was known for quark-quark mixed phases.
In this situation, the best one can do is to determine the 
maximal surface tension which is compatible with the existence of
a mixed phase and to compare this value with ``plausible'' estimates.
This is similar to what has been done in Ref.~\cite{ARRW01} for the interface 
between nuclear and CFL matter.

As an example we consider the 2SC-SC$_{us+ds}$ phase 
at $\mu = 407.5$~MeV where we have found a gain in bulk 
free energy of 3.7~MeV/fm$^3$. (The maximum of $|\delta\Omega|$
is located in a three-component mixed phase which cannot be addressed by the
formulae of Ref.~\cite{RPW83}.) 
At this point we have $x \equiv x^{(SC_{us+ds})} = 0.23$ and 
$\delta n_Q = 0.52\,\fm^{-3}$.
The resulting surface plus Coulomb energy density as a function of $\sigma$
is displayed in the right panel of \fig{figbulk}. 
It turns out that the $d=1$ solution (dotted) is somewhat higher in energy,
whereas the solutions for $d=2$ (dashed) and $d=3$ (solid) are practically 
identical. (The ``spaghetti phase'', $d=2$ is slightly favored, but this is 
hardly seen in the plot.) 
For these solutions the gain in bulk energy (indicated by the dash-dotted
line) is already weight out for a surface tension  
$\sigma\simeq 10$~MeV/fm$^2$. 

This sounds like a rather small number, which has originally been taken
as an indication against the mixed-phase scenario~\cite{NBO03}.
Very recently, however, Reddy and Rupak have {\it calculated} the 
surface tension between normal and 2SC phase in a two-flavor 
model~\cite{ReRu04}, employing Landau-Ginzburg theory to determine the 
gradient contribution to the free energy at the interface~\cite{BaLo84}.
The result was a surprisingly small surface tension of only a few MeV,
which led the authors to the conclusion that in their case the mixed
phase was in fact favored~\cite{ReRu04}.
It would be certainly very interesting to redo this kind of calculation
for the three-flavor case as well.

In the considerations above we have not yet included the color-electric 
energy. As an order-of-magnitude estimate we note that the corresponding 
contribution is of similar form as $\epsilon_{em}$ with 
$\alpha_{em}$ replaced by $\alpha_s$, and $\delta n_Q$ replaced by
the difference of excess blue quarks $\delta n_8/\sqrt{3}$.
In our example, the latter is about one order of magnitude smaller than
$\delta n_Q$ and has to be squared, whereas $\alpha_s$ should be about
two orders of magnitude larger than $\alpha_{em}$. We thus expect
the color-electric energy to be of the same order as the 
electromagnetic one. (Note that, because of the power $1/3$ in 
\eq{epsscmin}, we are not very sensitive to details.)
Anyway, it is clear that the color-electric energy will further lower the
maximum value of $\sigma$ compatible with a mixed phase.

On the other hand, if $\sigma$ is small, but color forces are strong, 
there could be a third 
scenario where color neutrality is realized locally whereas electric 
neutrality is realized only globally, leading to mixed phases of color neutral 
but electrically charged components~\cite{SHH03}.
In this case, we would be left with the more standard situation of only
two independent chemical potentials, $\mu$ and $\mu_Q$, and only mixed 
phases with two components would be possible. 

Finally, we should note that we have restricted our analysis to 
diquark condensates $s_{AA'}$ with $A=A'$. For homogeneous phases this
was motivated by the fact, that the two most important condensation patterns,
2SC and CFL, can always be brought into this form without loss of generality
(see \sect{patterns}). For mixed phases, this is in principle different. 
Here the color rotations are additional degrees of freedom which could 
be exploited to reduce the bulk energy of the mixed phase.
For instance, as discussed in \sect{results} we could construct
a color neutral mixed phase without applying color chemical potentials 
by combining several components of the same condensate, but rotated into
different color directions.
However, as we have seen in \fig{fig2scdens} the related gain in bulk free 
energy is small. Therefore it is unlikely that including color rotated 
condensates would strongly change the equation of state.

\section{Discussion: alternative pairing patterns}
\label{neutdisc}

The results of the previous section suggest that, in spite of the arguments
given in Ref.~\cite{AR02}, the 2SC phase could play an important role
under compact star conditions -- either as a homogeneous phase or as
the dominant component in a mixed phase.  
However, our analysis is still rather incomplete:

First, we need a better description of the hadronic phase, which 
so far is represented at best by a vacuum phase with spontaneously broken 
chiral symmetry. It is thus possible that a realistic hadronic phase is
much more stable and the deconfinement phase transition takes place at
much higher densities where the 2SC phase plays no role. 
In the next chapter, we will therefore use an alternative approach and
construct a hybrid equation of state with the hadronic part taken from 
other models.

In addition, there are other pairing patterns which we have not yet 
considered:

\subsection{CFL + Goldstone phase}

So far, we have neglected the role of the Goldstone bosons in the CFL phase.
      It is obvious that the charged Goldstone bosons ($\pi^\pm$, 
      $K^\pm$) are sensitive to $\mu_Q$. Moreover, the stress imposed
      by unequal quark masses acts as an additional effective chemical 
      potential. For instance, the reduced Fermi momentum
      $p_F^s = \sqrt{\mu^2 - m_s^2} \simeq \mu - m_s^2/(2\mu)$
      could be interpreted as a result of an effective strangeness chemical 
      potential $m_s^2/(2\mu)$.
      Combining $\mu_Q$ and mass effects, one finds the effective chemical
      potentials~\cite{KR02}
\beq
     \tilde \mu_{\pi^+} = \mu_Q + \frac{m_d^2-m_u^2}{2\mu}~,\quad
     \tilde \mu_{K^+} = \mu_Q + \frac{m_s^2-m_u^2}{2\mu}~,\quad
     \tilde \mu_{K^0} = \frac{m_s^2-m_d^2}{2\mu}~,
\eeq
      and the same with the opposite sign for $\pi^-$, $K^-$, and 
      $\bar K^0$.
      Hence, if one of these effective chemical potentials exceeds the
      mass of the corresponding Goldstone boson, these mesons condense.
   
      These processes can systematically be studied within high-density 
      effective field theory~\cite{CaGa99,SoSt00}. (For a recent overview, 
      see Ref.~\cite{Schaefer}.)
      Roughly speaking, this corresponds to $\chi PT$ in the CFL phase,
      but with the essential difference that at high densities 
      all coefficients can be calculated within high-density QCD,
      instead of being determined empirically.
      Assuming that $U_A(1)$-breaking effects can be neglected, this
      leads to rather small Goldstone masses of the order 
      $m_q\Delta/\mu$~\cite{SoSt00}. For instance, when the asymptotic results 
      are extrapolated down to 400~MeV, one finds kaon masses of about
      5-20~MeV~\cite{RaWi00}.
      As a consequence, meson condensation is expected to occur
      already for rather small values of $m_s \sim m_d^{1/3}\Delta^{2/3}$
      or $|\mu_Q| \sim \sqrt{m_d m_s} \Delta/p_F$~\cite{BS02}.   
      A phase diagram in the $m_s^2/(2\mu)-\mu_Q$ plane is 
      presented in Ref.~\cite{KR02}.       
      Further CFL phases containing $\eta$ condensates have recently been  
      discussed in Ref.~\cite{KKS04}.

      For $\mu_Q = 0$ and large enough $m_s$, the $K^0$ condense.
      Because of the condensation energy, the pressure in the 
      CFL$+K^0$ phase is higher than in the CFL phase (otherwise the
      kaons would not condense) and therefore the CFL$+K^0$ phase
      could be favored against neutral 2SC matter in regions where the
      CFL phase is not. This effect has been estimated in Ref.~\cite{SRP02}. 
      The authors found that the kaon condensate lowers the critical quark 
      chemical potential of the 2SC-CFL phase transition
      by about 16~MeV. This is not dramatic but could make a difference
      if, for instance, the transition point to the hadronic phase turns out 
      to be in the same region.   

      In principle, the CFL + Goldstone phases can also be studied within 
      NJL-type models. 
      In fact, since the applicability of the high-density effective
      Lagrangian method becomes questionable at moderate densities,
      it would be interesting to have alternative approaches to compare with.
      In particular, it would be interesting to investigate the role
      of non-vanishing quark-antiquark condensates and $U_A(1)$-breaking
      terms. 
      As mentioned in \sect{flamix}, NJL models have already been
      employed to study pion condensation at non-zero isospin chemical
      potential~\cite{MFrank} (see also Ref.~\cite{TK03}).
      An extention to the CFL phase is straight forward but technically 
      involved.

\subsection{Crystalline color superconductors}

      Another possibility is the formation of so-called crystalline
      color superconductors or ``LOFF phases'', named after 
      Larkin, Ovchinnikov, Fulde, and Ferrell who suggested this
      kind of pairing for electromagnetic superconductors~\cite{LO64,FF64}.
      In the QCD context this has been discussed first in Ref.~\cite{ABR01}.
      The essential feature of LOFF pairing is that the total momentum
      of a pair does not vanish, i.e., the momenta of the paired
      fermions are not opposite to each other.
      This has the advantage that both fermions can stay in the
      vicinity of their respective Fermi surfaces, even for rather
      different Fermi momenta, i.e., the pair can be formed without
      cost of free energy. 
      On the other hand, LOFF pairing is in general disfavored against BCS
      pairing by phase space.

      As an example, consider two quarks of type $a$ and $b$ with momenta 
\beq
     \vec k_a = \vec q + \vec p \quad\text{and}\quad 
     \vec k_b = \vec q - \vec p~, 
\label{kqp}
\eeq
      forming a diquark pair with total momentum $2\vec q$. 
      In the most simple case $\vec q$ is constant and the same for all
      pairs in the condensate.
      If we now restrict both quarks to their Fermi surfaces,
      $|\vec k_a| = k_a^F$ and $|\vec k_b| = k_b^F$, it follows 
      that the possible vectors $\vec p$ lie on a circle which corresponds
      to the crossing of two spherical shells with radii $k_a^F$ and 
      $k_b^F$ whose centers are displaced by $2\vec q$.
      Hence, in contrast to the standard BCS case where pairing takes
      place in the vicinity of the two-dimensional Fermi surface,
      it is now restricted to the region close to this one-dimensional
      circle. This explains why for equal Fermi momenta BCS pairing is
      favored. Moreover, because of the lower dimensionality of the
      regime where non-interacting LOFF pairs can be created (or destroyed)
      without cost of free energy, there is no divergence in the gap
      equation which guarantees the existence of a non-vanishing gap
      parameter for arbitrarily weak attractive interactions.  
 
      Nevertheless, for certain values of the chemical potential
      difference, LOFF pairing can be more favored than BCS pairing
      or no pairing at all. 
      If the analysis is restricted to condensates with a single fixed value 
      $2\vec q$ of the pair momentum (but $|\vec q|$ being a parameter
      which is varied to minimize the free energy for given chemical 
      potentials), this regime turns out to be very small~\cite{ABR01}.
      Note that in this case, rather than forming a real crystal,
      the gap function takes the form of a plane wave,
\beq
      \ave{q(\vec x)^T {\cal O}\, q(\vec x)} 
      \sim \Delta\, e^{2i\vec q\cdot \vec x}~.
\eeq  
      It is obvious that allowing for more complex structures could
      enlarge the available phase space, rendering LOFF phases
      more favorable. This has been analyzed by Bowers and 
      Rajagopal~\cite{BR02} within a Ginzburg-Landau approach. In this
      analysis, the authors considered superpositions of various
      plane waves, still with a single value of $|\vec q|$ but different 
      directions. It was found that crystalline phases can compete 
      with BCS pairing over a wide range of chemical potentials,
      with a face-centered cubic structure being most favored.  
      These findings led the authors to speculate that for neutral
      matter the regime between hadronic and CFL phase could completely 
      be occupied by a crystalline phase, with no 2SC phase at 
      all~\cite{BR02b}. The lower left phase diagram in 
      \fig{figschemphase} was inspired by this idea.

      It should be noted, however, that, in spite of being rather involved,
      the analysis of Ref.~\cite{BR02} has only been performed within a 
      simple toy model with two massless flavors. This leaves room for
      further investigations, in particular about mass effects and the 
      role of the strange quarks. Certainly, an NJL-model analysis of these 
      questions would be very interesting, although rather complicated.       
      In this context our analysis of mixed phases could be instructive.
      Since mixed phases are necessarily related to non-uniform structures
      in space and sometimes even to certain geometries, like rods or slabs,
      they might be viewed as some form of ``crystals'' as well. 
      Similar to the large number of different mixed phases which we have
      found in \sect{neutmix} we could thus imagine that the phase diagram
      contains a series of different crystal structures if masses and
      strange quarks are taken into account. 

      Of course, the mixed phases in \sect{neutmix} have been constructed
      under the simplifying assumption that each component is homogeneous 
      and infinitely extended. Obviously, this is only
      justified if the structures are much larger than the average distance
      between the quarks. 
      Unfortunately, we can not say much about the structure sizes in the 
      mixed phases as long as we do not know the surface tension.
      To get a rough idea we inspect the ``lasagna'' solution ($d=1$)
      of the example discussed in the right panel (dotted line) and take
      $\sigma = 10$~MeV/fm$^2$. For this case we find $r_0 = 0.8$~fm,
      meaning that the thickness of the $SC_{us+ds}$ slabs is 1.6~fm,
      while that of the $2SC$ slabs is 5.4~fm. (For $d=2$ or $d=3$ we get 
      somewhat larger structures, $r_0 = 1.8$~fm and $r_0 = 2.6$~fm, 
      respectively.) On the other hand, the average distance between the 
      quarks is 0.7~fm in the $SC_{us+ds}$ component and 0.8~fm in the 
      $2SC$ component. 
      This means, the geometric sizes are only a few times larger than  
      than the inter-quark distances  
      and our treatment of the mixed phases as consisting of infinite
      homogeneous components seems to be inappropriate.

      It is interesting that the sizes in the above example are almost 
      the same as in the LOFF phases. Taking again the $d=1$ solution 
      with  $\sigma = 10$~MeV/fm$^2$ we have a period of 7.0~fm. 
      For a LOFF phase described by a single plane wave, the authors 
      of Refs.~\cite{ABR01,BR02} find $|\vec q| \approx 1.2\,\delta\mu$,
      where $\delta\mu = \mu_Q/2$ (cf. \eq{mudef}). For the actual 
      value $\mu_Q \approx 155$~MeV, this translates into a periodicity 
      $L = \pi/(0.6 \mu_Q) = 6.7$~fm, i.e., almost the same as in the
      mixed phase. Since the latter depends on our (rather arbitrary)
      choice of the string tension, this coincidence might be completely
      accidental. On the other hand, it could also be taken as an
      indication that a small string tension of  $10$~MeV/fm$^2$ is not
      too unrealistic. 
       
      Before proceeding to the next point, we would like to mention
      that some authors have also investigated the possibility of
      particle-hole pairing with non-vanishing total momentum
      (``Overhauser pairing'')~\cite{Over,ShSo00,PRWZ00,RSZ01}.
      In weak coupling, it was found that this pairing scheme can
      compete with BCS only for an extremely large number of colors
      ($N_c \gtrsim 1000$)~\cite{ShSo00,PRWZ00}, but it could become
      competitive in the non-perturbative regime~\cite{RSZ01}.

\subsection{Gapless color superconductors}

      We have already mentioned the gapless color superconducting phases.
      Partially following an old idea of Sarma~\cite{Sa63}, this 
      has first been discussed by Shovkovy and Huang~\cite{ShHu03,HuSh03}
      in the context of two massless quark flavors.
      This phase does not correspond to an entirely new pairing pattern,
      but rather to the fact that in a certain interval of $\mu_Q$ 
      the 2SC gap equation has two branches of solutions. 
      The first branch is the continuation of the standard BCS solution
      at $\mu_Q=0$. Here $\Delta > |\mu_Q|/2$ and the 
      quasiparticle spectrum contains four gapped modes. 
      At $T=0$ the densities of the paired quark species are equal,
      as given in \eq{dens2sc}.
      For $\mu_Q \neq 0$, the four gapped modes are no longer completely
      degenerate, but they split into pairs of two with gaps
      $\Delta_\pm := \Delta \pm |\mu_Q|/2$. 
      In the second branch of solutions $\Delta < |\mu_Q|/2$. 
      Although, as before, four quark species participate in the
      condensate, only two of the corresponding dispersion laws are
      gapped, whereas the other two are not. This is the reason why
      this phase is called ``gapless 2SC'' (g2SC) 
      phase\footnote{However, it is possible that the gapless modes 
      experience another Cooper instability and, e.g., form a 
      secondary spin-1 condensate~\cite{HuSh03}.}. 
      Here the number densities of the paired quarks are not equal.

      Obviously, the g2SC solutions do not satisfy the stability
      criterion, \eq{stablegap}. In fact, at fixed chemical potentials
      this branch corresponds to local maxima of the thermodynamic
      potential and is indeed unstable.
      The important result of Refs.~\cite{ShHu03,HuSh03} is that this can
      change if local neutrality constraints $n_Q = n_8= 0$ are imposed,  
      instead of keeping $\mu_Q$ and $\mu_8$ at fixed values. In this case
      it depends on the coupling strength in the scalar diquark channel 
      whether the standard 2SC solution (strong coupling), the
      g2SC solution (intermediate) or the normal conducting solution
      (weak) is favored at given quark number chemical potential.  
      In their model calculation, the authors of Refs.~\cite{ShHu03,HuSh03} 
      find that the g2SC solution is favored at $\mu=400$~MeV if (in our 
      language) $0.7 \lesssim H:G  \lesssim 0.8$. 
      Although this is a relatively small interval, it is just centered
      by the ``standard'' value $H:G = 0.75$ and could therefore be
      quite relevant.
      A similar pairing pattern for the CFL phase with massive strange
      quarks has been discussed in Ref.~\cite{AKR04}.
 
      Very recently, Huang and Shovkovy have calculated the Meissner masses
      of the gluons in the g2SC phase~\cite{HuSh04}. It turned out that
      some of them become imaginary. While this seems to signal some kind of 
      instability, the final interpretation is not yet clear.

\subsection{Breached color superconductors}
        
      A similar mechanism has also been suggested for pairing
      one light and one heavy flavor with different Fermi 
      momenta~\cite{GLW03,LW02}. 
      For example, consider a system of ultra-relativistic up quarks
      and non-relativistic strange quarks with Fermi momenta
      $p_F^u > p_F^s$ and linearize the dispersion laws of the
      non-interacting particles near their Fermi surfaces, 
\beq
      E_-^i(p) \simeq V^i (p-p_F^i)~, \qquad  
      V^i =\frac{p_F^i}{\sqrt{{p_F^i}^2 + {M_F^i}^2}}~.
\eeq
      Obviously, $V^s \ll V^u \simeq 1$, which means that increasing
      the momentum of a strange quark is relatively inexpensive.
      In the presence of attractive interactions it could therefore
      be favorable to promote the strange quarks to pair with the
      up quarks around the up-quark Fermi surface, rather than deforming
      both Fermi surfaces. The result is also a gapless (``breached'')
      color superconductor with a similar characteristics as
      discussed above~\cite{GLW03}. Again, this scenario was found to
      be unstable for fixed chemical potentials, but it could be stable
      for fixed total or relative particle densities~\cite{GLW03}.
      Of course, in the thermodynamic limit this apparent difference
      between canonical and grand canonical treatment can only hold
      if a phase separation is again inhibited by long-range forces. 
      (See also Ref.~\cite{FGLW04} for a recent discussion of the stability
      problem.)

      The ``interior gap'' which has been discussed in Ref.~\cite{LW02} 
      is basically the same mechanism for the opposite case where the Fermi 
      momentum of the heavier species is larger than that of the lighter.
      Although this case is rather unlikely to play a role in quark matter,
      we have seen in the context of the ``exotic'' phase SC$_{us+ds}$
      that unusual orderings of the Fermi momenta should not be excluded
      completely.

\subsection{Deformed Fermi spheres}

      Yet another possibility to support the pairing of two quark species
      with different Fermi momenta is to deform their Fermi surfaces in
      an anisotropic way. Recently, this has been studied in 
      Ref.~\cite{MuSe03}. To that end, the authors considered a two-flavor
      system with an excess of down quarks.
      In this case the overlap of the two Fermi surfaces can be enhanced
      if the up- and down-quark Fermi spheres are deformed into oblate 
      and prolate ellipsoids, respectively.  
      While favoring the pairing near the equators of the ellipsoids,
      the deformation costs of course kinetic energy.     
      Nevertheless, the authors of Ref.~\cite{MuSe03} found that 
      the BCS state becomes unstable with respect to spontaneous quadrupole 
      deformations already at small asymmetries.

\chapter{Application: Neutron stars with color superconducting quark cores}
\label{stars}

In this chapter we apply our NJL quark matter equation of state to
investigate the possible existence of deconfined quark matter in
compact stars. 

We have already seen in \ref{neutresultshom}
that absolutely stable strange quark matter is unlikely to
exist within this  model, even in the presence of color-flavor locking 
diquark condensates. This rules out the existence of pure quark stars
(``strange stars''), but hybrid stars, consisting of hadronic matter in
the outer parts and a quark matter core, are not a priori excluded.

For a quantitative examination of this possibility we need a realistic 
equation of state for the hadronic phase. Since this cannot be obtained within
an NJL mean-field calculation we adopt hadronic equations of state from
other models (\sect{starshadron}). Comparing these with the NJL equation of 
state for the quark phase, we construct the phase transitions and
finally employ the resulting hybrid equations of state to calculate the
structure of compact stars.

Whereas most models of hybrid stars 
are based on bag-model descriptions of the quark matter phase,
two early investigations which employed the NJL model 
have been performed in Refs.~\cite{SLS99} and \cite{SPL00}. 
It was found by both groups that quarks exist at most in a small mixed 
phase regime but not in a pure quark core. Similar to our findings in 
the context with SQM this could be traced back 
to the relatively large values for the effective bag constant and the 
effective strange quark mass.

However, these calculations did not include diquark condensates.
As we have seen in \sect{neutresultshom} these give an extra contribution
to the pressure and therefore work in favor of a quark phase.
Recently it was shown within a bag model~\cite{AlRe03} that this
could have sizeable consequences for the hadron-quark phase transition and 
for the properties of compact stars.
On the other hand, as pointed out before, bag-model calculations
miss possible effects of the density and phase dependence of quark masses
and bag constants. 
It is therefore interesting to perform a similar investigation based on
an NJL model equation of state where these effects can be studied.
This has been done in Refs.~\cite{BBBNOS,BNOS04}. Below, we discuss the
results.

\section{Hadron-quark phase transition}
\label{hqphasetr}

\subsection{Hadronic equations of state}
\label{starshadron}

To put our analysis on a relatively broad basis, we employ four different
hadronic equations of state.
 
Two of them, taken from Refs.~\cite{bbb} and \cite{bbs},
are microscopic equations of state 
based on the non-relativistic Brueckner--Bethe--Goldstone 
(BBG) many-body theory, treated  in Brueckner--Hartree--Fock (BHF)
approximation. It has been shown
that the non-relativistic BBG expansion is well convergent  \cite{song,BaBu01},
and  the BHF level of approximation is accurate
in the density range relevant for neutron stars.  

The first equation of state, ``BHF(N,l)'' contains only nucleons and 
non-interacting leptons
and has been derived in Ref.~\cite{bbb} using the Paris potential
\cite{lac80} as two-nucleon interaction and the Urbana model as
three-body force \cite{CPW83,schi}.  
The resulting nuclear matter equation of state fulfills several 
requirements~\cite{bbb},
namely (i) it reproduces the correct nuclear matter saturation point
$\rho_0$,
(ii) the incompressibility is compatible with the values extracted
from phenomenology, (iii) the symmetry energy is compatible with
nuclear phenomenology, (iv) the causality condition is always
fulfilled. 

The second equation of state, ``BHF(N,H,l)'', is a recent extention of 
the first one which contains also hyperon ($\Sigma^-$ and $\Lambda$) degrees 
of freedom~\cite{bbs}. This is physically motivated by the fact that the
baryon chemical potentials reached in the interiors of neutron stars
are likely to be large enough that these particle states are populated.
Obviously, to include these additional degrees of freedom, the bare
nucleon-hyperon  and hyperon-hyperon interactions are needed.
For the nucleon-hyperon interaction 
the Nijmegen soft-core model \cite{mae89} has been adopted in Ref.~\cite{bbs}.
Unfortunately, because of lacking experimental data,
the hyperon-hyperon interaction is practically unknown. Therefore,
as a first approximation, the authors of Ref.~\cite{bbs} have decided to 
neglect this interaction completely.

In order to estimate the level of uncertainty caused by this neglect
we employ two other hadronic equations of state which also contain
hyperons but which are not based on microscopic interactions.
One of them, ``RMF240'', has been derived within relativistic mean-field 
theory and is tabulated in Ref.~\cite{Glendenning}.
It also reproduces the correct nuclear matter saturation point
and yields reasonable values for the symmetry energy and the
incompressibility $K$. Here we adopt the parameter set with $K = 240$~MeV.

Finally, we take an equation of state of Ref.~\cite{xsu3} 
(hereafter ``$\chi$SU(3)'') which has been 
derived within a QCD motivated hadronic model with a non-linear realization
of chiral $SU(3)$. 

\begin{figure}
\begin{center}
\epsfig{file=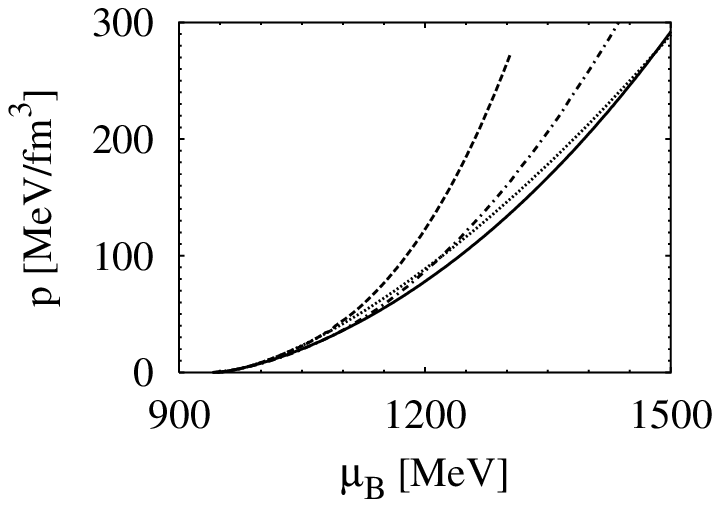,width=7.4cm}
\epsfig{file=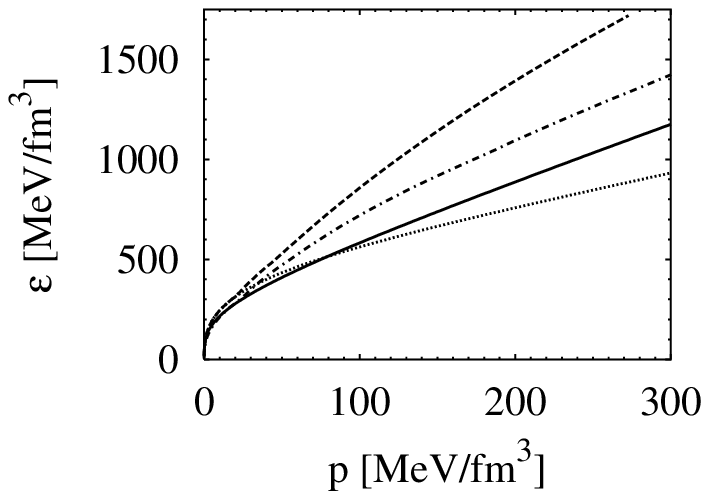,width=7.4cm}
\end{center}
\vspace{-0.5cm}
\caption{\small Properties of the hadronic equations of state employed
in our analysis: pressure as function of baryon chemical potential
(left) and energy density as function of pressure (right). 
The different lines correspond to the following equations of state
(see text): BHF(N,l)~\cite{bbb} (dotted), BHF(N,H,l)~\cite{bbs}
(dashed), $\chi$SU(3)~\cite{xsu3} (solid), and RMF240~\cite{Glendenning}
(dash-dotted).}
\label{figeosh}
\end{figure}

The key properties of the four hadronic equations of state are plotted
in \fig{figeosh}. In the left panel, the pressure is displayed as a 
function of the baryon chemical potential. 
The corresponding energy densities as functions of the pressure are
displayed in the right panel.
The curves correspond to neutral matter in beta
equilibrium. They agree fairly well at the lower end where the
equations of state are partially constrained by empirical data.
However, there are quite some differences if one moves up to higher 
densities, which may mainly reflect the uncertainties in the
hyperon sector. 

The two microscopic equations of state, (BHF(N,l) and BHF(N,H,l)),
are indicated by the dotted and the dashed lines, respectively.
They are of course identical below the hyperon threshold, but then
the presence of hyperons softens the equation of state considerably, 
which results in a steeper increase of pressure as a 
function of baryon chemical potential~\cite{bbs}.
It is therefore quite remarkable that the pressure predicted by the chiral
model $\chi$SU(3) (solid) remains even below the BHF(N,l) result up to
$\mu_B \simeq 1480$~MeV, although hyperons included. 
Finally, the relativistic mean-field equation of state RMF240 (dash-dotted)
plays some intermediate role, being softer than BHF(N,l) but still
considerably stiffer than BHF(N,H,l).

\subsection{Hadron-quark hybrid equation of state}
\label{hybrideos}

Having selected the hadronic equations of state we can now construct the
corresponding phase transitions to NJL quark matter.
Motivated by the conclusion of Ref.~\cite{ARRW01} that
a quark-hadron mixed phase is unlikely to be stable for reasonable 
values of the surface tension, and by the original conclusion of 
Ref.~\cite{NBO03} that this is also the case for quark-quark mixed phases
(see \sect{neutmixresults}), we restrict ourselves to analyze sharp phase
transitions from neutral hadronic matter to homogeneous neutral quark 
matter.

In order to study the influence of color superconductivity we consider
both, the equation of state without diquark condensates developed 
in \sect{sqm} and the equation of state with diquark condensates developed 
in \sect{neutresults}. Since the quark-quark coupling constant employed 
there, $H = G$, is likely to be an upper limit,
this could be considered as two extreme cases.

\begin{figure}
\begin{center}
\epsfig{file=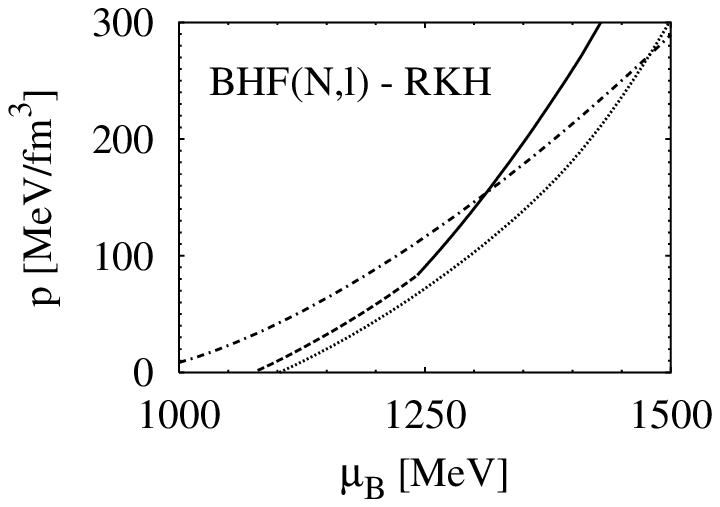,width=7.4cm}
\epsfig{file=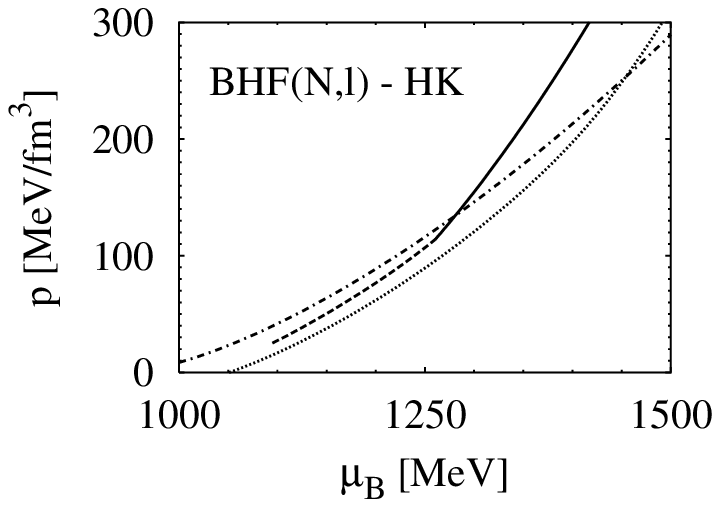,width=7.4cm}
\epsfig{file=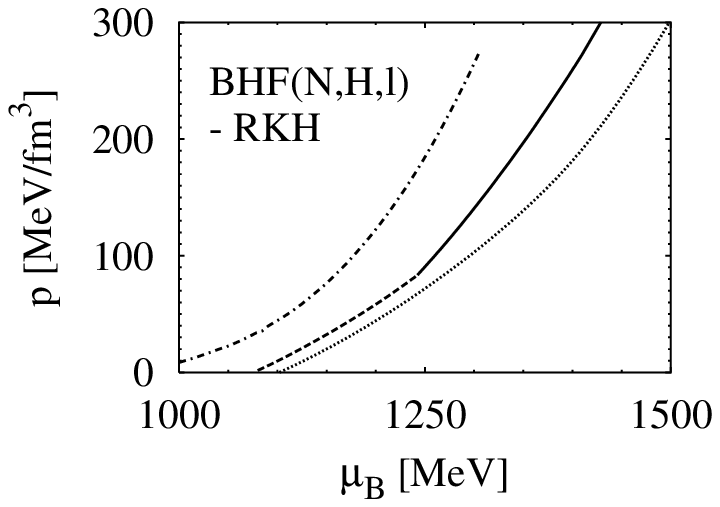,width=7.4cm}
\epsfig{file=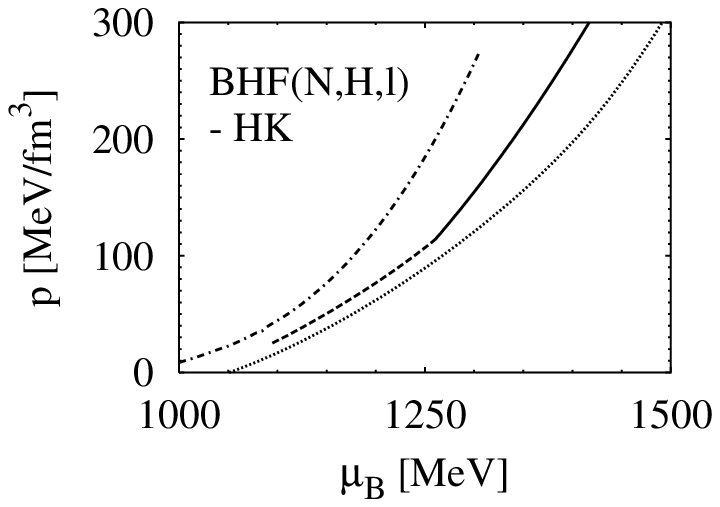,width=7.4cm}
\epsfig{file=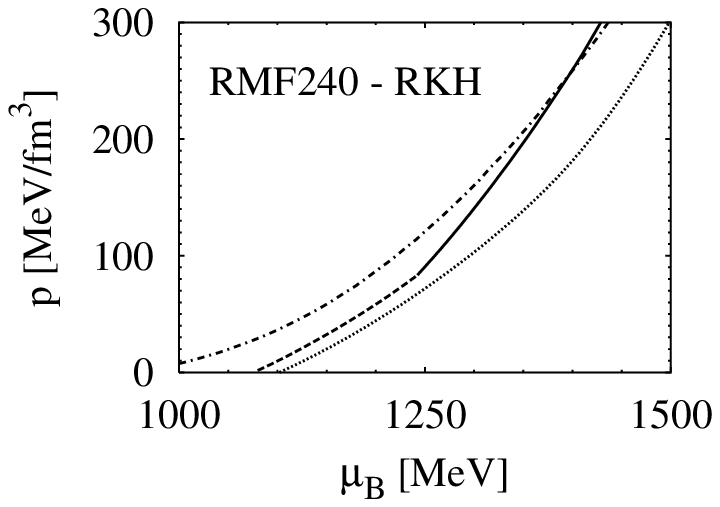,width=7.4cm}
\epsfig{file=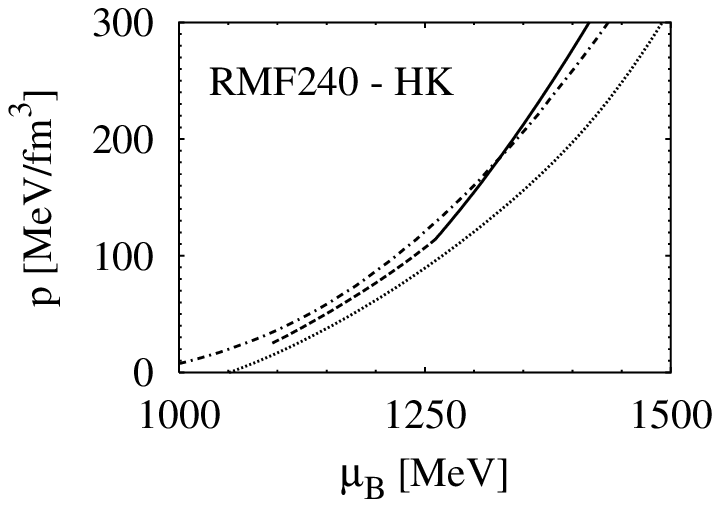,width=7.4cm}
\epsfig{file=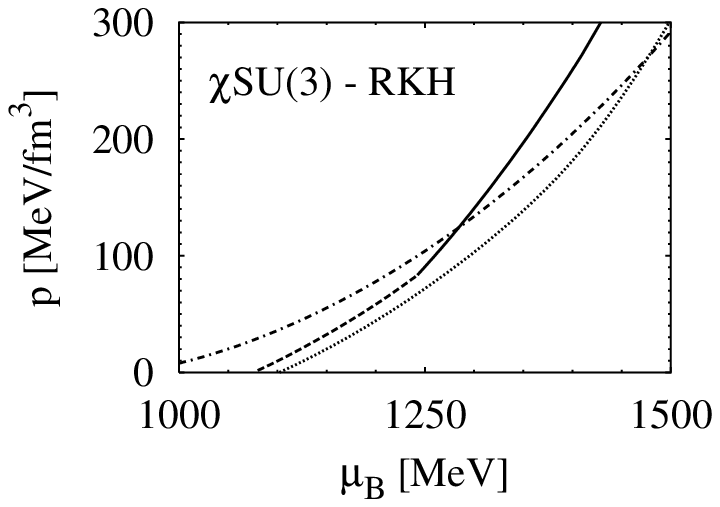,width=7.4cm}
\epsfig{file=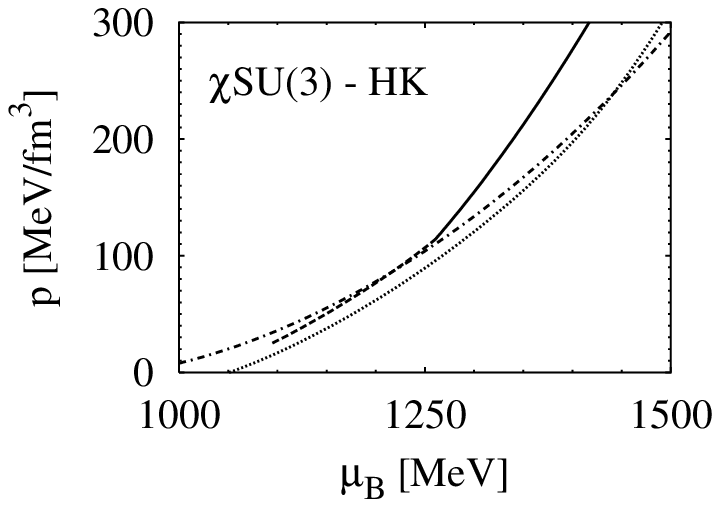,width=7.4cm}
\end{center}
\vspace{-0.5cm}
\caption{\small Pressure of neutral matter in beta
equilibrium as a function of baryon chemical potential. 
Dash-dotted lines: hadronic matter (BHF(N,l)~\cite{bbb}, 
BHF(N,H,l)~\cite{bbs}, RMF240~\cite{Glendenning}, $\chi$SU(3)~\cite{xsu3}).
The other lines correspond to NJL quark matter in the normal (dotted),
2SC (dashed) or CFL phase (solid), obtained with the parameter sets
RKH~\cite{Rehberg} (left panels) or HK~\cite{hatsuda} (right panels).
The quark-quark coupling constant was taken to be $H=G$.
The results of the upper three figures on the left have been presented in
a different form in Ref.~\cite{BBBNOS},
the two figures at the bottom have been adapted from Ref.~\cite{BNOS04}.}
\label{figqhphase}
\end{figure}

In \fig{figqhphase} the pressure of the various hadronic and quark matter
equations of state is displayed as a function of the baryon chemical potential.
The dash-dotted lines correspond to the hadronic equations of state.
These are the same curves as shown in \fig{figeosh}. 
The other lines correspond to NJL quark matter in the normal phase (dotted),
or in a color superconducting phase. Here we have indicated the part which
belongs to the 2SC phase by a dashed line and the part which belongs
to the CFL phase by a solid line.
Since the two solutions cross each other at the 2SC-CFL transition point,
the slope of the pressure increases discontinuously at this point.
Physically this is related to a sudden increase of the baryon number density
due to the fact that the number of strange quarks jumps from almost zero 
in the 2SC phase to 1/3 of the total quark number in the CFL phase. 
In contrast, there is no such behavior in the normal quark matter phase
where the strange quarks come in smoothly. (As we have seen earlier,
this can be different if the flavor mixing is weak.)

For given hadronic and quark equations of state, the hadron-quark phase 
transition point is now easily read off as the point of equal pressure, 
i.e., the point where the lines $p(\mu_B)$ cross. 
The results are summarized in \tab{tabletransqh}.
We begin our discussion with the four left panels of \fig{figqhphase}
where the quark equations of state are based on our ``standard''
parameter set RKH~\cite{Rehberg} which we have employed in most
calculations presented in this work.
In fact, the solid, dashed, and dotted lines are the same as the
solid, dashed, and dotted lines in \fig{figmuphom}, although displayed
on different scales. (Note that $\mu_B = 3\mu$.)

Depending on the hadronic equation of state, we find three different
cases: 
For the BHF calculation without hyperons (BHF(N,l))
and for the $\chi$SU(3) model we find a hadron-quark phase 
transition for both, normal and color superconducting quark matter.
Clearly, the presence of diquark condensates can lower
the critical chemical potential substantially.
On the other hand, the BHF calculation which includes hyperons
(BHF(N,H,l)) does not cross with the quark matter equation of state
up to large density, even if color superconducting phases are included.
In that case there would be no phase transition to quark matter at all,
at least at densities relevant for neutron stars. 
Finally, for the relativistic mean-field equation of state (RMF240),
we find an intermediate situation where a phase transition only
takes place if diquark pairing is taken into account.

In spite of these differences, the above results lead to the general 
observation
that a hadron-quark phase transition only takes place if strange quarks
play a non-negligible role in the quark-matter phase.
In the normal conducting phase it only happens (if it happens at all)
at relatively large chemical potentials, well above the strange quark
threshold, $\mu_B^{th} \simeq 1300$~MeV.
At the respective critical potentials we find about 25\% strange quarks.
For color superconducting quark matter we find no 
phase transition to the 2SC phase but only to the CFL phase.
Obviously, this is related to the kink at the 2SC-CFL transition
point which strongly accelerates -- or even enables -- the phase transition
from the hadronic phase. As discussed above, this increased slope corresponds
to a higher density which is mainly due to the sudden appearance
of a large amount of strange quarks in the CFL phase. 
On the contrary, in the 2SC phase the introduction of color 
superconductivity practically does not change the slope of the pressure 
curve, but
only leads to a moderate shift which can be attributed 
to some additional binding caused by the formation of Cooper pairs.
Thus, looking at the four left panels of \fig{figqhphase} 
there seems to be little chance for the presence of a 2SC phase in
neutral strongly interacting matter. This was also our conclusion in
Ref.~\cite{BBBNOS}.

\begin{table}[t]
\begin{center}
\begin{tabular}{|l| c c c c c|}
\hline
&&&&&\\[-3mm]
transition
&  $\mu_B$ [MeV] & $\rho_B^{(1)}/\rho_0$ &  $\rho_B^{(2)}/\rho_0$
&  $\epsilon^{(1)}$ [MeV/fm$^3$]&  $\epsilon^{(2)}$ [MeV/fm$^3$]
\\
&&&&&\\[-3mm]
\hline
&&&&&\\[-3mm]
BHF(N,l) $\rightarrow$ RKH(N)                & 1478 & 4.6 & 7.9 & \pho 884 & 1707 \\
\phantom{BHF(N,l) $\rightarrow$} RKH(CFL)    & 1312 & 3.7 & 6.4 & \pho 672 & 1282 \\
&&&&&\\[-2.5mm]
\phantom{BHF(N,l) $\rightarrow$} HK(N)       & 1454 & 4.5 & 6.9 & \pho 852 & 1446 \\
\phantom{BHF(N,l) $\rightarrow$} HK(CFL)     & 1280 & 3.6 & 6.1 & \pho 634 & 1186 \\
\hline
&&&&&\\[-3mm]
RMF240 \, $\rightarrow$ RKH(CFL)                  & 1397 & 6.5 & 7.7 & 1279 & 1567\\
&&&&&\\[-2.5mm]
\phantom{RMF240 \, $\rightarrow$} HK(CFL)         & 1326 & 5.4 & 6.8 & 1039 & 1345\\
\hline
&&&&&\\[-3mm]
$\chi$SU(3) \quad$\rightarrow$ RKH(N)             & 1477 & 5.4 & 7.9 & 1092 & 1701\\
\phantom{$\chi$SU(3) \quad$\rightarrow$} RKH(CFL) & 1284 & 3.6 & 6.0 & \pho 657 & 1188 \\
&&&&&\\[-2.5mm]
\phantom{$\chi$SU(3) \quad$\rightarrow$} HK(N)    & 1441 & 5.1 & 6.5 & 1000 & 1353\\
\phantom{$\chi$SU(3) \quad$\rightarrow$} HK(2SC)  & 1216 & 3.0 & 3.4 & \pho 536 & \pho 627 \\
HK(2SC) \,$\rightarrow$ HK(CFL)                & 1260 & 4.0 & 5.7 & \pho 746 &1117 \\
\hline
\end{tabular}
\end{center}
\caption{\small Various quantities related to the phase transitions
identified in \fig{figqhphase} (10 hadron-quark phase transitions 
and the 2SC-CFL phase transition for parameter set HK):
critical baryon chemical potential, and baryon and energy densities
below (1) and above (2) the phase transition.}
\label{tabletransqh}
\end{table}  

It turns out, however, that this statement depends strongly on the vacuum
constituent mass of the non-strange quarks.
This is quite obvious if we neglect the binding energy. In this case,
the point of zero pressure is just given by $\mu_B = 3M_u^\mi{vac}$.
Hence, a reduction of $M_u^\mi{vac}$ by, say, 50~MeV would approximately
shift the quark matter curves in \fig{figqhphase} by 150~MeV to the left,
and quark matter would become competitive to hadronic matter.  
(Note that the shift due to the formation of Cooper pairs in the 2SC phase
is only about 25~MeV.)
In fact, the authors of Ref.~\cite{SHH03} have been able to construct
a hadron-quark phase transition within a color superconducting 
two-flavor NJL model using a parametrization with 
$M^\mi{vac} = 314$~MeV~\cite{HZC03}. (In that calculation the hadronic phase 
was described by the $\chi$SU(3) equation of state.) 
However, since they did not include any strange quarks in their model, 
it was not clear whether there still would be a window for a stable 2SC 
phase if one allows for a CFL phase as well.

In Ref.~\cite{BNOS04}, we have studied the role of the non-strange quark 
mass in this context, employing -- besides the parameters used above --
the NJL-model parameters of Hatsuda
and Kunihiro, i.e., parameter set HK of \tab{tabnjl3fit}~\cite{hatsuda}.
With these parameters one finds $M_u^\mi{vac} = 335.5$~MeV, instead of
367.7~MeV we had before, i.e., the reduction is rather moderate.
For the quark-quark coupling constant we take again $H=G$.
 
The resulting pressure curves are displayed in the right panels of
\fig{figqhphase}.
As expected, the quark matter curves are shifted to lower chemical potentials,
but in most cases this does not lead to a qualitative change of the
behavior. The only exception is found for the $\chi$SU(3) equation of state.
In this case we indeed get a phase transition into the 2SC phase before
that is replaced by the CFL phase at a somewhat higher chemical potential
(see \tab{tabletransqh} for details). Although the window is rather
small, it could make a qualitative difference for compact stars,
as we will discuss below.

\begin{figure}
\begin{center}
\epsfig{file=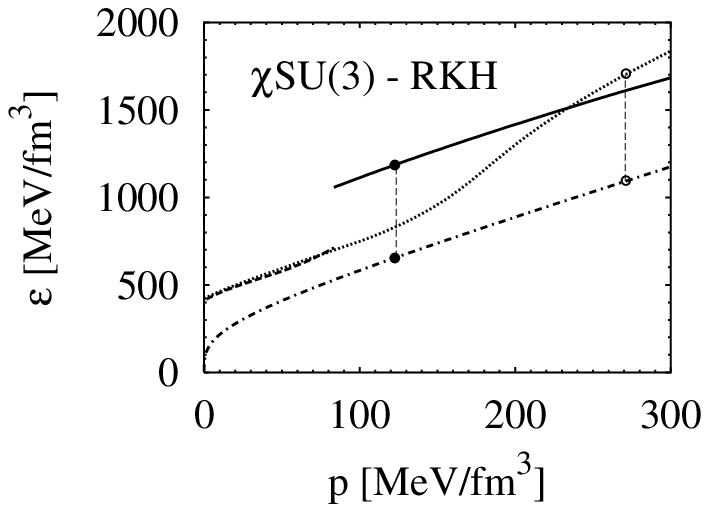,width=7.4cm}
\epsfig{file=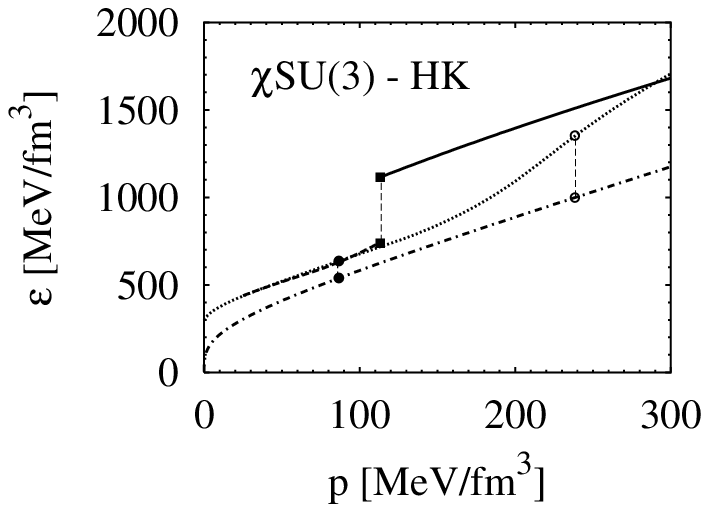,width=7.4cm}
\end{center}
\vspace{-0.5cm}
\caption{\small Energy density as a function of pressure for the hadronic
equation of state $\chi$SU(3)~\cite{xsu3} (dash-dotted lines) and 
NJL quark matter obtained with parameter sets RKH~\cite{Rehberg} 
(left panel) and HK~\cite{hatsuda} (right panel) and $H=G$:
normal phase (dotted), 2SC (dashed), CFL (solid). The points with the
thin vertical lines indicate the positions of the phase transitions.
Adapted from Ref.~\cite{BNOS04}.}
\label{figeoshq}
\end{figure}

To prepare for that discussion, let us have a look at the energy 
densities in the various phases. At the first-order phase boundaries
the density and, hence, the energy density increase discontinuously.
The corresponding values are listed in \tab{tabletransqh}. As one 
can see there, in most cases the discontinuity is rather large.
The only exception is the $\chi$SU(3)-2SC transition where the
energy density is only moderately increased.
This is further illustrated in \fig{figeoshq} where the energy density
is plotted as a function of pressure for the hadronic equation of state
$\chi$SU(3) and the various quark matter phases obtained with parameter
set RKH (left) and HK (right). 
The hadronic equation of state is again indicated by the dash-dotted lines,
while the dotted, dashed, and solid lines correspond to the normal,
2SC, and CFL phase, respectively. We have also indicated the positions
of the phase transitions. The open and closed circles mark the phase
transition from the hadronic phase to normal or color superconducting
quark matter, respectively, while the 2SC-CFL phase transition in the
right panel is indicated by the squares.
In the left panel, the 2SC-CFL phase transition is only ``virtually''
present because the hadronic phase is still favored in that region.
It is nevertheless an important effect, because it pushes the 
energy density in the quark phase to higher values at low pressure.
As a consequence, including color superconducting phases
in the NJL model is qualitatively different from the behavior in a 
bag model. This will be discussed below. 

\subsection{Comparison with bag model studies}
\label{stareosbag}

The effect of color superconductivity on the hadron-quark phase transition
and the structure of compact stars has been discussed in
Ref.~\cite{AlRe03} where the authors applied a bag model equation
of state to describe the quark phase. The corresponding contribution
to the pressure is given by~\cite{AlRe03}
\beq
    p_{BM}^{(CFL)}(\mu) \;=\; -\frac{6}{\pi^2} \int_0^\nu dp\, p^2\,(p-\mu)
         \;-\; \frac{3}{\pi^2} \int_0^\nu dp\, p^2\,(\sqrt{p^2+m_s^2}-\mu)
         \;+\; \frac{3\Delta^2\mu^2}{\pi^2} \;-\; B~,
\label{Omegabagcfl}
\eeq
where
\beq 
     \nu \;=\; 2\mu \;-\; \sqrt{\mu^2 + \frac{m_s^2}{3}}
\eeq 
is the common Fermi momentum which is related to the equal number densities
of the $u$, $d$, and $s$ quarks in the CFL phase,
$n_u = n_d = n_s = (\nu^3 + 2\Delta^2\mu)/\pi^2$.
These expressions are based on an expansion in $\Delta/\mu$, and the
masses of the up and down quarks have been neglected.
Like the bag constant $B$, the diquark gap $\Delta$ is treated as a
free parameter and is kept constant with varying $\mu$. 

Comparing \eq{Omegabagcfl} with the analogous expression for normal
quark matter in a bag model, \eq{OmegaBM}, one sees that the main effect
of the gap is the extra term $3\Delta^2\mu^2/\pi^2$.
It has been pointed out in Ref.~\cite{AlRe03} that, although this 
correction, being of order $\Delta^2\mu^2$, is formally small in comparison
with the total pressure (order $\mu^4$), it could have a strong impact
on the phase transition point. This is in agreement with our NJL model
results.

In the bag model the effect of the $\Delta^2$-term may be interpreted as an 
effective reduction of the bag constant, 
$B \rightarrow B - 3\Delta^2\mu^2/\pi^2$, enhancing the pressure and
reducing the energy density for a given chemical potential.
In other words, the function $\epsilon(p)$ is basically shifted downwards
and to the right by the diquark condensate, which both leads to a 
reduction of $\epsilon$ for a given $p$.
Obviously, this is quite different in the NJL model, as one can see in
\fig{figeoshq}. If we compare the solid line
with the dotted line we see that color superconductivity {\it enhances}
the energy density in a certain regime of pressure. The reason is again
the fact that the CFL phase always contains a large amount of (relatively
heavy) strange quarks which strongly contribute to the energy density,
whereas normal NJL quark matter contains no or very few
strange quarks up to much larger values of $p$.

In this context it is again useful to introduce an effective bag model
parametrization of the NJL model equation of state. 
To that end we insert the $\mu$ dependent values of the constituent
strange quark mass $M_s$ and the average diquark gap  
$\Delta = \sqrt{(\Delta_2^2 + \Delta_5^2 + \Delta_7^2)/3}$
(see \fig{figcondhom}) into \eq{Omegabagcfl} and
define an effective bag constant $B_\mi{eff}$
by equating the result with the pressure obtained in the NJL model. 
Alternatively, we may define an effective bag constant $B_\mi{eff}'$
in the analogous way, but using the current quark mass $m_s$ instead
of $M_s$\footnote{As pointed out in \sect{njlbag}, there is some arbitrariness
in defining effective bag constants. The above definitions of $B_\mi{eff}$
and $B_\mi{eff}'$ differ from those given in \eqs{Beff} and (\ref{Beffp})
by the fact that we now start from
the pressure as a function of the chemical potential instead of the 
energy density as a function of density. This is only a matter of convenience
but should not change the general picture. We repeat that we introduce
effective bag constants only for the {\it interpretation} of our results,
which are obtained in a well-defined way within the NJL model.}.

\begin{figure}
\begin{center}
\epsfig{file=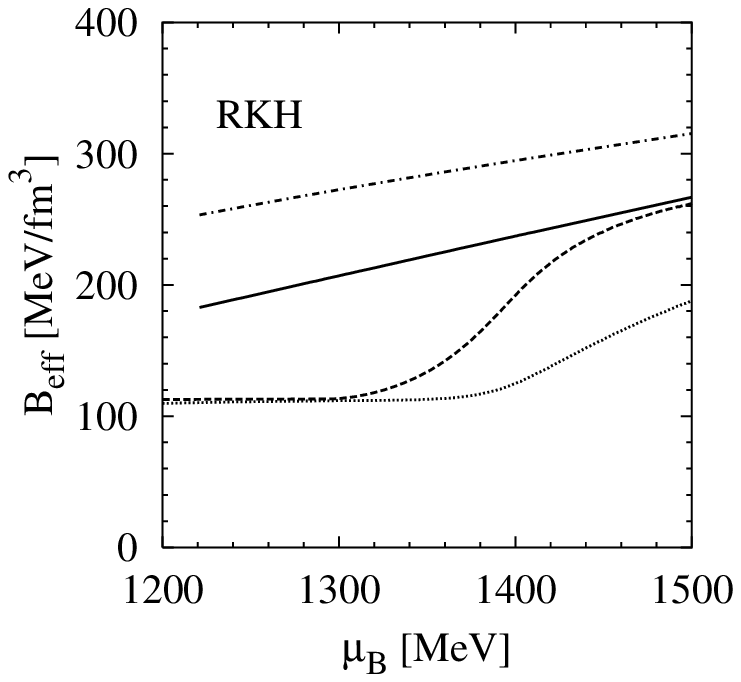,width=7.cm}
\epsfig{file=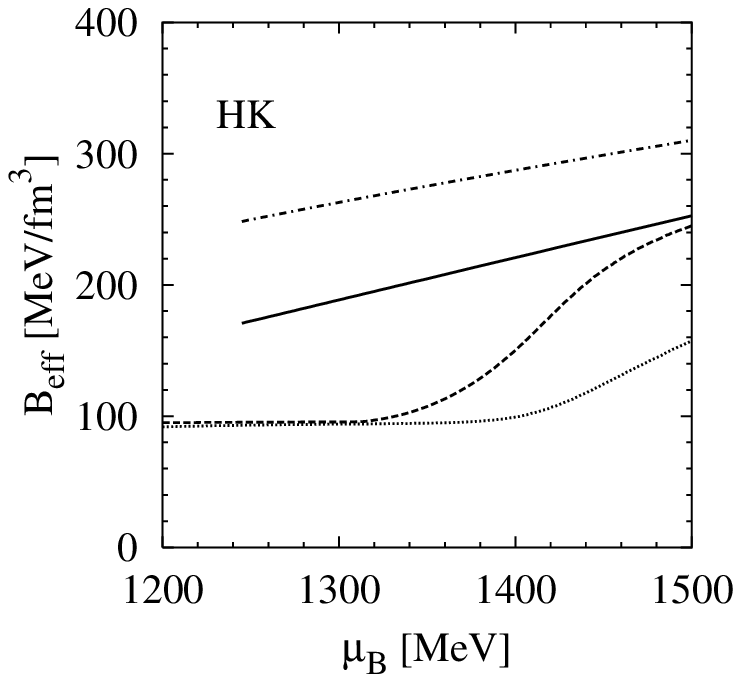,width=7.cm}
\end{center}
\vspace{-0.5cm}
\caption{\small Effective bag constants as functions of the 
baryon number chemical potential $\mu_B$:
$B_\mi{eff}$ (solid) and $B_\mi{eff}'$ (dash-dotted) 
for homogeneous neutral CFL matter, and
$B_\mi{eff}$ (dotted) and $B_\mi{eff}'$ (dashed) 
for homogeneous neutral normal quark matter (dotted).
Left: Parameter set RKH~\cite{Rehberg} with $H=G$.
Right: Parameter set HK~\cite{hatsuda} with $H=G$.
}
\label{figbeffmu}
\end{figure}

$B_\mi{eff}$ and $B_\mi{eff}'$ are displayed as function of $\mu_B$ 
in \fig{figbeffmu} (solid and dash-dotted lines, respectively). 
For comparison we also show the results for the normal quark phase 
(dotted and dashed).
In all cases the effective bag constants grow with $\mu_B$, and they are
always larger in the CFL phase than in normal quark matter.
Thus, although in the CFL phase the pressure is larger than
in the normal phase (see \fig{figmuphom}), the enhancement is smaller than 
one would naively expect from \eq{Omegabagcfl} if $B$ is kept constant. 

For the chemical potentials corresponding to the phase transitions
$M_s$, $B_\mi{eff}$  and $B_\mi{eff}'$ are listed in \tab{tabletransbag}.
These values are relatively large compared with ``typical'' bag model
parameters which are used in the context of neutron stars, e.g.,
Ref.~\cite{AlRe03}.
As a consequence, the energy densities are quite large as well.
As we will see below, this has important implications for the structure 
of compact stars.

\begin{table}[t]
\begin{center}
\begin{tabular}{|l| c c c c |}
\hline
&&&&\\[-3mm]
transition
&  $\Delta$ [MeV] &  $M_s$ [MeV] &  $B_\mi{eff}$ [MeV/fm$^3$] & 
$B_\mi{eff}'$ [MeV/fm$^3$] \\
&&&&\\[-3mm]
\hline
&&&&\\[-3mm]
BHF(N,l) $\rightarrow$ RKH(N)                & --- & 283 & 176 & 254 \\
\phantom{BHF(N,l) $\rightarrow$} RKH(CFL)    & 115 & 261 & 211 & 275 \\
&&&&\\[-2.5mm]
\phantom{BHF(N,l) $\rightarrow$} HK(N)       & --- & 346 & 127 & 215 \\
\phantom{BHF(N,l) $\rightarrow$} HK(CFL)     & 114 & 284 & 182 & 258 \\  
\hline
&&&&\\[-3mm]
RMF240 \, $\rightarrow$ RKH(CFL)             & 120 & 236 & 236 & 294 \\
&&&&\\[-2.5mm]
\phantom{RMF240 \, $\rightarrow$} HK(CFL)    & 118 & 267 & 197 & 269 \\
\hline
&&&&\\[-3mm]
$\chi$SU(3) \quad$\rightarrow$ RKH(N)             & --- & 284 & 175 & 254 \\  
\phantom{$\chi$SU(3) \quad$\rightarrow$} RKH(CFL) & 113 & 271 & 202 & 269 \\
&&&&\\[-2.5mm]
\phantom{$\chi$SU(3) \quad$\rightarrow$} HK(N)    & --- & 365 & 118 & 202 \\
HK(2SC) \,$\rightarrow$ HK(CFL)                   & 111 & 292 & 175 & 252 \\  
\hline
\end{tabular}
\end{center}
\caption{\small Average diquark gap 
$\Delta= \sqrt{(\Delta_2^2 + \Delta_5^2 + \Delta_7^2)/3}$,
constituent strange quark mass, and effective bag constants
in the strange quark matter phase at the phase transition points.}
\label{tabletransbag}
\end{table}

\section{Neutron star structure}
\label{starsstars}

The mass of a static compact star as a function of its
radius can be obtained by solving the Tolman-Oppenheimer-Volkoff 
equation~\cite{TOV} which reads
\beq
    \frac{d p}{d r} \;=\; 
    - \frac{[p(r) + \epsilon(r)][M(r) + 4\pi r^3 p(r)]}{r [r-2M(r)]}~.
\label{tov}
\eeq 
Here $p$ and $\epsilon$ are pressure and energy density, as before,
and
\beq
    M(r) \;=\; 4\pi \int_0^r {r'}^2\, dr'\, \epsilon(r')
\eeq
is the integrated energy inside a ``sphere'' of radius $r$ (in
Schwarzschild metric). We have used gravitational units, 
$G = c = 1$ ($G$ gravitational constant).

The radius $R$ and the gravitational mass $M_G$ of the star are 
given by the value of $r$ for which the pressure vanishes
and the corresponding value of $M(r)$, i.e., 
\beq
    p(R) = 0~, \quad M_G = M(R)~.
\eeq
For a given equation of state, $\epsilon = \epsilon(p)$, \eq{tov} can easily 
be integrated numerically, starting from the center of the star and moving 
outwards. In this way, if one continously varies the central 
pressure $p_c := p(0)$,
one obtains a curve $M_G(R)$ which relates masses and radii for that equation 
of state.  Note that stable branches of these curves have to fulfill the 
condition $d M_G/dp_c > 0$. Otherwise the solutions are unstable against small 
radial oscillations and collapse. 

Since we are mostly interested in 
quark-hadron hybrid stars we restrict our analysis to those hadronic 
equations of state for which we have found a transition to a quark phase in
\sect{hybrideos}. This was not the case for the microscopic equation of
state with hyperons, BHF(N,H,l).
Neutron star properties for this equation of state have been 
discussed in Ref.~\cite{bbs}. The authors have found a very low maximum
mass, $M_G^\mi{max} = 1.25 M_\odot$. Hence, since neutron star masses of 
$1.44 M_\odot$ have been observed~\cite{hulse,ThCh98},
it is clear that this equation of state cannot describe the whole interior of 
neutron stars correctly, i.e., it is probably too soft at high 
densities\footnote{Alternatively, this problem could in principle be cured
if a phase transition to quark matter takes place before the star becomes
unstable, provided the corresponding hybrid equation of state supports 
higher masses~\cite{BBSS02,MBBS04}.}. 

\begin{figure}
\begin{center}
\epsfig{file=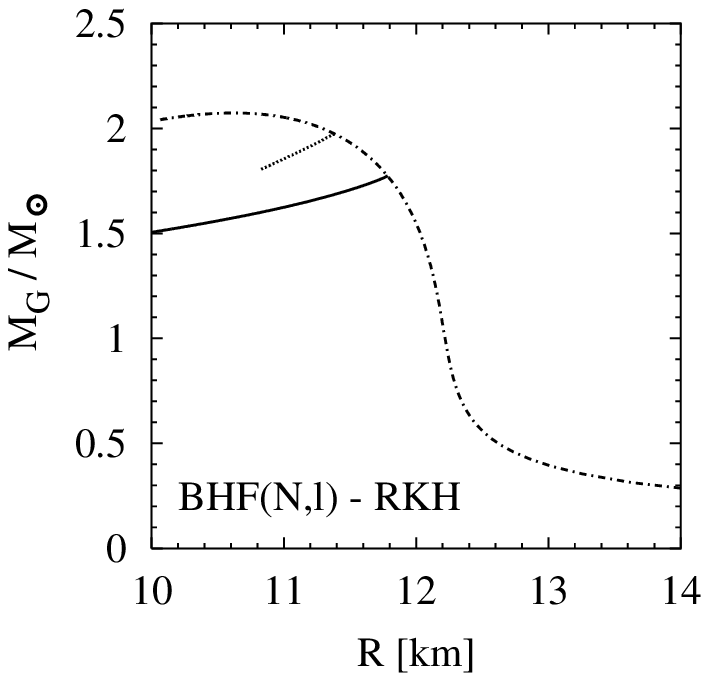,width=7.4cm}
\epsfig{file=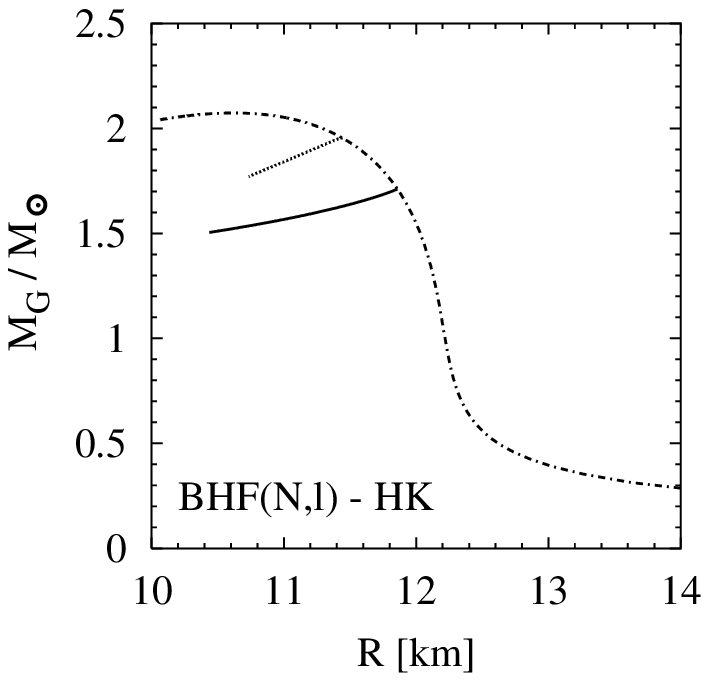,width=7.4cm}
\epsfig{file=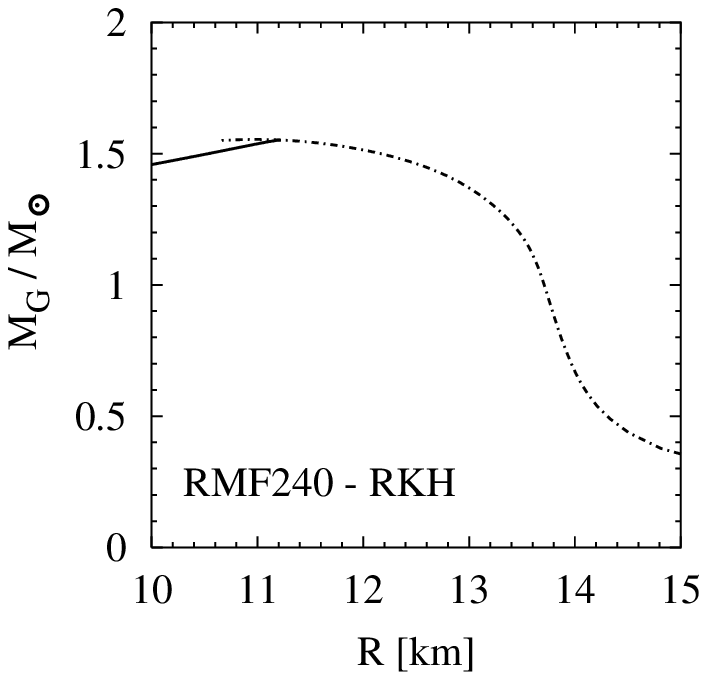,width=7.4cm}
\epsfig{file=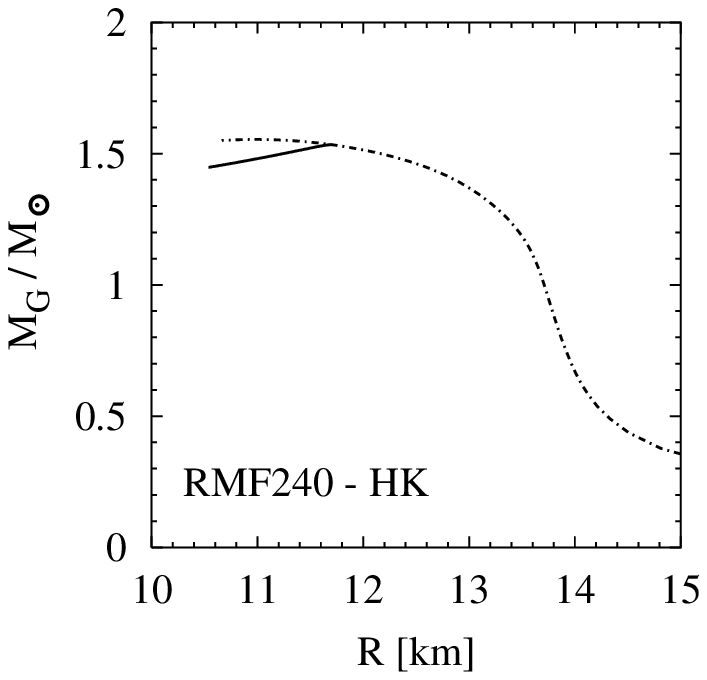,width=7.4cm}
\epsfig{file=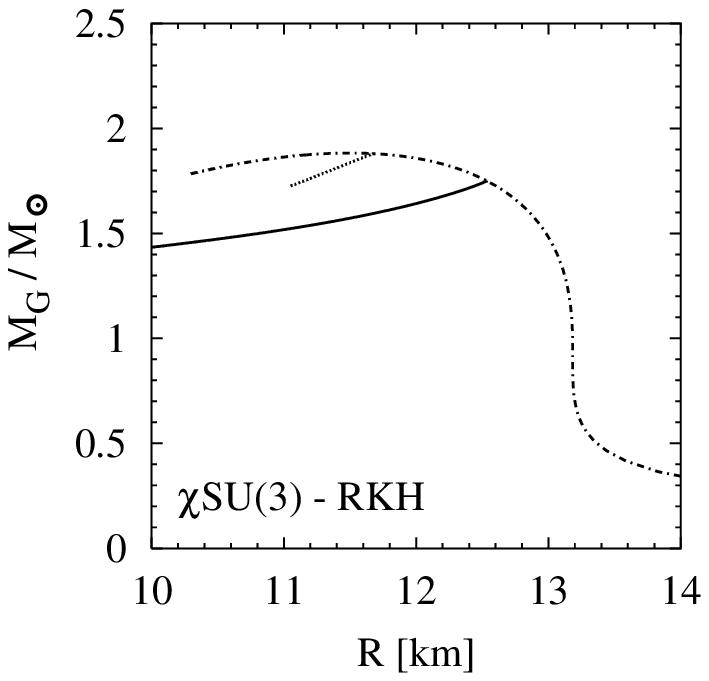,width=7.4cm}
\epsfig{file=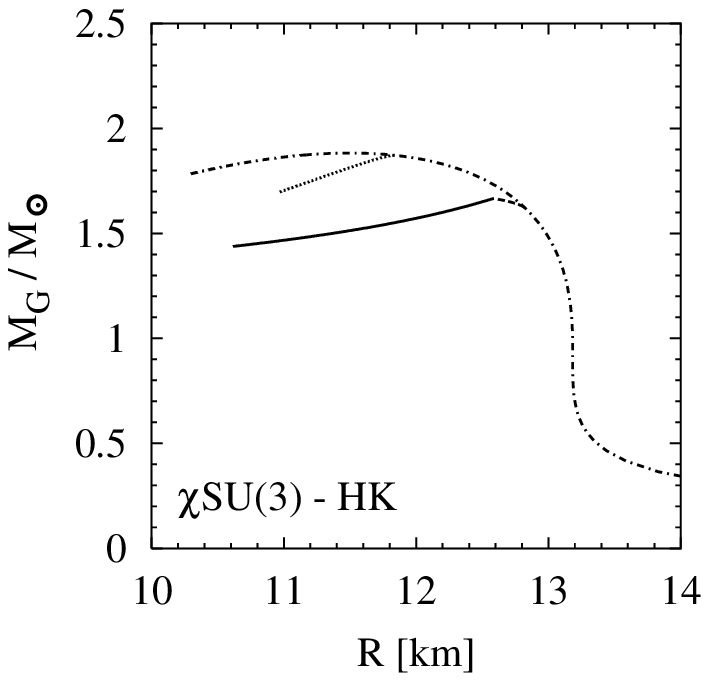,width=7.4cm}
\end{center}
\vspace{-0.5cm}
\caption{\small Gravitational masses of static compact stars as functions 
of the radius for the different equations of state. The dash-dotted lines
indicate the results for a purely hadronic star: BHF(N,l) (upper line),
RMF240 (central line), $\chi$SU(3) (lower line). The other lines indicate
the presence of a quark phase in the center: normal phase (dotted),
2SC dashed, CFL (solid). The quark phases in the three left panels have
been calculated with parameter set RKH with $H=G$,
those in the right panels with parameter set HK with $H=G$.
The results of the upper two figures on the left have been presented in
a different form in Ref.~\cite{BBBNOS},
the two figures at the bottom have been adapted from Ref.~\cite{BNOS04}.}
\label{figtov}
\end{figure}

\begin{figure}
\begin{center}
\epsfig{file=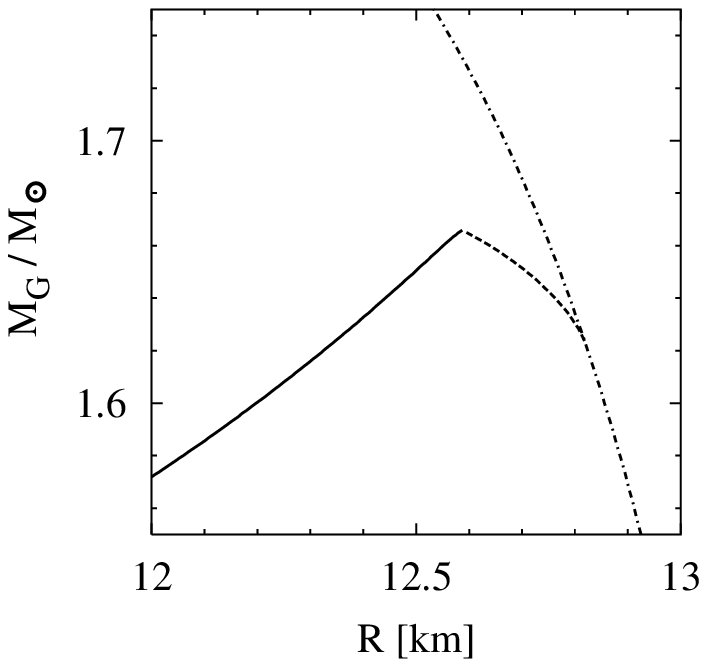,width=7.4cm}
\end{center}
\vspace{-0.5cm}
\caption{\small Enlarged detail of the lower right panel of \fig{figtov},
showing the emergence of a color superconducting quark core in a compact
star calculated with a $\chi$SU(3)-HK hybrid equation of state. 
The meaning of the various line types is the same as in \fig{figtov}.
Taken from Ref.~\cite{BNOS04}.
}
\label{figtov2}
\end{figure}

The curves $M_G(R)$ resulting for the three other hadronic and the related
hybrid equations of state are displayed in \fig{figtov}. 
The dash-dotted lines indicate the results for a purely hadronic star
(upper line: BHF(N,l), central line: RMF240, lower line: $\chi$SU(3)).
For a more realistic description of the crust, i.e., the region of 
subnuclear matter densities, we have employed the equations of state
of Baym, Pethick, and Sutherland~\cite{BPS71} for $\rho_B<0.001\,\fm^{-3}$
and of Negele and Vautherin~\cite{NV73} for 
$0.001\,\fm^{-3} <\rho_B <0.08\,\fm^{-3}$.
As typical for non-selfbound objects, large radii correspond to small
gravitational masses and thus small central pressures. Therefore with
increasing central pressure we have to follow the dash-dotted lines
from right to left. Eventually, the central pressure is large enough
that a phase transition to a quark phase takes place. The resulting
$M_G(R)$ above this point are indicated by dotted lines for a transition
to normal quark matter and by a solid line for a transition to quark 
matter in the CFL phase. 
As we have seen in \sect{hybrideos}, for the $\chi$SU(3) equation of
state in combination with the HK quark equation of state
(lower right panel), there is a phase transition from hadronic matter to 
the 2SC phase, followed by a second phase transition to the CFL phase.
Here we have indicated the part of the curve which corresponds to a
2SC phase (but not yet a CFL phase) in the center of the star by a
dashed line. This part of the figure is also presented in an enlarged
form in \fig{figtov2}.

The onset of a quark matter phase in the center of the star implies
of course a deviation from the corresponding hadronic matter curve. 
Because of the discontinuous energy density at the phase transition 
point (see \tab{tabletransqh} and \fig{figeoshq}) this is always related 
to a cusp in the mass-radius relation.
These cusps are clearly visible in Figs.~\ref{figtov} and \ref{figtov2}.
It turns out, however, that for all transitions to normal or to CFL quark matter 
the effect is so strong that the star is rendered unstable.
Only in the hadron-2SC phase transition (\fig{figtov2}) the star remains
stable but becomes unstable at the subsequent 2SC-CFL phase transition. 
Hence, if the correct equation of state lies in the range covered by our
model equations of state, hybrid stars, if they exist, should have 
a quark core in the 2SC phase and contain only a small 
fraction of strange quarks:
There is no stable star with a CFL or a normal quark matter core.   
This is at amusing variance with Ref.~\cite{AR02} who argued
that there is no 2SC phase in compact stars.

The essential question is thus, whether or not the equation of state
contains an interval where the 2SC phase is favored.
As we have demonstrated, the answer to this question can depend rather
sensitively on the value of the non-strange constituent quark mass in
vacuum. However, the over-all picture which emerges from our results is
that a stable 2SC phase and, hence, a stable hybrid star appear to be 
rather unlikely to exist. In all but one case we either found no hadron-quark
phase transition at all or a transition to normal or CFL matter, and
the star becomes unstable as soon as the phase transition occurs.
Only if we combine the very stiff $\chi$SU(3) equation of state
with the HK NJL model equation of state, hybrid stars can exist in a 
small mass window between 1.62 and 1.66~$M_\odot$.
Here we should also recall that we have chosen a relatively large
quark-quark coupling constant $H=G$. With a smaller coupling, 
there would be less binding in the 2SC phase and eventually the window
will shut. 
On the other hand, if we further decrease the constituent mass of the 
non-strange quarks, the window could become wider. Similarly,
a larger constituent mass of the strange quarks would render the CFL
phase less favored and thereby help stabilizing the 2SC phase. 
A systematic study in all these directions still remains to be done. 
This should also include the consideration of other color superconducting
phases which have not been taken into account here (see \sect{neutdisc}).

We should also note that we cannot exclude the existence of so-called
$3^{rd}$ family solutions of compact stars~\cite{Ger68} with pure quark 
matter cores.
This would be the case if the unstable branch ``recovers'' at some higher
density and rises again. Looking at our solutions, there seems to be some
tendency which points into this direction. Unfortunately, the densities
which are necessary to decide on this interesting point cannot be reached
in our model, since we would have to choose quark number chemical potentials
larger than the cut-off. 

In the above analysis we have assumed sharp phase transitions between homogeneous
neutral phases. In the case of a mixed phase the energy density would not jump,
but continously interpolate between the hadronic and the quark solution.
As a consequence, the cusps in \fig{tov} would be smoothed out and the
instability would not occur immediately at the onset of the mixed phase.
We expect, however, that in this case the star would become unstable 
before the mixed phase goes over into a pure quark phase if this is 
CFL or normal quark matter. 
Considering non-superconducting quark matter we thus confirm the
results of Ref.~\cite{SLS99}. Including color superconductivity in the 
CFL phase does not change these findings.  

Since this result is relatively insensitive to the choice of the hadronic
equation of state, it must mainly be attributed to the NJL-type
quark equation of state. 
In the bag-model investigation of Ref.~\cite{AlRe03}, the authors have 
employed a hadronic equation of state which is comparable to the
BHF equation of state without hyperons.
For a bag constant $B = 137$~MeV/fm$^3$ and a strange quark mass
$M_s = 200$~MeV they found that a star with a pure quark core is 
unstable without color superconductivity, but stable if a (CFL) diquark 
gap of 100~MeV is chosen. 
If we compare the above numbers with the effective quantities
listed in Table~\ref{tabletransbag}, we see that the essential differences
are again the relatively large values of the strange quark mass
and of the effective bag constant in the NJL model.
Therefore the energy densities just above the phase transition are considerably
larger in the NJL model than in the bag model, and this is finally responsible
for the instability\footnote{The authors of Ref.~\cite{AlRe03} have also
taken into account contributions of the ``pseudo''-Goldstone bosons of the 
broken chiral symmetry to the thermodynamic potential in the CFL phase.  
However, since in neutral CFL matter the electric charge chemical potential 
vanishes, no charged bosons are excited, while the contributions of a 
possible $K_0$-condensate are small~\cite{AlRe03}.}.

As pointed out many times before, our results rely on the
assumption that the NJL model parameters which have been fitted
in vacuum can be applied to dense matter. 
It is of course possible that this is not the case.
Thus our arguments could also be turned around:  
If there were strong hints, e.g. for the existence of a pure CFL quark
core in compact stars, this would indicate a considerable modification
of the effective NJL-type quark interactions in dense matter.
This observation could then be used to constrain the parameters
(provided the hadronic part is sufficiently well under control).

But even if there are no quark cores in compact stars and hence no
natural laboratories for color superconducting phases, the properties of
these phases can nevertheless influence the maximum mass of neutron
stars, as evident from \fig{figtov}.
The corresponding maximum masses are also listed in \tab{tablestars}.
Since the quark effects are larger if the 
phase transition occurs long before the would-be maximum mass of a
purely hadronic equation of state is reached,
the results become less sensitive to the hadronic equation of state
if quarks and in particular the effects of color superconductivity are 
included:
Whereas for the purely hadronic equations of state, we find
$M_G^\mi{max}$ between 1.55 and 2.07 solar masses, our results
scatter only between 1.53 and 1.77 solar masses if color superconducting
quark matter is included. In particular, the inclusion of color
superconducting quark matter keeps the neutron star maximum mass well
below two solar masses, independently of details of the hadronic
equation of state. Hence, the observation of a neutron star with a
mass well above two solar masses would seriously question our 
NJL-type description of the quark matter phase, using parameters which
are fixed in vacuum.

\begin{table}[t]
\begin{center}
\begin{tabular}{|l| c c c |}
\hline
equations of state  & \quad hadronic \quad & \quad hadr. + normal \quad 
& \quad hadr. + cs \quad \\ \hline 
&&&\\[-3mm]
BHF(N,l), RKH    & 2.07 & 1.97 & 1.77 \\
&&&\\[-4mm]
BHF(N,l), HK     &      & 1.96 & 1.71 \\
&&&\\[-2.5mm]
RMF240, RKH      & 1.55 & ---  & 1.55 \\
&&&\\[-4mm]
RMF240, HK       &      & ---  & 1.53 \\
&&&\\[-2.5mm]
$\chi$SU(3), RKH & 1.88 & 1.88 & 1.75 \\
&&&\\[-4mm]
$\chi$SU(3), HK  &      & 1.87 & 1.66 \\
\hline
\end{tabular}
\end{center}
\caption{\small Maximum gravitational mass $M_G^{max}$ (in terms of
         solar masses $M_\odot$) for a purely 
         hadronic equation of state and for hybrid equations of state
         with phase transitions to normal or color superconducting (cs)
         quark matter.}
\label{tablestars}
\end{table}  

\chapter{Summary and discussion}
\label{conclusions}

In this report we have explored the properties of deconfined quark matter,
focusing on the regime of low temperatures and ``moderate'' densities,
which cannot be accessed by perturbative or present-day lattice
calculations. 
Main issue was the investigation of color-superconducting phases
and the influence of dynamically generated effective quark masses on
these phases. To that end we employed NJL-type models, where both, 
diquark pairing and dynamical mass generation, can be treated on the 
same footing.
This turned out to be crucial, in particular for the understanding of
the transition from two-flavor (2SC) to three-flavor (CFL) 
color superconductors. We found that this transition is basically 
triggered by a strong discontinuous drop of the effective strange quark 
mass which has its roots in the corresponding chiral phase transition. 

To work this out more clearly, we started with a detailed study of the 
chiral phase transition for color non-supercon\-ducting quark matter.
A central point was the 
comparison of the NJL model with the MIT bag model, which is the
most commonly used model to describe the equation of state of
deconfined quark matter. 
In the NJL model, the bag constant arises dynamically as a consequence 
of spontaneous chiral symmetry breaking. 
Whereas for vanishing quark masses (and zero
temperature) the two models behave almost identically, differences
arise when finite current quark masses are introduced. 
Naturally, these effects become more pronounced in the three-flavor
case. Here we found that both, the effective strange quark mass
and the effective bag constant, are density dependent quantities
which are large compared with ``typical'' bag model parameters. 
One important consequence is that
the model does not support the hypothesis of absolutely
stable strange quark matter, as long as the model parameters do not 
differ drastically from a vacuum fit.

As pointed out above, the consideration of dynamical quark masses
remains crucial when we turn to color superconducting phases.
We also analyzed a spin-1 diquark condensate as a possible pairing
channel for those quarks which are left over from the standard
spin-0 condensate in the 2SC phase.
Another important question was the effect of neutrality constraints,
which must be imposed, e.g., to describe possible quark matter cores
of neutron stars. 
Generally, these constraints tend to disfavor the 2SC phase. 
Nevertheless, whereas estimates based on the assumption of small strange
quark masses predict that the 2SC phase is never the most favored neutral
phase, NJL model calculations reveal that large values of $M_s$ could
again stabilize the 2SC solution. 
In this context we have also investigated the possibility of electrically 
and color neutral mixed phases. Neglecting Coulomb and surface effects
we found a rich structure of mixed phases with up to four components. 
The gain in bulk free energy is, however, quite small. It is therefore
unclear whether the mixed phases survive if surface and Coulomb 
effects are included. 

Finally, we applied the NJL model to
study the possibility of color superconducting
quark matter cores in neutron stars.
To that end, we combined different hadronic equations of state 
with the NJL model ones to construct a sharp hadron-quark
phase transition. The resulting hybrid equations of state have then
been employed in a Tolman-Oppenheimer-Volkoff equation.
In most cases we found a phase transition from hadronic matter to the
CFL phase. It turned out that in these cases the
energy density in the quark phase is too large to support
a stable star with a quark matter core. This result could again be
traced back to the large effective strange quark masses.
So far we found a single example where the hadron-quark
phase transition went to the 2SC phase. In this case a stable 
quark matter core was possible.  
However, a more systematic investigation of the parameter dependence
of these results still needs to be done. 

In fact, for all topics we discussed in this work, there are
details which deserve further investigation. Many of them have been
discussed in the corresponding sections and will not be repeated here.
In the following we concentrate on the major questions.

Presumably the most severe limitation of our analysis is the problem
of parameter fixing, which has been addressed in \sect{interaction} and 
several other places.
At present, there is practically no other way than relying on vacuum
fits, which could be a rather bad assumption at large densities.
Therefore, most of our analysis must primarily be taken as
qualitative hints, e.g., about the role of the strange quark mass
or the effect of neutrality constraints.
In this situation, further improvements on the more fundamental side,
allowing to evolve the weak-coupling approaches down to lower densities,
are highly desirable.
To some extent we may also hope to learn something about the interaction
by comparing our predictions with observational facts. 
Unfortunately, since so far most of the predictions are negative ones,
e.g., the non-existence of strange quark matter or CFL matter cores
in neutron stars, this can at best be a falsification. 
This underlines ones more the necessity to work out more phenomenological
details. For instance data on neutron star cooling could provide severe 
constraints on the minimal pairing gaps in a possible quark-matter 
core~\cite{BVG04}. 

Returning to \fig{figschemphase} in the Introduction, we note that
we have mostly elaborated on the phases shown in the upper right diagram
(plus some extra phases which might appear if we add further axes to the
diagram, see \fig{figneutphasemuq}). 
On the other hand, phases like the CFL$+K$ phase or crystalline phases,
suggested in the two lower phase diagrams of \fig{figschemphase}, 
have not yet been studied in the framework of NJL-type models.
As pointed out in \sect{neutdisc}, this could be interesting,
since $\mu$-dependent quark masses could have important effects
on these phases as well. 
The description of crystalline phases is probably more difficult, 
whereas in order to study CFL$+K$ (or more general CFL$+$Goldstone) phases 
one only has to add further condensates with the corresponding quantum 
numbers. 

In general, the role and the properties of Goldstone bosons in the 
various color superconducting phases should be worked out in more
detail.
On the RPA-level, the Goldstone bosons can be constructed generalizing
the Bethe-Salpeter equation shown in \fig{fignjlrpa} to Nambu-Gorkov space. 
Because of baryon number non-conservation, mesons and diquarks can mix.
This is formally described by the presence of off-diagonal Nambu-Gorkov
components\footnote{A similar but technically simpler issue are 
possible precursor effects of color superconductivity at temperatures 
{\it above} $T_c$, which have been described in Ref.~\cite{KKKN02}.
Employing an NJL-type model, the authors calculated the spectral function 
in the scalar color-antitriplet diquark channel and monitored its behavior 
when the temperature approaches $T_c$ from above. 
At $T_c$ both, mass and width, of this mode go to zero but already 
at $T \simeq 1.2\, T_c$ the authors find a relatively sharp low-lying
peak which indicates precritical fluctuations.
The authors suggest that this could have observable effects in
heavy-ion collisions.}. 

A theoretically interesting case is the anisotropic spin-1 condensate 
discussed in \sect{aniso}.
Since general arguments predict rather different Goldstone spectra
for $M_J = 0$ and for $M_J = \pm 1$~\cite{Ho98,OM98} 
it would be instructive to see how this comes about by explicit calculation.
In this context it is interesting to note that quite recently,
an abnormal number of Goldstone modes has also been reported for the case of
the usual scalar diquark condensate in a two-flavor NJL model~\cite{BEKVY}.
Although an academic example (since these modes correspond to the spontaneously
broken {\it global} $SU(3)_c$ of the NJL model whereas in QCD, they are ``eaten'' 
by the gluons), it nicely illustrates that the counting of Goldstone bosons
becomes highly non-trivial in Lorentz-noninvariant systems~\cite{NiCh76}.

The Goldstone spectrum arising from chiral symmetry breaking in the CFL phase
has been determined in leading-order QCD at asymptotic 
densities~\cite{SoSt00,RWZ00}. 
The most interesting features are that the excitations 
have reversed mass ordering (i.e., $m_K < m_\pi$) and are very light.
For instance, if one extrapolates the resulting expressions down to 
``moderate'' chemical potentials, $\mu \sim 500$~MeV, one typically
finds $m_K \sim 15$~MeV. 
These small masses are a consequence of the fact that the symmetry breaking
terms in the corresponding effective Lagrangian must be quadratic,
rather than linear, in the quark masses in order to be consistent with 
the ${Z_2}_A$ symmetry of the CFL phase~\cite{ARW99}.  
As discussed in \eq{cflsymmetries} the
latter is a residual part of the $U_A(1)$ symmetry which is not explicitly
broken at asymptotic energies where instanton effects vanish. 
It would be interesting to compare these results with an
NJL-model calculation with ``realistic'' parameters.
In particular the effects of non-vanishing $\ave{\bar q q}$ condensates
and of $U_A(1)$ breaking terms ('t Hooft interaction) should be analyzed.

A partially related problem is the question whether the BCS mean-field
approach we have used throughout this work is appropriate to describe
quark matter at ``moderate'' densities. 
It is well known that BCS theory is a weak-coupling theory
which works best if the correlation length $\xi_c$ is much larger than 
the average distance $d$ between the fermions. 
In this case many Cooper pairs 
-- which are only temporary correlations between changing partners -- 
overlap each other and phase fluctuations average away.
On the other hand, for $\xi_c/d \ll 1$, the system is better described
as a Bose-Einstein condensate, consisting of strongly coupled  pairs with 
fixed partners\footnote{We are in not necessarily restricted to 
pair correlations. 
For instance, one might think of superfluid nuclear matter as 
three quarks strongly coupled to a nucleon, which is then Cooper paired
with a second three-quark cluster.
It is possible that similar correlations still exist above the 
deconfinement phase transition. From this point of view it might be
worthwhile to revisit the idea of a six-quark condensate which has
been suggested long ago by Barrois~\cite{Ba77}. 
}. The transition region is more difficult to describe. 
However, it is known from metallic superconductors that
deviations from BCS behavior, e.g., in the relation between $T_c$ and the 
gap at $T=0$ (\eq{Tcapp}), are visible already if $\xi_c/d$ becomes 
smaller than about 100.

In Ref.~\cite{AHI02}, $\xi_c/d$ has been studied for a two-flavor color 
superconductor within a QCD-like model.
This model basically corresponds to solving a Dyson-Schwinger equation
of the weak-coupling type, \eq{sigmabcsoge}, but with a running coupling 
constant which has been modified to regulate the low-momentum 
behavior. 
Keeping the momentum dependence of the gap, the authors analyzed the
spatial structure of the Cooper pairs via a Fourier transform and
calculated the coherence length.
The result was that $\xi_c/d$ drops below 10 for $\mu < 1$~GeV. 
Since in our NJL-model approach the gaps are momentum independent
we cannot perform the same kind of analysis. 
However, as a rule of thumb we may identify the coherence length
with the inverse of the gap, $\xi_c \sim 1/\Delta$.
Inserting typical numbers one finds $\xi_c/d$ of the order 2 to 3 
at $\mu = 500$~MeV, which clearly calls into question the applicability
of the BCS treatment. 

It should be noted that the analogous estimate 
applied to the vacuum gap equation looks even more worrisome. 
If we identify the constituent quark mass with $1/\xi_c$ and the
quark condensate with a density, we get 
$\xi_c/d \sim |\ave{\bar q q}|^{1/3} / M \lesssim 1$.   
In fact, it has been argued some time ago that chiral symmetry is
{\it not} broken in the NJL model because of strong phase fluctuations
due to pionic modes~\cite{KVdB00}. 
This point has subsequently been studied by various authors, explicitly
taking into account meson loops, e.g., via a $1/N_c$ expansion. 
It was found that meson-loop effects could be large 
but do not necessarily restore chiral symmetry~\cite{OBW00,OBW00a,RipkaNc}.
(Similar conclusions have been drawn in Refs.~\cite{Baba00, Baba01,LeOh00}) 
within different approaches.)
However, as discussed in \sect{nc}, the importance of the meson loops is
in general controlled by a new cut-off parameter which has to be
introduced because of the non-renormalizability of the NJL model. 
Therefore no definite statement can be made without fixing this 
parameter~\cite{OBW00a}. 

This analogy tells us that fluctuation effects could be, but do not need to
be dangerous for the phases we have discussed in this report.
In this context it is interesting to see that recent lattice studies 
of the NJL model at finite chemical potential find non-vanishing
diquark condensates which are in rather good agreement with the
corresponding large-$N_c$ limit~\cite{HaWa02}. 
Also, NJL-model results seem to be in good agreement with lattice QCD
for $N_c = 2$~\cite{RaWe04},
which is accessible by standard methods even at $\mu\neq0$.
On the other hand, an explicit investigation of fluctuation effects
within the (continuum) NJL model does not seem to be a rewarding effort.

Here we have more or less reached the limits of the model.
After all, NJL-type models are schematic models 
which are motivated to major extent by their simplicity. 
They become questionable
when the merits of the simplified interaction get lost by using
highly complicated approximation schemes. 
In that case one should think about other approaches.
To study fluctuation effects, Lagrangians with bosonic degrees of freedom,
like the effective Lagrangians mentioned in \sect{neutdisc},
are probably more appropriate. 
Of course, final answers must come from QCD or -- if available -- from 
empirical observations, but model calculations
may help to bring the ideas on the right track.


\begin{appendix}

\chapter{Fierz transformations}
\label{fierz}

\section{General aim}
\label{fierzfocus}

We consider a local four-point interaction of the form
\beq
    {\cal L}_\mi{int} \;=\; g_I (\bar q\, \hat \Gamma^{(I)} q)^2 \;=\;
    g_I\,\Gamma^{(I)}_{ij} \,\Gamma^{(I)}_{kl}\;\bar q_i\,q_j\,\bar q_k\,q_l~.
\eeq  
Taking into account the anticommutation rules for fermions, this leads to the 
identities
\beq
    {\cal L}_\mi{int} \;=\; 
    -g_I\,\Gamma^{(I)}_{ij} \,\Gamma^{(I)}_{kl}\;\bar q_i\,q_l\,\bar q_k\,q_j 
     \;=:\; {\cal L}_\mi{ex} 
\eeq  
and
\beq
    {\cal L}_\mi{int} \;=\; g_I\,\Gamma^{(I)}_{ij} \,\Gamma^{(I)}_{kl}\;
    \bar q_i\bar q_k\,q_l\,q_j
    \;=:\; {\cal L}_\mi{qq}~.
\eeq  
So far, ${\cal L}_\mi{int}={\cal L}_\mi{ex}={\cal L}_\mi{qq}$.
However, if we restrict ourselves to Hartree-type approximations where
the first field is contracted with the second,  and the third one with
the fourth, ${\cal L}_\mi{ex}$ yields the exchange diagrams (Fock terms)
of ${\cal L}_\mi{int}$ while ${\cal L}_\mi{qq}$ yields the particle-particle
and antiparticle-antiparticle contributions. To that end we wish to
rewrite the operators as
\beq
    \Gamma^{(I)}_{ij}\,\Gamma^{(I)}_{kl} \;=\; 
    \sum_{M} c^I_M\, \Gamma^{(M)}_{il}\,\Gamma^{(M)}_{kj}
\label{cim}
\eeq 
to get
\beq
    {\cal L}_\mi{ex}  
    \;=\; -g_I\,\sum_{M} c^I_M\;(\bar q\, \hat \Gamma^{(M)} q)^2~.
\eeq  
Combining this with the Hartree Lagrangian 
${\cal L}_\mi{dir} \equiv {\cal L}_\mi{int}$ we get for the total
effective quark-antiquark interaction  
\beq
    {\cal L}_{q\bar q} \;=\; {\cal L}_\mi{dir} + {\cal L}_\mi{ex } \;=\;
    \sum_{M} G_M\,(\bar q \, \hat \Gamma^{(M)} q)^2~,
\eeq  
with $G_M = c^I_M g_I$ for $M\neq I$ and $G_I = (1- c^I_I)g_I$.

In the same way one can employ
\beq
    \Gamma^{(I)}_{ij}\,\Gamma^{(I)}_{kl} \;=\; 
    \sum_{D} d^I_D\, (\hat\Gamma^{(D)}\,C)_{ik}\,(C\,\hat\Gamma^{(D)})_{lj}~, 
\label{did}
\eeq 
to write the quark-quark interaction as
\beq
    {\cal L}_{qq} \;=\; 
\sum_{D} H_D\,(\bar q\,\hat\Gamma^{(D)}\,C \,\bar q^T)
    (q^T\,C\,\hat\Gamma^{(D)} q)~,    
\eeq 
with $H_D = d^I_D g_I$.

By construction, ${\cal L}_{q\bar q}$ and ${\cal L}_{qq}$ are to be used in 
Hartree approximation only, to avoid double counting.

\section{Fierz identities for local four-point operators}
\label{operators}

In this section we list the coefficients $c^I_M$ and $d^I_D$ defined
in \eqs{cim} and (\ref{did}) for various operators.

\subsection{Operators in Dirac space}
\label{diracoperators}

\begin{itemize}
\item[(a)] \underline{quark-antiquark channel (exchange diagrams):}

\beq
    \left(\begin{array}{c}
             (\unity)_{ij}\,(\unity)_{kl} \\[1mm]
             (i\gamma_5)_{ij}\,(i\gamma_5)_{kl} \\[1mm]
             (\gamma^\mu)_{ij}\,(\gamma_\mu)_{kl} \\[1mm]
             (\gamma^\mu\gamma_5)_{ij}\,(\gamma_\mu\gamma_5)_{kl} \\[1mm]
             (\sigma^{\mu\nu})_{ij}\,(\sigma_{\mu\nu})_{kl}
    \end{array}\right)
\;=\;
    \left(\begin{array}{ccccc}
    \phm\frac{1}{4}&  -\frac{1}{4} &\phm\frac{1}{4}&-\frac{1}{4}&\phm\frac{1}{8}\\[1mm]
      -\frac{1}{4} &\phm\frac{1}{4}&\phm\frac{1}{4}&-\frac{1}{4}&  -\frac{1}{8} \\[1mm]
       \phm 1      &  \phm 1       &  -\frac{1}{2} &-\frac{1}{2}&     \phm 0    \\[1mm]
          -1       &      -1       &  -\frac{1}{2} &-\frac{1}{2}&     \phm 0    \\[1mm]
       \phm 3      &      -3       &    \phm 0     &  \phm  0   &  -\frac{1}{2}
    \end{array}\right)
\;
    \left(\begin{array}{c}
             (\unity)_{il}\,(\unity)_{kj} \\[1mm]
             (i\gamma_5)_{il}\,(i\gamma_5)_{kj} \\[1mm]
             (\gamma^\mu)_{il}\,(\gamma_\mu)_{kj} \\[1mm]
             (\gamma^\mu\gamma_5)_{il}\,(\gamma_\mu\gamma_5)_{kj} \\[1mm]
             (\sigma^{\mu\nu})_{il}\,(\sigma_{\mu\nu})_{kj}
    \end{array}\right)
\eeq

\begin{alignat}{3}
    (\gamma^0)_{ij}\,(\gamma^0)_{kl} 
\;=\;\frac{1}{4}\,\Big\{\;&\,(\unity)_{il}\,(\unity)_{kj} && 
&+\,&(i\gamma_5)_{il}\,(i\gamma_5)_{kj}
\nonumber\\
+\,&(\gamma^0)_{il}\,(\gamma^0)_{kj} &&-(\gamma^m)_{il}\,(\gamma_m)_{kj} 
&+\,&(\gamma^0\gamma_5)_{il}\,(\gamma^0\gamma_5)_{kj} 
-(\gamma^m\gamma_5)_{il}\,(\gamma_m\gamma_5)_{kj} 
\nonumber\\
-\,&(\sigma^{0n})_{il}\,(\sigma_{0n})_{kj}
&&+ \frac{1}{2} (\sigma^{mn})_{il}\,(\sigma_{mn})_{kj} &&\;\Big\}
\end{alignat}

\item[(b)] \underline{quark-quark channel:}

\beq
    \left(\begin{array}{c}
             (\unity)_{ij}\,(\unity)_{kl} \\[1mm]
             (i\gamma_5)_{ij}\,(i\gamma_5)_{kl} \\[1mm]
             (\gamma^\mu)_{ij}\,(\gamma_\mu)_{kl} \\[1mm]
             (\gamma^\mu\gamma_5)_{ij}\,(\gamma_\mu\gamma_5)_{kl} \\[1mm]
             (\sigma^{\mu\nu})_{ij}\,(\sigma_{\mu\nu})_{kl}
    \end{array}\right)
\;=\;
    \left(\begin{array}{ccccc}
    \phm\frac{1}{4}&  -\frac{1}{4} &\phm\frac{1}{4}&  -\frac{1}{4} &  -\frac{1}{8} \\[1mm]
      -\frac{1}{4} &\phm\frac{1}{4}&\phm\frac{1}{4}&  -\frac{1}{4} &\phm\frac{1}{8}\\[1mm]
       \phm 1      &  \phm 1       &  -\frac{1}{2} &  -\frac{1}{2} &     \phm 0    \\[1mm]
       \phm 1      &  \phm 1       &\phm\frac{1}{2}&\phm\frac{1}{2}&     \phm 0    \\[1mm]
          - 3      &  \phm 3       &    \phm 0     &    \phm  0    &  -\frac{1}{2}
    \end{array}\right)
\;
    \left(\begin{array}{c}
             (i\gamma_5\,C)_{ik}\,(C\,i\gamma_5)_{lj} \\[1mm]
             (C)_{ik}\,(C)_{lj} \\[1mm]
             (\gamma^\mu\gamma_5\,C)_{ik}\,(C\,\gamma_\mu\gamma_5)_{lj} \\[1mm]
             (\gamma^\mu\,C)_{ik}\,(C\,\gamma_\mu)_{lj} \\[1mm]
             (\sigma^{\mu\nu}\,C)_{ik}\,(C\,\sigma_{\mu\nu})_{lj}
    \end{array}\right)
\eeq

\begin{alignat}{2}
    (\gamma^0)_{ij}\,(\gamma^0)_{kl} 
\;=\;\frac{1}{4}\,\Big\{\;&
\,(i\gamma_5\,C)_{ik}\,(C\,i\gamma_5)_{lj} 
&&\quad+\;(C)_{ik}\,(C)_{lj}
\nonumber\\
+\,&(\gamma^0\gamma_5\,C)_{ik}\,(C\,\gamma^0\gamma_5)_{lj} 
&&-\, (\gamma^m\gamma_5\,C)_{ik}\,(C\,\gamma_m\gamma_5)_{lj}
\nonumber\\
+\,&(\gamma^0\,C)_{ik}\,(C\,\gamma^0)_{lj} 
&&-\,(\gamma^m\,C)_{ik}\,(C\,\gamma_m)_{lj}
\nonumber\\
-\,&(\sigma^{0n}\,C)_{ik}\,(C\,\sigma_{0n})_{lj}
&&+\, \frac{1}{2} (\sigma^{mn}\,C)_{ik}\,(C\,\sigma_{mn})_{lj}\quad\Big\} 
\end{alignat}

\end{itemize}

\subsection{Generators of ${\mathbf U(N)}$}
\label{unoperators}

We use the following notation:\\
$\tau_a$, $a=1,\dots,N^2-1$:\quad generators of $SU(N)$, 
normalized as $\tr{\tau_a\tau_b} = 2 \delta_{ab}$,\\
$\unity$: \quad $N \times N$ unit matrix,\;
$\tau_0 = \sqrt{2/N}\,\unity$,\\
$\tau_S$:\quad symmetric generators (including $\tau_0$),\;
$\tau_A$:\quad antisymmetric generators.

\begin{itemize}
\item[(a)] \underline{quark-antiquark channel (exchange diagrams):}

\beq
    \left(\begin{array}{c}
             (\unity)_{ij}\,(\unity)_{kl} \\[1mm]
             (\tau_a)_{ij}\,(\tau_a)_{kl}
    \end{array}\right)
\;=\;
    \left(\begin{array}{cc} \frac{1}{N} & \frac{1}{2} \\[1mm]
                            2\,\frac{N^2-1}{N^2} & -\frac{1}{N} \end{array}\right)
\;
    \left(\begin{array}{c}
             (\unity)_{il}\,(\unity)_{kj} \\[1mm]
             (\tau_a)_{il}\,(\tau_a)_{kj}
    \end{array}\right)
\eeq

\item[(b)] \underline{quark-quark channel:}

\beq
    \left(\begin{array}{c}
             (\unity)_{ij}\,(\unity)_{kl} \\[1mm]
             (\tau_a)_{ij}\,(\tau_a)_{kl}
    \end{array}\right)
\;=\;
    \left(\begin{array}{cc} \frac{1}{2} & \frac{1}{2} \\[1mm]
                            \frac{N-1}{N} & -\frac{N+1}{N} \end{array}\right)
\;
    \left(\begin{array}{c}
             (\tau_S)_{ik}\,(\tau_S)_{lj} \\[1mm]
             (\tau_A)_{ik}\,(\tau_A)_{lj}
    \end{array}\right)
\eeq

\end{itemize}

\section{Specific examples}
\label{fierzexamples}

In the following, $\tau_a$ and $\lambda_a$ denote operators in 
$SU(N_f)$ flavor space or $SU(N_c)$ color space, respectively.
Repeated indices are summed over.
Flavor or color indices run from $1$ to $N_{f,c}^2-1$,
unless explicitly stated otherwise. For Dirac indices 
$(a^\mu)^2 \equiv a^\mu a_\mu$, etc..

\subsection{Color current interaction}
\label{fierzhge}

\begin{itemize}
\item[(a)] {Lorentz-invariant interaction:}
\beq
    {\cal L}_\mi{int} \;=\; -g (\bar q\, \gamma^\mu\,\lambda_a q)^2 
\eeq  

\begin{alignat}{6}
\Rightarrow \; {\cal L}_\mi{ex} 
\;&=\;&\frac{2(N_c^2-1)}{N_f N_c^2}\,g\,
&\Big[\, (\bar qq)^2 & &+ (\bar qi\gamma_5 q)^2&
&- \frac{1}{2}(\bar q \gamma^\mu q)^2 &
&- \frac{1}{2}(\bar q \gamma^\mu\gamma_5 q)^2& &\Big]
\nonumber\\
&&+\;\frac{(N_c^2-1)}{N_c^2}\,g\,
&\Big[\, (\bar q\tau_aq)^2 & &+ (\bar q i\gamma_5\tau_a  q)^2&
&- \frac{1}{2}(\bar q \gamma^\mu\tau_a  q)^2 & 
&- \frac{1}{2}(\bar q \gamma^\mu\gamma_5\tau_a  q)^2& &\Big]
\nonumber\\
&&\;-\;\frac{1}{N_f N_c}\,g\,
&\Big[\, (\bar q\lambda_{a} q)^2 & &+ (\bar q i\gamma_5\lambda_{a} q)^2&
&- \frac{1}{2}(\bar q \gamma^\mu\lambda_{a} q)^2 & 
&- \frac{1}{2}(\bar q \gamma^\mu\gamma_5\lambda_{a} q)^2& &\Big]
\nonumber\\
&&-\;\frac{1}{2N_c}\,g\,
&\Big[\, (\bar q\tau_a\lambda_{a'}q)^2 & 
&+ (\bar q i\gamma_5\tau_a\lambda_{a'}  q)^2&
&- \frac{1}{2}(\bar q \gamma^\mu\tau_a\lambda_{a'}  q)^2 &
&- \frac{1}{2}(\bar q \gamma^\mu\gamma_5\tau_a\lambda_{a'}  q)^2& &\Big]
\nonumber\\[2mm]
&=\;&\frac{(N_c^2-1)}{N_c^2}\,g\,\sum_{a=0}^{N_f^2-1}
&\Big[\, (\bar q\tau_aq)^2 & &+ (\bar q i\gamma_5\tau_a  q)^2&
&- \frac{1}{2}(\bar q \gamma^\mu\tau_a  q)^2 & 
&- \frac{1}{2}(\bar q \gamma^\mu\gamma_5\tau_a  q)^2& &\Big]
\nonumber\\
&&-\;\frac{1}{2N_c}\,g\,\sum_{a=0}^{N_f^2-1}
&\Big[\, (\bar q\tau_a\lambda_{a'}q)^2 & 
&+ (\bar q i\gamma_5\tau_a\lambda_{a'}  q)^2&
&- \frac{1}{2}(\bar q \gamma^\mu\tau_a\lambda_{a'}  q)^2 &
&- \frac{1}{2}(\bar q \gamma^\mu\gamma_5\tau_a\lambda_{a'}  q)^2& &\Big]
\label{Lexli}
\end{alignat}

\begin{alignat}{4}
{\cal L}_\mi{qq} \;=\;\frac{N_c+1}{2N_c}\,g\,
&\Big[ & &(\bar q i\gamma_5C \tau_A\lambda_{A'}\bar q^T)
       (q^T C i\gamma_5 \tau_A\lambda_{A'} q) & &+\;  
       (\bar q C \tau_A\lambda_{A'}\bar q^T)
       (q^T C  \tau_A\lambda_{A'} q) & &  
\nonumber\\
& & -\frac{1}{2}&(\bar q \gamma^\mu\gamma_5C \tau_A\lambda_{A'}\bar q^T)
       (q^T C \gamma_\mu\gamma_5 \tau_A\lambda_{A'} q) & &-\frac{1}{2} 
       (\bar q\gamma^\mu C \tau_S\lambda_{A'}\bar q^T)
       (q^T C\gamma_\mu  \tau_S\lambda_{A'} q) & &\Big]
\nonumber\\
-\frac{N_c-1}{2N_c}\,g\,
&\Big[ & &(\bar q i\gamma_5C \tau_S\lambda_{S'}\bar q^T)
       (q^T C i\gamma_5 \tau_S\lambda_{S'} q) & &+\;  
       (\bar q C \tau_S\lambda_{S'}\bar q^T)
       (q^T C  \tau_S\lambda_{S'} q) & &  
\nonumber\\
& & -\frac{1}{2}&(\bar q \gamma^\mu\gamma_5C \tau_S\lambda_{S'}\bar q^T)
       (q^T C \gamma_\mu\gamma_5 \tau_S\lambda_{S'} q) & &-\frac{1}{2} 
       (\bar q\gamma^\mu C \tau_A\lambda_{S'}\bar q^T)
       (q^T C\gamma_\mu  \tau_A\lambda_{S'} q) & &\Big]
\label{Lqqli}
\end{alignat}

In particular we have
\begin{alignat}{2}
   G :&= \text{coeff}\Big(\,(\bar q \tau_a q)^2\,\Big) &
    \;&=\; \frac{(N_c^2-1)}{N_c^2}\,g
\\
    H :&= \text{coeff}\Big(\,(\bar q i\gamma_5C \tau_A\lambda_{A'}\bar q^T)
       (q^T C i\gamma_5 \tau_A\lambda_{A'} q)\,\Big) &
       \;&=\; \frac{N_c+1}{2N_c}\,g\,
\end{alignat}
and thus
\beq
    H \;:\; G \;=\; \frac{N_c}{2(N_c-1)} \;=\; \frac{3}{4}~,
\eeq
where the last equality holds for $N_c = 3$.

\item[(b)] {Electric and magnetic gluon exchange:}

\beq
     {\cal L}_\mi{int} \;=\; 
              \;-\; g_E \,(\bar q \gamma^0 \lambda_a q)^2
              \;+\; g_M \,(\bar q \vec\gamma \lambda_a q)^2
\label{hgeem}
\eeq

We want to derive to derive the six effective coupling constants of 
\eq{geff}. 

To this end, we first consider electric gluons only:
\beq
    {\cal L}_\mi{int}^{(E)} \;=\; -g_E (\bar q\, \hat \gamma^0\,\lambda_a q)^2 
\eeq  

\begin{alignat}{4}
\Rightarrow \quad {\cal L}_\mi{ex}^{(E)} 
\;&=\;&\frac{2(N_c^2-1)}{N_f N_c^2}\,g_E\,
&\Big[\, \frac{1}{4}(\bar qq)^2 &
&+ \frac{1}{4}(\bar q \gamma^0 q)^2 & \;+ \dots\;&\Big]\hspace{28mm}
\nonumber\\
&&\;-\;\frac{1}{N_f N_c}\,g_E\,
&\Big[\, \frac{1}{4}(\bar q\lambda_{a} q)^2 & 
&+ \frac{1}{4}(\bar q \gamma^0\lambda_{a} q)^2 & \;+ \dots\;&\Big]
\nonumber\\
&&+\;\dots\hspace{10mm}&
\label{Lexel}
\end{alignat}

\begin{alignat}{1}
{\cal L}_\mi{qq}^{(E)} \;=\;\frac{N_c+1}{2N_c}\,g_E\,
\Big[\;&\frac{1}{4}(\bar q i\gamma_5C \tau_A\lambda_{A'}\bar q^T)
       (q^T C i\gamma_5 \tau_A\lambda_{A'} q)
\nonumber\\
+&\frac{1}{4}(\bar q \gamma^0\gamma_5C \tau_A\lambda_{A'}\bar q^T)
       (q^T C \gamma_0\gamma_5 \tau_A\lambda_{A'} q) \;+\; \dots\;\Big]
\nonumber\\
+\;\dots\hspace{10mm}&~,
\label{Lqqel}
\end{alignat}
where only those terms have been listed explicitly which are
relevant for \eq{geff}. 

Now we rewrite \eq{hgeem} as
\beq
     {\cal L}_\mi{int} \;=\; 
              \;-\; (g_E-g_M) \,(\bar q \gamma^0 \lambda_a q)^2
              \;-\; g_M \,(\bar q \gamma^\mu \lambda_a q)^2~.
\eeq
and combine the results of \eq{Lexli} with \eq{Lexel}, and of
\eq{Lqqli} with \eq{Lqqel}.

\begin{alignat}{4}
\Rightarrow \quad {\cal L}_\mi{ex} 
\;&=\;&\frac{N_c^2-1}{2N_f N_c^2}\,
&\Big[\, (g_E+3g_M)(\bar qq)^2 &
&+ (g_E-g_M)(\bar q \gamma^0 q)^2 & \;+ \dots\;&\Big]\hspace{13mm}
\nonumber\\
&&\;-\;\frac{1}{4N_f N_c}\,
&\Big[\, (g_E+3g_M)(\bar q\lambda_{a} q)^2 & 
&+ (g_E-g_M)(\bar q \gamma^0\lambda_{a} q)^2 & \;+ \dots\;&\Big]
\nonumber\\
&&+\;\dots\hspace{7mm}&
\end{alignat}

\begin{alignat}{1}
{\cal L}_\mi{qq} \;=\;\frac{N_c+1}{8N_c}\,
\Big[\;&(g_E+3g_M)(\bar q i\gamma_5C \tau_A\lambda_{A'}\bar q^T)
       (q^T C i\gamma_5 \tau_A\lambda_{A'} q)
\nonumber\\
+& (g_E-g_M)(\bar q \gamma^0\gamma_5C \tau_A\lambda_{A'}\bar q^T)
       (q^T C \gamma_0\gamma_5 \tau_A\lambda_{A'} q) \;+\; \dots\;\Big]
\nonumber\\
+\;\dots\hspace{7mm}&
\end{alignat}

Adding the Hartree term ${\cal L}_\mi{dir}$, and taking $N_f=2$ and
$N_c=3$ we reproduce the coefficient given in \eq{geff}. 
\end{itemize}

\subsection{Two-flavor instanton induced interaction}
\label{fierzinst}

The two-flavor instanton induced interaction reads~\cite{SS98,RSSV00}:
\begin{alignat}{1}
{\cal L}_\mi{int} \;=\;\frac{g}{4(N_c^2-1)}\,\Big\{
\frac{2N_c-1}{2N_c}\,&\Big[(\bar q q)^2 - (\bar q \,i\gamma_5 \,q)^2 
               - (\bar q \, \tau_a \,q)^2 + (\bar q \,i\gamma_5 \tau_a \,q)^2 \Big]
\nonumber\\
-\frac{1}{4N_c}\,&\Big[(\bar q \,\sigma^{\mu\nu} \,q)^2 -
                    (\bar q \,\sigma^{\mu\nu} \tau_a \,q)^2 \Big] 
\hspace{30mm} \Big\}~. 
\end{alignat}

\begin{alignat}{1}
\Rightarrow \; {\cal L}_\mi{\bar qq} \;=\; g\,\Big\{\hspace{19mm}
\frac{1}{4N_c^2}\,&\Big[ (\bar q q)^2 - (\bar q \,i\gamma_5 \,q)^2 
               - (\bar q \,\tau_a \,q)^2 + (\bar q \,i\gamma_5 \tau_a \,q)^2 \Big] 
\nonumber\\
+\,\frac{N_c-2}{16N_c(N_c^2-1)}
&\Big[ (\bar q \,\lambda_{a'} \,q)^2 - (\bar q \,i\gamma_5 \lambda_{a'} \,q)^2 
- (\bar q \,\tau_a \lambda_{a'} \,q)^2 
+ (\bar q \,i\gamma_5 \tau_a \lambda_{a'} \,q)^2 \Big]\hspace{8mm}
\nonumber\\
+\frac{1}{32(N_c^2-1)}\,&\Big[(\bar q \,\sigma^{\mu\nu}\lambda_{a'} \,q)^2 -
                    (\bar q \,\sigma^{\mu\nu} \tau_a\lambda_{a'} \,q)^2 \Big] 
\hspace{20mm} \Big\}
\end{alignat}

\begin{alignat}{1}
{\cal L}_\mi{qq} \;=\; g\,\Big\{
\frac{1}{8N_c(N_c-1)}\,&\Big[ (\bar q \,i\gamma_5C \tau_2\lambda_{A}\,\bar q^T)
                              (q^T \,C i\gamma_5 \tau_2\lambda_{A} \,q)
-(\bar q \,C \tau_2\lambda_{A}\,\bar q^T)(q^T \,C \tau_2\lambda_{A} \,q)\Big]
\nonumber\\
-\frac{1}{16N_c(N_c+1)}\,&(\bar q \,\sigma^{\mu\nu} C \tau_2\lambda_{S}\,\bar q^T)
                              (q^T \,C \sigma_{\mu\nu} \tau_2\lambda_{S} \,q)
\quad\Big\}
\end{alignat}

Comparing the coefficients with \eqs{Lqqbar} and (\ref{Lqq}) we can
identify:
\beq
    G = \frac{1}{4N_c^2}~,\qquad H_s = \frac{1}{8N_c(N_c-1)}~,
    \qquad H_t = \frac{1}{16N_c(N_c+1)}~,
\eeq
and thus
\beq
    G \;:\; H_s \;:\; H_t 
    \;=\; 1 \;:\; \frac{N_c}{2(N_c-1)} \;:\; \frac{N_c}{4(N_c+1)}
    \;=\; 1 \;:\; \frac{3}{4} \;:\; \frac{3}{16}~,
\eeq
where the last equality holds for $N_c=3$.

\chapter{Two-flavor Nambu-Gorkov propagator}
\label{prop}
 
In this appendix we sketch the derivation of the
two-flavor Nambu-Gorkov propagator $S(p)$ which corresponds to the
inverse propagator
\beq
    S^{-1}(p)
\;=\; \left(\begin{array}{cc} \psl + \hat\mu\gamma^0 - \hat M 
    & (\Delta + \Delta_0\gamma^0)\,\gamma_5\tau_2\lambda_2
\\
      (-\Delta^* + \Delta_0^*\gamma^0)\,\gamma_5\tau_2\lambda_2
    & \psl - \hat\mu\gamma^0 - \hat M \end{array}\right)~,
\eeq
given in \sect{2SCHF}, \eq{NGSinv}.
The inverse is defined by
\beq
    S^{-1}(p)\,S(p) \;=\; \unity~.
\eeq
Writing the Nambu-Gorkov components of $S(p)$ explicitly,
\beq
    S(p)
\;=\; \left(\begin{array}{cc} S_{11}(p) & S_{12}(p)
\\
    S_{21}(p) & S_{22}(p) \end{array}\right)~,
\eeq
this yields
\begin{alignat}{3}
 (\psl + \hat\mu\gamma^0 - \hat M)\, &S_{11}(p) &
\;&+\; & (\Delta + \Delta_0\gamma^0)\,\gamma_5\tau_2\lambda_2\,S_{21}(p) &\;=\;1~,
\label{b1}\\
  (-\Delta^* + \Delta_0^*\gamma^0)\,\gamma_5\tau_2\lambda_2\, &S_{11}(p) &
\;&+\; & (\psl - \hat\mu\gamma^0 - \hat M)\,S_{21}(p) &\;=\;0~,
\label{b2}
\end{alignat}
and two analogous equations which result from the above ones if one replaces
\beq
    S_{11} \;\rightarrow\; S_{22}~, \quad
    S_{21} \;\rightarrow\; S_{12}~, \quad
    \hat\mu \;\rightarrow\; -\hat\mu~, \quad
    \Delta \;\rightarrow\; -\Delta^*~, \quad
    \Delta_0 \;\rightarrow\; \Delta_0^*~.
\label{replace}
\eeq
Employing \eq{b2}, we can eliminate $S_{21}$,
\beq
    S_{21}(p) \;=\; \frac{\psl_r^- + M_r}{p_r^{-\,2} - M_r^2} \;
    (\Delta^* \,-\, \Delta_0 \gamma^0)\;\gamma_5\tau_2\lambda_2\;S_{11}(p)~,
\label{b3}
\eeq
where we have defined
\beq
    p^\pm \;=\; \left(\begin{array}{c} p^0 \pm \hat \mu \\ \vec p
    \end{array}\right)~,
\eeq
and the indices $r$ and $b$ denote the red and blue color components,
respectively. 
Then \eq{b1} becomes
\begin{alignat}{1}
    \Big[\; \Big(\,\psl_r^+ &- M_r  
     \;-\; |\Delta|^2 \frac{\psl_r^- - M_r}{p_r^{-\,2} - M_r^2}
     \;-\; \Delta\Delta_0^* \frac{\psl_r^- - M_r}{p_r^{-\,2} - M_r^2}\gamma^0
\nonumber \\
     \;&-\; \Delta^*\Delta_0 \gamma^0\frac{\psl_r^- - M_r}{p_r^{-\,2} - M_r^2}
     \;-\; |\Delta_0|^2 \gamma^0\frac{\psl_r^- - M_r}{p_r^{-\,2} - M_r^2}
     \gamma^0
     \,\Big)\,\hat P_{12}^{(c)} \;+\;  \Big(\,\psl_b^+ - M_b\,\Big)\,
\hat P_{3}^{(c)}
\;\Big] \; S_{11}(p) \;=\;1~.
\end{alignat}
Here $P_3^{(c)} = 1/3 - 1/\sqrt{3}\,\lambda_8$ and 
$P_{12}^{(c)} = 1 - P_3^{(c)}$ are the projectors on the blue 
and the red/green sector in color space, respectively.

The important observation is that the operator in front of 
$S_{11}$ is diagonal in color space and does not depend on flavor.
Moreover, the ``blue'' part, i.e., the term proportional to $\hat P_3^{(c)}$,
takes the standard form and can easily be inverted. 
The problem has thus been reduced to inverting an, admittedly complicated,
expression in Dirac space for the ``red'' part of the propagator.

The final result reads
\begin{alignat}{1}
S_{11}(p) \;=&\; \frac{1}{(p_0^2 - \omega_-^2)(p_0^2 - \omega_+^2)}
\;\Big[\;({p_r^{-\,2}} - M_r^2)(\psl_r^+ + M_r)
                \,-\,|\Delta|^2 (\psl_r^- + M_r) 
 \,+\,\Delta\Delta_0^* \gamma^0(\psl_r^- + M_r)
\nonumber \\
&\hspace{50mm}+\,  \Delta^*\Delta_0(\psl_r^- + M_r)\gamma^0
 \,-\,|\Delta_0|^2 \gamma^0(\psl_r^- + M_r)\gamma^0 \;\Big]\;\hat P_{12}^{(c)}
\nonumber \\
&\,+\; \frac{\psl_b^+ + M_b}{(p_0 - E_-)(p_0 + E_+)
}\;\hat P_{3}^{(c)}~,
\end{alignat}
where $\omega_\mp$ and $E_\mp$ are the dispersion laws given in 
\eqs{2scepm} and (\ref{epspm}), respectively.
Inserting this into \eq{b3}, one finds
\begin{alignat}{1}
S_{21}(p) \;=\; \frac{1}{(p_0^2 - \omega_-^2)(p_0^2 - \omega_+^2)}
\;\Big[\;&-\Delta^*\,\Big((\psl_r^- + M_r)(\psl_r^+ - M_r) 
\,-\, |\Delta|^2 \,\,- \Delta^*\Delta_0 \gamma^0\Big)
\nonumber\\
&+\Delta_0^*\,\Big((\psl_r^- + M_r)\gamma^0(\psl_r^+ - M_r) 
\,-\,\Delta\Delta_0^* \,\,- |\Delta_0|^2 \gamma^0\Big)
\;\Big]\;\gamma_5\tau_2\lambda_2~.
\end{alignat}
The two remaining Nambu-Gorkov components are easily obtained by
the symmetry relations \eq{replace}.

Note that in all expressions given above, $p^0$ has to be interpreted
as a short-hand notation for a Matsubara frequency $i\omega_n$.

\end{appendix}




\newpage

\begin{center}{\Large {\bf Acknowledgements}} \end{center}
\bigskip\noindent

Most of the work reported here has been performed at the
Institut f\"ur Kernphysik at the TU Darmstadt. 
I would like to thank the entire theory group for a stimulating 
atmosphere and in particular 
Jochen Wambach for constant support and encouragement.
I also acknowledge financial support by GSI Darmstadt between 2001 and 2003.
Early roots go back to my time at the 
State University of New York in Stony Brook, and I would like
to thank Gerry Brown and the Alexander von Humboldt Foundation
for support. 
The final write-up of the report has been completed while I was participating 
in the INT program on ``QCD and Dense Matter: From Lattices to Stars'' 
(INT-04-1) in Seattle.

I am most indebted to my long-term collaborator Micaela Oertel who 
has actively contributed to most topics discussed in this work.  
I am also grateful to her and to Bernd-Jochen Schaefer for 
carefully reading the manuscript.

It was a great pleasure for me
to collaborate with Ji\v r\'{\i} Ho\v sek
who taught me the basics of color superconductivity and always opened
my mind for new ideas. 
I also thank him for his repeated hospitality in Prague. 

Work related to neutron stars has been performed in collaboration with
Marcello Baldo, Fiorella Burgio, and Hans-Josef Schulze. 
I thank them for their repeated hospitality in Catania and for 
providing me with the data of their equations of state~\cite{bbb,bbs}.
I also thank Matthias Hanauske and Detlev Zschiesche for supplying me with
the data of their equation of state~\cite{xsu3}.

I also would like to thank Martin Frank and especially Frederik Neumann
for fruitful collaborations.  
Part of the present work has overlap with their diploma theses.

I have benefitted a lot from various workshops, where I had illuminating
discussions with
Mark Alford,
Burkhard K\"ampfer,
Daniel Litim, 
Maria-Paola Lombardo,
Krishna Rajagopal,
Georges Ripka,
Dirk Rischke,
Sanjay Reddy,
Thomas Sch\"afer,
J\"urgen Schaffner-Bielich,
Igor Shovkovy,
and
Edward Shuryak.
In this context I would like to thank
ECT$^*$ in Trento, INT in Seattle, and KITP in Santa Barbara
for financial support. 
I also acknowledge stimulating discussions during the weekly meetings
of the Frankfurt-Darmstadt Color Superconductivity Group of the
Virtual Institute for Dense Hadronic Matter and QCD Phase Transitions
by the Helmholtz Association. 

Special thanks go to David Blaschke for many stimulating discussions and
his hospitality at various places.

Useful discussions with
Zoheir Aouissat,
Peter Braun-Munzinger,
Bengt Friman,
Axel Maas, 
Ralf Rapp,
Robert Roth,
Bernd-Jochen Schaefer,
Michael Urban, 
Jochen Wambach,
Verena Werth, 
and
Andreas Wirzba
are also gratefully acknowledged.

I also would like to thank Norbert Grewe for illuminating discussions
and for providing me with his lecture notes on superconductivity.

Finally, I thank Carsten Isselhorst, Alexander Mai, and especially
Thomas Roth for their help with computer related problems.

\end{document}